\documentclass[11pt, a4paper]{article}
\usepackage{jheppub}
\usepackage{epsfig}
\usepackage{amsfonts}
\usepackage{amsmath}
\usepackage{amssymb}
\usepackage{dsfont}
\usepackage{hyperref}
\usepackage{color}
\usepackage[usenames,dvipsnames]{xcolor}
\usepackage{bbm} 
\usepackage{bm}
\usepackage{cleveref}
\usepackage[normalem]{ulem}

\allowdisplaybreaks

\newcommand{\si}{\sigma}

\newcommand{\vL}{\ensuremath{\mathcal{L}}}
\newcommand{\vp}{\varphi}

\newcommand{\dslash}[1]{#1 \llap{/\kern-0.5pt}}
\newcommand{\Dslash}[1]{#1 \llap{/\kern+1.5pt}}
\newcommand{\DDslash}[1]{#1 \llap{/\kern+2.3pt}}
\newcommand{\dslashh}[1]{#1 \llap{/\kern+1pt}}
\newcommand{\abs}[1]{\left\lvert#1\right\rvert}

\newcommand{\beq}{\begin{equation}}
\newcommand{\eeq}{\end{equation}}

\newcommand{\hsigma}{\hat{\sigma}}

\newcommand{\bea}{\begin{eqnarray}}
\newcommand{\eea}{\end{eqnarray}}
\newcommand{\bma}{\begin{pmatrix}}
\newcommand{\ema}{\end{pmatrix}}
\newcommand{\nn}{\nonumber}

\newcommand\PythiaEight{{\tt Pythia8}}

\newcommand{\met}{\not{\!\!{\rm E}}_{T}}

\newcommand{\eq}[1]{Eq.~\eqref{eq:#1}}
\newcommand{\eqs}[2]{Eqs.~\eqref{eq:#1}~and~\eqref{eq:#2}}

\newcommand{\ssec}[1]{Sec.~\ref{ssec:#1}}
\newcommand{\bra}[1]{\left\langle #1\right\rvert}
\newcommand{\ket}[1]{\left\lvert #1\right\rangle}
\DeclareMathOperator{\Tr}{Tr}

\DeclareMathOperator{\Disc}{Disc}
\def\kslash{k\!\!\!\slash}

\def\bnslash{\bar n\!\!\!\slash}
\def\Pslash{P\!\!\!\!\slash\,}
\def\qslash{q\!\!\!\slash}
\newcommand{\mcdot}{\!\cdot\!}
\newcommand{\DLR}{\smash{\overset{\text{\small$\leftrightarrow$}}{\smash{D}
\vphantom{+}}}}
\newcommand{\DL}{\smash{\overset{\text{\small$\leftarrow$}}{\smash{D}\vphantom{+}}}}

\newcommand{\nocontentsline}[3]{}
\newcommand{\tocless}[2]{\bgroup\let\addcontentsline=\nocontentsline#1{#2}\egroup}

\title{Charged Lepton Flavor Violation at the EIC}

\author{Vincenzo Cirigliano,}
\author{Kaori Fuyuto,}
\author{Christopher Lee,}
\author{Emanuele Mereghetti,}
\author{\\ and Bin Yan}

\affiliation{Theoretical Division, Los Alamos National Laboratory,
Los Alamos, NM 87545, U.S.A.}
\emailAdd{cirigliano@lanl.gov}
\emailAdd{kfuyuto@lanl.gov}
\emailAdd{clee@lanl.gov}
\emailAdd{emereghetti@lanl.gov}
\emailAdd{binyan@lanl.gov}

\abstract{We present a comprehensive analysis of the potential sensitivity of the Electron-Ion Collider (EIC) to  charged lepton flavor violation (CLFV)  
in the channel $ep\to \tau X$,  within the model-independent framework of the Standard Model Effective Field Theory (SMEFT). 
We compute the relevant cross sections to leading order in QCD and electroweak corrections and 
perform simulations of signal and SM background events in various $\tau$ decay channels, 
suggesting simple cuts to enhance the associated estimated efficiencies. 
To assess the discovery potential of the EIC in $\tau$-$e$ transitions, we  
study the sensitivity of other probes of this physics across a broad range of energy scales,  
from $pp \to e \tau X$ at the Large Hadron Collider to decays of  $B$ mesons and  $\tau$ leptons, 
such as $\tau \to e \gamma$, $\tau \to e \ell^+ \ell^-$, and crucially the hadronic modes $\tau \to e Y$ with $Y \in \{ \pi, K, \pi \pi,  K \pi, ...\}$.  
We find that electroweak dipole and four-fermion semi-leptonic operators involving light quarks 
are already strongly constrained by $\tau$ decays, while operators involving the $c$ and $b$ quarks present more promising discovery potential for the EIC. An analysis of three models of leptoquarks confirms the expectations based on the SMEFT results. We also  identify  future directions needed to maximize the reach of the EIC in CLFV searches: 
these include an optimization of the $\tau$ tagger in hadronic channels, an exploration of background suppression through tagging  $b$ and $c$ jets in the final state, and a global fit by turning on all SMEFT couplings,  which  will  likely reveal new 
discovery windows  for the EIC. }

\begin{document}
{\flushright LA-UR-21-20531 \\[-9ex]}

\maketitle

\section{Introduction}

Processes involving charged lepton flavor violation (CLFV) are very powerful tools to search for  
new physics beyond the Standard Model (BSM)  for a number of reasons. 
First, the observation of CLFV  at experiments in the foreseeable future  would immediately point to new physics beyond the minimal extension of the SM that only 
includes neutrino mass (so-called $\nu$SM).  This is  because in the $\nu$SM, CLFV amplitudes are proportional to $(m_\nu/m_W)^2$~\cite{Petcov:1976ff,Marciano:1977wx,Lee:1977qz,Lee:1977tib}, 
leading to rates forty orders of magnitude below current sensitivity.  Furthermore,  current and future CLFV  searches are sensitive 
to new mediator particles with masses that can be well above the scales directly accessible in current and near-future high-energy colliders. 
Classic examples include supersymmetric models~\cite{Lee:1984kr,Lee:1984tn,Borzumati:1986qx,Barbieri:1995tw}. 
Finally, CLFV processes play a special role in probing  extensions  of the Standard Model (SM) connected to the generation of neutrino mass. 
Correlations between neutrino mass models and   signatures 
in CLFV processes have  been highlighted in the literature 
(e.g.  TeV see-saw mechanisms~\cite{Abada:2007ux,Alonso:2012ji}  
or in minimally flavor-violating GUT scale see-saw models~\cite{Cirigliano:2005ck}). 
In a nutshell, CLFV processes offer a great discovery tool for BSM physics 
as well as the possibility to ``diagnose'' the underlying new physics and its effect on neutrino mass generation. 
There is a vast literature on the subject, and for reviews we refer the reader to Refs.~\cite{Raidal:2008jk,deGouvea:2013zba,Bernstein:2013hba,Calibbi:2017uvl}.

Probes of CLFV exist across a broad spectrum of energy scales. 
Low-energy probes include decays of the   $\mu$ and $\tau$ leptons, 
decays of the $B$ and $K$ mesons and quarkonia. 
High-energy probes include searches for SM-forbidden events  
such as $p p \to \ell_\alpha \bar \ell_\beta + X$ at the Large Hadron Collider (LHC) 
or  $ e p \to \ell + X$ at electron-hadron colliders such as HERA and the upcoming Electron-Ion Collider (EIC). 
Currently, the most stringent limits on CLFV come from 
searches for $\mu\leftrightarrow e$  processes, 
e.g. the branching ratio ${\rm BR} (\mu^+ \to  e^+ \gamma)  < 4.2 \times 10^{-13}$ at 90\% CL~\cite{TheMEG:2016wtm}.  
The constraints on $\tau \leftrightarrow e$ 
transitions, however, are much weaker, with 
${\rm BR} (\tau^\pm \to  e^\pm Y)  <  {\rm few}  \times 10^{-8}$~\cite{Tanabashi:2018oca}, with $Y \in \{ \gamma, \pi \pi, ...\}$. 
Although Belle-II~\cite{Kou:2018nap} is expected to improve these $\tau$ BR constraints, and High-Luminosity LHC~\cite{Cerri:2018ypt} to extend its reach in $pp \to \bar{e} \tau + X$, both by an order of magnitude, there remains nevertheless a competitive opportunity for colliders to search for events $e p \to \tau X$, with hadronic final states $X$.

In the recent past,  HERA was able to put competitive constraints on   $\tau\leftrightarrow e$ transitions~\cite{Aaron:2011zz}.   
The EIC will collide $e$'s and $p$'s at center-of-mass energy $\gtrsim 100$ GeV, smaller than HERA, but at vastly higher luminosity,  
reaching  10--100 fb$^{-1}$  per year~\cite{Accardi:2012qut}.
Thus its reach to find CLFV may be a  thousand  times greater than HERA~\cite{Aaron:2011zz} 
and possibly competitive with improved searches for $\tau \to e Y$ at Belle-II~\cite{Kou:2018nap}.
The promise of the EIC as a probe of CLFV  was highlighted by the early study
of Ref.~\cite{Gonderinger:2010yn}, which estimated that an EIC with a collision energy of 90 GeV could probe currently allowed CLFV interactions 
in the context  of  leptoquark models. 

In this paper we perform a first comprehensive analysis of the CLFV physics reach of the EIC in the general framework of the 
Standard Model Effective Field Theory (SMEFT)
\cite{Weinberg:1979sa,Wilczek:1979hc,Buchmuller:1985jz,Grzadkowski:2010es,Jenkins:2013zja,Jenkins:2013wua,Alonso:2013hga,Crivellin:2013hpa}, 
which captures new potential sources of CLFV above the electroweak scale $v_\text{ew}$ in a model-independent way.
SMEFT encodes  new physics originating at energies higher than $v_\text{ew}$
in  operators of dimension greater than four  built out of SM fields.  
The SMEFT framework is  applicable to processes in which the center-of-mass energy is well below the expected scale of new physics. 
Given the null results so far for new physics searches at LHC, the SMEFT is perfectly applicable at an EIC with center-of-mass energy $\sqrt{S} < v_{\rm ew} \sim 200$~GeV. In fact, the effect of {\it any} new physics model with   particle  masses above the electroweak scale  will reduce to the SMEFT operators, with a model-specific pattern of effective couplings.  Therefore, the SMEFT framework allows one 
to assess the discovery potential and model diagnosing power of the EIC 
in full generality, also allowing a consistent comparison with probes 
at lower energies, such as  $\tau \to e X$  and LFV $B$ meson decays.
Our work  considerably  improves on the current state of the art, in two ways. 
First, for EIC itself, we  account for all 
leading (dimension-six) CLFV operators, including heavy quark operators,  in computing EIC's reach in inclusive and differential $ep\to\tau X$ searches. 
Second, we compare this reach with all existing CLFV probes today, at both high and low energy, within the model-independent framework of SMEFT. These include  searches for $pp \to e \tau$ at the LHC 
and decays of  the $\tau$ lepton ($\tau \to e Y$) and $B$ meson. 
Concerning the $\tau$ decays, we will consider  
not only radiative ($\tau \to e \gamma$) and 
leptonic modes ($\tau \to e  \ell^+ \ell^-$), but also hadronic modes such as $\tau \to e  \pi$, 
$\tau \to e \pi \pi$~\cite{Arganda:2008jj,Petrov:2013vka,Celis:2013xja,Celis:2014asa}, 
which have so far not been considered in studies of CLFV at EIC (e.g.~\cite{Gonderinger:2010yn}).
The inclusion of hadronic  channels is very relevant because 
(i) the current and prospective sensitivity 
in $\tau$ BRs for radiative and hadronic modes are  at the same level, namely $\sim 10^{-8}-10^{-9}$; (ii) the hadronic modes 
provide  the strongest constraints on CLFV operators involving quarks and gluons~\cite{Celis:2013xja,Celis:2014asa}. 
Through this analysis,  we will also identify  synergies and complementarity of CLFV searches at the EIC and in $\tau$ decays. 

In the recent literature, studies of $e \to \tau$  transitions have appeared in various contexts. 
Ref.~\cite{Antusch:2020vul}  discusses  $e \to \tau (\mu)$  
at a future LHeC, using a small subset of 
BSM operators, namely vector and scalar vertex corrections.  
Ref.~\cite{Husek:2020fru} focuses on  $e  N \to \tau N$ transitions at a fixed target experiment such as NA64~\cite{Gninenko:2018num} 
within  the SMEFT framework, performing a comparative study of this process with CLFV $\tau$ decays.  
Ref.~\cite{Abada:2016vzu} studies  $e N \to \tau N$ transitions at a fixed target experiment within minimal SM extensions with sterile fermions. 
Ref.~\cite{Takeuchi:2017btl} discusses $e \to \tau$ transitions mediated by  gluonic operators  at both fixed target experiments and LHeC. 
In the context of  this rich literature, our work introduces several new elements: the use of the full set of SMEFT operators, 
the study of a larger set of probes (including LHC and $B$ meson decays besides all CLFV  $\tau$ decays) 
and  the focus on the EIC sensitivity and reach.

The paper is organized as follows.
In Section~\ref{sect:sensitivities} we present  a high-level 
discussion of the relative sensitivity of collider and lepton decays in probing  CLFV. 
This analysis will provide the minimum luminosity requirements for    $e p$  colliders to be competitive with 
CLFV lepton decays and will show that the EIC will be competitive only for $e  \leftrightarrow \tau$  
and not for $e  \leftrightarrow \mu$ transitions.   
Specializing to $e  \leftrightarrow \tau$,  in Section~\ref{basis}  we present the basis of relevant CLFV operators
at dimension-six  in the SMEFT. 
In Section~\ref{sect:CLFVDIS}  we present our results for the CLFV deep inelastic scattering (DIS) process $e p \to \tau X$  
mediated by all dimension-six operators in SMEFT,  
and in  Section~\ref{efficiency}  we discuss the EIC sensitivity to CLFV couplings. 
In Section~\ref{sect:LHC} we discuss complementary high-energy probes of CLFV, 
such as CLFV decays of the top quark, Higgs boson,  $Z$ boson and LFV Drell-Yan at the LHC. 
Going down in energy scale, 
in Section~\ref{low-energy} we discuss the connection between SMEFT and the low-energy effective theory (LEFT) 
and study the constraints from  CLFV decays of the $\tau$ lepton and $B$ meson. 
Indirect low-energy probes of CLFV involving charged-current processes and neutrinos are discussed in  Section~\ref{sect:indirect}. 
In Section~\ref{sect:summary} we summarize the single-coupling constraints, and identify the classes of operators for which the EIC is competitive 
with other high- and low-energy probes.
Finally, in  Section~\ref{sect:leptoquarks} we apply our EFT formalism to the analysis of  three different leptoquark models 
and compare our findings with the existing literature. 
Our conclusions and outlook are given in Section~\ref{sect:concl}.
The appendices contain technical details of our analysis. 

\section{Comparing collider and decay sensitivities}
\label{sect:sensitivities}

Historically, very strong  constraints on CLFV couplings have  been obtained by  
studying  decays of $\mu$ and $\tau$ leptons, 
with current upper limits on the BRs in the $10^{-13}$ and $10^{-8}$ ballpark, respectively. 
Given an underlying LFV scenario (e.g. represented by one or more CLFV operators in the SMEFT), 
the lepton decay BR limits translate into requirements on the luminosity, energy, and efficiency 
for a collider search to be competitive.  
We formulate the criterion as follows: 
for  $\ell= \tau, \mu$, we require that the number of expected signal events in  
a given decay channel $\ell  \to eY$,  denoted by $N_S^{decay}$,   
and  in a collider process,  denoted by $N_S^{scatt}$,   be  comparable.  
For definiteness, we will phrase our discussion in terms of the collider process  $e p \to \ell X$, relevant for the EIC, 
but we will  also consider $pp \to e \ell X$, relevant for the LHC. 

Searches for $\ell \to e Y$  typically analyze a sample of $N_\ell$ charged leptons produced either at $e^+ e^-$ machines 
or by hadronic decays in a fixed target experiment. 
These searches are also characterized by a signal efficiency $\epsilon_d$, 
so that 
\beq
N_S^{decay} 
=  \epsilon_d \, N_\ell \, {\rm BR}_{\ell \to eY} 
= \epsilon_d \, N_\ell \, \Gamma_{\ell \to eY}  \, \tau_\ell~, 
\eeq
where $\tau_\ell$ is the $\ell$ lepton lifetime. 
For example, in the case of both BaBar and Belle, $N_\tau \sim 10^9$ and $\epsilon_d$ is in the $2.5\% \to 6\%$ range depending on the decay channel 
considered~\cite{Aubert:2009ag,Miyazaki:2013yaa}.
Currently, from experimental analyses one can infer only $O(1)$ upper limits on  $N_S^{decay}$, from which one deduces 
upper limits (UL) on the BRs
\beq
 {\rm BR}^{UL}_{\ell \to eY}  
 \sim \frac{1}{\epsilon_d \, N_\ell}~, 
 \label{eq:comp0}
\eeq
where the symbol  $\sim$ is used to indicate that analysis-dependent  $O(1)$ factors are missing on the RHS. 

Conversely, in a collider setup  the relevant quantities are 
the  integrated luminosity ${\cal L}$, 
the total  signal efficiency $\epsilon_s$ (including selection and reconstruction) and the cross section $\sigma_{ep \to \ell X}$,  
leading to
\beq
N_S^{scatt} =  \epsilon_s   \, \sigma_{ep \to \ell X} \, {\cal L}  ~. 
\eeq

Equating  $N_S^{scatt}$  and $N_S^{decay}$ one gets 
\beq
\epsilon_s \, {\cal L} =    \left(  \epsilon_d \, N_\ell  \right)  \,  \tau_\ell   \,   \frac{\Gamma_{\ell \to e Y}}{\sigma_{ep \to \ell X}} 
\sim        \frac{1}{   {\rm BR}^{UL}_{\ell \to eY}  }    \,  \tau_\ell   \,     \frac{\Gamma_{\ell \to e Y}}{\sigma_{ep \to \ell X}} 
~,
\label{eq:comp1}
\eeq
where in the last step we used \eqref{eq:comp0}. 
In  Eq.~\eqref{eq:comp1} the  ratio  $\Gamma_{\ell \to e Y} / \sigma_{ep \to \ell X}$ depends in principle on 
the underlying new physics parameters. 
However, when considering a single dominant source of LFV (i.e.~one SMEFT operator at  a time), 
the dependence on new physics parameters cancels completely in the ratio, 
which then depends only on the relevant masses,  collider energy, phase space factors and non-perturbative matrix elements.  
We will consider below a few  benchmark scenarios, in which the dominant new physics 
is either in $\ell  \to e \gamma$ dipole operators or in $\ell q \,\leftrightarrow\, e q$ four-fermion interactions. 

Denoting the new physics scale by $\Lambda$, for dipole operators  dimensional considerations lead to 
\beq
\Gamma_{\tau \to e \gamma} \sim \frac{m_\tau^3 v^2}{\Lambda^4}~, 
\quad 
\sigma_{ep \to \tau  X} \sim \frac{v^2}{\Lambda^4}~, 
\quad 
 \frac{\Gamma_{\tau \to e \gamma}}{\sigma_{ep \to \tau  X}} =  \kappa_D   m_\tau^3  =  \kappa_D \cdot 2.2 \cdot 10^{52} {\rm cm}^{-2} s^{-1}~, 
   \label{eq:comp2}
\eeq
where $\kappa_D \sim O(1)$. Explicit calculations to be presented later in the manuscript  show that  $\kappa_D = 0.33$ for $\sqrt{S} = 100$~GeV.
Similarly, for pseudo-scalar and axial-vector operators involving first-generation quarks  one can estimate 
\beq
\Gamma_{\tau \to e \pi} \sim \frac{m_\tau^3 \Lambda_{\rm QCD}^2}{\Lambda^4}\,, 
\ \ 
\sigma_{ep \to \tau  X} \sim \frac{S}{\Lambda^4}\,, 
\ \ 
 \frac{\Gamma_{\tau \to e  \pi}}{\sigma_{ep \to \tau  X}} 
 =  \kappa_{A,P}   \,   \frac{m_\tau^3  \Lambda_{\rm QCD}^2}{S} = 
  \kappa_{A,P}  \cdot 2.0 \cdot 10^{47} {\rm cm}^{-2} s^{-1} \,,
  \label{eq:comp3}
\eeq
where $\kappa_{A,P} \sim O(1)$ and  explicit calculation shows that  $\kappa_P = 2.7$ and $\kappa_A= 0.95$
for $\sqrt{S} = 100$~GeV.
An analogous estimate for scalar and vector operators leads to 
\beq
\Gamma_{\tau \to e \pi \pi} \sim \frac{m_\tau^5}{\Lambda^4}\,, 
\ \ 
\sigma_{ep \to \tau  X} \sim \frac{S}{\Lambda^4}\,, 
\ \ 
 \frac{\Gamma_{\tau \to e \pi \pi}}{\sigma_{ep \to \tau  X}} 
 =  \kappa_{S,V}    \frac{m_\tau^5}{(2 \pi)^2 S} = 
  \kappa_{S,V}  \cdot 1.7 \cdot 10^{47} {\rm cm}^{-2} s^{-1} \,,
  \label{eq:comp3v2}
\eeq
where the extra $(2 \pi)^2$  in $\Gamma/\sigma$ accounts for the mismatch in phase space factors between 
decay and collider process.   Numerically we find $\kappa_S = 0.3$ and $\kappa_V=0.1$.

Using  the above estimates in Eq.~\eqref{eq:comp1},  we can make the following observations: 

\begin{itemize}
\item  For the dipole operator, using the   current limit  ${\rm BR}^{UL}_{\tau \to e\gamma}    \sim 10^{-8}$~\cite{Aubert:2009ag}, 
Eq.~\eqref{eq:comp1} implies that to match the $\tau \to e \gamma$ sensitivity  one would need 
an EIC with  integrated luminosity  satisfying $ \epsilon_s {\cal L}_D \sim 10^{8} \, {\rm fb}^{-1}$.  
This is out of reach for the current EIC design. 

\item  For (pseudo)scalar and (axial) vector contact interactions involving first-generation quarks, 
using  ${\rm BR}^{UL}_{\tau \to e\pi \pi}    \sim 10^{-8}$~\cite{Miyazaki:2013yaa}  
one needs at $\sqrt{S} = 100$~GeV  an integrated luminosity of 
 $\epsilon_s {\cal L}_{S,V} \sim 10^{3} \,  {\rm fb}^{-1}$, 
 which could be within reach of the current EIC design under optimal conditions after several years of running~\cite{Accardi:2012qut}.
Therefore, the EIC should be  competitive in constraining contact CLFV interactions and in probing  the many  directions in the SMEFT parameter space that are left unconstrained 
by low-energy probes of CLFV.   It is also worth noting that in the case of contact interactions 
Eq. \eqref{eq:comp3}  implies that 
the constraining power of the EIC  grows linearly with $S$. 

\item From the above discussion one also sees that 
for new physics patterns that involve more than a single dominant operator,  
the ratio  $\Gamma_{\tau \to e Y} / \sigma_{ep \to \tau X}$ could be suppressed due to cancellations  
and  therefore  even for flavor-conserving light quark operators  
the EIC could be more competitive than the simplest scenarios suggest.

\item  Importantly,  considering operators involving heavy  quark flavors $Q=c,b$ 
makes the analysis more favorable for the EIC.  
As an example consider vector operators:  
the  cross section   $ \sigma_{ep \to \tau  X_Q}$ is suppressed with respect to the 
light flavor case by about one order of magnitude, due to the heavy flavor PDFs. 
On the other hand the heavy flavor operators can contribute to $\tau$  decays such as 
$\tau \to e \pi \pi$ only through loop amplitudes suppressed by 
 a factor of a few $\times 10^{-3}$. 
 In turn, this implies a suppression of  about $\approx 10^{-5}$ in the decay rate, 
 much larger than the suppression in the cross section. 
Putting the ingredients together we find that  $ \Gamma_{\tau \to e \pi \pi}  / \sigma_{ep \to \tau  X_Q}$ 
is suppressed by a factor of $\approx 10^{-4}$ compared to the light flavor case. 
Therefore, the requirement on the luminosity is only
 $\epsilon_s {\cal L}\sim 0.1 \,  {\rm fb}^{-1}$, 
 well within the reach of the current EIC design, even with realistic $\epsilon_s \sim O(\%)$. 
 This analysis suggests that the largest discovery potential at the EIC is in the 
 DIS processes involving production of heavy quark  flavors in the final state. 

\item For the LHC-relevant process $pp \to e \ell X$,  the cross section scaling   
given in Eqs. \eqref{eq:comp2}-\eqref{eq:comp3v2}  for $ep \to \ell X$  is still valid, with $\sqrt{S}$ 
replaced by the $\tau$-$e$ invariant mass, $m_{\tau e}$. Existing analyses reach $m_{\tau e}$ of a few TeV \cite{Aaboud:2018jff}.
As a consequence, for dipole operators and vertex corrections one does not expect particularly great sensitivity at the LHC. 
On the other hand,  for four-fermion operators the larger $m_{\tau e}$ brings the luminosity requirement for the LHC 
to the realistic levels  $\epsilon_s {\cal L}_{S,V} \sim 50\, {\rm fb}^{-1}$. 
Taking into account the numerical factors and PDF integrations, this brings the LHC constraints on dimension-six Wilson coefficients 
to within an order of magnitude of the constraints from $\tau$ decays. 
We also note the recent study \cite{Boughezal:2020uwq} comparing LHC and EIC constraints on lepton flavor-conserving vector four-fermion operators, showing the potential power of the EIC to lift degeneracies or flat directions in the space of SMEFT operators that would remain using LHC alone.

\end{itemize}

Finally, we note that one could repeat the above analysis for the case of $e \leftrightarrow \mu$ transitions, 
using ${\rm BR}^{UL}_{\mu \to eY}    \sim 10^{-13}$ and the appropriate changes $m_\tau \to m_\mu$ and 
$\tau_\tau \to \tau_\mu$.   Taking these effects into account we find  that the integrated luminosity 
required for EIC to be competitive  in $e \to \mu$ transitions 
would be  eight orders of magnitude larger than the one required for $e \to \tau$ transitions. 
This result implies  that for these transitions the EIC cannot compete with low-energy muon processes, 
in agreement with the findings of Ref.~\cite{Gonderinger:2010yn}. 
Therefore, in what follows we will focus on  $e \leftrightarrow \tau$ transitions.

\section{The operator basis}\label{basis}

We consider in this paper CLFV at the EIC, LHC and in low-energy $\tau$ and meson decays.
At the center-of-mass energies reached at the EIC, it is appropriate to integrate out the degrees of freedom that induce CLFV, and to work in the framework of the SMEFT.
We will also frame the analysis of LHC data in the SMEFT, even though in this case the limits we obtain should be interpreted with some care.

\subsection{The SMEFT Lagrangian} 
The dimension-six SMEFT Lagrangian was constructed in Refs.\ \cite{Buchmuller:1985jz,Grzadkowski:2010es},
and it contains the most general set of operators that are invariant under the Lorentz group, the gauge group $SU(3)_c \times SU(2)_L \times U(1)_Y$, 
and that have the same field content as the SM. We consider here the SM in its minimal version, with three families of leptons and quarks, and one scalar doublet. In particular, we do not introduce a light sterile neutrino $\nu_R$. 
The left-handed quarks and leptons transform as doublets under $SU(2)_L$
\begin{equation}
q_L  = \Biggl( \begin{array}{c}
u_L \\
d_L
\end{array}
\Biggr), \qquad \ell_L = \Biggl( \begin{array}{c}
\nu_L \\
e_L
\end{array}
\Biggr),
\end{equation}
while the right-handed quarks, $u_R$ and $d_R$, and charged leptons, $e_R$,
are singlets under $SU(2)_L$. 
The scalar field $\varphi$ is a doublet under $SU(2)_L$. In the unitary gauge we have
\begin{equation}
\varphi = \frac{v}{\sqrt{2}} U(x) \Biggl( 
\begin{array}{c}
0 \\
1 +  \frac{h}{v}
\end{array}\Biggr),
\end{equation}
where $v=246$ GeV is the scalar vacuum expectation value (vev), $h$ is the physical Higgs field and $U(x)$ is a unitary matrix that encodes the Goldstone bosons. We will denote by  $\tilde \vp$ the combination $\tilde \vp = i\tau_2 \vp^*$. 
The gauge interactions are determined by the covariant derivative
\begin{equation}
D_\mu =  \partial_\mu + i g_1 {\rm y} B_\mu + i  \frac{g_2}{2} \tau^I  W^I_\mu   + i g_s G^a_\mu t^a  
\end{equation}
where $B_\mu$, $W^I_{\mu}$ and $G^a_{\mu}$ are the $U(1)_Y$, $SU(2)_L$ and $SU(3)_c$ gauge fields, respectively, and $g_1$, $g_2$, and $g_s$ are their gauge couplings. Furthermore, 
$\tau^I/2$ and $t^a$ are the $SU(2)_L$ and $SU(3)_c$ generators, in the representation of the field on which the derivative acts.
In the SM, the gauge couplings $g_1$ and $g_2$ are related to the electric charge and the Weinberg angle by $g_2 s_w = g_1 c_w = e$, where $e > 0$ is the charge of the positron
and $s_w = \sin\theta_W$, $c_w = \cos\theta_W$. These relations are affected by SMEFT dimension-six operators,
but these corrections are subleading for the processes considered here, which have no SM background.
Similarly, at the order we are working, we can interchangeably use $v$ or the Fermi constant $G_F$, using the SM relation $\sqrt{2} G_F = v^{-2}$.
The values of the couplings $g_s$, $g_1$, $g_2$ and of the quark masses, and the hypercharge assignments of the SM fields are given in Table \ref{Tab:SM} and in Eq.~\eqref{hypc}.

In the SM, lepton flavor is exactly conserved. There is a single, gauge-invariant dimension-five operator  \cite{Weinberg:1979sa} 
\begin{equation}\label{Weinberg}
\mathcal L_5 =  C_5 \left(\tilde{\varphi} \ell_L \right)^T  C (\tilde\varphi^{\dagger} \ell_L),
\end{equation}
where $C$ is the charge conjugation matrix.
When the Higgs takes its vev, $\mathcal L_5$ gives rise to the neutrino Majorana masses and mixings, and thus to LFV in the neutral sector. 
The operator in Eq.~\eqref{Weinberg} violates lepton number, and thus two insertions of  $C_5$ are needed to induce CLFV at the loop level. 
While formally dimension-six, the resulting CLFV is proportional to the masses of the light neutrinos and thus negligible~\cite{Petcov:1976ff,Marciano:1977wx}.

CLFV processes are affected by many dimension-six operators. Following the notation of Ref.\ \cite{Grzadkowski:2010es}, we classify the relevant operators 
according to their gauge (denoted by $X$), fermion ($\psi$), and scalar field ($\varphi$) content. The operators that contribute  at tree level fall in the following four classes:
\begin{eqnarray}\label{eq:basis1}
\mathcal L =    \mathcal L_{\psi^2 \varphi^2 D} + \mathcal L_{\psi^2 X \varphi} + \mathcal L_{\psi^2 \varphi^3} + \mathcal L_{\psi^4}.   
\end{eqnarray}
The first three classes contain fermion bilinear operators. 
$\psi^2 \varphi^2 D$ contains corrections to the SM couplings of quarks and leptons to the $Z$ and $W$ bosons, 
$\psi^2 X \varphi$ contains dipole couplings to the $U(1)_Y$, $SU(2)_L$ and $SU(3)_c$ gauge bosons, and  $\psi^2 \varphi^3$ contains non-standard Yukawa interactions. Focusing on purely leptonic operators,
we consider
\begin{eqnarray}
\vL_{\psi^2 \varphi^2 D} &=&   -  \frac{\varphi^{\dagger} i \DLR_{\mu} \varphi}{v^2} \, \left(   \bar \ell_L   \gamma^{\mu} \, c^{(1)}_{L\varphi} \ell_L +   
  \bar e_R  \gamma^{\mu}\, c_{e\varphi}  e_R\right)  -  \frac{ \varphi^{\dagger}  i \DLR^I_{\mu} \varphi}{v^2}\,  \bar \ell_L \tau^I  \gamma^{\mu} c^{(3)}_{L\varphi} \ell_L, \label{eq:Z}\\
\mathcal L_{\psi^2 X \varphi} &=&  -\frac{1}{\sqrt{2}}\bar \ell_L \si^{\mu\nu}(g_1\Gamma_{B}^e B_{\mu\nu}+g_2\Gamma_{W}^e {\tau}^I  {W}^I_{\mu\nu})\frac{\varphi}{v^2}  e_R + \mathrm{h.c.}\ , \label{eq:dipole}\\
\mathcal L_{\psi^2 \varphi^3} &=& - \sqrt{2} \frac{\varphi^{\dagger} \varphi}{v^2} \bar \ell_L Y^\prime_e \varphi e_R  + \textrm{h.c.}, \label{eq:Y}
\end{eqnarray}
where $\DLR_{\mu}=  D_\mu-\DL_\mu$, $\DLR^I_{\mu}= \tau^I D_\mu-\DL_\mu \tau^I$.\footnote{Here, $\varphi^{\dagger}\overleftarrow D_{\mu}\varphi \equiv \left(D_{\mu}\varphi \right)^{\dagger}\varphi$.}
The couplings $c_{L\vp}^{(1)}$, $c_{L\vp}^{(3)}$, $c_{e\vp}$ are hermitian, 3 $\times$ 3 matrices in lepton-flavor space.
Expanding the covariant derivative in \eq{Z}, these operators induce CLFV $Z$ couplings, so that the $Z$ vertices are given by
\begin{eqnarray}
\mathcal L_Z &=& -\frac{g_2}{c_w} Z_\mu \bigg\{ \left( z_{e_L}\delta_{p r} + \frac{1}{2} \left[c^{(1)}_{L\varphi} + c^{(3)}_{L\varphi} \right]_{pr} \right)\bar e^p_L \gamma^\mu e^r_L
+ \left( z_{e_R} \delta_{pr} + \frac{1}{2} \left[c^{}_{e\varphi}  \right]_{pr}\right) \bar e^p_R \gamma^\mu e^r_R \nonumber  \\
& & +  z_{u_L}  \bar u^p_L \gamma^\mu u^p_L + z_{u_R}  \bar u^p_R \gamma^\mu u^p_R
+ z_{d_L}  \bar d^{\, p}_L \gamma^\mu d^{\,p}_L + z_{d_R}  \bar d^{\,p}_R \gamma^\mu d^p_R
\bigg\}, 
\end{eqnarray}
with $p,r$ being lepton flavor or quark family indices. The couplings $z_{f_L}$ and $z_{f_R}$ are
\begin{eqnarray}\label{zcouplings}
z_{f_L}  = T_{3f} - Q_f s_w^2, \qquad  z_{f_R}  = - Q_f s_w^2,
\end{eqnarray}
where $T_{3f}$ and $Q_f$ are the fermion isospin and charge. 

Meanwhile, $\Gamma_W^{e}$ and $\Gamma_B^{e}$ in \eq{dipole} are generic $3\times 3$ matrices in flavor space,  which we find convenient to trade for dipole couplings to the $Z$ and photon field
\begin{equation}
\label{eq:gZdipole}
\Gamma^e_\gamma = \Gamma^e_B  - \Gamma^e_W, \qquad  \Gamma^e_Z = -c_w^2 \Gamma^e_W - s_w^2 \Gamma^{e}_B.
\end{equation}
Finally, $Y_e^\prime$ in \eq{Y} is a dimension-six Yukawa coupling, which corrects the dimension-four SM Yukawa
\begin{equation}
\mathcal L_{\psi^2 \varphi} = - \sqrt{2}  \bar \ell_L Y^{(0)}_e \varphi e_R + \textrm{h.c.}
\end{equation}
When the Higgs gets its vev, we can write
\begin{equation}
\label{eq:hYukawa}
\mathcal L_{\rm yuk} = -  v \bar e_L Y_e  e_R  \left(1 + \frac{h}{v}\right) - \bar e_L Y_e^\prime e_R h + \ldots + \textrm{h.c.}, \qquad Y_e = Y_e^{(0)} + \frac{1}{2} Y_e^\prime,
\end{equation}
where the dots denote higher-order terms in $h$.
We can always diagonalize the first term, so that the charged lepton masses are given by $M_e = v Y_e$. The second term can in general be off-diagonal.
For both quark and lepton SM Yukawa couplings we will use the convention $M_f = v Y_f$. The quark Yukawa interactions are the same as the $Y_e$ term in \eq{hYukawa} with each $e\to q$.

$\mathcal L_{\psi^4}$ includes four-fermion operators. The most relevant for collider searches are semileptonic four-fermion operators,
\begin{align}\label{eq:fourfermion}
&\mathcal L_{\psi^4} = -\frac{4 G_F}{\sqrt{2}}  \bigg\{ 
C^{(1)}_{LQ}\, \bar \ell_L \gamma^\mu \ell_L \, \bar q_L \gamma_\mu q_L + C^{(3)}_{LQ} \, \bar \ell_L \tau^I \gamma^\mu \ell_L \, \bar q_L  \tau^I \gamma_\mu q_L \\
& \qquad\qquad\qquad + C_{eu} \, \bar e_R \gamma^\mu e_R \, \bar u_R \gamma_\mu u_R  +\
 C_{ed} \, \bar e_R \gamma^\mu e_R \, \bar d_R \gamma_\mu d_R \nn \\
 & \qquad\qquad\qquad + C_{Lu}\, \bar \ell_L \gamma^\mu \ell_L \, \bar u_R \gamma_\mu u_R +  C_{Ld}\, \bar \ell_L \gamma^\mu \ell_L \, \bar d_R \gamma_\mu d_R  +\ C_{Qe}  \, \bar e_R \gamma^\mu e_R \, \bar q_L \gamma_\mu q_L   \bigg\}  \nn \\
&\quad -\frac{4 G_F}{\sqrt{2}} \bigg\{  C_{LedQ}\, \bar \ell^i_L e_R\, \bar d_R q_L^i + C^{(1)}_{LeQu}\, \varepsilon^{ij} \bar \ell^i_L e_R\, \bar q_L^j u_R +  C^{(3)}_{LeQu}\, \varepsilon^{ij} \bar \ell^i_L \sigma^{\mu \nu} e_R\, \bar q_L^j \sigma_{\mu \nu} u_R \, + \textrm{h.c.}
  \bigg\}. \nonumber
\end{align}
Here, $i,j$ represent $SU(2)_L$ indices.
Of these operators, only a few affect charged currents, introducing new Lorentz structures, such as scalar-scalar and tensor-tensor interactions. All of the above operators modify neutral currents and the couplings are, in general, four-index tensors in flavor. We allow the operators to have a generic structure in quark flavor. We follow the flavor conventions of Ref.~\cite{Alioli:2018ljm} and assign operator labels to the neutral current components
with charged leptons, after rotating to the $u$ and $d$ quark mass basis. 
This induces  factors of the SM CKM matrix $V_{\rm CKM}$ in the charged-current and in the neutral current neutrino components, which play a minimal role here.
For example, introducing 
\begin{equation}
C_{LQ, U} = \left(U^u\right)_L^{\dagger} \left( C_{LQ}^{(1)} - C_{LQ}^{(3)} \right) U_L^{u}, \qquad  C_{LQ, D} = \left(U^d_L\right)^{\dagger} \left( C_{LQ}^{(1)} + C_{LQ}^{(3)} \right) U^d_L,
\end{equation}
where $U_{L,R}^{u,d}$ are unitary matrices that diagonalize the quark mass matrices, the first two terms in the four-fermion Lagrangian \eq{fourfermion} become
\begin{eqnarray}
\mathcal L &=&
-\frac{4 G_F}{\sqrt{2}}  \bigg\{ 
   \left[ C^{}_{LQ, U} \right]_{p r s t}\, \bar e^p_L \gamma^\mu e^r_L \, \bar u^s_L \gamma_\mu u^t_L 
+  \left[C^{}_{LQ, D} \right]_{pr st}\, \bar e^p_L \gamma^\mu e^r_L \, \bar d^s_L \gamma_\mu d^t_L \nonumber \\
& & +\left[ V_{\rm CKM} C_{LQ, D} V^{\dagger}_{\rm CKM } \right]_{p r s t} \, \bar \nu^{p}_L \gamma^\mu \nu^r_L \, \bar u^s_L \gamma_\mu u^t_L 
+ \left[V_{\rm CKM}^{\dagger} C^{}_{LQ, U} V_{\rm CKM} \right]_{p r s t}\, \bar \nu^p_L \gamma^\mu \nu^r_L \, \bar d^s_L \gamma_\mu d^t_L  \nonumber\\
& &+ \left( \left[C_{LQ, D} V^{\dagger}_{\rm CKM} - V_{\rm CKM}^{\dagger} C^{}_{LQ, U} \right]_{p r s t} \bar \nu^p_L \gamma^\mu e^r_L  \bar d^s_L \gamma_\mu u^t_L  + \textrm{h.c.}\right)
\bigg\},
\end{eqnarray}
where $p$, $r$, $s$, $t$ are flavor indices in quark/lepton mass bases.
With these conventions, all semileptonic operators have naturally either $u$-type or $d$-type quark flavor indices. The only exception is $C_{Qe}$, which we choose to be $d$-type,
 leading to neutral current vertices of the form
\begin{equation}\label{CQe}
\mathcal L = - \frac{4 G_F}{\sqrt{2}}    \, \bar e_R \gamma^\mu e_R \, \left(  \bar d_L C_{Qe} \gamma_\mu d_L +  \bar u_L V_{\rm CKM} C_{Qe} V^{\dagger}_{\rm CKM}  \gamma_\mu u_L \right) .
\end{equation}

As we discuss in Appendix \ref{RGEs}, the renormalization group evolution of the operators in Eq. \eqref{eq:fourfermion} also induces  purely leptonic operators
\begin{equation}\label{eq:fourleptons}
\mathcal L_{\psi^4} = -\frac{4 G_F}{\sqrt{2}}  \bigg\{ 
C^{}_{LL}\, \bar \ell_L \gamma^\mu \ell_L \, \bar \ell_L \gamma_\mu \ell_L  + C_{ee} \, \bar e_R \gamma^\mu e_R \, \bar e_R \gamma_\mu e_R  
 + C_{Le}\, \bar \ell_L \gamma^\mu \ell_L \, \bar e_R \gamma_\mu e_R  \bigg\}.
\end{equation}
These operators could be probed at the EIC by looking for final states with multiple leptons. At low energy, they can be sensitively probed by the process $\tau \rightarrow e \bar\ell \ell$.

\subsection{Running to the electroweak scale}

The Lagrangians in Eqs.~\eqref{eq:Z}, \eqref{eq:dipole},  \eqref{eq:Y} and \eqref{eq:fourfermion} are defined just below the new physics scale $\Lambda \gg v$.
For the study of DIS at the EIC, of LHC constraints and of low-energy processes we first evolve the Lagrangian to a scale $\mu$ close to the electroweak scale. 
The renormalization group equations (RGEs) in the SMEFT were derived in Refs.  \cite{Jenkins:2013zja,Jenkins:2013wua,Alonso:2013hga}
and we report them for convenience in Appendix \ref{RGEs}, where we also provide the numerical solutions of the RGEs at leading logarithmic accuracy.

We comment here on the most important qualitative effects:
\begin{itemize}
\item The scalar and tensor operator coefficients $C^{(1,3)}_{Le Q u}$ and $C^{}_{L e d Q}$ run in QCD. The running from $\Lambda \sim 1$ TeV to 
$\mu = m_t$ increases (decreases) the coefficient of the scalar (tensor) operators by roughly 10\% (5\%).

\item $Z$ dipoles, scalar and tensor operators mix into the photon dipole $\Gamma^e_\gamma$ at leading log \cite{Jenkins:2013zja,Jenkins:2013wua,Alonso:2013hga}.
We show the relevant RGEs in Eqs. \eqref{rge_st0}--\eqref{rge_st4}.
The mixing of the $Z$ dipole is at the $10^{-2}$ level, as expected from a weak loop correction.
The mixing of the tensor operator is proportional to the quark Yukawa coupling and thus it is particularly important for the $tt$ component of $C^{(3)}_{LeQu}$. 
The strong constraints on flavor-changing dipoles imply that this mixing is also non-negligible for the charm component of the tensor operator.
$C^{(1)}_{LeQu}$ mixes with $C^{(3)}_{LeQu}$ via an electroweak loop. For the $tt$ component of the scalar operator, the resulting contribution to $\Gamma^e_\gamma$ is sizable.
The coefficients of photon and $Z$ dipoles, scalar and tensor four-fermion operators at the scale $\mu = m_t$, as a function of
top, bottom and charm scalar and tensor operators at the scale $\mu_0 = 1$ TeV, are given in Table \ref{ScalarTensorRGE}.

\begin{figure}
\vspace{-10pt}
\center
\includegraphics[width=\textwidth]{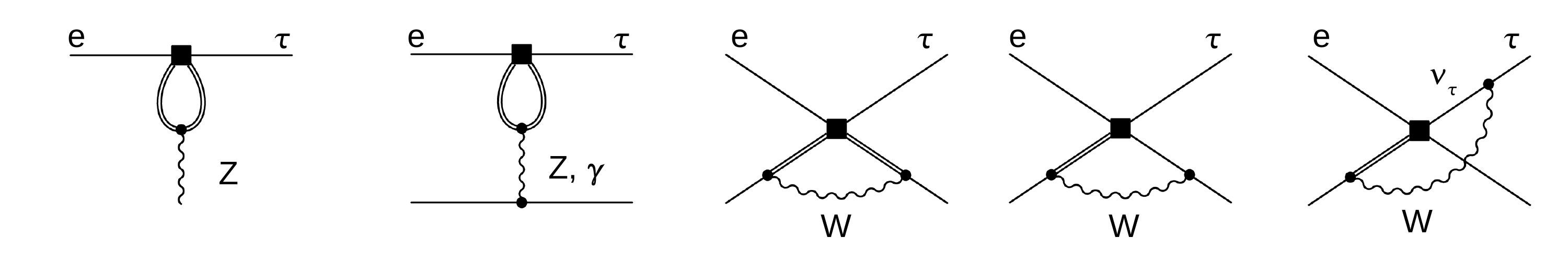}
\vspace{-24pt}
\caption{
One loop diagrams contributing to the running of heavy flavor operators onto operators that can be tested at the EIC and in $\tau$ decays.  
Plain lines denotes leptons and light quarks, double lines heavy quarks, a square an insertion of a CLFV operator and dots SM vertices.
The first two diagrams represent penguin contributions of heavy flavor vector operators to CLFV Z couplings, leptonic operators and semileptonic operators with light quarks. The latter also receive contributions from $W$ exchanges, as shown in the last three diagrams. Tensor operators run into $\Gamma^{e}_{\gamma, Z}$
via a diagram with the same topology as the first. 
}\label{Penguin1}
\end{figure}

\item Vector-like four-fermion operators with heavy quarks mix onto $Z$-boson vertices and four-fermion operators with light quarks and leptons via the penguin diagrams shown in Fig. \ref{Penguin1}.
As shown in Eqs.~\eqref{cLphi} and \eqref{cephi}, the mixing with the CLFV $Z$ couplings has a component proportional to the quark Yukawa coupling and one to the gauge couplings. For top-quark operators, the Yukawa component dominates, 
and induces a very sizable mixing
\begin{equation}
\left(c^{(1)}_{L\varphi} + c^{(3)}_{L\varphi} \right)(\mu_t)\sim 0.1 \left( C_{Lu }  - C_{LQ, U} \right)_{tt}(\mu_0), \quad c_{e \varphi}(\mu_t) \sim 0.1 \left( C_{eu} - C_{Qe}  \right)_{tt}(\mu_0), \label{top-rge}
\end{equation}
where  $(C_{Qe})_{tt} = V_{t j} (C_{Qe})_{j k} V^*_{t k}$,
$\mu_t \sim m_t$ and $\mu_0 \sim \Lambda \sim  1$ TeV. For operators with $b$ and $c$ quarks, the gauge component dominates, and gives percent level corrections to the $Z$ couplings.
The mixing with light-quark and lepton four-fermion operators, driven by the RGEs in Eqs. \eqref{peng1}--\eqref{peng7}
and \eqref{peng8}--\eqref{peng13}, is the same for all the flavor components of $u$- or $d$-type operators, and these mixing coefficients are at the $10^{-3}$ level. The coefficients of $Z$ couplings, leptonic and semileptonic four-fermion operators at the scale $\mu = m_t$, as a function 
of heavy quark operators at the scale $\mu_0 = 1$ TeV, are given in Table \ref{topRGE}.

\item The mixing of quark-flavor off-diagonal four-fermion operators onto flavor diagonal is suppressed by small CKM and/or Yukawa couplings, 
see Eqs. \eqref{fc1}--\eqref{fc6}, and it is in most cases negligible.

\end{itemize}

In addition to the running effects, integrating out heavy flavors induces gluonic operators.
The EIC is sensitive to the CLFV Yukawa $Y^\prime_e$ in \eq{hYukawa} via the couplings of the Higgs bosons to quarks
and the  effective Higgs-gluon coupling induced at the top threshold,
\begin{equation}
\label{eq:hgg}
\mathcal L_{h gg} = \frac{\alpha_s}{12 \pi v} h\,  G^a_{\mu \nu} G^{a\, \mu \nu}.
\end{equation}
In addition, CLFV SMEFT operators with heavy quarks can induce dimension-seven gluonic operators of the form
\begin{align}
\label{eq:Lgluonic}
\mathcal L_g = C_{GG}  \frac{1}{v^3} \frac{\alpha_s}{4\pi} \left( G^a_{\mu \nu} G^{a \mu \nu} \right)   \bar e_L  e_R  & + C_{G\widetilde G} \frac{1}{v^3} \frac{\alpha_s}{4\pi} \left( G^a_{\mu \nu} \widetilde G^{a\,\mu \nu} \right)  \bar e_L  e_R   + \textrm{h.c.} \\
\text{where}\quad \widetilde G^{a\mu\nu} &= \frac{1}{2} \epsilon^{\mu\nu\alpha\beta} G^{a}_{\alpha\beta}\,. \nn
\end{align}
At the top threshold, $C_{GG,G \widetilde G}$ are induced by the scalar operators with matching coefficients
\begin{eqnarray}
\left[C_{GG} \right]_{\tau e} &=& -\frac{1}{3} \frac{v}{m_t} \left[ C^{(1)}_{L e Q u} \right]_{\tau e tt}, \qquad 
\left[C_{GG} \right]_{e \tau} =   - \frac{1}{3} \frac{v}{m_t} \left[ C^{(1)}_{L e Q u} \right]_{e \tau tt}, \label{matching_GG}\\
\left[C_{G\widetilde G} \right]_{\tau e} &=& -\frac{i}{2} \frac{v}{m_t} \left[ C^{(1)}_{L e Q u} \right]_{\tau e tt}, \qquad
\left[C_{G\widetilde G} \right]_{e \tau} =  -\frac{i}{2} \frac{v}{m_t} \left[ C^{(1)}_{L e Q u} \right]_{e \tau tt}. \label{matching_GGtilde}
\label{eq:gluonic}
\end{eqnarray}
Notice that both sides of Eqs. \eqref{matching_GG} and \eqref{matching_GGtilde} are renormalization-scale-independent, at one loop in QCD.

\section{CLFV Deep Inelastic Scattering}
\label{sect:CLFVDIS}

We obtain in this Section the expressions for deep inelastic scattering (DIS) cross sections in the presence of CLFV SMEFT operators.
In \ssec{factorization} we factorize the generic DIS cross section into leptonic and hadronic structures, matching the latter onto partonic hard matching coefficients convolved with parton distribution functions (PDFs), reviewing the standard derivation in QCD, followed by generalization to contributions from arbitrary SMEFT operators. We simplify to tree-level cross sections for the remainder of the analysis, and in \ssec{CLFV_partonic} we collect the tree-level partonic cross sections induced by all the CLFV SMEFT operators we consider. In \ssec{partonic_numerical}, we provide numerical values of the cross sections multiplying the SMEFT operator coefficients, and obtain initial estimates of EIC sensitivity to each coupling based on the partonic cross sections. In Sec.~\ref{efficiency} we will go to the more realistic case of detector-level cross sections.

\subsection{Factorization of the cross section}
\label{ssec:factorization}

\tocless\subsubsection{General cross section}

The generic cross section differential in the momentum transfer $q = k-k'$ in the scattering $\ell(k) p(P) \to \ell'(k') X(p_X)$ is
\begin{equation}
\frac{d\sigma}{d^4q} = \frac{1}{2S} \int d\Phi_L \sum_X \abs{\mathcal{M}(\ell p \to \ell' X)}^2 (2\pi)^4\delta^4(P + q - p_X)\delta^4(q-k+k')\,.
\end{equation}
where $S=(k+P)^2$, $\Phi_L$ is the outgoing lepton $\ell'$ phase space, and the sum is over all other final state particles $X$. We do not yet specify whether we sum over $\ell, p$ spins, allowing for the possibility of polarized beams. We sum over $\ell'$ spins.
We will use the standard DIS kinematic variables,
\begin{equation}
Q^2\equiv - q^2\,,\quad x\equiv \frac{Q^2}{2P\mcdot q} \,,\quad y \equiv \frac{2P\mcdot q}{2P\mcdot k}\,,\qquad xyS = Q^2\,.
\end{equation}
To form the cross section differential in the DIS variables $x,y$, we insert the delta functions defining these variables,
\begin{equation}
\label{eq:CSxy}
\frac{d\sigma}{dx\,dy} = \int d^4 q\frac{d\sigma}{d^4q} \delta\Bigl(x + \frac{q^2}{2P\mcdot q}\Bigr)\delta\Bigl(y -  \frac{2P\mcdot q}{2P\mcdot k}\Bigr)\,.
\end{equation}
It is convenient to pick a particular frame to perform the integrals with the delta functions, though the result is still Lorentz-invariant. For example, in the Breit or CM frames, the proton can be put in the $+\hat z$ direction, and $P,q$ take the forms
\begin{equation}
\label{eq:Pqmomenta}
P = \bar n_z\mcdot P\frac{n_z}{2}\,, \quad q= \bar n_z\mcdot q \frac{n_z}{2} + n_z\mcdot q \frac{\bar n_z}{2} + q_T\,,
\end{equation}
where $n_z = (1,\hat z)$, $\bar n_z = (1,-\hat z)$. Then we use the delta functions in \eq{CSxy} to integrate over $n_z\mcdot q,\bar n_z\mcdot q$. To do the $q_T$ integrals, we express the $\ell'$ phase space integral to leading order in electroweak interactions:
\begin{equation}
\int d\Phi_L = \int \frac{d^4k'}{(2\pi)^3} \delta\bigl((q-k)^2\bigr)\,,
\end{equation}
which will let us do the $q_T$ integral (using also azimuthal symmetry). In the end, our formula \eq{CSxy} becomes
\begin{equation}
\label{eq:DIScs}
\frac{d\sigma}{dx\,dy} = \frac{y}{32\pi^2}\sum_X \abs{\mathcal{M}(\ell p \to \ell' X)}^2 (2\pi)^4\delta^4(P + q - p_X)\,,
\end{equation}
where the value of $q$ has been fixed by the above delta function integrals, e.g. in frames where $P$ takes the form in \eq{Pqmomenta}, we have
\begin{equation}
q = \frac{yS}{\bar n_z\mcdot P} \frac{\bar n_z}{2} - xy \bar n_z\mcdot P\frac{n_z}{2} + Q\sqrt{1-y}\,\hat n_T\,,
\end{equation}
where 
$\hat n_T$ is a unit vector in any direction transverse to $n_z$ (azimuthally symmetric). \eq{DIScs} is our basic starting formula for a DIS cross section. 

\begin{figure}
\vspace{-1em}
\centering
\includegraphics[width=0.45\textwidth]{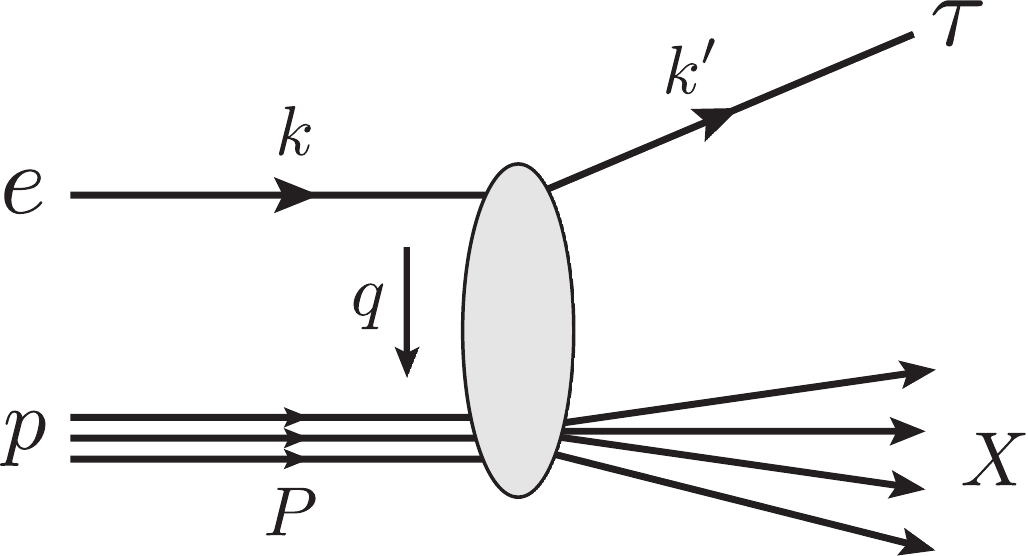}
\vspace{-6pt}
\caption{The DIS process induced by CLFV SMEFT operators. The gray blob represents arbitrary CLFV interactions mediating $ep\to\tau X$.}
\label{DIS}
\end{figure}

The bulk of our calculations will come in evaluating the squared amplitudes $\abs{\mathcal{M}}^2$ in the presence of arbitrary SMEFT operators that can mediate the process $\ell P \to \ell' X$, where primarily we shall be interested in $\ell = e $ and $\ell' = \tau$ as in Fig. \ref{DIS}. All of the operators or channels we consider give amplitudes that can be expressed in a form,
\begin{equation}
\label{eq:DISamplitude}
\mathcal{M}(\ell p \to \ell' X) = \sum_I C_I \bra{\ell'(k')}\mathcal{O}_\text{lep}^I \ket{\ell(k)}\bra{X}\mathcal{O}_\text{had}^I\ket{p(P)}\,,
\end{equation}
where each operator is factored into a leptonic and hadronic part, the two parts containing the relevant leptonic and hadronic fields:
\begin{equation}
\label{eq:OlepOhad}
\mathcal{O}_\text{lep} \sim \bar \ell' \Gamma_l^I \ell \,,\quad \mathcal{O}_\text{had}  \sim \bar q_{f'}\Gamma_h^I \bar q_f \,,G_{\alpha\beta}G_{\mu\nu}\,,
\end{equation}
and in general we will lump constant prefactors into $\mathcal{O}_\text{lep}$.
Here $\Gamma_{l,h}$ are any allowed Dirac matrix structures, and the gluon field indices may be contracted in different ways, e.g. $GG,G\widetilde G$. These effective operators may also arise from contractions of other operators, in which case relevant propagators or other factors are lumped into the coefficients. In the sum over operator structures $I$, any appropriate contractions over Dirac or flavor indices are understood.

With amplitudes of the form \eq{DISamplitude}, the cross section \eq{DIScs} also factors into leptonic and hadronic structures,
\begin{equation}
\label{eq:DISLW}
\frac{d\sigma}{dx\,dy} = \sum_{IJ} L_{IJ}\otimes W_{IJ}\,,
\end{equation}
where 
\begin{subequations}
\label{eq:LW}
\begin{align}
L_{IJ} &= \frac{y}{32\pi^2}C_I C_J^* \bra{\ell(k)} \bar\ell\bar\Gamma_l^{J} \ell' \ket{\ell'(k')}\bra{\ell'(k')} \bar\ell' \Gamma_l^I\ell\ket{\ell(k)} \\
W_{IJ}&= \sum_X (2\pi)^4\delta(P+q-p_X) \bra{p(P)}\bar{\mathcal{O}}_\text{had}^{J}\ket{X}\bra{X}\mathcal{O}_\text{had}^I\ket{p(P)}\,,
\end{align}
\end{subequations}
where $\bar\Gamma = \gamma^0\Gamma^\dag \gamma^0$, $\bar{\mathcal{O}} = \mathcal{O}^\dag$, and the $\otimes$ in \eq{DISLW} represents any appropriate index contractions. We have assumed that the inclusive state $X$ is purely hadronic, appropriate for us working at tree level in electroweak interactions.

At this point we have not yet specified whether the incoming lepton and proton are polarized or spin-averaged. In the leptonic part, we can pick out right- or left-handed polarizations by summing over spins but including projection operators in the leptonic Dirac structures $\Gamma_l^{I,J}$, i.e., again at tree level,
\begin{align}
\label{eq:LIJ}
L_{IJ} &= \frac{y}{32\pi^2} C_I C_J^* \sum_\text{spins} \bar u(k) \frac{1-\lambda_\ell\gamma_5}{2} \bar \Gamma_l^J u(k') \bar u(k') \Gamma_l^I\frac{1+\lambda_\ell\gamma_5}{2}u(k) \\
&=  \frac{y}{32\pi^2} C_I C_J^* \Tr \Bigl( \kslash\bar\Gamma_l^J \kslash' \Gamma_l^I \frac{1+\lambda_\ell\gamma_5}{2}\Bigr) \,, \nn
\end{align}
where $\lambda_\ell = \pm 1$ for $R,L$-handed incoming $\ell$. For the case of SM electroweak interactions, with photon and $Z$ boson exchanges, expressions for the traces in \eq{LIJ} can be found in, e.g., \cite{Kang:2013nha}. In the simplest case of tree-level photon exchange in the SM, we will relabel $I,J \to \gamma f f'$ indicating the photon coupling to quark flavors $f,f'$ in the hadronic part, and the tensor \eq{LIJ} takes the value
\begin{equation}
\label{eq:Lff}
L_{\gamma f f'}^{\mu\nu} = -\frac{\alpha_\text{em}^2e_f e_{f'}}{2xS} (g_T^{\mu\nu} - i\lambda_\ell \epsilon_T^{\mu\nu})\,, 
\end{equation}
where $e_f$ is the electric charge of quark flavor $f$ in units of $e$, and $\alpha_\text{em}\equiv  e^2/(4\pi)$. The tensor structures appearing in \eq{Lff} are:
\begin{equation}
g_T^{\mu\nu} \equiv g^{\mu\nu} - 2\frac{k^\mu {k'}^\nu + k^\nu {k'}^\mu}{Q^2}\,,\quad \epsilon_T^{\mu\nu} = \frac{2}{Q^2} \epsilon^{\alpha\beta\mu\nu}k_\alpha k'_\beta\,.
\end{equation}
When $W_{IJ}\to W_{\gamma ff'}$ in \eq{LW} is evaluated for partonic initial states, at tree level, we will simply obtain the Born cross section for \eq{DISLW}. In general we need to match $W_{IJ}$ onto quark and gluon PDFs (polarized and unpolarized) in the proton state. We sketch this matching procedure in the next subsection.

\tocless\subsubsection{Hadronic tensor}

\begin{figure}
\vspace{-1em}
\centering
\includegraphics[width=.9\textwidth]{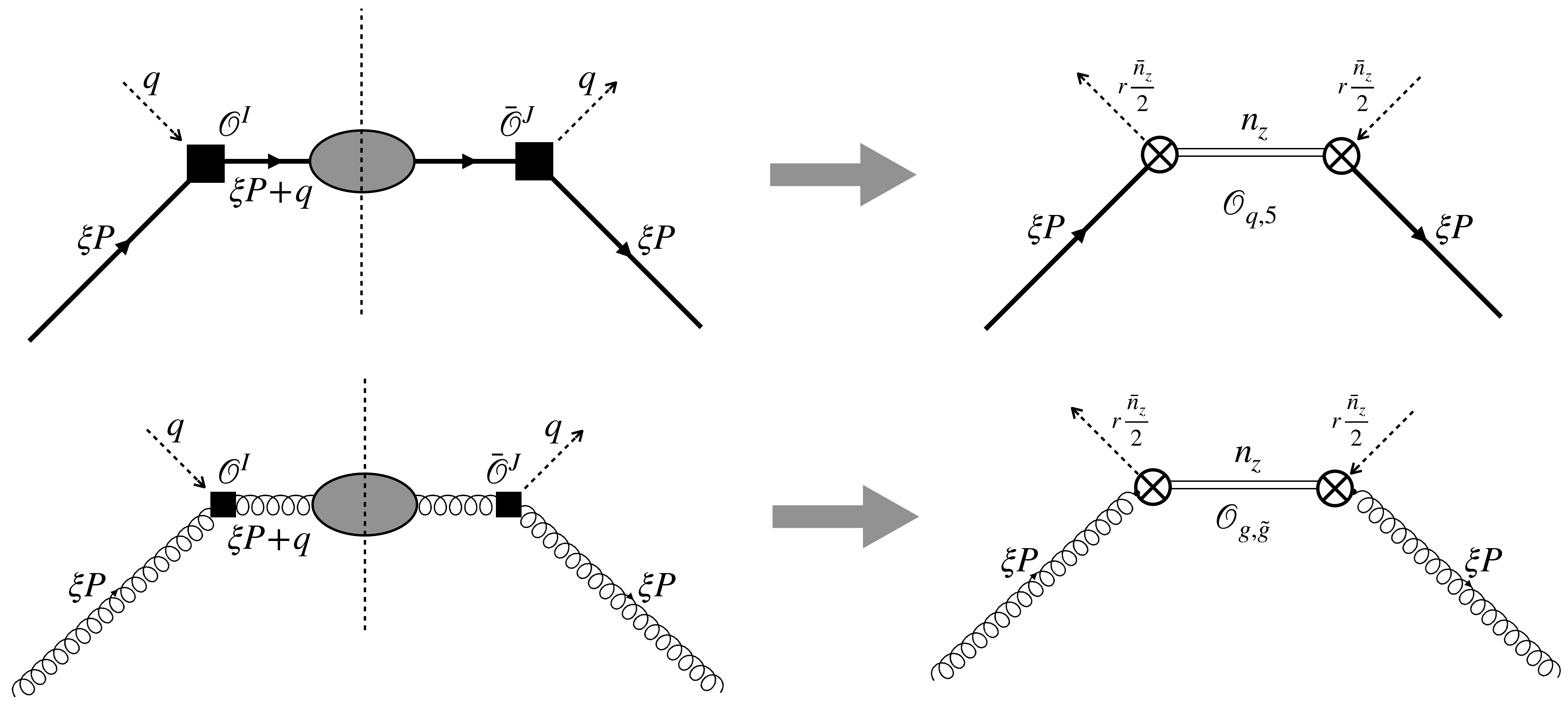}
\vspace{-1em}
\caption{Matching products of operators in hadronic tensor \eq{WIJ} onto quark or gluon bilinear operators in \eq{OqO5} or \eq{OgOg} at tree level, using partonic matrix elements in external quark or gluon states with momentum $\xi P$. The operators $\mathcal{O}^{I,J}$ are the hadronic part of generic SM or SMEFT operators or amplitudes, and proton matrix elements of $\mathcal{O}_{q,5}$ or $\mathcal{O}_{g,\tilde g}$ give (un)polarized quark and gluon PDFs in the proton, see \eqs{qpdf}{gpdf}.}
\label{pdf_matching}
\end{figure}

The hadronic part of the amplitude $W_{IJ}$ in \eq{LW} can be expressed, as in usual DIS, as convolutions of perturbative matching coefficients and PDFs. Using the delta function to translate one of the operators, and summing over $X$, we obtain:
\begin{equation}
\label{eq:WIJ}
W_{IJ} = \int d^4 x \, e^{iq\cdot x} \bra{p(P)} \bar{\mathcal{O}}_\text{had}^J (x) \mathcal{O}_\text{had}^I(0)\ket{p(P)}\,.
\end{equation}
This forward matrix element of the product of operators can be related to twice the imaginary part or the discontinuity of the matrix element of the time-ordered product of the operators (e.g. \cite{Bauer:2002nz,Manohar:2003vb}):
\begin{equation}
\label{eq:WT}
W_{IJ} = \Disc T_{IJ} \,, \quad T_{IJ}\equiv i\int d^4 x\,e^{iq\cdot x}\bra{p(P)} T\bigl\{ \bar{\mathcal{O}}_\text{had}^J (x) \mathcal{O}_\text{had}^I(0)\bigr\}\ket{p(P)}\,,
\end{equation}
which can be evaluated from ordinary Feynman diagrams. This operator product typically contains two pairs of quark or gluon bilinears, separated by $x$. We will perform an operator product expansion (OPE) to match onto products of a single bilinear operator containing quark or gluon fields, separated only along the light-cone direction $n_z$ conjugate to the proton momentum $P$.
In general, the product of operators in \eq{WT} will match, at leading power (twist) onto:
\begin{align}
\label{eq:Tpdf_matching}
W_{IJ}  \longrightarrow \int dr  \bigl[\mathcal{C}_{q}^{IJ}(r)\mathcal{O}_q(r) + \mathcal{C}_{5}^{IJ}(r)\mathcal{O}_5(r) + \mathcal{C}_{g}^{IJ} \mathcal{O}_g(r) + \mathcal{C}_{\tilde g}^{IJ} \mathcal{O}_{\tilde g}(r) \bigr] \,, 
\end{align}
where $\mathcal{O}_{q,5}$ are quark bilinear operators:
\begin{subequations}
\label{eq:OqO5}
\begin{align}
\mathcal{O}_q (r) &=  \int\frac{dz}{2\pi}e^{-iz r}\bar q( z \bar n_z) \frac{\bnslash_z}{2} W(z \bar n_z,0)q(0) \\
\mathcal{O}_5(r)  &= \int\frac{dz}{2\pi}e^{-iz r}\bar q( z \bar n_z) \frac{\bnslash_z \gamma_5}{2} W(z \bar n_z,0)q(0)
\end{align}
\end{subequations}
and $\mathcal{O}_{g,\tilde g}$ are gluon bilinear operators:
\begin{subequations}
\label{eq:OgOg}
\begin{align}
\mathcal{O}_g(r) &=  -\int\! \frac{dz}{4\pi r}e^{-izr} \bar n_z^\mu \bar n_z^\alpha G_{\mu\lambda}(z\bar n_z) Y(z \bar n_z,0) {G^\lambda}_\alpha(0) - (r\to -r) \\
\mathcal{O}_{\tilde g} (r) &=  i \int\frac{dz}{4\pi r}e^{-iz r} \bar n_z^\mu \bar n_z^\alpha G_{\mu\lambda}(z\bar n_z) Y(z \bar n_z,0) {\widetilde G^\lambda}_{\ \alpha}(0) + (r\to -r) \,.
\end{align}
\end{subequations}
In \eqs{OqO5}{OgOg}, each pair of quark or gluon fields are separated only along the light-cone direction $\bar n_z$ conjugate to the large proton momentum along $n_z$, and the $W,Y$ are fundamental or adjoint Wilson line gauge links along $\bar n_z$ ensuring gauge invariance (in this paper, we can take $W=Y=1$). Matrix elements of these bilinear operators in the proton state give the unpolarized and polarized PDFs \cite{Collins:1981uw,Soper:1996sn,Manohar:1990jx,Manohar:1990kr}:
\begin{subequations}
\label{eq:qpdf}
\begin{align}
f_q (\xi) &=  \frac{1}{2}\sum_\lambda \bra{p,\lambda} \mathcal{O}_q(\xi \bar n_z\mcdot P) \ket{p,\lambda}\,, \quad 
\lambda \Delta f_q (\xi) =  \bra{p,\lambda}\mathcal{O}_5(\xi \bar n_z\mcdot P)\ket{p,\lambda}  \,, \\
\label{eq:gpdf}
f_g(\xi) &= \frac{1}{2}\sum_\lambda\bra{p,\lambda} \mathcal{O}_g(\xi \bar n_z\mcdot P) \ket{p,\lambda} \,,\quad  \lambda\Delta f_g(\xi) =  \bra{p,\lambda} \mathcal{O}_{\tilde g}(\xi \bar n_z\mcdot P)\ket{p,\lambda}.
\end{align}
\end{subequations}
The matching coefficients $\mathcal{C}_{q,5,g,\tilde g}^{IJ}$ in \eq{Tpdf_matching} are computed by matching partonic matrix elements of the operators on either side of the equation, with hard propagators between extra fields on the left-hand side contracted or integrated out. This procedure is illustrated in Fig.~\ref{pdf_matching}. 
At tree level we will not encounter mixing of quark and gluon operators, but at higher orders they will mix.

\leavevmode
\paragraph{SM QCD:} In the usual case of QCD in the SM, for photon exchange diagrams, we obtain for the hadronic tensor in \eq{LW} that contracts with the leptonic tensor in \eq{Lff},
\begin{equation}
\label{eq:WSM}
W_{\gamma ff'}^{\mu\nu} = \Disc i \int d^4 x\, e^{iq\cdot x} \bra{p(P)} T\{ \bar q_{f'} \gamma^\nu q_{f'}(x) \bar q_f\gamma^\mu q_f (0) \} \ket{p(P)}\,.
\end{equation}
To match the operator in this matrix element onto those on the RHS of \eq{Tpdf_matching}, we compute matrix elements of each in a quark state (see Fig.~\ref{pdf_matching}):
\begin{align}
\label{eq:WSMq}
W_{\gamma ff'}^{\mu\nu(q)} &= \Disc i\int d^4 x\,e^{iq\cdot x} \bra{q(\xi P;\lambda)} T\{\bar q_{f'} \gamma^\nu q_{f'}(x) \bar q_f\gamma^\mu q_f (0)\ket{q(\xi P;\lambda)} \\
& = -2\pi x\delta(\xi-x) \Bigl[ g_\perp^{\mu\nu} - \frac{4x^2 P^\mu P^\nu}{Q^2} + i\lambda \epsilon_\perp^{\mu\nu}\Bigr] \delta_{ff'}\,, \nn
\end{align}
with momentum $\xi P$ and spin $\lambda$, and where the transverse tensor structures are:
\begin{equation}
g_\perp^{\mu\nu} \equiv g^{\mu\nu} - \frac{P^\mu q^\nu + P^\nu q^\mu}{P\cdot q}\,,\qquad \epsilon_\perp^{\mu\nu} \equiv \frac{1}{P\cdot q} \epsilon^{\mu\nu\alpha\beta}P_\alpha q_\beta\,.
\end{equation}
Meanwhile, the quark matrix elements of $\mathcal{O}_q,\mathcal{O}_5$ in \eq{OqO5} are:
\begin{subequations}
\label{eq:OqO5part}
\begin{align}
\bra{q(\xi P;\lambda)} \mathcal{O}_q(r) \ket{q(\xi P;\lambda)} &= \delta\Bigl(\frac{r}{\xi\bar n_z\mcdot P}-1\Bigr) \\
\bra{q(\xi P;\lambda)} \mathcal{O}_5(r) \ket{q(\xi P;\lambda)} &= \lambda \delta\Bigl(\frac{r}{\xi\bar n_z\mcdot P}-1 \Bigr).
\end{align}
\end{subequations}
This tells us that the matching coefficients in \eq{Tpdf_matching} for the operator in \eq{WSM} are:
\begin{subequations}
\begin{align}
\mathcal{C}_{q,\gamma ff'}^{\mu\nu}(r) &= -2\pi \delta(r- x\bar n_z\cdot P)  \Bigl[ g_\perp^{\mu\nu} - \frac{4x^2 P^\mu P^\nu}{Q^2} \Bigr] \delta_{ff'} \\
\mathcal{C}_{5,\gamma ff'}^{\mu\nu} &= -2\pi \delta(r- x\bar n_z\cdot P) i\epsilon_\perp^{\mu\nu} \delta_{ff'} \,.
\end{align}
\end{subequations}
Using these matching conditions in \eq{WSM} and the PDF operator definitions in \eq{qpdf}, and contracting the perturbative matching coefficients in $W_{\gamma ff'}^{\mu\nu}$ with the leptonic tensor in \eq{Lff}, we obtain for the cross section \eq{LW} for the photon channel in SM QCD DIS at LO:
\begin{equation}
\label{eq:DISSMQCD}
\frac{d\sigma^\text{LO}_{\lambda_e\lambda_T}}{dx\,dy} = 2\sigma_0 \sum_{q}e_q^2 \biggl\{ [ 1 + (1-y)^2] f_q(x)  - \lambda_e\lambda_T y(2-y)\Delta f_q(x)  \biggr\}\,,
\end{equation}
for incoming $e$ and proton target spins $\lambda_{e,T}$, and
where
\begin{equation}
\label{eq:Bornfactor}
\sigma_0 \equiv \frac{\alpha^2_{\rm em} \pi S x}{Q^4}.
\end{equation}
Averaged over incoming spins, we obtain the familiar unpolarized DIS cross section at LO in QCD:
\begin{equation}
\label{eq:DISLOcs}
\frac{d\sigma_\text{un}^\text{LO}}{d x d y}  = 2 \sigma_0  \sum_q e_q^2  f_q(x)  \left[1 + \left( 1 - y\right)^2 \right]\,. 
\end{equation}

\paragraph{SMEFT four-fermion operators:}
We can generalize the above derivation in the SM for generic four-fermion operators in SMEFT. The leptonic tensor for a given operator still takes the form \eq{LIJ}, and the hadronic tensor \eq{WIJ} will take the form:
\begin{equation}
\label{eq:W4fermi}
W_{\psi^4}^{IJ} = \Disc i \int d^4 x\, e^{iq\cdot x} \bra{p(P)} T\{ \bar q_{f} \bar\Gamma_h^J q_{f'}(x) \bar q_{f'}\Gamma_h^I q_{f} (0) \} \ket{p(P)}\,,
\end{equation}
where $f,f'$ are the particular quark flavors appearing in a given operator from, e.g., \eq{fourfermion}. The operator in the matrix element matches onto $\mathcal{O}_{q,5}$ in \eq{OqO5} in similar manner as in the SM above and illustrated in Fig.~\ref{pdf_matching}, with the partonic matrix element analogous to \eq{WSMq} now given by:
\begin{align}
\label{eq:W4fermion}
W_{\psi^4}^{IJ(q)} &= \Disc i\int d^4 x\,e^{iq\cdot x} \bra{q_f(\xi P;\lambda)} T\{\bar q_{f} \bar\Gamma_h^J q_{f'}(x) \bar q_{f'}\Gamma_h^I q_{f} (0)\ket{q_f(\xi P;\lambda)} \\
&= 2\pi \delta(\xi-x) \frac{1}{2P\mcdot q} \Tr\Bigl[ \xi\Pslash \bar\Gamma_h^J (\xi\Pslash + \qslash)\Gamma_h^I \frac{1+\lambda\gamma_5}{2}\Bigr] \,, \nn
\end{align}
where any $L,R$ projections of the quark fields $q_{f,f'}$ are understood to be contained in $\Gamma_h^{I,J}$. The matching coefficients in \eq{Tpdf_matching}  for these operators onto $\mathcal{O}_{q,5}$ are:
\begin{subequations}
\label{eq:CqC54fermion}
\begin{align}
\mathcal{C}_{q,\psi^4}^{IJ}(r) &= 2\pi \delta(r-xn_z\mcdot P) \frac{1}{4P\mcdot q} \Tr\Bigl[ x\Pslash \bar\Gamma_h^J (x\Pslash + \qslash)\Gamma_h^I \Bigr]  \\
\mathcal{C}_{5,\psi^4}^{IJ}(r) &= 2\pi \delta(r-xn_z\mcdot P) \frac{1}{4P\mcdot q} \Tr\Bigl[ x\Pslash \bar\Gamma_h^J (x\Pslash + \qslash)\Gamma_h^I \gamma_5\Bigr] \,.
\end{align}
\end{subequations}
These perturbative coefficients contracted with the leptonic tensor in \eq{LIJ} will give the partonic cross sections when plugged into \eq{DISLW}, and we can write, similar to the SM formula \eq{DISSMQCD}, the four-fermion operator contribution to the full cross section:
\begin{equation}
\frac{d\sigma^{\psi^4}_{\lambda_e\lambda_T}}{dx\,dy} = L_{IJ}^{\lambda_e} \int dr\, \Bigl[ \mathcal{C}_{q,\psi^4}^{IJ}(r) f_q(x)  + \lambda_T \mathcal{C}_{5,\psi^4}^{IJ}(r)  \Delta f_q(x)  \Bigr]\,,
\end{equation}
using \eq{OqO5part} for the PDFs. At tree level the integral over $r$ here just removes the delta function in \eq{CqC54fermion}. The contraction of $L_{IJ}$ with each of $C_{q,5}^{IJ}$ gives the tree-level partonic cross section from each operator.

For example, for the scalar ${LeQu}$ operator in \eq{fourfermion}, the operator contributing to $ep\to \tau X$ is 
\begin{align}
\label{eq:scalarop}
&(C^{(1)}_{LeQu})_{ij}(\mathcal{O}^{(1)}_{LeQu})_{ij} \equiv   \mathcal{O}_\text{lep}^S \mathcal{O}_\text{had}^S\,, \nn \\
&\quad \text{where} \quad \mathcal{O}_\text{lep}^S =  \frac{4G_F}{\sqrt{2}} (C_{LeQu}^{(1)} )_{ij}\bar\tau_L e_R \,, \ \mathcal{O}_\text{had}^S  = \bar u_L^i u_R^j = \bar u^i\frac{1+\gamma_5}{2} u^j\,,
\end{align}
where here $i,j=u,c$ label the quark flavors. 
So the leptonic ``tensor'' \eq{LIJ} for initial electron spin $\lambda_e$ and hadronic ``tensor'' in \eq{W4fermion} in a quark state of momentum $\xi P$ and spin $\lambda_q$ are:
\begin{equation}
L_S^{(ij)} = \frac{G_F^2 y Q^2}{4\pi^2} \frac{1+\lambda_e}{2} \abs{(C_{LeQu}^{(1)})_{ij}}^2  \,,\quad W_S^{(ij)} = 2\pi x \delta(\xi-x)\frac{1+\lambda_q}{2}\,.
\end{equation}
The matching coefficients  \eq{CqC54fermion} for the hadronic tensor are then
\begin{equation}
\mathcal{C}_{q}^S(r) = \mathcal{C}_{5}^S(r) = \pi \delta(r-xn_z\mcdot P)\,,
\end{equation}
and at tree level the gluon coefficients are zero. Thus the contribution of the operators \eq{scalarop} to the cross section is
\begin{equation}
\frac{d\sigma^S_{\lambda_e\lambda_T}}{dx\,dy} = \frac{G_F^2 y Q^2}{4\pi}   \frac{1+\lambda_e}{2} \sum_{i,j=u,c} \abs{(C_{LeQu}^{(1)})_{ij}}^2\bigl[ f_{u_j}(x) + \lambda_T \Delta f_{u_j}(x)\bigr]\,.
\end{equation}
and similarly for $\bar u_j$ antiquark contibutions. The procedure for other four-fermion operator contributions is also similar. (Many of the resulting cross sections have been given recently in Ref.~\cite{Boughezal:2020uwq}.) The contribution of dipole and Higgs operators follows substantially the same procedure as SM QCD or SMEFT four-fermion operator matching as well. We collect all relevant partonic cross sections in Sec.~\ref{ssec:CLFV_partonic}.

\paragraph{SMEFT gluon operators:}

The matching procedure for products of gluon operators from \eq{Lgluonic} is similar: 
\begin{align}
\label{eq:LGlephad}
\mathcal{L}_g &= \mathcal{O}^{G}_\text{lep} \mathcal{O}^{G}_\text{had} + \mathcal{O}^{\widetilde G}_\text{lep} \mathcal{O}^{\widetilde G}_\text{had} + \text{h.c.} \,, \\
\text{where}\quad \mathcal{O}^{G}_\text{lep} &= \bigl[C_{GG}\bigr]_{\tau e} \frac{1}{v^3}\frac{\alpha_s}{4\pi }\bar\tau_L e_R \,, \quad \mathcal{O}^{\widetilde G}_\text{lep} =\bigl[C_{G\widetilde G}\bigr]_{\tau e} \frac{1}{v^3}\frac{\alpha_s}{4\pi} \bar\tau_L e_R\,, \nn \\
\mathcal{O}^{G}_\text{had} &= G^a_{\mu\nu}G^{a\mu\nu}\,, \quad \mathcal{O}^{\widetilde G}_\text{had} = G^a_{\mu\nu}\widetilde G^{a\mu\nu}\,. \nn
\end{align}
The cross section in \eq{DISLW} then takes the form
\begin{equation}
\frac{d\sigma^G}{dx\,dy} = L_{G} W_{G} + L_{\widetilde G} W_{\widetilde G} + L_{G\widetilde G} W_{G\widetilde G} + L_{\widetilde G G} W_{\widetilde G G} \,,
\end{equation}
where 
\begin{align}
L_G &= \frac{yQ^2}{32\pi^2 v^6} \Bigl(\frac{\alpha_s}{4\pi}\Bigr)^2 \Bigl\{ \frac{1+\lambda_e}{2} \abs{\bigl[C_{GG}\bigr]_{\tau e}}^2 + \frac{1- \lambda_e}{2} \abs{\bigl[C_{GG}\bigr]_{e \tau}}^2\Bigr\}  \,, \\
 L_{\widetilde G} &= \frac{yQ^2}{32\pi^2 v^6} \Bigl(\frac{\alpha_s}{4\pi}\Bigr)^2  \Bigl\{ \frac{1+\lambda_e}{2} \abs{\bigl[C_{G\widetilde G}\bigr]_{\tau e}}^2 + \frac{1-\lambda_e}{2} \abs{\bigl[C_{G\widetilde G}\bigr]_{e\tau}}^2\Bigr\} \nn \\
L_{G\widetilde G} &= L_{\widetilde G G}^* = \frac{yQ^2}{32\pi^2 v^6}\Bigl(\frac{\alpha_s}{4\pi}\Bigr)^2 \Bigl\{ \frac{1+\lambda_e}{2} \bigl[C_{GG}\bigr]_{\tau e}^*\bigl[C_{G\widetilde G}\bigr]_{\tau e} + \frac{1-\lambda_e}{2} \bigl[C_{GG}\bigr]_{e\tau}^*\bigl[C_{G\widetilde G}\bigr]_{e\tau} \Bigr\} \nn\,,
\end{align}
and
\begin{align}
\label{eq:WG}
W_G &= \int d^4x\,e^{iq\cdot x} \bra{p(P)} G^a_{\mu\nu}G^{a\mu\nu}(x) G^b_{\alpha\beta}G^{b\alpha\beta}(0)\ket{p(P)} \\
W_{\widetilde G} &= \int d^4x\,e^{iq\cdot x} \bra{p(P)} G^a_{\mu\nu}\widetilde G^{a\mu\nu}(x) G^b_{\alpha\beta}\widetilde G^{b\alpha\beta}(0)\ket{p(P)}  \nn \\
W_{G\widetilde G} &= \int d^4x\, e^{iq\cdot x}\bra{p(P)}G^a_{\mu\nu} G^{a\mu\nu}(x) G^b_{\alpha\beta}\widetilde G^{b\alpha\beta}(0)\ket{p(P)}  = W_{\widetilde G G}(q\to -q)\,. \nn
\end{align}
The matrix elements of the gluon PDF operators in \eq{gpdf} in a partonic gluon state wtih polarization $\lambda$ (Fig.~\ref{pdf_matching}) are
\begin{align}
\bra{g(\xi P;\lambda) } \mathcal{O}_g(r)\ket{g(\xi P;\lambda)} &= r [\delta(r-\xi\bar n_z\mcdot P) + \delta(r+\xi\bar n_z\mcdot P) ] \varepsilon_\lambda^{*\alpha} \varepsilon_{\lambda\alpha}\,, \\
\bra{g(\xi P;\lambda) } \mathcal{O}_{\tilde g}(r)\ket{g(\xi P;\lambda)} &=r [\delta(r-\xi\bar n_z\mcdot P) - \delta(r+\xi\bar n_z\mcdot P) ]  \frac{i}{2}\epsilon_{\mu\nu\alpha\beta}\bar n_z^\mu n_z^\nu \varepsilon_\lambda^{*\alpha} \varepsilon_{\lambda}^{\beta}\,,  \nn
\end{align}
where $\varepsilon_\lambda$ is the polarization vector for the gluon in polarization state $\lambda$.
Meanwhile, the tree level matrix elements of the operators in \eq{WG} in a gluon state (Fig.~\ref{pdf_matching}) are
\begin{align}
W_G^{(g)} = W_{\widetilde G}^{(g)}  = 8\pi xQ^2\delta(\xi-x)  \varepsilon_\lambda^{*\alpha} \varepsilon_{\lambda\alpha} \,,\quad W_{G\widetilde G}^{(g)} = 0\,.
\end{align}
Thus the gluon matching coefficients in \eq{Tpdf_matching} are:
\begin{equation}
\mathcal{C}_{g}^G(r) = \mathcal{C}_{g}^{\widetilde G}(r) = 8\pi Q^2 \delta(r-x\bar n_z\mcdot P)\,,\quad \mathcal{C}_{\tilde g}^{G,\widetilde G} = 0\,,
\end{equation}
and at tree level the quark coefficients are zero.
The contribution of the gluon operators \eq{LGlephad} to the cross section \eq{DISLW} is then
\begin{align}
\frac{d\sigma^G_{\lambda_e\lambda_T}}{dx\,dy} = \frac{yQ^4}{4\pi v^6} \Bigl(\frac{\alpha_s}{4\pi}\Bigr)^2 f_g(x) \Bigl\{  \frac{1+\lambda_e}{2} ( \abs{\bigl[C_{GG}\bigr]_{\tau e}}^2 &+ \abs{\bigl[C_{G\widetilde G}\bigr]_{\tau e}}^2 ) \\
+ \frac{1- \lambda_e}{2} ( \abs{\bigl[C_{GG}\bigr]_{e\tau}}^2 &+ \abs{\bigl[C_{G\widetilde G}\bigr]_{e\tau}}^2 )\Bigr\}  \nn
\end{align}

For other possible SMEFT operator channels $I,J$ in the hadronic tensor \eq{WIJ}, we can compute the matching of the $T$-ordered products of operators in \eq{WT} onto quark and gluon bilinears \eq{qpdf} in the same way as we hve illustrated above. With more exclusive measurements on final states we may even become sensitive to more general parton distributions in the proton. In this paper, we shall limit ourselves to tree-level results in QCD, which will always yield the na\"\i ve Born-level parton model prediction, which we collect in \ssec{CLFV_partonic} for all the SMEFT operators we consider.

\tocless\subsubsection{Tree-level cross section}
\label{sect:treeDIS}

At LO in QCD, following the steps in the previous subsection at tree level in the matching onto PDFs in hadronic tensor, we obtain the DIS cross sections induced by the CLFV operators introduced in Section \ref{basis}  in terms of the partonic cross sections $\sigma^{a}_{i j}$
where $i, j \in \{ L, R\}$ denote the helicity of the electron and quark/gluon, respectively, and $a = q,g$ (where $q = u,d,s,c,b$ or their antiquarks) denotes the partonic species. For beams with electron and proton polarizations $\lambda_{e,T}$, we obtain the generic cross section
\begin{align}
\label{eq:DISpolarized_cs}
\frac{1}{\sigma_0} \frac{d\sigma_{\lambda_e\lambda_T}}{dx \,dy} &=   \frac{1}{2} \sum_a\left[ \frac{1 - \lambda_e }{2} \left( \hat{\sigma}^{a}_{\rm LL } + \hat{\sigma}^{a}_{\rm LR } \right) 
+ \frac{1 + \lambda_e }{2} \left( \hat{\sigma}^{a}_{\rm RL } + \hat{\sigma}^{a}_{\rm RR } \right)
\right]  f_{a}(x,Q^2). 
 \\
&+ \frac{1}{2}  \sum_a  \left[ \frac{1 - \lambda_e }{2} \left( - \hat{\sigma}^{a}_{\rm LL } +
 \hat{\sigma}^{a}_{\rm LR } \right) + \frac{1 + \lambda_e }{2} \left(-  \hat{\sigma}^{a}_{\rm RL }
+  \hat{\sigma}^{a}_{\rm RR }\right)
\right]  \lambda_T \Delta f_{a}(x,Q^2) \,,\nn
\end{align}
where $\lambda_{e,T} = \pm 1$ for $R,L$ polarizations, respectively. Each individual $\hat\sigma_{ij}^a$ on the right-hand side is the cross section for the specified incoming polarizations, normalized by $\sigma_0$ in \eq{DISLOcs}, i.e.
\begin{align}
\label{eq:partonic_cs}
\hat\sigma_{ij}^a = \frac{y^2 Q^2}{32\pi^3\alpha_\text{em}^2}\sum_{a'} \abs{\mathcal{M}\bigl(\ell(k,\lambda_i) a(xP,\lambda_j) \to \ell'(k') a'\bigr)}^2 (2\pi)^4\delta^4(xP + q - p_{a'})\,, 
\end{align}
with the incoming parton $a=q,g$ having the momentum fraction $x$ of the proton momentum $P$. The spin and flavor of the outgoing parton $a'$ are determined by the SMEFT operator(s) mediating the amplitude $\mathcal{M}$.

In the case of the unpolarized cross section, the dependence on polarized PDFs $\Delta f$ in \eq{DISpolarized_cs} drops out, and we obtain the familiar spin-averaged unpolarized cross section:
\begin{eqnarray}
\frac{d \sigma_\text{un}}{\sigma_0 d x \, d y} &=& \frac{1}{4} \sum_i  \left(\hat{\sigma}^{i}_{\rm LL } + \hat{\sigma}^{i}_{\rm LR } + \hat{\sigma}^{i}_{\rm RL } + \hat{\sigma}^{i}_{\rm RR } \right) f_{i}(x,Q^2) \,.
\label{eq:sigma}
\end{eqnarray}
Since the absolute value of polarized PDFs are always smaller and have a larger uncertainty than their unpolarized counterparts~\cite{Ball:2013lla,Nocera:2014gqa,Boughezal:2020uwq}, we will focus on unpolarized targets in this work and defer the impact of nonzero $\lambda_T$ to future work. For example, single-spin asymmetries could be used to study the polarized beam effects since the PDF uncertainties cancel to a good degree. In the next subsection we give the expressions for the partonic cross sections corresponding to different operators.

\leavevmode

\subsection{CLFV partonic cross sections}
\label{ssec:CLFV_partonic}

\tocless\subsubsection{Vertex corrections and vector-axial four-fermion operators}

The $Z$ couplings $c^{(1,3)}_{L\varphi}$ and $c_{e\varphi}$, and the four-fermion operators
$C_{LQ, U}$, $C_{LQ, D}$, $C_{Lu}$, $C_{Ld}$, $C_{eu}$, $C_{ed}$ and $C_{Qe}$, which are the product of a quark and lepton left- or right-handed vector current,
induce DIS cross sections whose $x$ and $y$ dependence are similar to neutral current DIS in the SM.
For example, defining the prefactor $F_Z$ as
\begin{equation}\label{FZ}
F_Z = \frac{1}{c_w^4 s_w^4} \frac{Q^4}{(Q^2 + m_Z^2)^2},
\end{equation}
we find that the partonic cross sections for $u$-type quarks are given by
\begin{eqnarray}\label{eq:up1}
\hat{\sigma}^{u_i}_{LL} &=& F_Z   \bigg\{ \! \left|\left[c^{(1)}_{L\varphi} \! + \! c^{(3)}_{L\varphi}\right]_{\! \tau e} \! z_{u_L} +  \frac{Q^2 \!+\! m_Z^2}{ m_Z^2} 
\! \left[C^{}_{LQ,\, U}\right]_{\! \tau e u_i u_i}   \right|^2  +  \sum_{j \neq i }\left|   \frac{Q^2 \!+\! m_Z^2}{ m_Z^2} 
\left[C^{}_{LQ,\, U}\right]_{\! \tau e u_j u_i}    \right|^2 \! \bigg\} \nn \\
\hat\sigma^{u_i}_{RR} &=& F_Z   \biggl\{  \left|  \left[c_{e\varphi}\right]_{\tau e} z_{u_R} +  \frac{Q^2 + m_Z^2}{ m_Z^2} \left[C_{e u} \right]_{\tau e u_i u_i} \right|^2  + \sum_{j \neq i} \left|   \frac{Q^2 + m_Z^2}{ m_Z^2} 
 \left[C_{e u} \right]_{\tau e u_j u_i}   \right|^2 \biggr\}  \\
\hat\sigma^{u_i}_{LR} &=& F_Z  (1 -y)^2  \bigg\{  \left| \left[ c^{(1)}_{L\varphi}  +  c^{(3)}_{L\varphi}\right]_{\tau e} z_{u_R} + \frac{Q^2 + m_Z^2}{m_Z^2} \left[C_{Lu}\right]_{\tau e u_i u_i}   \right|^2   \nonumber \\ & & +  \sum_{j \neq i }\left|   \frac{Q^2 + m_Z^2}{ m_Z^2} 
\left[C^{}_{Lu}\right]_{\tau e u_j u_i}    \right|^2 \bigg\}  \nn \\
\hat\sigma^{u_i}_{RL} &=& F_Z  (1-y)^2  \biggl\{  \left|  \left[ c_{e \varphi}\right]_{\tau e} z_{u_L}  + \frac{Q^2 + m_Z^2}{m_Z^2} \left[C_{Q e}\right]_{\tau e u_i u_i}  \right|^2  + \sum_{j \neq i} \left|   \frac{Q^2 + m_Z^2}{ m_Z^2} 
 \left[C_{Q e}\right]_{\tau e u_j u_i}   \right|^2 \biggr\}, \nonumber
\end{eqnarray} 
where $u_i = \{u,\, c\}$,  and $\left[C_{Q e}\right]_{u_j u_i}$ includes factors of the CKM matrix as in Eq. \eqref{CQe}.
The partonic cross sections for $\bar{u},d,\bar{d}$-type (anti)quarks are given in Appendix \ref{partonic}. 

The $Z$ couplings $c^{(1)}_{L \varphi} + c^{(3)}_{L \varphi}$ and $c_{e\varphi}$ induce contributions that are diagonal in quark flavor, and, as seen in the relevant terms of \eq{up1}, can interfere with the quark-flavor-diagonal components of the semileptonic operators in \eq{fourfermion}. Note the $Z$ coupling contributions and four-fermion operator contributions have different dependences on $Q^2$, which, as we will discuss in Section \ref{efficiency}, will lead to different transverse momentum and rapidity distributions for the  $\tau$ decay products, and thus to different efficiencies.

\tocless\subsubsection{Dipole operators}

In the case of dipole operators given by \eqs{dipole}{gZdipole}, we factor out the prefactor
\begin{eqnarray}\label{Fdip}
F_{\rm dip} = \frac{4 Q^2}{v^2}.
\end{eqnarray}
For the $\tau e$ coefficient of the dipole operators, the electron is right-handed, and the $u$-type quark contribution to the cross section is
\begin{eqnarray}
\label{eq:udipole}
\hat \sigma_{LL}^{u}&=&\hat \sigma_{LR}^{u} = 0, \nn\\
\hat \sigma_{RR}^{u}&=& F_{\rm dip} (1-y) \biggl \lvert \left[\Gamma_\gamma^e\right]_{\tau e} Q_u  + \frac{z_{u_R}}{c_w^2 s_w^2} \frac{Q^2}{(Q^2 + m_Z^2)} \left[ \Gamma_Z^e \right]_{\tau e}  \biggr\rvert^2, \nn \\
\hat \sigma_{RL}^{u}&=& F_{\rm dip} (1-y) \biggl\lvert \left[\Gamma_\gamma^e\right]_{\tau e} Q_u  + \frac{z_{u_L}}{c_w^2 s_w^2} \frac{Q^2}{(Q^2 + m_Z^2)} \left[ \Gamma_Z^e \right]_{\tau e}  \biggr\rvert^2. 
\end{eqnarray}
The $d$-type quark contribution is obtained by the following replacements
\beq
\label{eq:dipoleud}
z_{u_L}\to z_{d_L},\quad z_{u_R}\to z_{d_R}, \quad Q_u\to Q_d,
\eeq
and, since the helicity of massless antiparticles is opposite to their chirality, the antiquark contributions can be obtained from the quarks by the replacement
\beq
\hsigma_{RR}^{\bar u,\bar d}\leftrightarrow\hsigma_{RL}^{u,d}.
\eeq
The expressions for $\bigl[\Gamma^{e}_{\gamma, Z}\bigr]_{e \tau}$ are identical, upon replacing the lepton helicity label $R \rightarrow L$. For completeness, we give the expressions in Appendix \ref{partonic}.
Notice that for the photon dipole $\Gamma^{e}_\gamma$, the power of $Q^2$ in $F_{\rm dip}$ is not sufficient to cancel the divergence at $Q^2 \rightarrow 0$ seen in \eq{Bornfactor}.
The terms proportional to $|\Gamma^{e}_Z|^2$ and to the $\Gamma^{e}_\gamma$--$\Gamma^{e}_Z$ interference are, on the other hand, finite at $Q^2 \rightarrow 0$.

\leavevmode

\tocless\subsubsection{Higgs, scalar and tensor four-fermion operators}

\noindent 
In the case of Higgs Yukawa operators \eq{hYukawa} and scalar and tensor operators in the last line of \eq{fourfermion}, we define the prefactor
\begin{eqnarray}\label{FS}
F_{S} = \frac{Q^4}{4 c_w^4 s_w^4 m_Z^4}.
\end{eqnarray}
Starting from the $\tau e$ component of the operator coefficients, the $e$ is right-handed. 
The partonic cross sections initiated by $u$-type quarks receive contributions from both scalar and tensor operators. In both, the $u$ is right-handed. In addition, Higgs exchanges contribute to this channel, and the Higgs couples to both right- and left-handed $u$ quarks. The total contributions of all these operators to the partonic cross sections are:
\begin{eqnarray}
\label{eq:HiggsST}
\hsigma_{LL}^{u}&=& \hsigma_{LR}^{u} =0, \nn \\
\hsigma_{RR}^{u_i}&=& F_S\, y^2 \bigg\{ \left|\left[C^{(1)}_{LeQu}\right]_{\tau e u_i u_i} + 4 \left( 1- \frac{2}{y}  \right) \left[C^{(3)}_{LeQu}\right]_{\tau e  u_i u_i} 
+  \frac{Y_{u_i}}{2} \left[  Y_e^\prime\right]_{\tau e} \frac{v^2}{m_H^2 + Q^2} \right|^2 \nn \\
& &+  \sum_{j \neq i}  \left|\left[C^{(1)}_{LeQu}\right]_{\tau e u_j u_i} + 4 \left( 1- \frac{2}{y}  \right) \left[C^{(3)}_{LeQu}\right]_{\tau e  u_j u_i} \right|^2 \bigg\} \nn \\
\hsigma_{RL}^{u_i}&=& F_S\, y^2\left| \frac{Y_{u_i}}{2} \left[  Y_e^\prime\right]_{\tau e} \frac{v^2}{m_H^2 + Q^2} \right|^2.
\end{eqnarray}
The partonic cross sections for the $\bar u$-type antiquarks are given in Appendix \ref{partonic}.
For $d$-type quarks, the main difference is the absence of a tensor operator, and the chirality of the incoming $d$ quark, which is now left-handed. The relevant expressions are given in Appendix  \ref{partonic}.
For $e\tau$ operators, the results are the same, but the electron is left-handed.

\leavevmode

\tocless\subsubsection{Gluonic operators}

\noindent
We finally consider gluonic operators. These come from two sources, first,  through \eq{hgg}, which talks to $e\tau$ through the Yukawa interaction \eq{hYukawa}; and second, from \eq{Lgluonic}, induced by scalar and tensor operators below the top threshold.
The left-handed and right-handed gluon will give same results,
\begin{align}
\hsigma_{LL}^{g}=\hsigma_{LR}^g&=F_G y^2\biggl\{ \left| \left[C_{GG}\right]_{e \tau} - \frac{1}{3} \frac{v^2}{Q^2 + m_H^2} \left[ Y^\prime_{e \tau}\right]  \right|^2
+ \left| \left[C_{G\tilde G}\right]_{e \tau}\right|^2 \biggr\},\nn\\
\hsigma_{RR}^{g}=\hsigma_{RL}^g&=F_G y^2\biggl\{\left| \left[C_{GG}\right]_{\tau e} - \frac{1}{3} \frac{v^2}{Q^2 + m_H^2} \left\{Y^\prime_{\tau e}\right]  \right|^2
+ \left| \left[C_{G\tilde G}\right]_{\tau e}\right|^2 \biggr\}
\end{align}
where here the factor $F_G$ is
\beq\label{FG}
F_G = \frac{1}{4}  \left(\frac{\alpha_s}{4\pi}\right)^2  \frac{1}{c_w^4 s_w^4} \frac{Q^6}{m_Z^4 v^2} \,.
\eeq

\subsection{Numerical results for partonic EIC cross sections and sensitivity}
\label{ssec:partonic_numerical}

\begin{table}
\centering
\small
\begin{tabular}{||c||c | c | c || c|| c | c | c||}
\hline
\small $\sqrt{S}$ &\small 63 GeV     & \small 100 GeV &\small 141 GeV &\small$ \sqrt{S} $&\small  63 GeV     &  \small 100 GeV & \small 141 GeV \\
\hline\hline
 $\tau_L$  & $\sigma_1$ (pb)     &  $\sigma_{2}$ (pb) & $\sigma_3$ (pb) & $\tau_R$ & $\sigma_1$ (pb) & $\sigma_2$ (pb) & $\sigma_3$ (pb)\\
\hline 
      $c_{L \varphi}^{(1)} + c_{L \varphi}^{(3)}$    & $1.86(4)$ & $4.2(1)$ & $ 7.6(2) $  & $c_{e \varphi}$  & $1.30(3)$ & $3.1(1)$ & $5.6(2)$  \\
\hline
\hline
 $\tau_L$  & $\sigma_1$ (pb)     &  $\sigma_{2}$ (pb) & $\sigma_3$ (pb) & $\tau_L$ & $\sigma_1$ (pb) & $\sigma_2$ (pb) & $\sigma_3$ (pb)\\
 \hline
 $(C_{L Q,\, U}^{} )_{uu}$               & 8.0(4)   & 20(1)    & 38(2)       &  $(C_{Lu})_{uu}$   & 3.9(2)    & 9.5(4)   & 19(1) \\
 $(C_{L Q, U}^{} )_{cu}$                 & 7.8(4)   & 20(1)    & 37(2)       &  $(C_{Lu})_{cu}$   & 3.1(3)    & 7.8(7)   & 15(1) \\
 $(C_{L Q,\, U}^{} )_{uc}$               & 1.0(2)   & 2.5(6)   & 5.2(1.1)    &  $(C_{Lu})_{uc}$   & 1.4(2)    & 3.7(4)   & 7.5(8)                     \\   
 $(C_{L Q, U}^{} )_{cc}$                 & 0.7(3)   & 1.9(7)   & 4.0(1.4)    &  $(C_{Lu})_{cc}$   & 0.7(3)    & 1.9(7)   & 4.0(1.4)\\
 $(C_{L Q,\, D }^{})_{dd}$               & 4.4(2)   & 10.8(4)  & 21(1)       &  $(C_{Ld})_{dd}$   & 2.8(1)    & 7.1(3)   & 14(1)                  \\
 $(C_{L Q,\, D}^{} )_{sd}$               & 3.9(2)   & 9.7(4)   & 19(1)       &  $(C_{Ld})_{sd}$   & 1.6(2)    & 3.9(6)   &  7.8(1.2)                 \\
 $(C_{L Q, D}^{} )_{bd}$                 & 3.9(1)   & 9.5(3)   & 19(1)       &  $(C_{Ld})_{bd}$   & 1.4(1)   & 3.4(1)   &  7.0(3)                 \\
 $(C_{L Q, D}^{} )_{ds}$                 & 0.8(3)   & 2.0(8)   & 4.1(1.5)    &  $(C_{Ld})_{ds}$   & 1.6(2)    & 4.1(4)   &  8.3(9)                  \\
 $(C_{L Q, D}^{} )_{ss}$                 & 0.35(31) & 1.0(8)   & 2.0(1.7)    &  $(C_{Ld})_{ss}$   & 0.33(27)  & 0.9(7)   &  1.9(1.5)                \\
 $(C_{L Q,\, D}^{} )_{bs}$               & 0.28(26) & 0.8(7)   & 1.7(1.4)    &  $(C_{Ld})_{bs}$   & 0.14(10)   & 0.5(3)   &  1.1(6)                 \\
 $(C_{L Q,\, D }^{} )_{db}$              & 0.57(7)  & 1.6(2)   & 3.2(3)      &  $(C_{Ld})_{db}$   & 1.6(1)    & 4.0(2)   &  8.0(5)                 \\
 $(C_{L Q, D}^{} )_{sb}$                 & 0.13(7)  & 0.4(2)   & 1.1(5)      &  $(C_{Ld})_{sb}$   & 0.26(19)  & 0.7(5)   &  1.6(1.1)                 \\
 $(C_{L Q,\, D}^{} )_{b b}$              & 0.07(4)  & 0.3(2)   & 0.8(2)      &  $(C_{Ld})_{bb}$   & 0.07(6)   & 0.3(1)   & 0.8(0.5)                 \\
 \hline
 \hline 
 $\tau_R$  & $\sigma_1$ (pb)     &  $\sigma_{2}$ (pb) & $\sigma_3$ (pb) & $\tau_R$ & $\sigma_1$ (pb) & $\sigma_2$ (pb) & $\sigma_3$ (pb)\\
 \hline
  $(C_{Qe}^{})_ {dd}$                    & 7.5(3)  & 19(1)   & 37(2)    & $(C_{Qe}^{} )_{ds}$   & 5.7(5)   & 14(1)  & 29(2) \\
  $(C_{Qe}^{} )_{sd}$                    & 4.1(2)  & 10.3(5) & 21(1)    & $(C_{Qe}^{} )_{ss}$   & 2.3(2)   & 5.8(5) & 12(1)\\
  $(C_{Qe}^{} )_{bd}$                    & 1.4(6) & 3.7(1)  & 7.4(3)   & $(C_{Qe}^{} )_{bs}$   & 0.20(11) & 0.6(3) & 1.4(7) \\
  $(C_{Qe}^{} )_{db}$                    & 1.7(1)  & 4.3(3)  & 8.7(5)   & $(C_{Qe}^{} )_{sb}$   & 0.32(19) & 0.9(5) & 2.0(1.1) \\ 
  $(C_{Qe}^{} )_{bb}$                    & 0.07(6) & 0.3(1) & 0.8(5)  & & & & \\ 
 
 \hline
 \end{tabular}
\caption{Numerical coefficients $a_{iJ}$ that control the 
cross sections  $\sigma_i = a_{i J} |C_J|^2$ 
for the CLFV process $e p \rightarrow \tau X$  induced by CLFV $Z$ couplings, vector and axial four-fermion operators.
The subscript $i = \{1,2,3\}$ denotes each of the three benchmark points discussed in the text,  at $\sqrt{S}=63,100,141$ GeV, respectively,
while $J$ is the operator label. Here we omit interference terms between $Z$ couplings and four-fermion operators.
The labels $\tau_{L,R}$ denote the polarization of the $\tau$ lepton emitted by the effective operators. 
The cross section is computed with the  \texttt{NNPDF31\_lo\_as\_0118} PDF set \cite{Ball:2017nwa}. 
The error estimates includes PDF and scale uncertainties. 
Terms quadratic in $C_{eu}$ and $C_{ed}$ are identical to $C_{LQ,\, U}$ and $C_{LQ, D}$, respectively, and are not given explictly.  }
\vspace{2em}
 \label{CrossSection:AV}
\end{table}

\begin{table}
\centering
\small
\begin{tabular}{||c||c | c | c || c|| c | c | c||}
\hline
\small $\sqrt{S}$ &\small 63 GeV     & \small 100 GeV &\small 141 GeV &\small$ \sqrt{S} $&\small  63 GeV     &  \small 100 GeV & \small 141 GeV \\
\hline\hline
 $\tau_L$  & $\sigma_1$ (pb)    &  $\sigma_{2}$ (pb) & $\sigma_3$ (pb) & $\tau_L$ & $\sigma_1$ (pb) & $\sigma_2$ (pb) & $\sigma_3$ (pb)\\
\hline 
      $Y^\prime_{\tau e}$ $(10^{-4})$    &   0.22(2)        &    $0.73(5)$      &    $1.7(1) $         & $C_{GG}$ ($10^{-5}$)  & 0.103(5)  & 0.32(1) &  1.77(7)  \\
      $\Gamma_\gamma^e$                  &   26(2)          &    35(3)          &    $43.6(4.5)$       & $\Gamma_Z^{e}$        & 0.0174(3) & 0.088(2) &  0.276(5) \\
\hline
\hline
 $\tau_L$  & $\sigma_1$ (pb)     &  $\sigma_{2}$ (pb) & $\sigma_3$ (pb) & $\tau_L$ & $\sigma_1$ (pb) & $\sigma_2$ (pb) & $\sigma_3$ (pb)\\
 \hline

 $(C^{(1)}_{L e Q u} )_{uu}$             & 0.72(3)  & 1.78(6)  &   3.5(1)         &  $(C^{(3)}_{L e Q u} )_{uu}$   & 83(3)  & 203(7) & 399(15) \\
 $(C^{(1)}_{L e Q u} )_{cu}$             & 0.67(2)  & 1.63(7)  &   3.2(1)         &  $(C^{(3)}_{L e Q u} )_{cu}$   & 76(3)  & 186(5) & 367(13) \\
 $(C^{(1)}_{L e Q u} )_{uc}$             & 0.16(2)  & 0.40(6)  &   0.8(1)         &  $(C^{(3)}_{L e Q u} )_{uc}$   & 17(3)  &  43(7)  &  90(12)                             \\   
 $(C^{(1)}_{L e Q u} )_{cc}$             & 0.09(3)  & 0.25(8)  &   0.5(2)         &  $(C^{(3)}_{L e Q u} )_{cc}$   & 10(4)  &  26(9) &  55(19)\\
 $(C^{(1)}_{L e d Q } )_ {dd}$           & 0.44(1)  & 1.10(3)  &   2.2(1)         &  $(C^{(1)}_{L e d Q } )_{ds}$  & 0.15(2)  &  0.39(5) & 0.8(1) \\
 $(C^{(1)}_{L e d Q } )_{sd}$            & 0.34(2)  & 0.84(5)  &   1.7(1)         &  $(C^{(1)}_{L e d Q } )_{ss}$  & 0.046(38)  &  0.12(9) & 0.26(21)\\
 $(C^{(1)}_{L e d Q } )_{bd}$            & 0.32(1)  & 0.80(3)  &   1.6(1)         &  $(C^{(1)}_{L e d Q } )_{bs}$  & 0.031(23)  &  0.09(6) & 0.19(11) \\
 $(C^{(1)}_{L e d Q } )_{db}$            & 0.14(8)  & 0.35(6)  &   0.7(1)         &  $(C^{(1)}_{L e d Q } )_{sb}$   & 0.028(9)  &  0.09(5) & 0.19(9) \\ 
 $(C^{(1)}_{L e d Q } )_{bb}$            & 0.013(1) & 0.05(2) &   0.13(5)         & & & & \\ 
 
 \hline
 \end{tabular}
\caption{
Numerical coefficients $a_{iJ}$ that control the 
cross sections  $\sigma_i = a_{i J} |C_J|^2$ 
for the CLFV process $e p \rightarrow \tau X$, 
induced by CLFV Higgs couplings, photon and $Z$ dipoles and scalar and tensor four-fermion operators. 
The subscript $i = \{1,2,3\}$ denotes each of the three benchmark points discussed in the text,  at $\sqrt{S}=63,100,141$ GeV, respectively,
while $J$ is the operator label. Here we omit interference terms between photon and $Z$ dipoles and between Higgs couplings, scalar and tensor four-fermion operators.
The cross section is computed with the  \texttt{NNPDF31\_lo\_as\_0118} PDF set. 
The error estimates includes PDF and scale uncertainties. 
We give here the cross section for the $\tau e$ component of the operators, in which the $\tau$ lepton is left-handed. The results are identical for the $e\tau$ components, with the  difference that a right-handed $\tau$ is emitted.
}\label{CrossSection:ST}
\end{table}

To get an idea of the number of CLFV events that can be produced at the EIC, we calculate here the total DIS cross section
from different SMEFT operators, obtained by integrating \eq{sigma} over $x$ and $y$ in the range $x,y \in [0,1]$.
To illustrate the $S$ dependence of the SMEFT cross sections, we use a few benchmark points,
\begin{enumerate}
\item $E_e=20~{\rm GeV}$,  $E_p=50~{\rm GeV}$, $\sqrt{S} = 63$ GeV,
\item $E_e=10~{\rm GeV}$,  $E_p=250~{\rm GeV}$, $\sqrt{S} =100$ GeV,
\item $E_e=20~{\rm GeV}$,  $E_p=250~{\rm GeV}$, $\sqrt{S} = 141$ GeV.
\end{enumerate}
These are typical beam energies of EIC~\cite{Accardi:2012qut,Aschenauer:2017jsk}, with the last point corresponding to the maximum $\sqrt{S}$ the EIC plans to achieve.
The renormalization and factorization scales are chosen as $\mu_F=\mu_R=Q$, and we assess the scale uncertainty by varying $\mu_F = \mu_R$ between $Q/2$ and $2 Q$.
We use the  \texttt{NNPDF31\_lo\_as\_0118} PDF set \cite{Ball:2017nwa}, and we evaluate the PDF errors by calculating the cross section for the 100 members of this PDF set.  
Furthermore, we have compared the results of our numerical calculations with those obtained using \texttt{MadGraph5}~\cite{Alwall:2014hca} and found excellent agreement. 
We show the cross section from various CLFV operators with $\lambda_e=0$ in Tables~\ref{CrossSection:AV} and \ref{CrossSection:ST}.
It is straightforward to include the polarization of the electron beam, see \eq{DISpolarized_cs}.

The cross section for SMEFT operators grows as $\sqrt{S}$ increases, with more marked increase for the dimension-7 gluonic operators.
The CLFV $Z$ couplings and four-fermion operators induce cross sections that are comparable to the $Z$ boson contributions to standard DIS,
multiplied by the square of the operator coefficients, scaling as $(v/\Lambda)^4$. Operators with a sea quark in the initial state are suppressed by the PDF of the $s$, $c$ or $b$ quark.
The suppression is not too severe, but notice that the PDF and scale errors become sizable, especially in the case of operators dominated by the $s$ and $b$ contribution. For these operators, it will be important to extend the analysis beyond leading order. We stress that we use the PDF and scale errors only as a rough estimate of the theoretical error, a more robust assessment requires extending the calculation to next-to-leading order (NLO).

The scalar and tensor four-fermion operators induce contributions of similar size as vector operators, with some enhancement in the case of the $C^{(3)}_{LeQu}$.
The photon dipole $\Gamma^e_\gamma$ gives a large contribution to the cross section, but, as we will discuss in Section \ref{efficiency}, the
divergence at $Q^2 \rightarrow 0$ implies that 
the shape of the $p_T$ distributions of the $\tau$ decay products is hardly distinguishable from the SM backgrounds. 
The Yukawa operator $Y^\prime_e$ contributes to DIS via the Higgs coupling to light quarks and the effective gluon-Higgs coupling 
induced by top loops. At the EIC, the dominant contribution arises from the Higgs coupling to $b$ quarks. The cross section is however too small to provide 
bounds on $Y^\prime_e$ that are competitive with the LHC or low-energy probes.

We can use the cross sections in Tables \ref{CrossSection:AV} and \ref{CrossSection:ST} to provide a first estimate of the EIC sensitivity to CLFV operators,
as a function of a selection efficiency $\epsilon_{n_b}$, defined as the number of signal events that pass the cuts required to reduce the SM background to $n_b$ events.
We consider separately the three decays channel $\tau^- \rightarrow e^- \bar{\nu}_e \nu_\tau$, $\tau^- \rightarrow \mu^- \bar{\nu}_\mu \nu_\tau$
and $\tau^- \rightarrow X_h \nu_\tau$, where $X_h$ denotes an hadronic final state. The branching ratios in these channels are \cite{Zyla:2020zbs}
\begin{eqnarray}
& &\textrm{BR} (\tau^- \rightarrow  e^-   \bar\nu_e \nu_\tau  ) = 17.82 \pm 0.04 \%, \\ 
& &\textrm{BR} (\tau^- \rightarrow  \mu^- \bar\nu_\mu \nu_\tau  ) = 17.39 \pm 0.04 \%, \\ 
& & \textrm{BR} (\tau^- \rightarrow  X_h \nu_\tau  ) = 64.8 \%. 
\end{eqnarray}
Assuming the backgrounds are known with negligible errors, we can estimate the 
upper limit on the CLFV coefficients at the  $1-\alpha$ credibility level, 
when $n$ events have been observed and $n_b$ events are expected, 
by solving the equation~\cite{Zyla:2020zbs}
\begin{equation}
1- \alpha =   1 - \frac{\Gamma\left(1 + n, n_b + n_s\right)}{\Gamma\left(1 + n, n_b\right)}, 
\label{eq:CL}
\end{equation}
where $n_s$ is a function of the SMEFT coefficient, of the decay channel and of the selection efficiency
\begin{equation}\label{poissonmean}
n_s(C_i, \epsilon_{n_b}, X_j ) = \mathcal L \times ( \sigma \epsilon_{n_b} |C_i|^2  ) \times \textrm{BR}(\tau \rightarrow X_j \nu_\tau ),
\end{equation}
with $\mathcal L$ the integrated luminosity. 
For the cross section $\sigma$ we use the central values given in Tables \ref{CrossSection:AV} and \ref{CrossSection:ST}.
We however notice that processes initated by sea quarks have large theoretical uncertainties, which can significantly shift the upper bound on the SMEFT coefficients.

\begin{table}
\center
\begin{tabular}{||c || c | c || c |c  || c |c ||}
\hline
              &  \multicolumn{2}{|c||}{$\tau \rightarrow e \bar\nu_e \nu_\tau$ or $\tau \rightarrow \mu \bar\nu_\mu \nu_\tau$  }   &   \multicolumn{2}{|c||}{$\tau \rightarrow X_h \nu_\tau$}  \\ \hline
              & $n_b= 0$ & $n_b=100$ &  $n_b= 0$ & $n_b=100$  \\
\hline 
 $\left|(c_{L \varphi}^{(1)} + c_{L \varphi}^{(3)} ) \sqrt{\epsilon_{n_b}}\right|$        & $4.2 \cdot 10^{-3}$    &  $1.2 \cdot 10^{-2}$  & $2.2 \cdot 10^{-3}$ & $6.1 \cdot 10^{-3}$  \\
  $|c_{e \varphi} \sqrt{\epsilon_{n_b}} |    $                                            & $4.9 \cdot 10^{-3}$    &  $1.4 \cdot 10^{-2}$  & $2.6 \cdot 10^{-3}$ & $7.0 \cdot 10^{-3}$ \\                        
\hline
 $|(C_{L Q,\, U}^{} )_{uu}  \sqrt{\epsilon_{n_b}}  |$        & $1.9 \cdot 10^{-3}$  &  $5.2 \cdot 10^{-3}$  & $1.0 \cdot 10^{-3}$ & $2.7 \cdot 10^{-3}$ \\
 $|(C_{L Q, U}^{} )_{cu}    \sqrt{\epsilon_{n_b}}  |$        & $1.9 \cdot 10^{-3}$  &  $5.2 \cdot 10^{-3}$  & $1.0 \cdot 10^{-3}$ & $2.8 \cdot 10^{-3}$ \\
 $|(C_{L Q,\, U}^{} )_{uc}  \sqrt{\epsilon_{n_b}}  |$        & $5.1 \cdot 10^{-3}$  &  $1.4 \cdot 10^{-2}$  & $2.6 \cdot 10^{-3}$ & $7.3 \cdot 10^{-3}$\\
 $|(C_{L Q, U}^{} )_{cc}    \sqrt{\epsilon_{n_b}}  |$        & $5.8 \cdot 10^{-3}$  &  $1.6 \cdot 10^{-2}$  & $3.0 \cdot 10^{-3}$ & $8.3 \cdot 10^{-3}$ \\
 $|(C_{L Q,\, D }^{})_{dd}  \sqrt{\epsilon_{n_b}}  |$        & $2.5 \cdot 10^{-3}$  &  $6.9 \cdot 10^{-3}$  & $1.3 \cdot 10^{-3}$ & $3.6 \cdot 10^{-3}$\\ 
 $|(C_{L Q,\, D}^{} )_{sd}  \sqrt{\epsilon_{n_b}}  |$        & $2.6 \cdot 10^{-3}$  &  $7.3 \cdot 10^{-3}$  & $1.4 \cdot 10^{-3}$ & $3.8 \cdot 10^{-3}$\\ 
 $|(C_{L Q, D}^{} )_{bd}    \sqrt{\epsilon_{n_b}}  |$        & $2.6 \cdot 10^{-3}$  &  $7.4 \cdot 10^{-3}$  & $1.4 \cdot 10^{-3}$ & $3.8 \cdot 10^{-3}$\\ 
 $|(C_{L Q, D}^{} )_{ds}    \sqrt{\epsilon_{n_b}}  |$        & $5.7 \cdot 10^{-3}$  &  $1.6 \cdot 10^{-2}$  & $2.9 \cdot 10^{-3}$ & $8.1 \cdot 10^{-3}$\\ 
 $|(C_{L Q, D}^{} )_{ss}    \sqrt{\epsilon_{n_b}}  |$        & $8.2 \cdot 10^{-3}$  &  $2.3 \cdot 10^{-2}$  & $4.2 \cdot 10^{-3}$ & $1.2 \cdot 10^{-2}$\\  
 $|(C_{L Q,\, D}^{} )_{bs}  \sqrt{\epsilon_{n_b}}  |$        & $8.9 \cdot 10^{-3}$  &  $2.5 \cdot 10^{-2}$  & $4.6 \cdot 10^{-3}$ & $1.3 \cdot 10^{-2}$\\ 
 $|(C_{L Q,\, D }^{} )_{db} \sqrt{\epsilon_{n_b}}  |$        & $6.4 \cdot 10^{-3}$  &  $1.8 \cdot 10^{-2}$  & $3.3 \cdot 10^{-3}$ & $9.3 \cdot 10^{-3}$\\   
 $|(C_{L Q, D}^{} )_{sb}    \sqrt{\epsilon_{n_b}}  |$        & $1.1 \cdot 10^{-2}$  &  $3.1 \cdot 10^{-2}$  & $5.8 \cdot 10^{-3}$ & $1.6 \cdot 10^{-2}$\\   
 $|(C_{L Q,\, D}^{} )_{bb}  \sqrt{\epsilon_{n_b}}  |$        & $1.3 \cdot 10^{-2}$  &  $3.7 \cdot 10^{-2}$  & $6.9 \cdot 10^{-3}$ & $1.9 \cdot 10^{-2}$\\   
 \hline
 \end{tabular}
\caption{EIC sensitivity to CLFV $Z$ couplings and vector four-fermion operators with left-handed quark and leptons, from the $\tau$ electronic, muonic and hadronic decay channels. We assume $\sqrt{S} = 141$ GeV and $\mathcal L = 100$ fb$^{-1}$.
The two sets of  90\% CL bounds are obtained assuming that the EIC will observe $n = n_b = 0$ and $n = n_b = 100$ events.  
$\epsilon_0$ and $\epsilon_{100}$ accounts for the selection cuts that ensure 0 and 100 background events, respectively, and are functions of the decay channel and of the operator structure.
Bounds on the right-handed operators  $C_{eu}$ and $C_{ed}$ are the same as $C_{LQ, U}$ and $C_{LQ, D}$.
}\label{EIC_sens_LL}
\end{table}

\begin{table}[t]
\center
\begin{tabular}{||c || c | c || c |c  || c |c ||}
\hline
              &  \multicolumn{2}{|c||}{$\tau \rightarrow e \bar\nu_e \nu_\tau$ or $\tau \rightarrow \mu \bar\nu_\mu \nu_\tau$  }   &   \multicolumn{2}{|c||}{$\tau \rightarrow X_h \nu_\tau$}  \\ \hline
              & $n_b= 0$ & $n_b=100$ &  $n_b= 0$ & $n_b=100$  \\
\hline
 $|(C_{L u})_{uu}  \sqrt{\epsilon_{n_b}}  |$        & $2.6 \cdot 10^{-3}$  &  $7.4 \cdot 10^{-3}$  & $1.4 \cdot 10^{-3}$ & $3.8 \cdot 10^{-3}$ \\
 $|(C_{L u})_{cu}  \sqrt{\epsilon_{n_b}}  |$        & $2.9 \cdot 10^{-3}$  &  $8.2 \cdot 10^{-3}$  & $1.5 \cdot 10^{-3}$ & $4.2 \cdot 10^{-3}$ \\
 $|(C_{L u})_{uc}  \sqrt{\epsilon_{n_b}}  |$        & $4.2 \cdot 10^{-3}$  &  $1.2 \cdot 10^{-2}$  & $2.2 \cdot 10^{-3}$ & $6.1 \cdot 10^{-3}$\\
 $|(C_{L u})_{cc}  \sqrt{\epsilon_{n_b}}  |$        & $5.8 \cdot 10^{-3}$  &  $1.6 \cdot 10^{-2}$  & $3.0 \cdot 10^{-3}$ & $8.3 \cdot 10^{-3}$ \\
 $|(C_{L d})_{dd}  \sqrt{\epsilon_{n_b}}  |$        & $3.1 \cdot 10^{-3}$  &  $8.5 \cdot 10^{-3}$  & $1.6 \cdot 10^{-3}$ & $4.4 \cdot 10^{-3}$\\ 
 $|(C_{L d})_{sd}  \sqrt{\epsilon_{n_b}}  |$        & $4.1 \cdot 10^{-3}$  &  $1.2 \cdot 10^{-2}$  & $2.1 \cdot 10^{-3}$ & $5.9 \cdot 10^{-3}$\\ 
 $|(C_{L d})_{bd}  \sqrt{\epsilon_{n_b}}  |$        & $4.4 \cdot 10^{-3}$  &  $1.2 \cdot 10^{-2}$  & $2.3 \cdot 10^{-3}$ & $6.3 \cdot 10^{-3}$\\ 
 $|(C_{L d})_{ds}  \sqrt{\epsilon_{n_b}}  |$        & $4.0 \cdot 10^{-3}$  &  $1.1 \cdot 10^{-2}$  & $2.1 \cdot 10^{-3}$ & $5.8 \cdot 10^{-3}$\\ 
 $|(C_{L d})_{ss}  \sqrt{\epsilon_{n_b}}  |$        & $8.3 \cdot 10^{-3}$  &  $2.3 \cdot 10^{-2}$  & $4.3 \cdot 10^{-3}$ & $1.2 \cdot 10^{-2}$\\  
 $|(C_{L d})_{bs}  \sqrt{\epsilon_{n_b}}  |$        & $1.1 \cdot 10^{-2}$  &  $3.1 \cdot 10^{-2}$  & $5.7 \cdot 10^{-3}$ & $1.6 \cdot 10^{-2}$\\ 
 $|(C_{L d})_{db}  \sqrt{\epsilon_{n_b}}  |$        & $4.1 \cdot 10^{-3}$  &  $1.1 \cdot 10^{-2}$  & $2.1 \cdot 10^{-3}$ & $5.9 \cdot 10^{-3}$\\   
 $|(C_{L d})_{sb}  \sqrt{\epsilon_{n_b}}  |$        & $9.1 \cdot 10^{-3}$  &  $2.5 \cdot 10^{-2}$  & $4.7 \cdot 10^{-3}$ & $1.3 \cdot 10^{-2}$\\   
 $|(C_{L d})_{bb}  \sqrt{\epsilon_{n_b}}  |$        & $1.3 \cdot 10^{-2}$  &  $3.7 \cdot 10^{-2}$  & $6.8 \cdot 10^{-3}$ & $1.9 \cdot 10^{-2}$\\   
 \hline
 $|(C_{Qe})_{dd}  \sqrt{\epsilon_{n_b}}  |$        & $1.9 \cdot 10^{-3}$  &  $5.3 \cdot 10^{-3}$  & $1.0 \cdot 10^{-3}$ & $2.7 \cdot 10^{-3}$\\ 
 $|(C_{Qe})_{sd}  \sqrt{\epsilon_{n_b}}  |$        & $2.5 \cdot 10^{-3}$  &  $7.0 \cdot 10^{-3}$  & $1.3 \cdot 10^{-3}$ & $3.6 \cdot 10^{-3}$\\ 
 $|(C_{Qe})_{bd}  \sqrt{\epsilon_{n_b}}  |$        & $4.2 \cdot 10^{-3}$  &  $1.2 \cdot 10^{-2}$  & $2.2 \cdot 10^{-3}$ & $6.1 \cdot 10^{-3}$\\ 
 $|(C_{Qe})_{ds}  \sqrt{\epsilon_{n_b}}  |$        & $2.2 \cdot 10^{-3}$  &  $6.0 \cdot 10^{-3}$  & $1.1 \cdot 10^{-3}$ & $3.1 \cdot 10^{-3}$\\ 
 $|(C_{Qe})_{ss}  \sqrt{\epsilon_{n_b}}  |$        & $3.3 \cdot 10^{-3}$  &  $9.3 \cdot 10^{-3}$  & $1.7 \cdot 10^{-3}$ & $4.8 \cdot 10^{-3}$\\  
 $|(C_{Qe})_{bs}  \sqrt{\epsilon_{n_b}}  |$        & $9.7 \cdot 10^{-3}$  &  $2.7 \cdot 10^{-2}$  & $5.0 \cdot 10^{-3}$ & $1.4 \cdot 10^{-2}$\\ 
 $|(C_{Qe})_{db}  \sqrt{\epsilon_{n_b}}  |$        & $3.9 \cdot 10^{-3}$  &  $1.1 \cdot 10^{-2}$  & $2.0 \cdot 10^{-3}$ & $5.6 \cdot 10^{-3}$\\   
 $|(C_{Qe})_{sb}  \sqrt{\epsilon_{n_b}}  |$        & $8.2 \cdot 10^{-3}$  &  $2.3 \cdot 10^{-2}$  & $4.3 \cdot 10^{-3}$ & $1.2 \cdot 10^{-2}$\\   
 $|(C_{Qe})_{bb}  \sqrt{\epsilon_{n_b}}  |$        & $1.3 \cdot 10^{-2}$  &  $3.6 \cdot 10^{-2}$  & $6.8 \cdot 10^{-3}$ & $1.9 \cdot 10^{-2}$\\   
 \hline
 \end{tabular}
\caption{EIC sensitivity to CLFV four-fermion operators with left(right)-handed leptons and right(left)-handed quarks, from the $\tau$ electronic, muonic and hadronic decay channels. 
The two sets of  90\% CL bounds are obtained assuming that the EIC will observe $n = n_b = 0$ and $n = n_b = 100$ events.  
$\epsilon_0$ and $\epsilon_{100}$ accounts for the selection cuts that ensure 0 and 100 background events, respectively, and are functions of the decay channel and of the operator structure.
}\label{EIC_sens_LR}
\end{table}

\begin{table}[t]
\center
\begin{tabular}{||c || c | c || c |c  || c |c ||}
\hline
              &  \multicolumn{2}{|c||}{$\tau \rightarrow e \bar\nu_e \nu_\tau$ or $\tau \rightarrow \mu \bar\nu_\mu \nu_\tau$  }   &   \multicolumn{2}{|c||}{$\tau \rightarrow X_h \nu_\tau$}  \\ \hline
              & $n_b= 0$ & $n_b=100$ &  $n_b= 0$ & $n_b=100$  \\
\hline
 $\left|\Gamma^e_\gamma \sqrt{\epsilon_{n_b}} \right|$          & $1.8 \cdot 10^{-3}$                &  $5.0 \cdot 10^{-3}$                &  $9.5 \cdot 10^{-4}$            & $2.6 \cdot 10^{-3}$                \\
 $\left|\Gamma^e_Z \sqrt{\epsilon_{n_b}} \right|$               & $2.2 \cdot 10^{-2}$                &  $6.1 \cdot 10^{-2}$                &  $1.1 \cdot 10^{-2}$            & $3.2 \cdot 10^{-2}$                \\
 $\left|Y^\prime_{te} \sqrt{\epsilon_{n_b}} \right|$            & $0.90$                             &  $2.5$                &  $0.47$            & $1.3$                \\           
 $\left|C_{GG} \sqrt{\epsilon_{n_b}} \right|$                   & $2.8$                              &  $7.7$                &  $1.4$            & $4.0$                \\
 \hline
 $\left|(C^{(1)}_{L e Q u})_{uu}  \sqrt{\epsilon_{n_b}}  \right|$        & $6.1 \cdot 10^{-3}$  &  $1.7 \cdot 10^{-2}$  & $3.2 \cdot 10^{-3}$ & $8.8 \cdot 10^{-3}$ \\
 $\left|(C^{(1)}_{L e Q u})_{cu}  \sqrt{\epsilon_{n_b}}  \right|$        & $6.4 \cdot 10^{-3}$  &  $1.8 \cdot 10^{-3}$  & $3.3 \cdot 10^{-3}$ & $9.2 \cdot 10^{-3}$ \\
 $\left|(C^{(1)}_{L e Q u})_{uc}  \sqrt{\epsilon_{n_b}}  \right|$        & $1.3 \cdot 10^{-2}$  &  $3.6 \cdot 10^{-2}$  & $6.6 \cdot 10^{-3}$ & $1.8 \cdot 10^{-2}$\\
 $\left|(C^{(1)}_{L e Q u})_{cc}  \sqrt{\epsilon_{n_b}}  \right|$        & $1.6 \cdot 10^{-2}$  &  $4.4 \cdot 10^{-2}$  & $8.2 \cdot 10^{-3}$ & $2.3 \cdot 10^{-2}$ \\
 $\left|(C_{L e d Q})_{dd}  \sqrt{\epsilon_{n_b}}  \right|$              & $7.8 \cdot 10^{-3}$  &  $2.2 \cdot 10^{-2}$  & $4.0 \cdot 10^{-3}$ & $1.1 \cdot 10^{-2}$\\ 
 $\left|(C_{L e d Q})_{sd}  \sqrt{\epsilon_{n_b}}  \right|$              & $8.9 \cdot 10^{-3}$  &  $2.5 \cdot 10^{-2}$  & $4.6 \cdot 10^{-3}$ & $1.3 \cdot 10^{-2}$\\ 
 $\left|(C_{L e d Q})_{bd}  \sqrt{\epsilon_{n_b}}  \right|$              & $9.1 \cdot 10^{-3}$  &  $2.5 \cdot 10^{-2}$  & $4.7 \cdot 10^{-3}$ & $1.3 \cdot 10^{-2}$\\ 
 $\left|(C_{L e d Q})_{ds}  \sqrt{\epsilon_{n_b}}  \right|$              & $1.3 \cdot 10^{-2}$  &  $3.6 \cdot 10^{-2}$  & $6.7 \cdot 10^{-3}$ & $1.9 \cdot 10^{-2}$\\ 
 $\left|(C_{L e d Q})_{ss}  \sqrt{\epsilon_{n_b}}  \right|$              & $2.2 \cdot 10^{-2}$  &  $6.2 \cdot 10^{-2}$  & $1.2 \cdot 10^{-2}$ & $3.2 \cdot 10^{-2}$\\  
 $\left|(C_{L e d Q})_{bs}  \sqrt{\epsilon_{n_b}}  \right|$              & $2.6 \cdot 10^{-2}$  &  $7.2 \cdot 10^{-2}$  & $1.3 \cdot 10^{-2}$ & $3.7 \cdot 10^{-2}$\\ 
 $\left|(C_{L e d Q})_{db}  \sqrt{\epsilon_{n_b}}  \right|$              & $1.4 \cdot 10^{-2}$  &  $3.8 \cdot 10^{-2}$  & $7.0 \cdot 10^{-3}$ & $2.0 \cdot 10^{-2}$\\   
 $\left|(C_{L e d Q})_{sb}  \sqrt{\epsilon_{n_b}}  \right|$              & $2.6 \cdot 10^{-2}$  &  $7.3 \cdot 10^{-2}$  & $1.4 \cdot 10^{-2}$ & $3.8 \cdot 10^{-2}$\\   
 $\left|(C_{L e d Q})_{bb}  \sqrt{\epsilon_{n_b}}  \right|$              & $3.2 \cdot 10^{-2}$  &  $9.0 \cdot 10^{-2}$  & $1.7 \cdot 10^{-2}$ & $4.7 \cdot 10^{-2}$\\   
 \hline
 $\left|(C^{(3)}_{L e Q u})_{uu}  \sqrt{\epsilon_{n_b}}  \right|$        & $5.8 \cdot 10^{-4}$ &  $1.6 \cdot 10^{-3}$  & $3.0 \cdot 10^{-4}$ & $8.4 \cdot 10^{-4}$ \\
 $\left|(C^{(3)}_{L e Q u})_{cu}  \sqrt{\epsilon_{n_b}}  \right|$        & $6.0 \cdot 10^{-4}$ &  $1.7 \cdot 10^{-3}$  & $3.1 \cdot 10^{-4}$ & $8.7 \cdot 10^{-4}$ \\
 $\left|(C^{(3)}_{L e Q u})_{uc}  \sqrt{\epsilon_{n_b}}  \right|$        & $1.2 \cdot 10^{-3}$  &  $3.4 \cdot 10^{-3}$  & $6.4 \cdot 10^{-4}$ & $1.8 \cdot 10^{-3}$\\
 $\left|(C^{(3)}_{L e Q u})_{cc}  \sqrt{\epsilon_{n_b}}  \right|$        & $1.6 \cdot 10^{-3}$  &  $4.3 \cdot 10^{-3}$  & $8.1 \cdot 10^{-4}$ & $2.2 \cdot 10^{-3}$ \\
 \hline
 \end{tabular}
\caption{EIC sensitivity to CLFV $\gamma$ and $Z$ dipole couplings, Higgs couplings, gluon couplings and scalar and tensor four-fermion operators, from the $\tau$ electronic, muonic and hadronic decay channels. We assume $\sqrt{S} = 141$ GeV and $\mathcal L = 100$ fb$^{-1}$. The two sets of  90\% CL bounds are obtained assuming that the EIC will observe $n = n_b = 0$ and $n = n_b = 100$ events.  
$\epsilon_0$ and $\epsilon_{100}$ accounts for the selection cuts that ensure 0 and 100 background events, respectively, and are functions of the decay channel and of the operator structure.
}\label{EIC_sens_S}
\end{table}

In Tables \ref{EIC_sens_LL},  \ref{EIC_sens_LR} and \ref{EIC_sens_S}  we give the $90\%$ CL bounds on the product of the operator coefficients
and the efficiency $\epsilon$, assuming $n = n_b$ and for two choices, $n_b =0$ and $n_b = 100$. We consider a center of mass energy of $\sqrt{S} = 141$ GeV,
and assume an integrated luminosity of $100$ fb$^{-1}$.
In the case of $Z$ couplings and four-fermion operators with valence quarks, the EIC could reach better than percent sensitivities with $\epsilon_0 \sim 10\%$
in the $\tau$ leptonic or hadronic decay channels. Flavor-changing operators and operators with heavy quarks could also be probed at the few percent level.
In these cases, however, theoretical uncertainties cannot be neglected. Considering, e.g., the extreme case of the operator $[C_{Ld}]_{bb}$, varying the cross section in the uncertainty range given in Table \ref{CrossSection:AV} 
causes the $90\%$ CL upper limit to vary between $5.7 \cdot 10^{-3}$ and   $12 \cdot 10^{-3}$. This large range can be narrowed by including NLO QCD corrections.
We will present a detailed comparison of sensitivities of EIC with  other collider and low-energy probes  in 
Section \ref{sect:summary}.  Here we anticipate that the EIC can be quite competitive for four-fermion semileptonic operators, 
both diagonal and non-diagonal in quark flavor.
We will present an estimate of the selection efficiencies $\epsilon_{n_b}$ in Section \ref{efficiency}.

\section{EIC sensitivity to CLFV}
\label{efficiency}

Next we perform a detailed Monte Carlo simulation to explore the potential of probing CLFV effects via $e^-p\to \tau^-X$ at the EIC with collider energy $E_e=20~{\rm GeV}$ and $E_p=250~{\rm GeV}$ (benchmark point 3 at $\sqrt{S}=141$ GeV in \ssec{partonic_numerical}).  It is straightforward to generalize our analysis  to other collider energies.
The main challenges for the identification of $\tau$ CLFV at the EIC are, first of all, that, differently from muons, the $\tau$ leptons decay very quickly
inside the detector and, secondly, that all decay channels involve missing energy, complicating the reconstruction of the $\tau$ momentum and thus of the DIS variables $x$ and $y$.
It is therefore necessary to identify distinctive features of the signal events, in order to disentangle them from the SM background.
Based on the $\tau$ decay modes, there are three classes of final states: (1) $e^- p \rightarrow \tau^- X \to  e^- \bar{\nu}_e {\nu}_\tau X$; (2) $e^- p \rightarrow \tau^- X \to \mu^-\bar{\nu}_\mu {\nu}_\tau X$; (3) $e^- p \rightarrow \tau^- X\to {\nu}_\tau\,X_h\, X$. In the first case, signal events are characterized by an electron and missing energy recoiling against at least one jet. In the second case, the electron is replaced by a muon, which, as we will see, largely suppresses the SM background. Finally, in the hadronic channels the signal events have missing energy, at least two jets and no charged leptons.
The major backgrounds from SM processes include neutral current ($e^-p\to e^-j$) and charged current  ($e^-p\to \nu_e j$) DIS. Other backgrounds, such as lepton pair production ($e^-p\to e^-\ell^+\ell^-j$) and real $W$ boson production ($e^-p\to e^- W^{\pm}j$), can at this stage be ignored due to the small cross sections. 

We use \PythiaEight~\cite{Sjostrand:2007gs} to generate $10^8$ and $10^7$ events for the background and signals, respectively.
A  transverse momentum cut on the final states transverse momentum $p_T>10~{\rm GeV}$ is applied to the DIS background generation.
The {\tt Delphes} package is used to simulate the detector smearing effects~\cite{deFavereau:2013fsa}.
We use in this analysis the EIC input card developed by M. Arratia and S. Sekula, based on parameters in Ref.~\cite{Arratia} and used and provided in \cite{Arratia:2020nxw,Arratia:2020azl}. As the EIC handbook does not specify muon identification parameters \cite{Arratia}, 
we assumed the same performance for muons and electrons, and we modified the EIC Delphes card accordingly. 
This assumption relies on having a dedicated muon detector in the EIC design, which is currently being discussed\footnote{We thank M. Arratia for 
clarifying this point.}.  
The anti-$k_t$ jet algorithm with jet cone size $R=1$ and $p_T>5~{\rm GeV}$ will be used to define the observed jets.

\begin{figure}[t]
\vspace{-1em}
\centering
\includegraphics[width=0.49\textwidth]{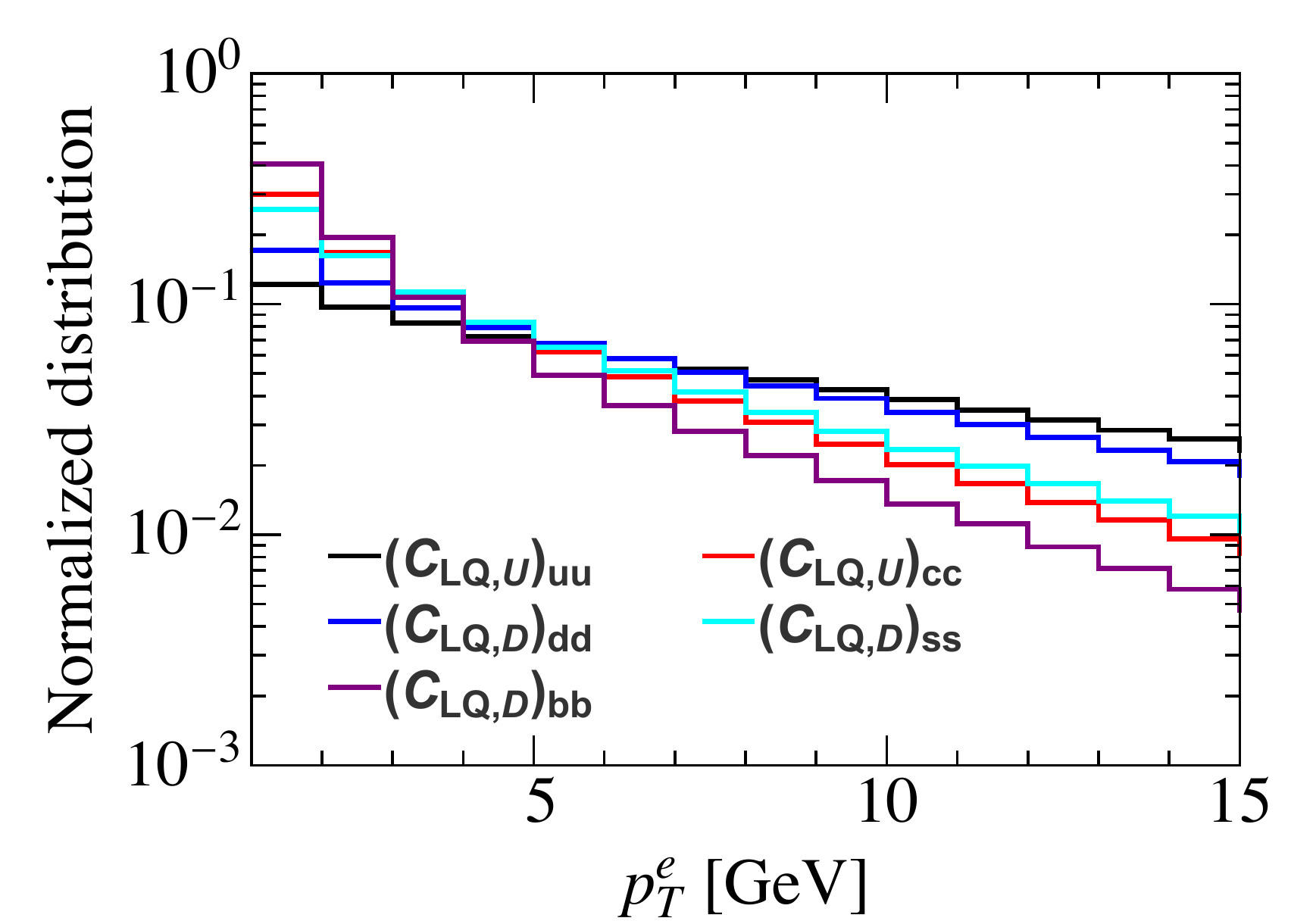}
\includegraphics[width=0.49\textwidth]{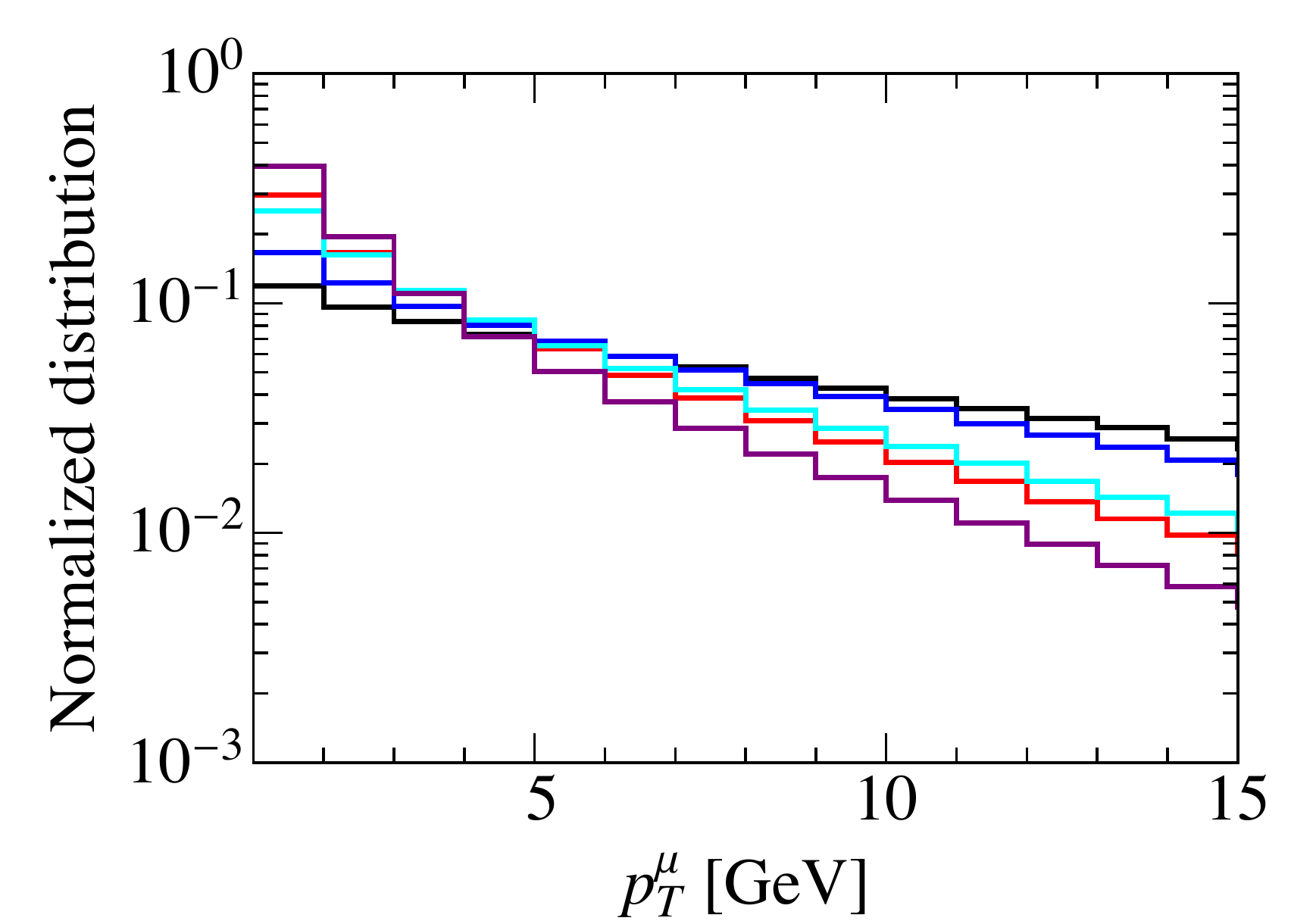} 
\includegraphics[width=0.49\textwidth]{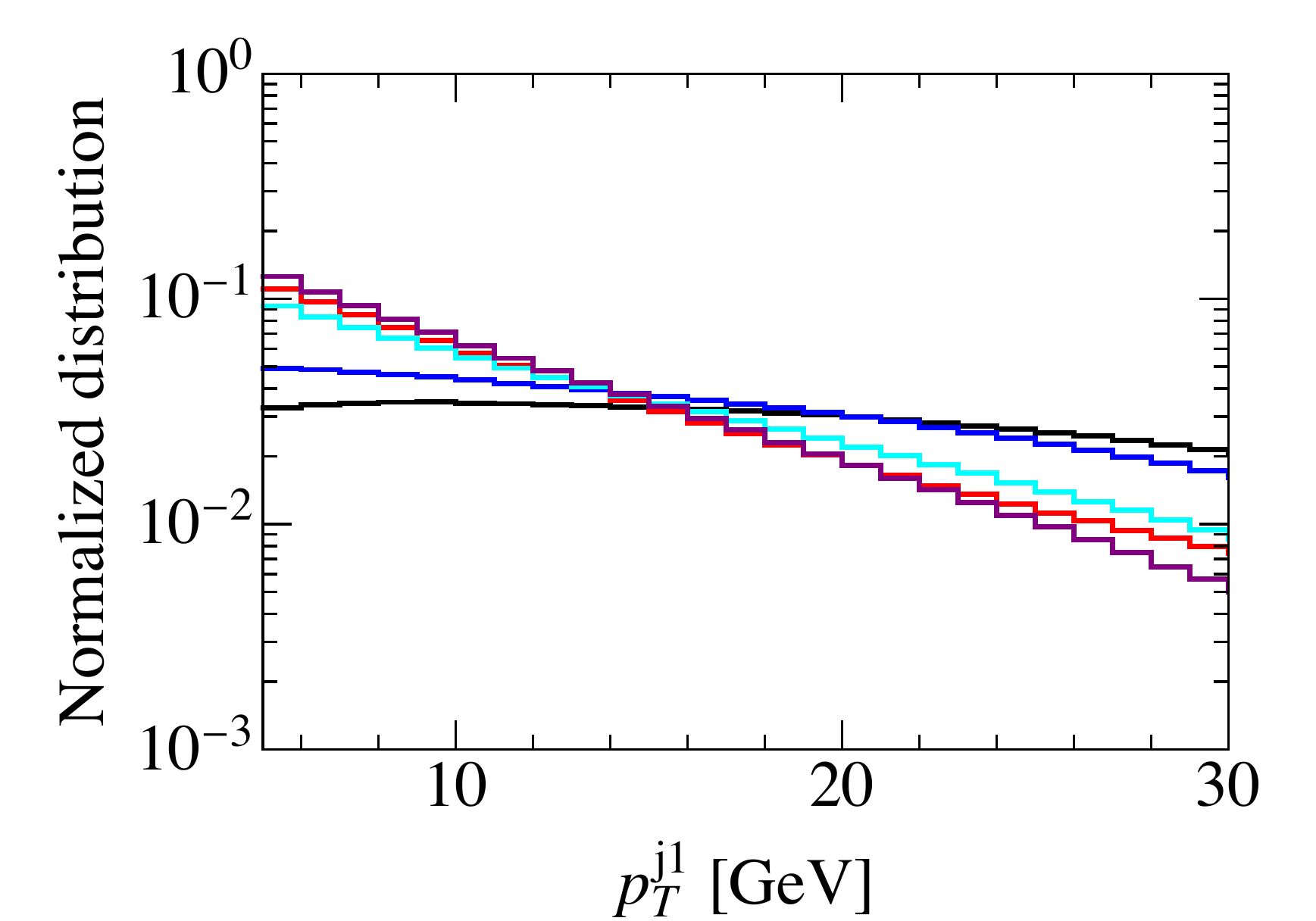}
\includegraphics[width=0.49\textwidth]{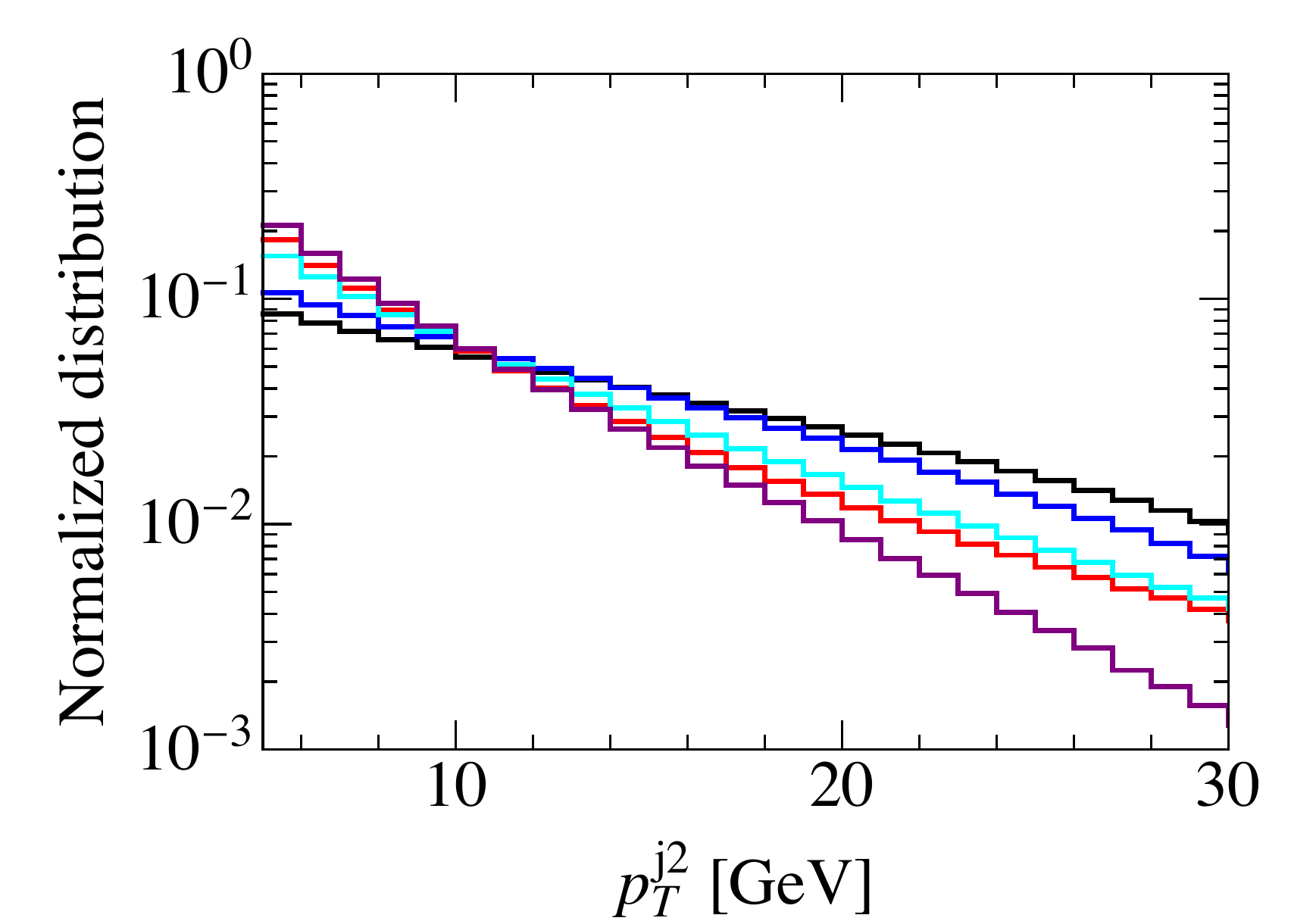}
\includegraphics[width=0.49\textwidth]{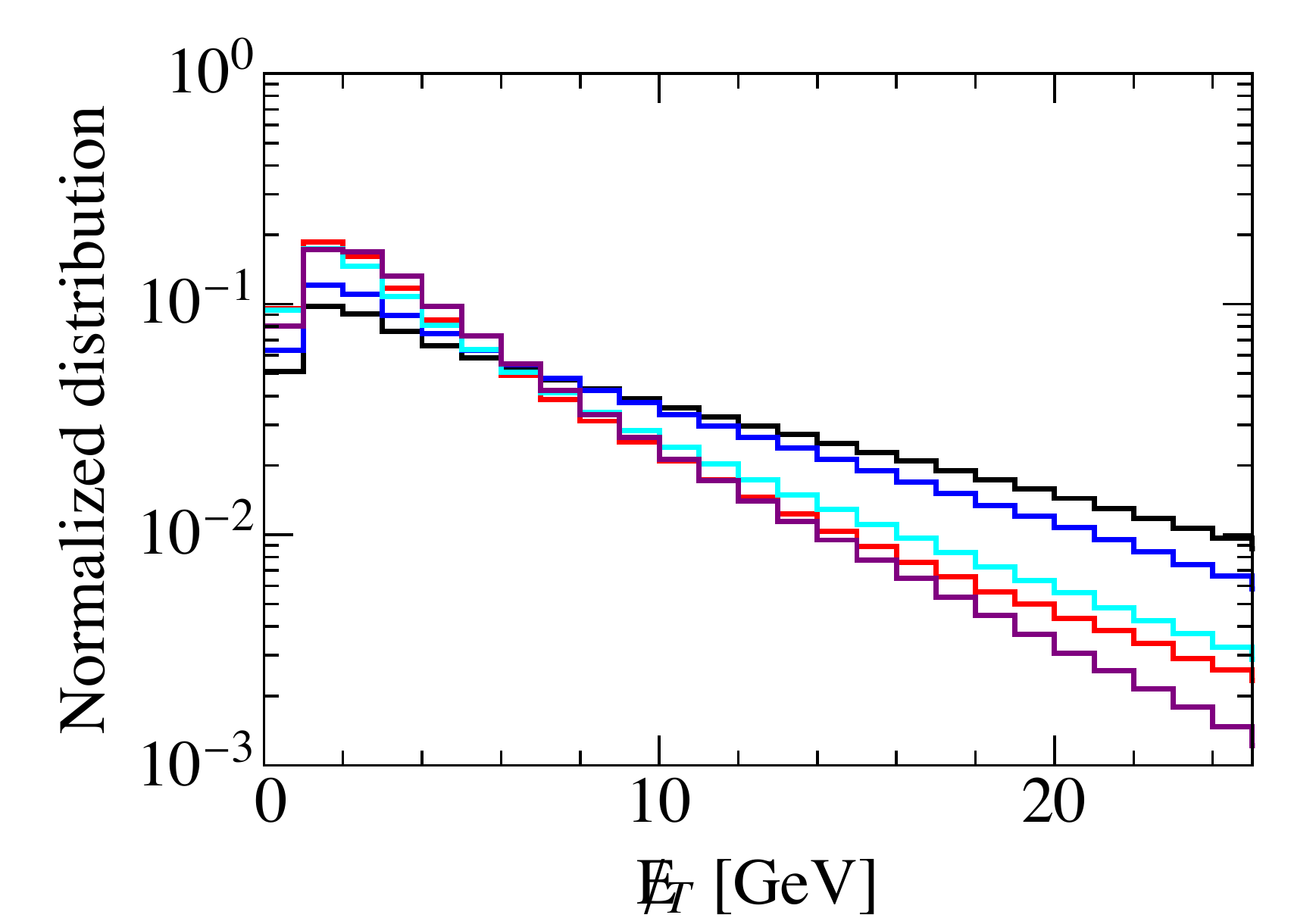}
\vspace{-1ex}
\caption{
Electron, muon, leading and subleading jet $p_T$ distributions
and missing energy distribution induced by four-fermion operators with different flavor components at the EIC, with $E_e=20~{\rm GeV}$ and $E_p=250~{\rm GeV}$ ($\sqrt{S}=141$ GeV). }
\label{Fig:kinF}
\end{figure}

\begin{figure}[t]
\vspace{-1em}
\centering
\includegraphics[width=0.49\textwidth]{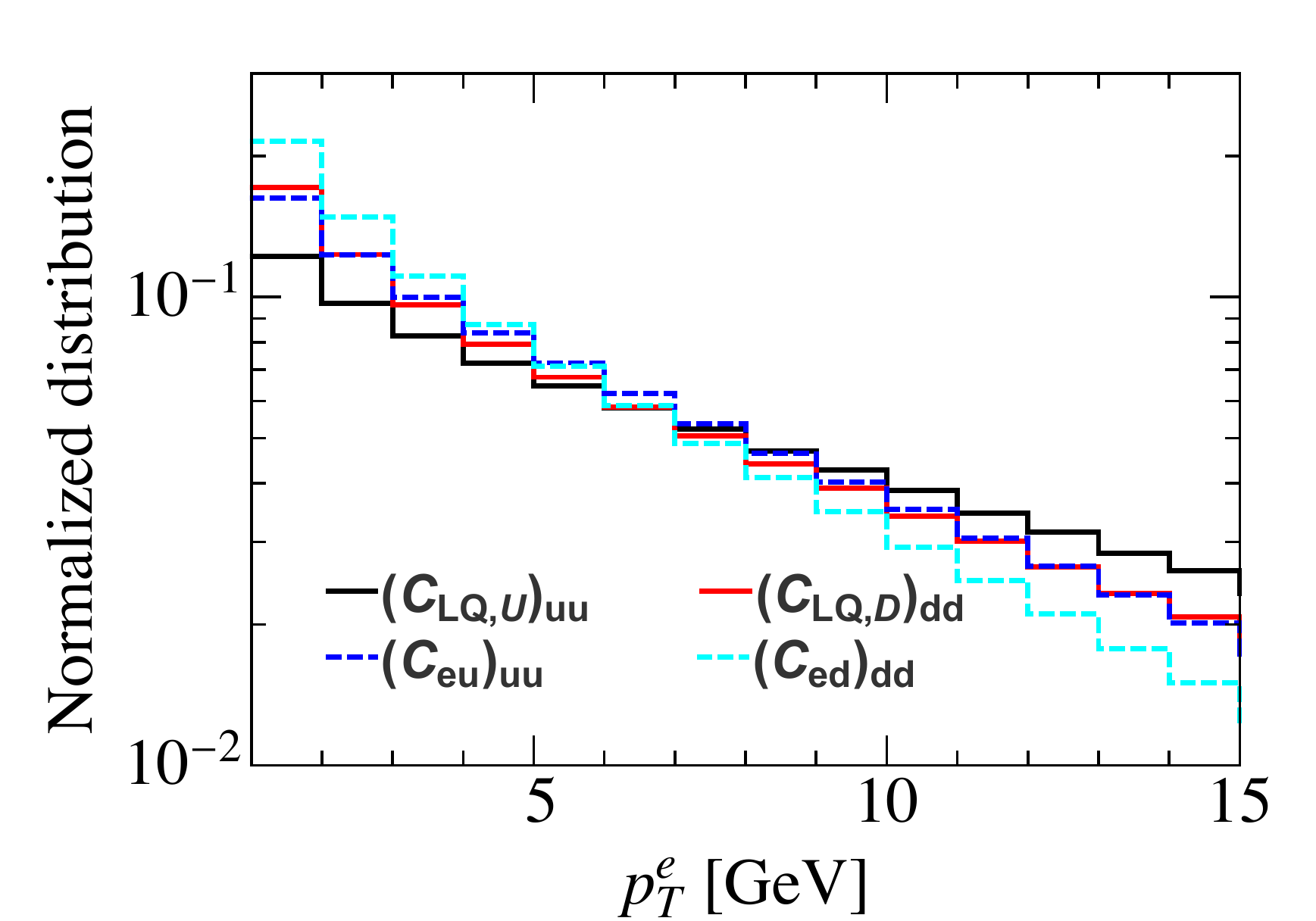}
\includegraphics[width=0.49\textwidth]{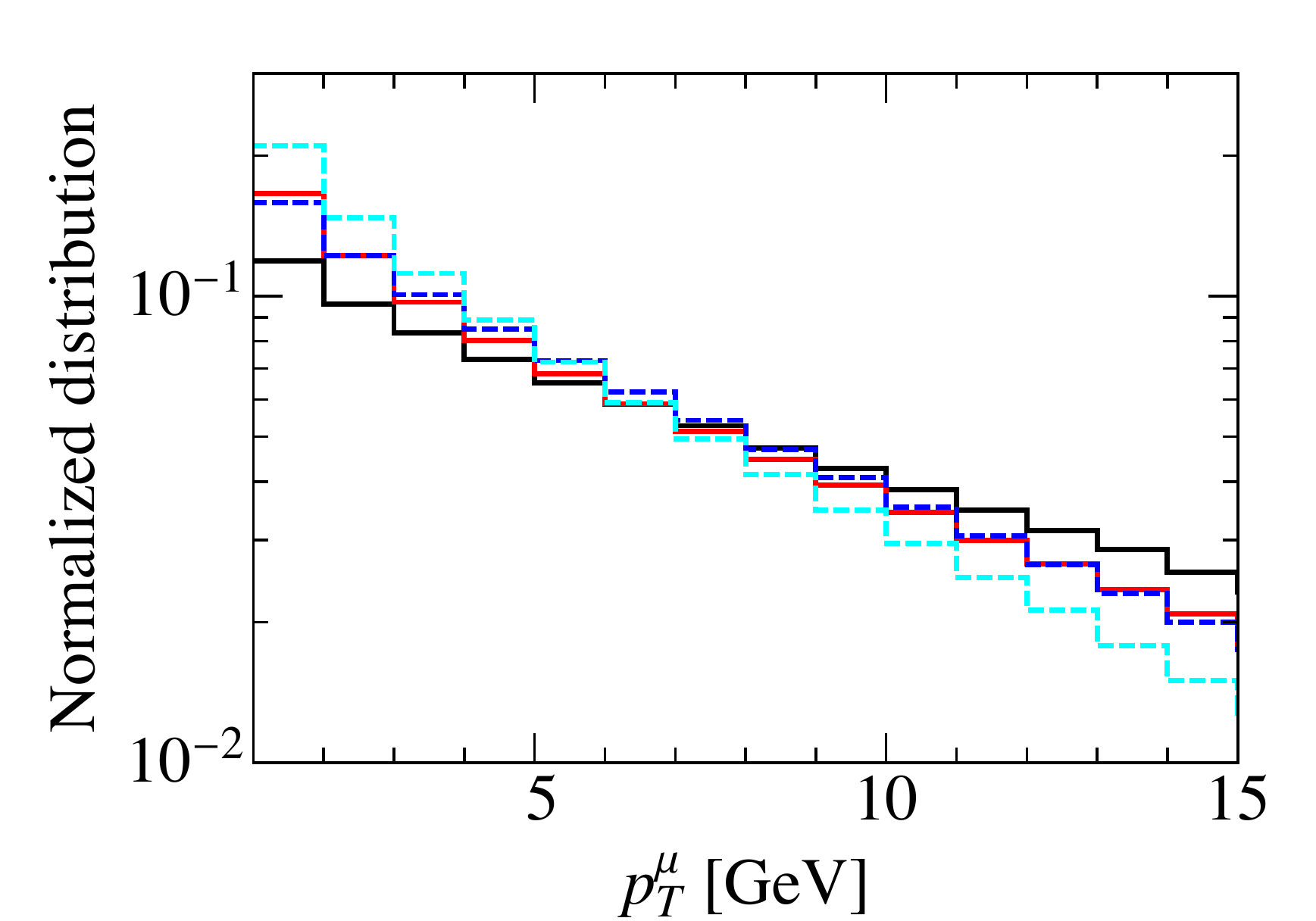} 
\includegraphics[width=0.49\textwidth]{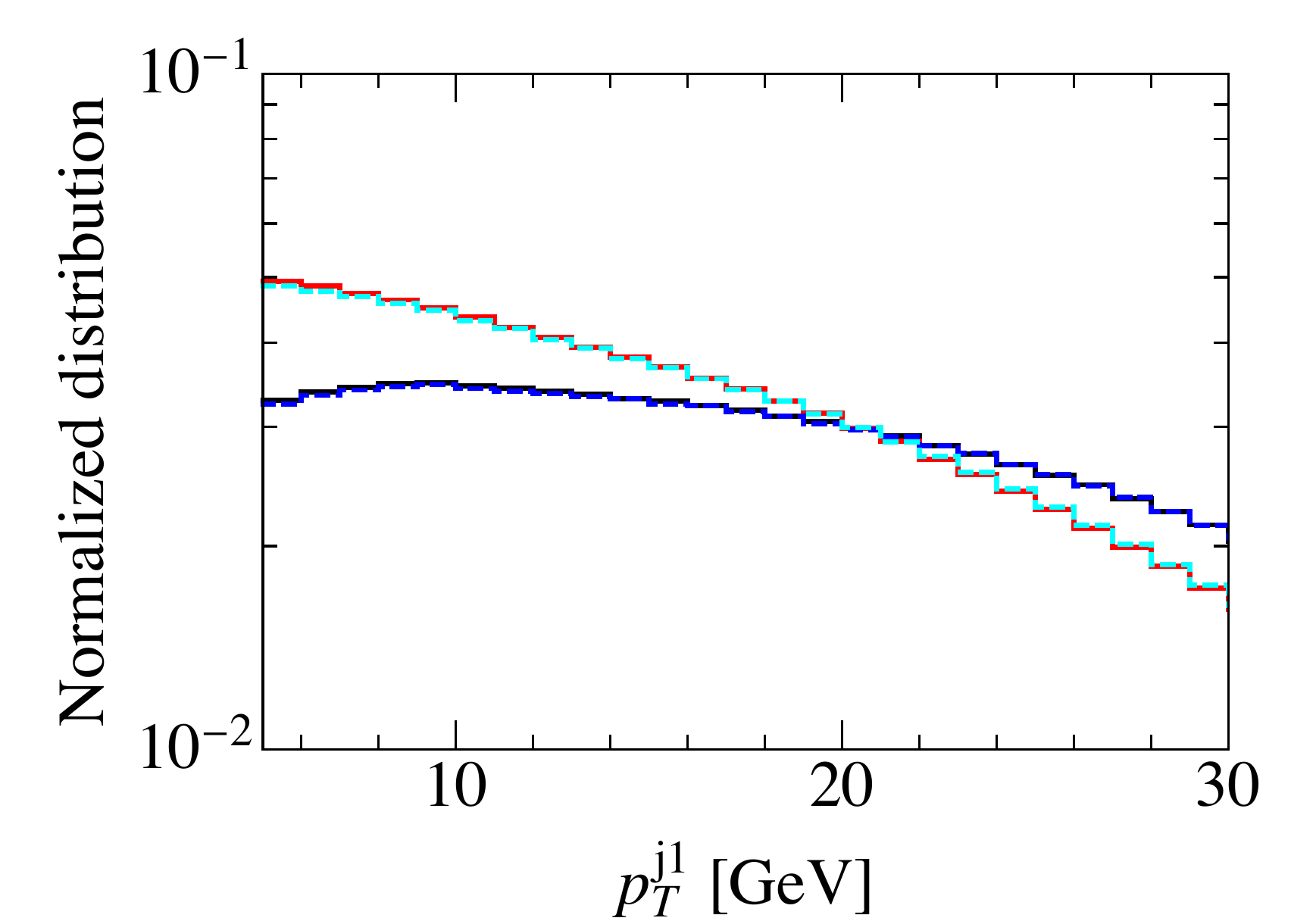}
\includegraphics[width=0.49\textwidth]{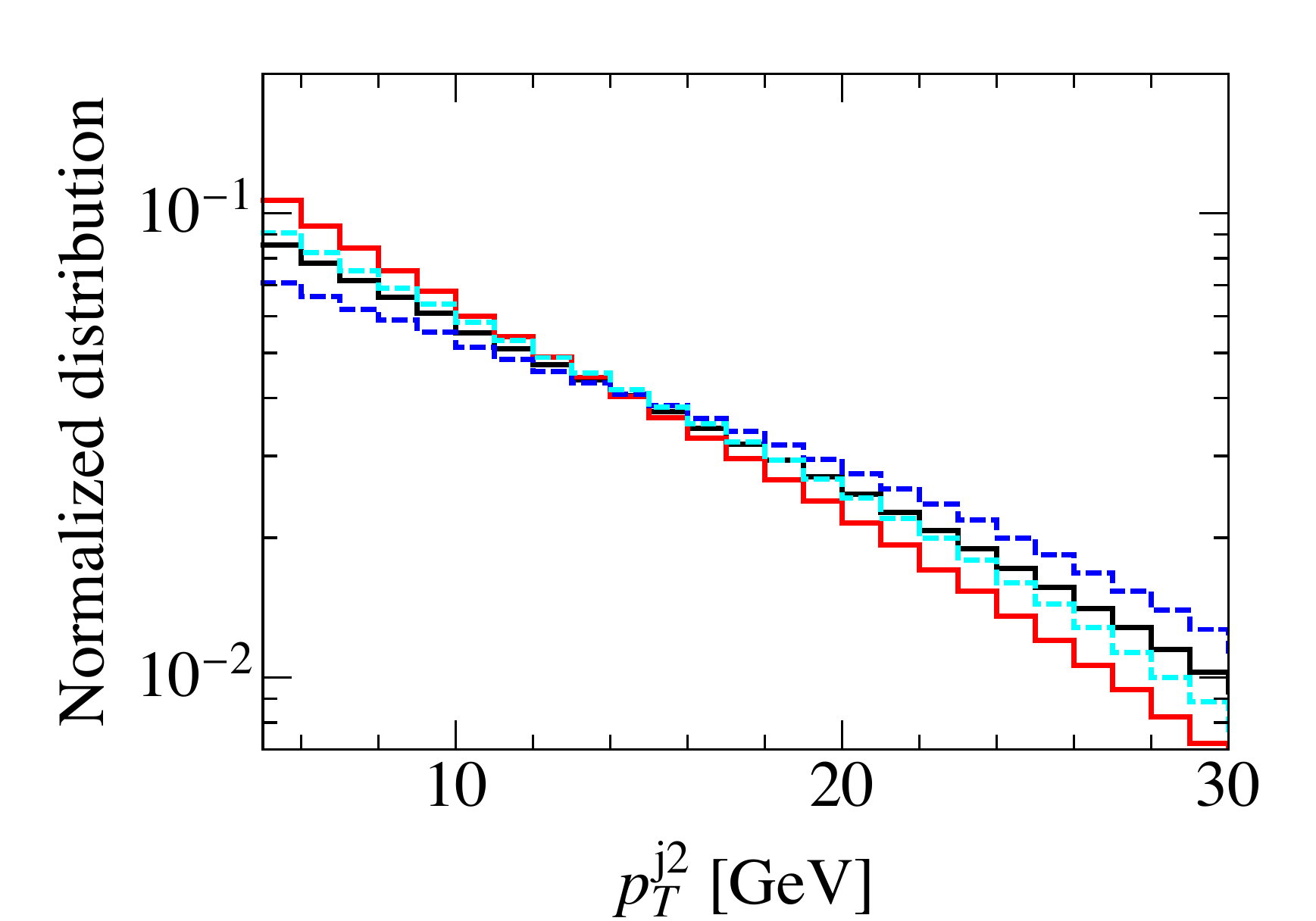}
\includegraphics[width=0.49\textwidth]{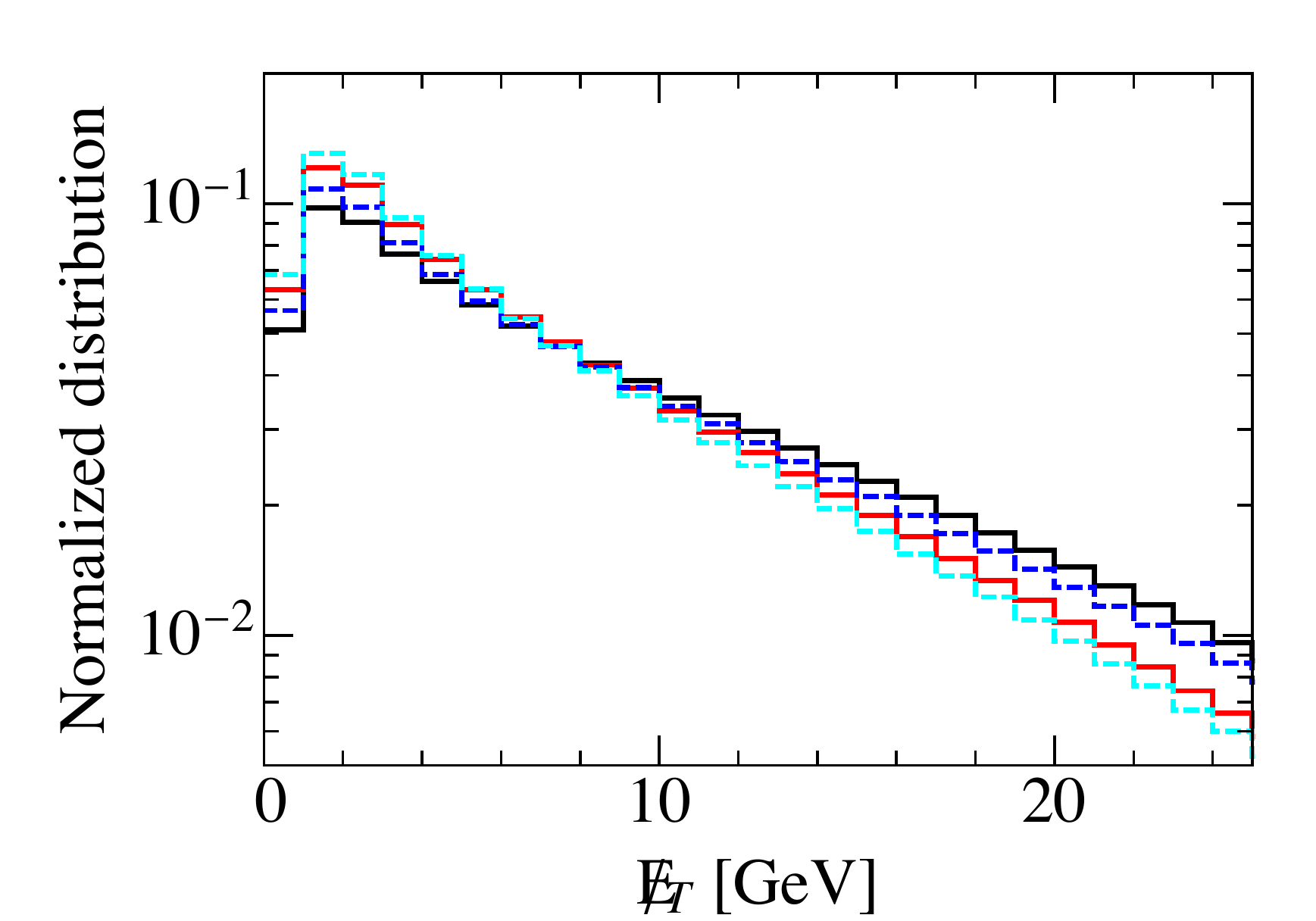}
\vspace{-1ex}
\caption{Electron, muon, leading and subleading jet $p_T$ distributions
and missing energy distribution induced by four-fermion operators with different $\tau$ polarization at the EIC, with $E_e=20~{\rm GeV}$ and $E_p=250~{\rm GeV}$ ($\sqrt{S}=141$ GeV). }
\label{Fig:kinpol}
\end{figure}

In Figs. \ref{Fig:kinF} and \ref{Fig:kinpol} we show the transverse momentum distributions of the hardest electron ($p_T^e$), muon ($p_T^\mu$) and of the leading ($p_T^{j1}$), sub-leading ($p_T^{j2}$) jets
and the missing energy ($\met$) distribution induced by various four-fermion SMEFT operators. The distributions are normalized by the total cross section for each individual contribution, i.e. normalized to a total integral of 1. (Thus these figures compare the shapes but not relative sizes of individual cross sections.).
We note that these  distributions are very sensitive to the flavor of the quark in the initial state,  while they do not strongly depend on the polarization  of the $\tau$ lepton. 
In Fig. \ref{Fig:kinF} we consider the purely left-handed operators $(C_{LQ,U})_{ii}$ and $(C_{LQ,D})_{jj}$, where $i=u,c$ and $j=d,s,b$.
In the massless limit, these operators create a left-handed $\tau$ and the different kinematic behaviors in Fig. \ref{Fig:kinF} are solely due to the flavor
of the quark in the initial state.
The strange and heavy quark components $(C_{LQ,D})_{ss}$, $(C_{LQ,U})_{cc}$ and $(C_{LQ,D})_{bb}$ would favor  small $p_T$ or $\met$, due to the suppression of the sea quark PDFs at large transverse momenta, while the valence components $(C_{LQ,U})_{uu}$ and $(C_{LQ,D})_{dd}$ have significant tails at large $p_T$ and $\met$.
Fig.~\ref{Fig:kinpol} shows the  same distributions for the left-handed operators $(C_{LQ,U})_{uu}$ and $(C_{LQ,D})_{dd}$,
and the right-handed operators $(C_{eu})_{uu}$ and $(C_{ed})_{dd}$. In the massless limit, the $\tau$ lepton is left-handed polarized for $(C_{LQ,U})_{uu}$, and $(C_{LQ,D})_{dd}$ (solid lines), and right-handed polarized for $(C_{eu})_{uu}$ and $(C_{ed})_{dd}$ (dashed lines).  Fig. \ref{Fig:kinpol} shows that the kinematical distributions we are considering in this work are not sensitive to the $\tau$ polarization. This is true in particular for the $p_T$ of the leading jet, which, being produced in the  hard scattering $e^-p\to \tau^- j$, does not depend on the $\tau$ polarization.
In Figs. \ref{Fig:kinF} and \ref{Fig:kinpol} we only show vector and axial operators. We verified that scalar, pseudoscalar and tensor four-fermion operators give rise to similar distributions.

\begin{figure}[t]
\vspace{-1em}
\centering
\includegraphics[width=0.49\textwidth]{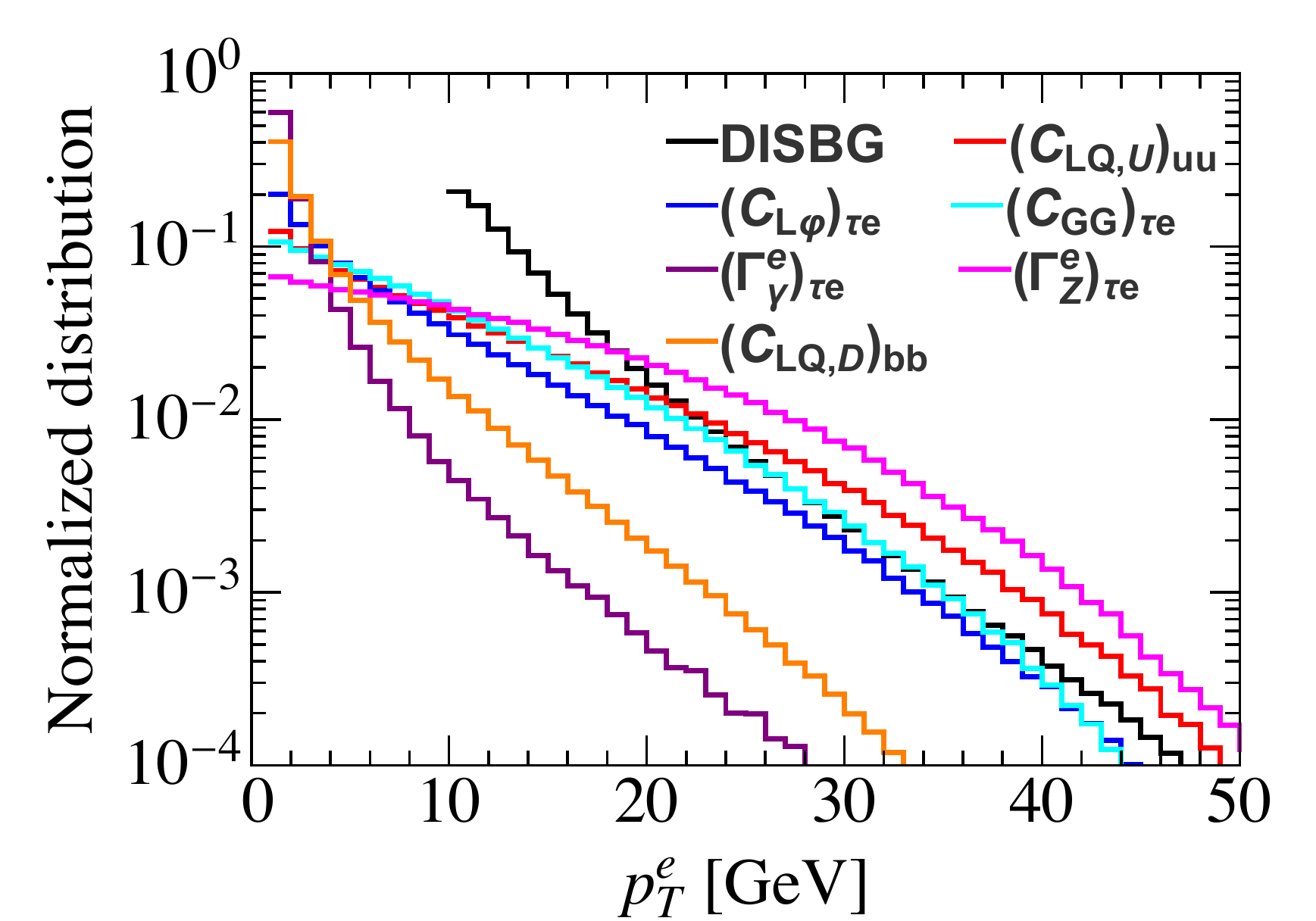}
\includegraphics[width=0.49\textwidth]{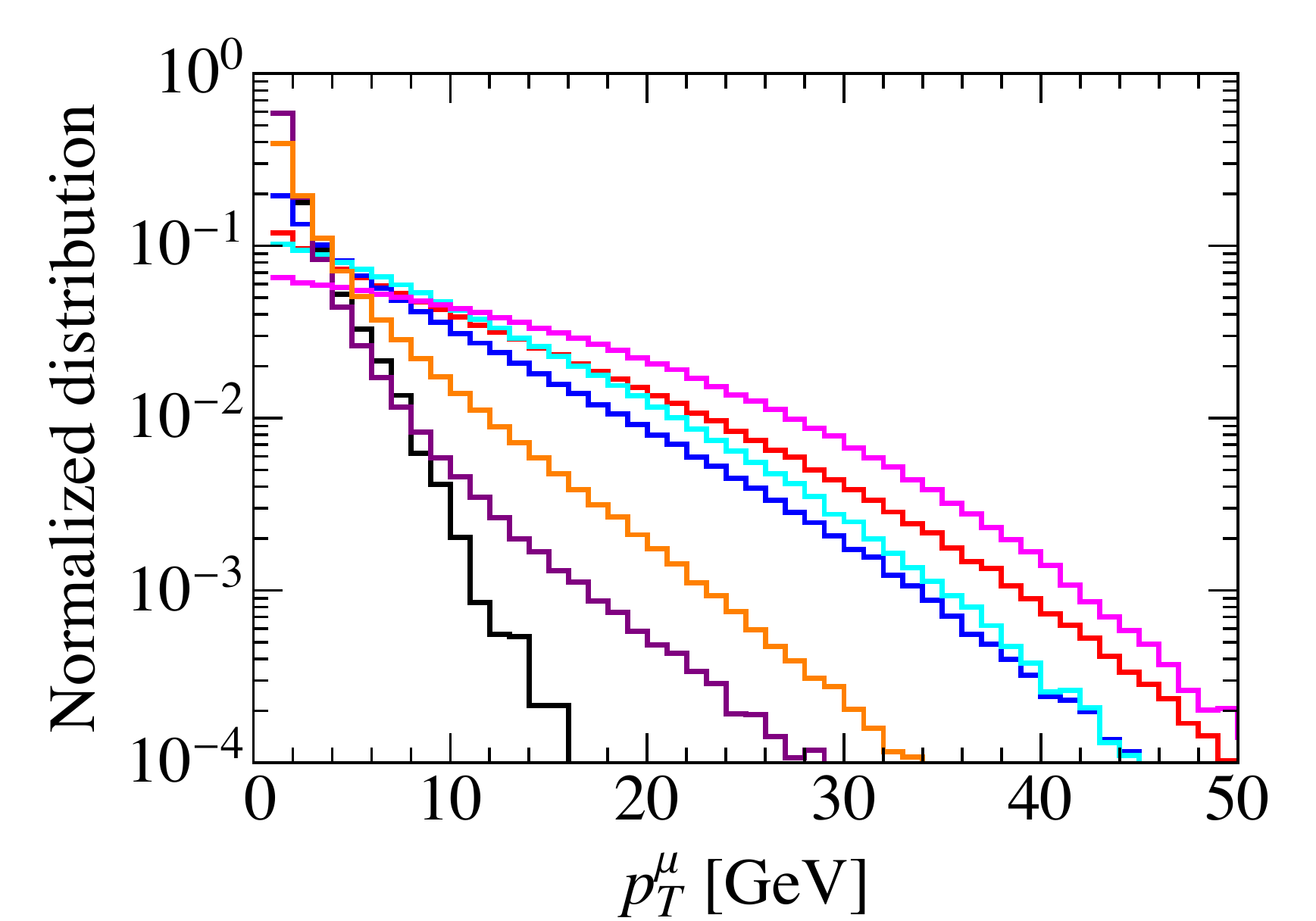} 
\includegraphics[width=0.49\textwidth]{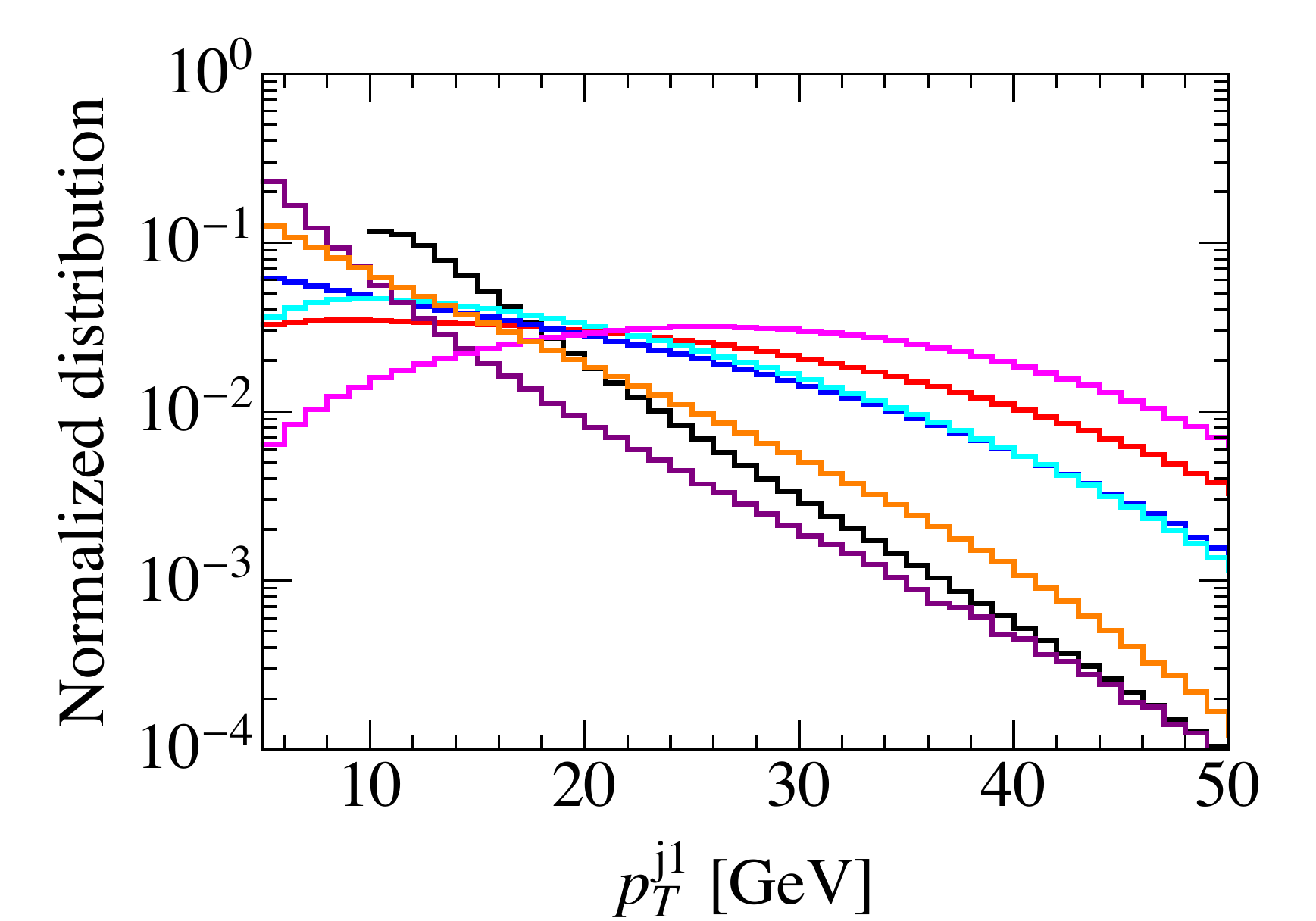}
\includegraphics[width=0.49\textwidth]{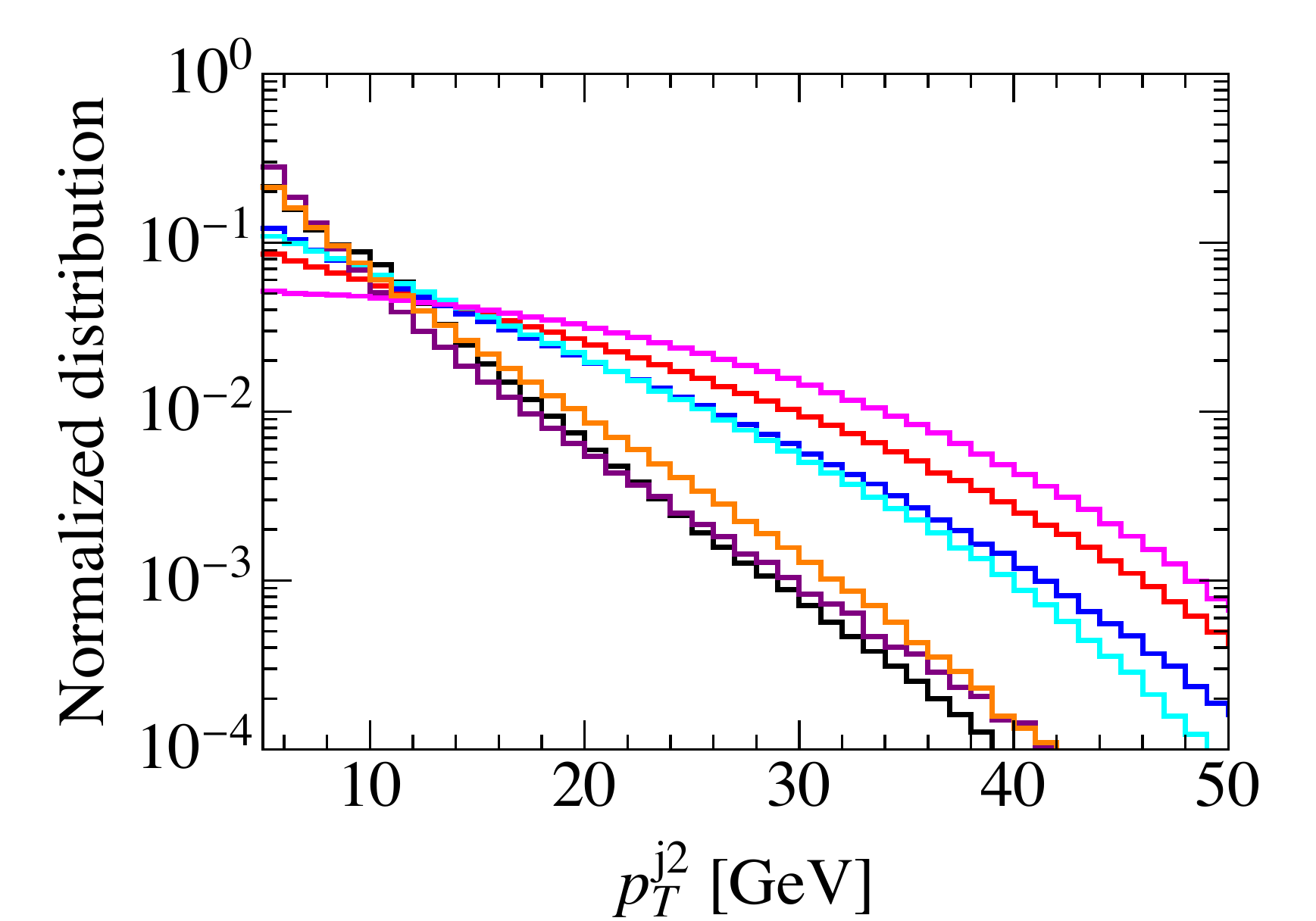}
\includegraphics[width=0.49\textwidth]{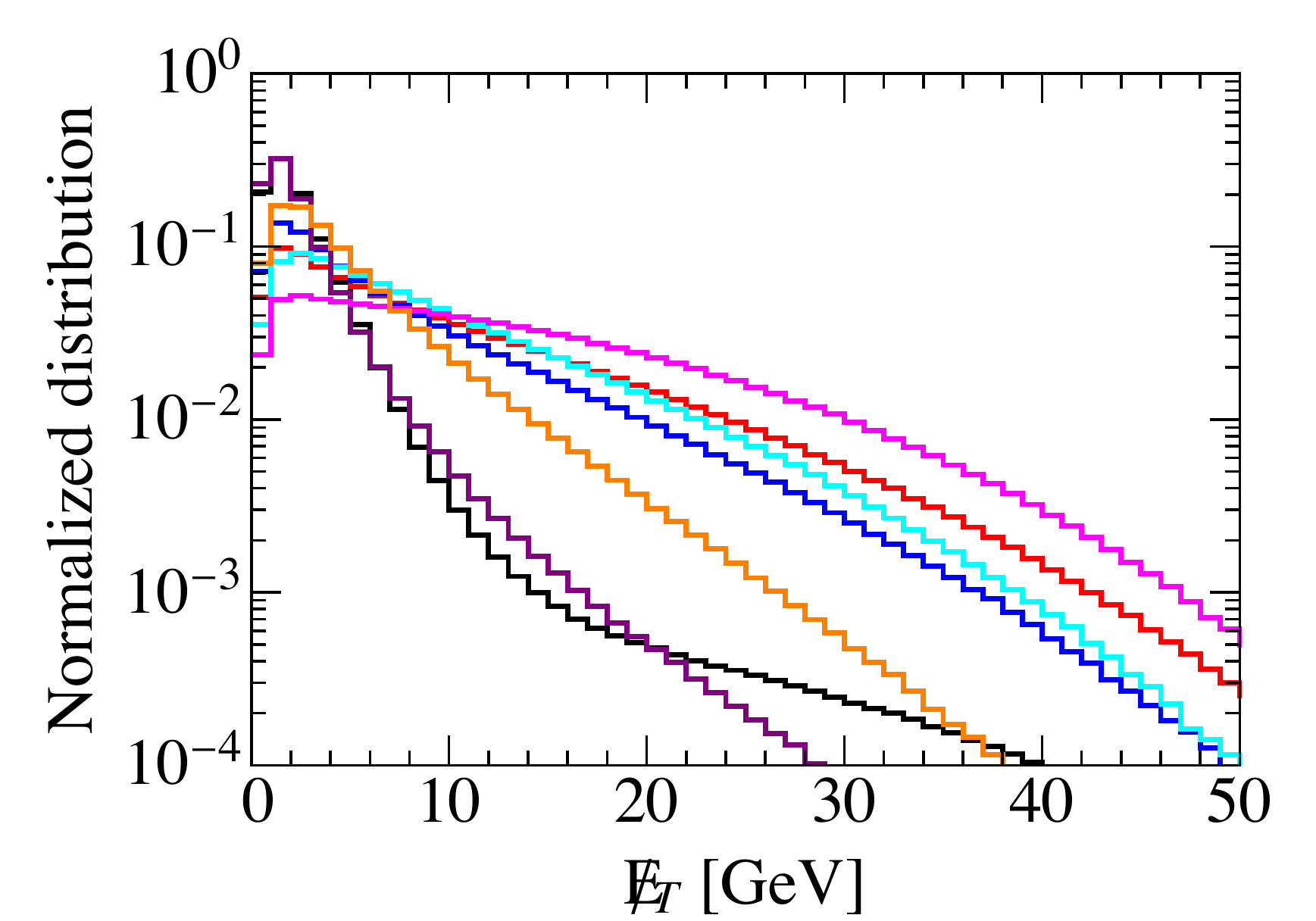}
\vspace{-1ex}
\caption{Electron, muon, leading and subleading jet $p_T$ distributions
and missing energy distribution induced by SMEFT operators with left-handed $\tau$ leptons and by the SM background (DISBG) at the EIC, with $E_e=20~{\rm GeV}$ and $E_p=250~{\rm GeV}$ ($\sqrt{S}=141$ GeV). }
\label{Fig:kinL}
\end{figure}

As discussed in Section \ref{sect:CLFVDIS}, flavor-changing $Z$ couplings, photon and $Z$ dipoles, and gluonic operators induce DIS cross sections with 
different dependence on $Q^2$ with respect to four-fermion operators.
As a consequence, also the $p_T$ and $\met$ distributions  show different features. 
In Fig.~\ref{Fig:kinL}, we show kinematic distributions for the SM background and  SMEFT operators with left-handed $\tau$ leptons. 
The distributions induced by operators with right-handed $\tau$ are similar to those with left-handed $\tau$, and will not be shown here. We use $(C_{LQ,U})_{uu}$ and $(C_{LQ,D})_{bb}$ as  examples of four-fermion operators, and, in addition, we show the signal from the left-handed $Z$ coupling $c_{L\varphi}\equiv c_{L\varphi}^{(1)}+c_{L\varphi}^{(3)}$,
from the photon and $Z$ dipoles $\Gamma^e_\gamma$ and $\Gamma^e_Z$, and from the CP-even gluonic operator $C_{GG}$.
All distributions are again normalized to area one.

With the results depicted in Fig.~\ref{Fig:kinL}, several comments are in order:
\begin{itemize}
\item The SM distributions tend to peak/grow at small values of $p_T$ and $\met$. In the case of the electron and leading jet 
$p_T$  distributions, we begin plotting the DIS background only at 10 GeV, in order to limit the number of events we had to simulate, as the SM cross section blows up rapidly as these $p_T\to 0$.

\item The electron $p_T$ distribution induced by valence four-fermion operators, $Z$ couplings, $\Gamma^e_Z$ and gluonic operators shows a slower 
decrease at high $p_T$ compared to the SM. Still the very large SM background implies that even imposing hard cuts on the electron $p_T$ is not sufficient to fully suppress the SM background.

\item Muons in the background sample are generated by the parton shower and by the decay of hadrons. Therefore, most background muons have 
very small $p_T$. For signal events, the muon spectrum is similar to the electron $p_T$ spectrum.

\item The $p_T$ spectra of the two leading jets induced by four-fermion operators with valence quarks, $Z$ couplings, $\Gamma^e_Z$ and gluonic operators 
are harder than for the SM background. For heavy-quark operators, the shape of the signal is similar to the SM background.

\item $\met$ in the background sample is generated by charged-current DIS, by the parton shower and by the decay of hadrons. The background distribution
is peaked at small $\met$, but, differently from the muon $p_T$ distributions, charged-current DIS causes a sizable tail at larger values of $\met \gtrsim 20$ GeV.

\item There is a collinear enhancement for the $p_T$  of leptons and jets from the photon dipole operator $(\Gamma_\gamma^e)_{\tau e}$. Consequently, the  distributions from $\Gamma_\gamma^e$ are similar to the DIS background.
\end{itemize}

\begin{figure}[t]
\vspace{-1em}
\centering
\includegraphics[width=0.62\textwidth]{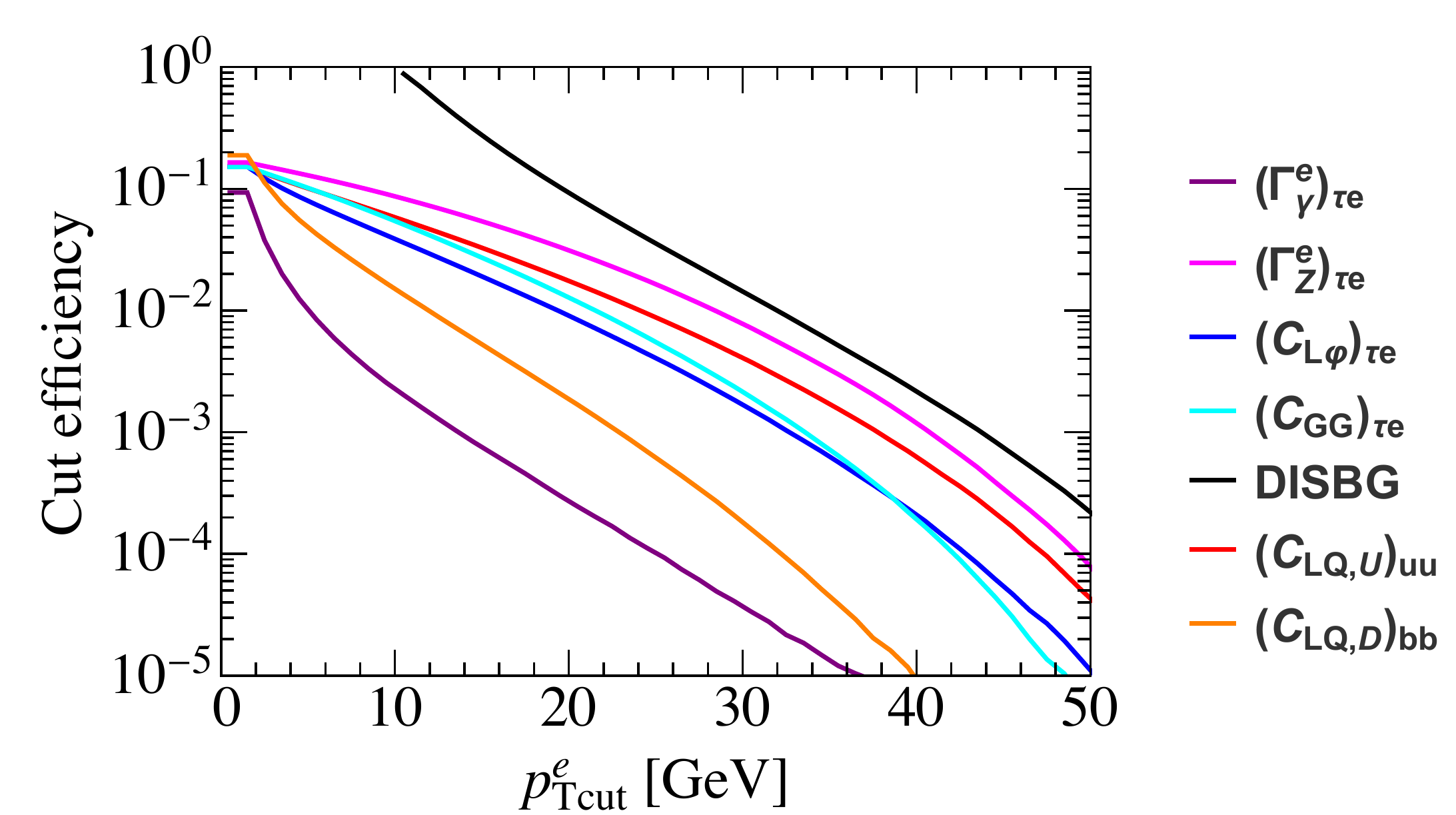}\\
\includegraphics[width=0.49\textwidth]{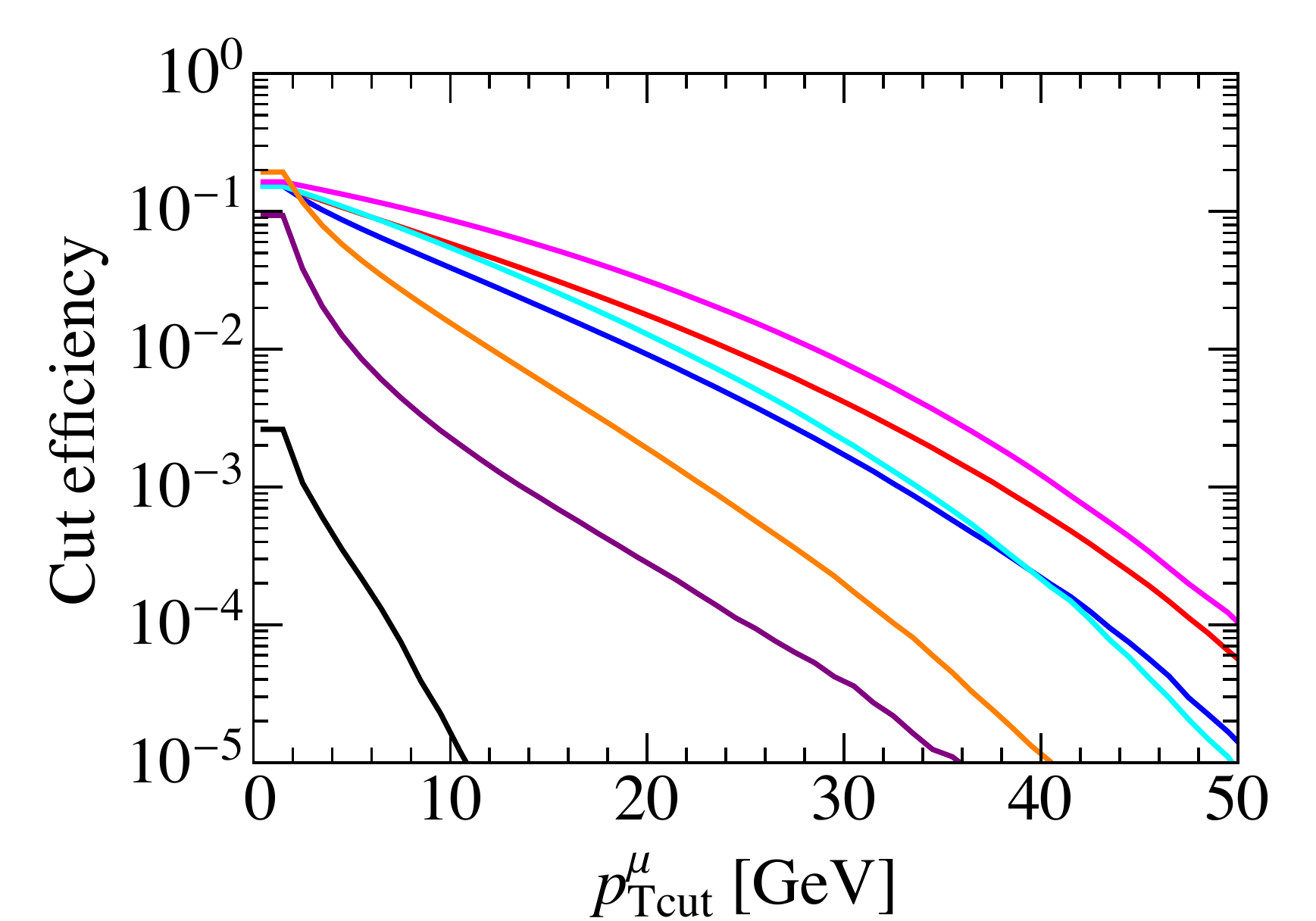} 
\includegraphics[width=0.49\textwidth]{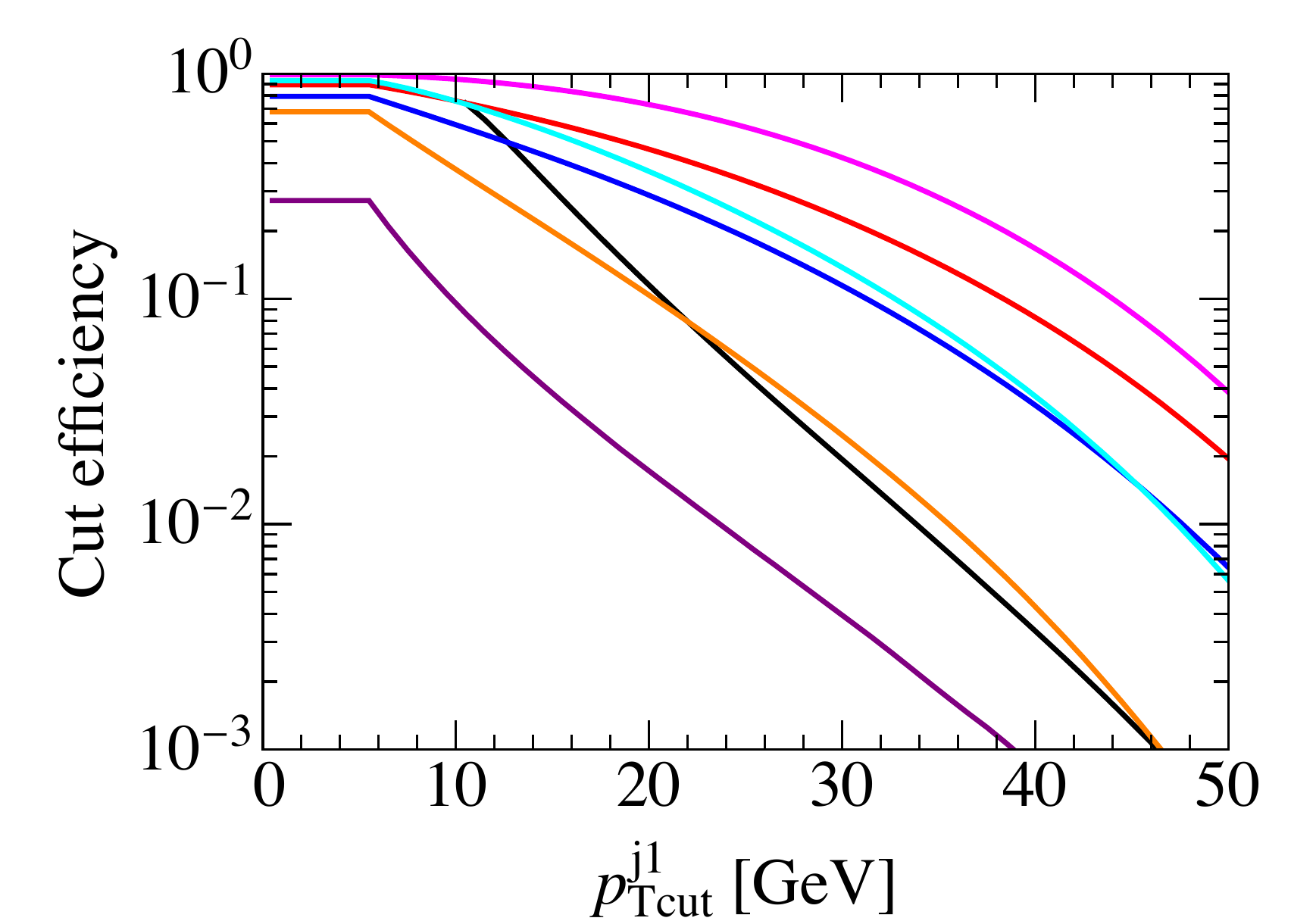}
\includegraphics[width=0.49\textwidth]{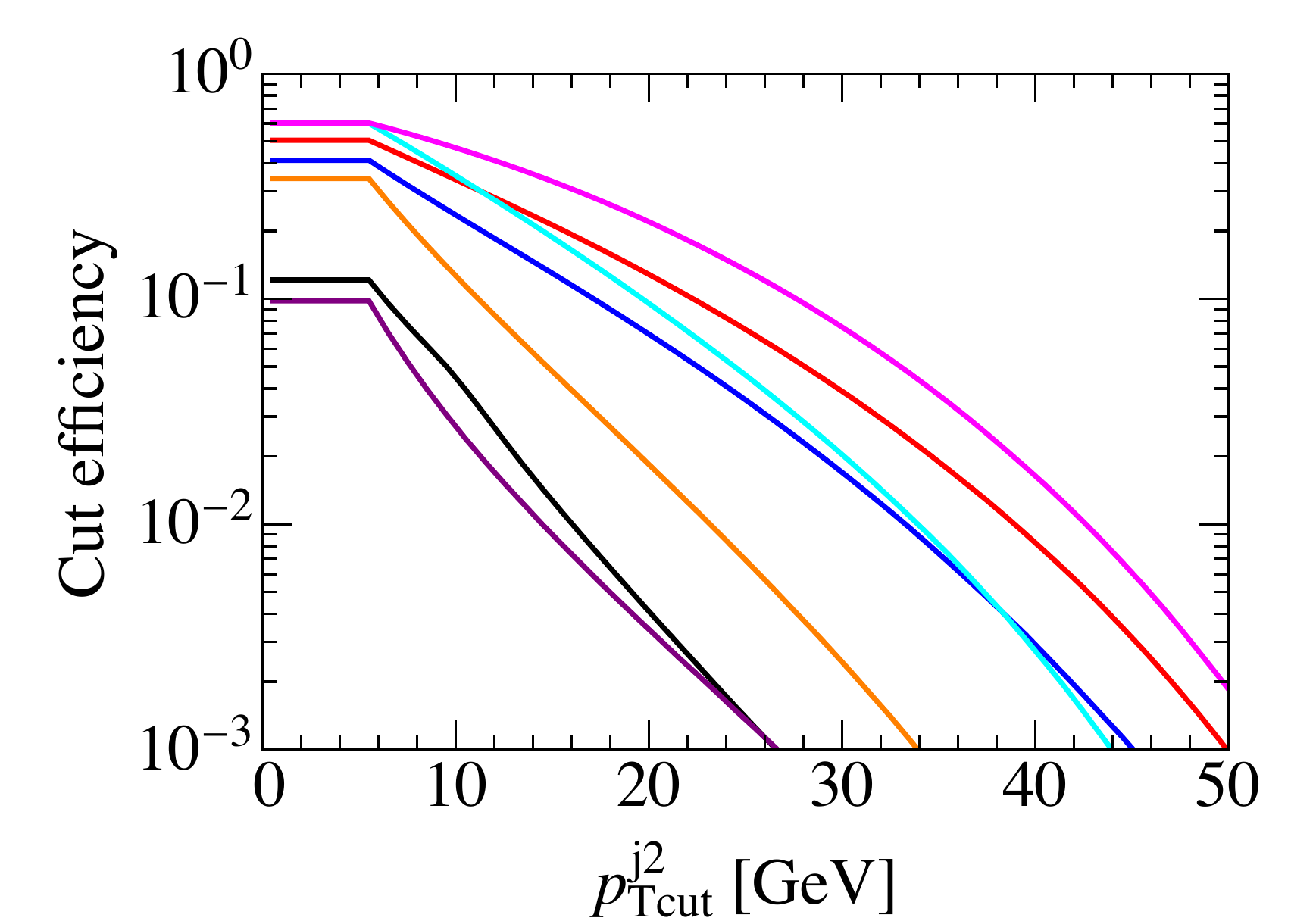}
\includegraphics[width=0.49\textwidth]{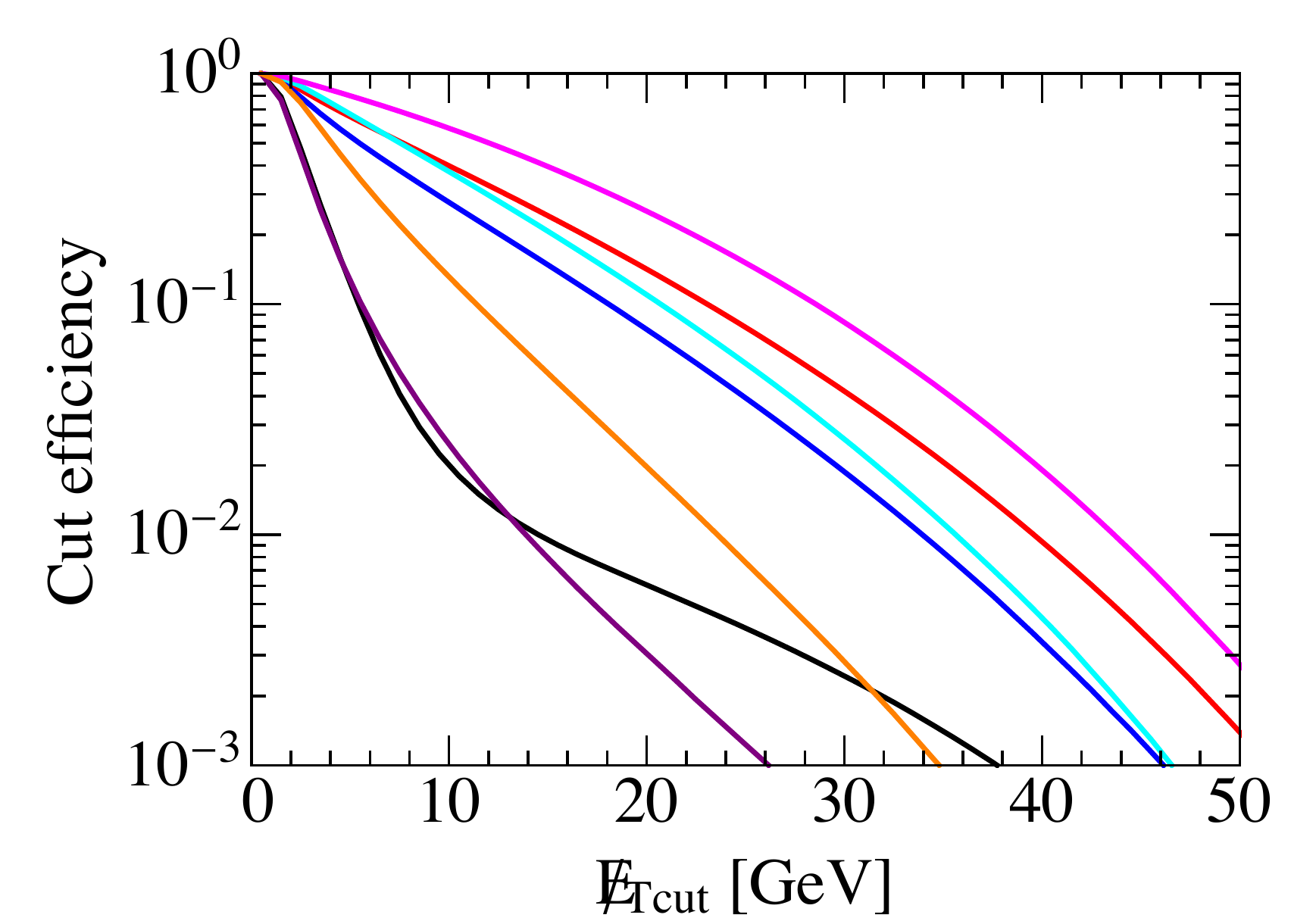}
\vspace{-1ex}
\caption{Cut efficiency for the SM background (DISBG) and the signal induced by SMEFT operators with left-handed $\tau$ leptons, as a function of the cut
on the electron, muon, leading and subleading jet $p_T$ or on the missing energy. We only implement one cut at a time.
}
\label{Fig:cutL}
\end{figure}

\begin{figure}[t]
\vspace{-1em}
\centering
\includegraphics[width=0.49\textwidth]{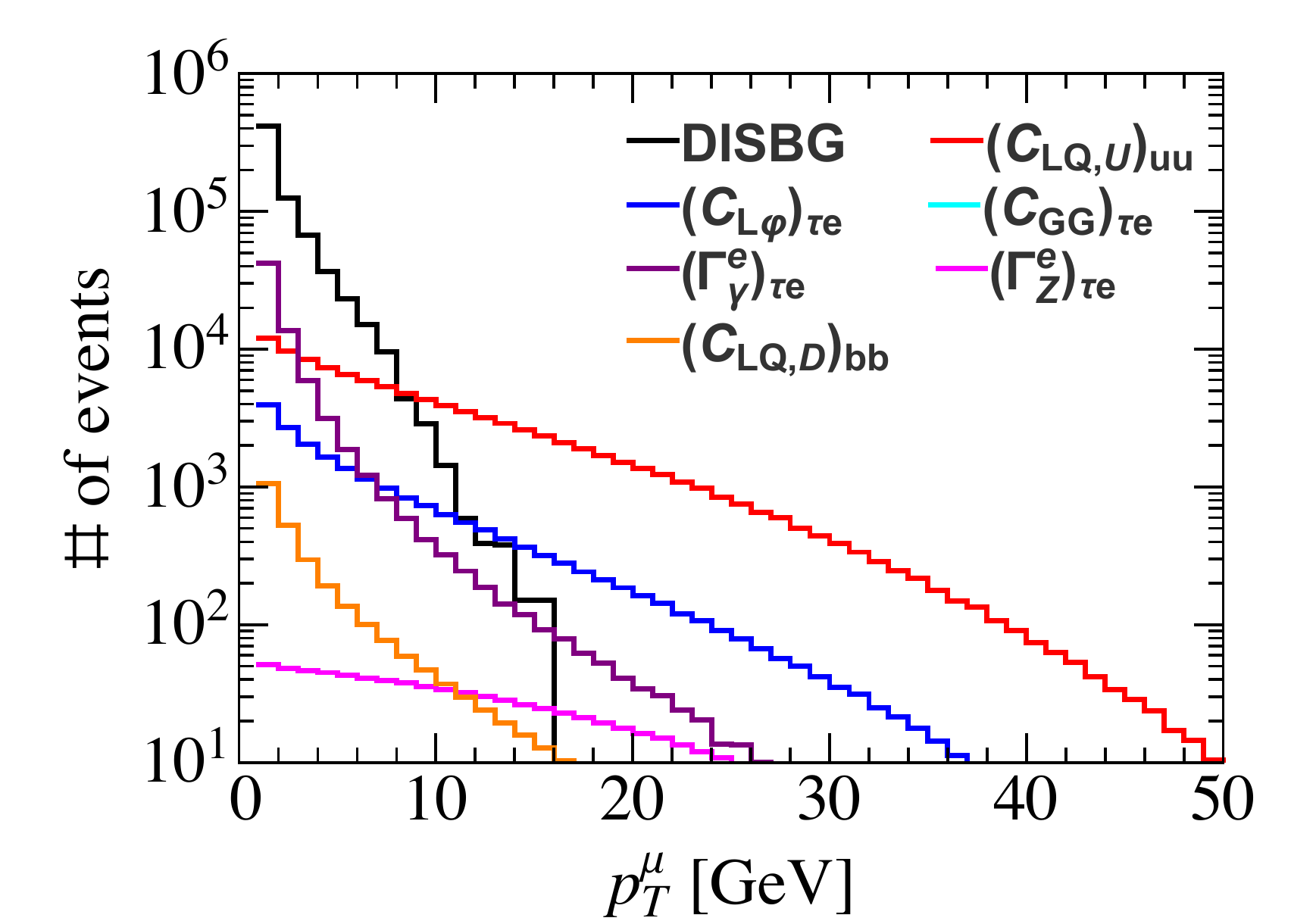}
\includegraphics[width=0.49\textwidth]{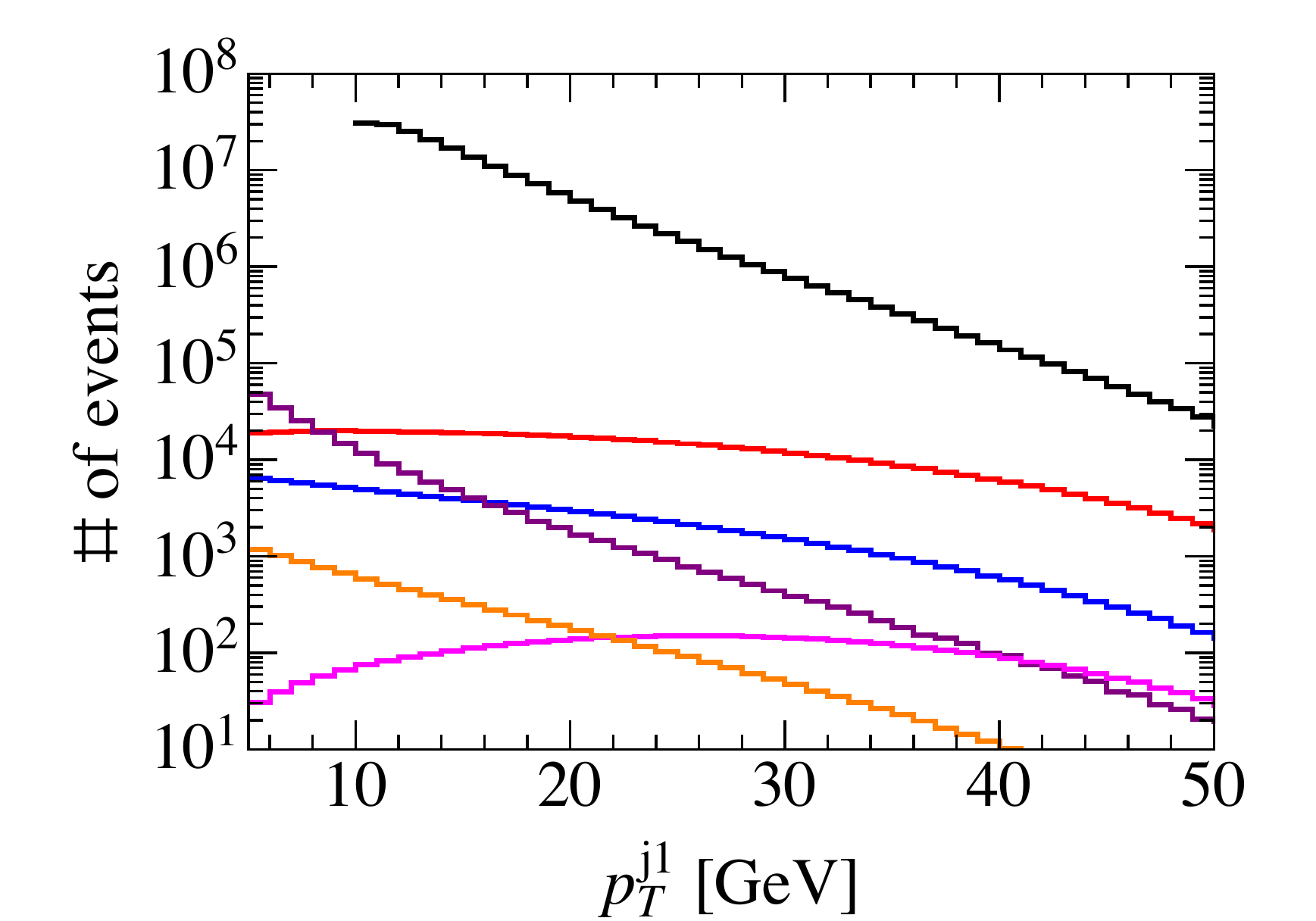} 
\includegraphics[width=0.49\textwidth]{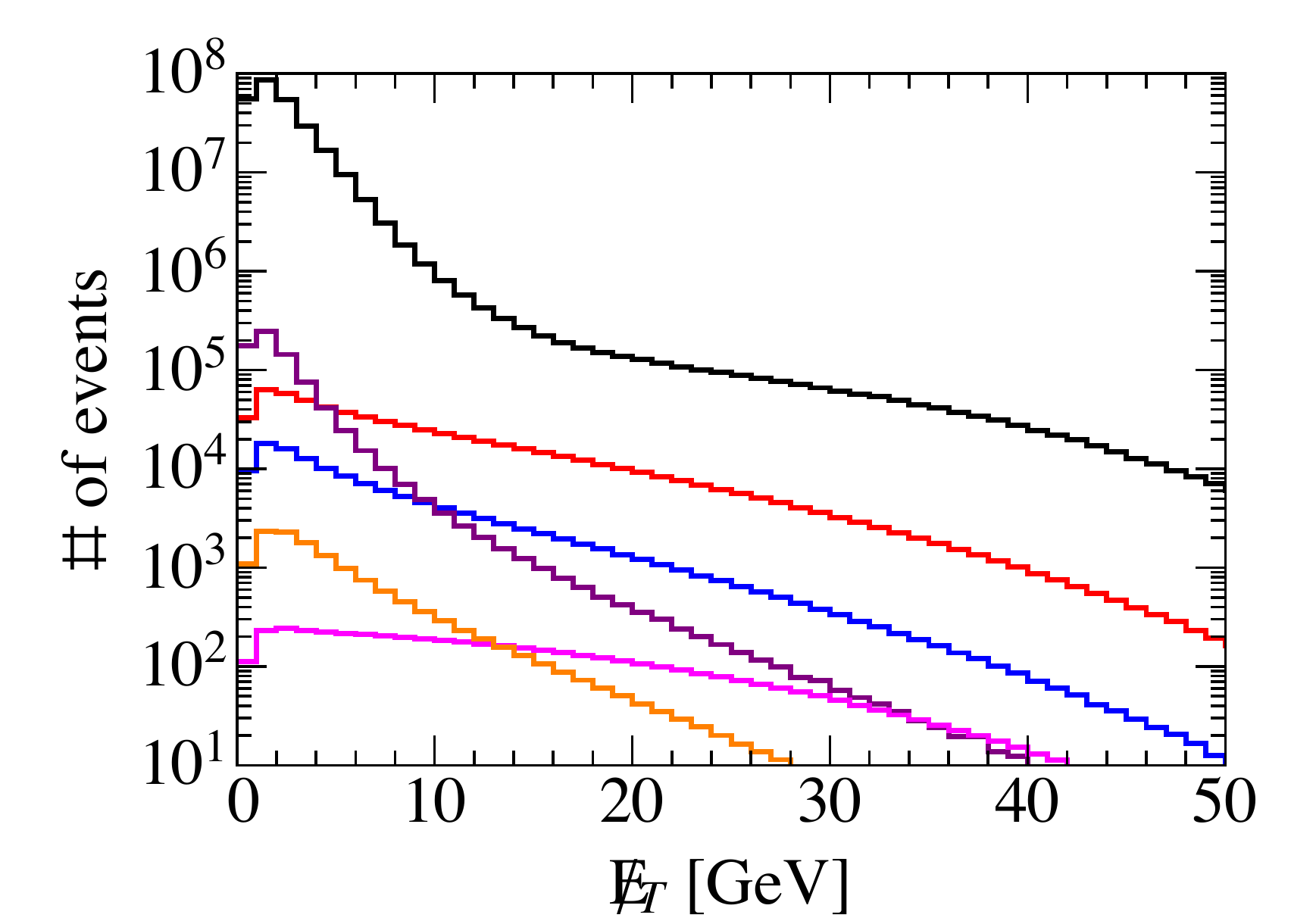}
\vspace{-1ex}
\caption{The distributions of $p_T^\mu,~p_T^{j1}$ and $\met$ from $ep\to \tau (\rightarrow  \mu \bar\nu_\mu \nu_\tau)+X$  at the EIC with $E_e=20~{\rm GeV}$, $E_p=250~{\rm GeV}$ ($\sqrt{S}=141$ GeV) and $\mathcal{L}=100~{\rm fb}^{-1}$. The Wilson coefficients of the CLFV operators is set to $C_i=1$.}
\label{Fig:event}
\end{figure}
\begin{figure}
\vspace{-1ex}
\centering
\includegraphics[width=0.65\textwidth]{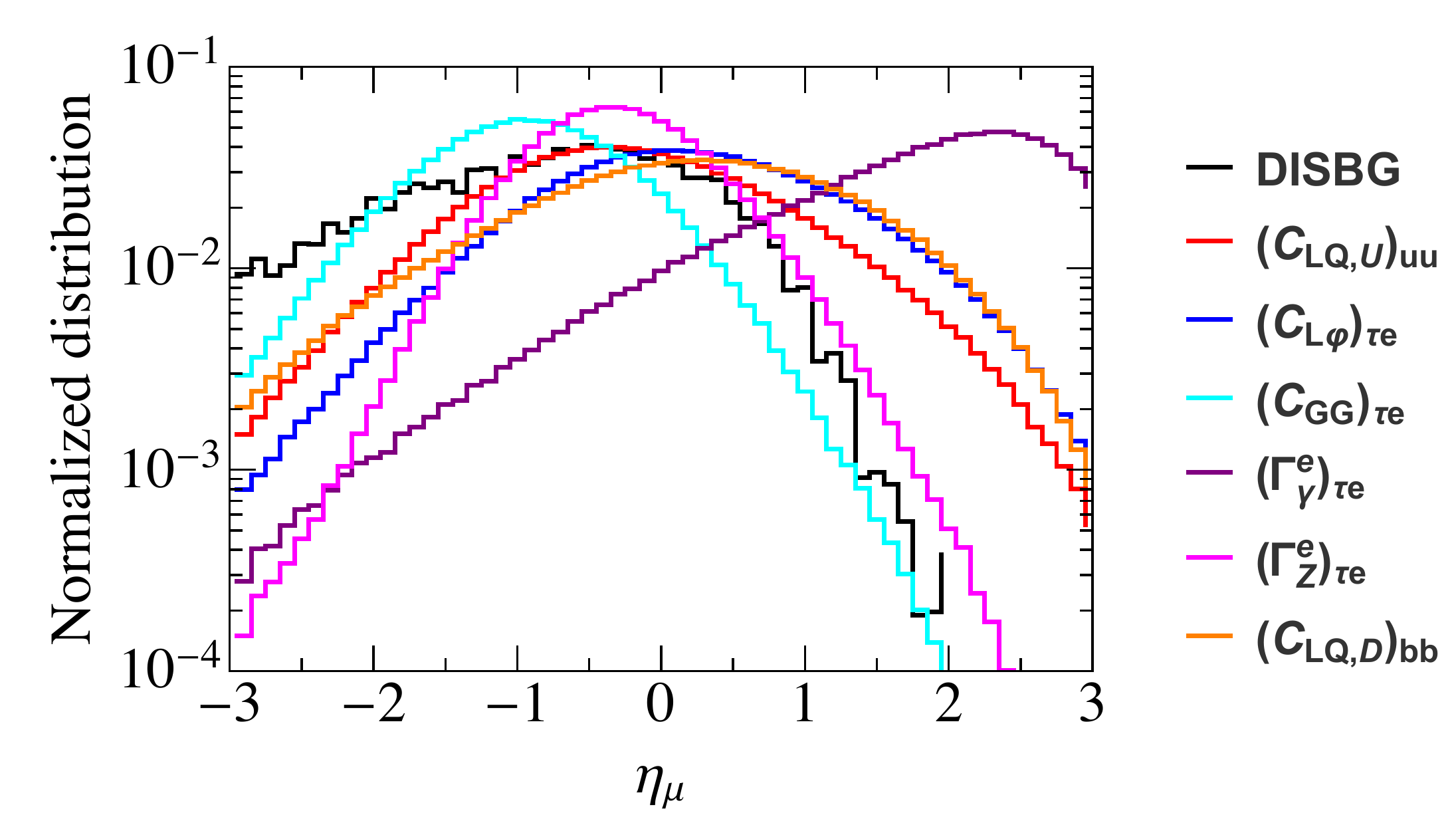}
\vspace{-1em}
\caption{Muon $\eta$ distribution induced by SMEFT operators with left-handed $\tau$ leptons and by the SM background (DISBG) at the EIC, with $E_e=20~{\rm GeV}$ and $E_p=250~{\rm GeV}$ ($\sqrt{S}=141$ GeV).}
\label{Fig:rap}
\end{figure}

These observations are summarized in Fig.~\ref{Fig:cutL}, where we show the cut efficiency as a function of the kinematic cut for both the signals and background. In these plots, we consider one observable at a time. 
Fig.~\ref{Fig:cutL} shows that the cut efficiencies for the SM background and for the $\gamma$ dipole operator $\Gamma_\gamma^e$ drop quickly as  we increase the $p_T$ or $\met$ cut. This is particularly true for the muon channel. Here, asking for a muon in the final state already suppresses the SM background by a factor of about $10^{-3}$, and requiring that $p_T^\mu>10~{\rm GeV}$ brings the suppression to $10^{-5}$. The same $p_T$ cut reduces the signal events by about  $\sim 10\%$, corresponding to the $\tau$ branching ratio in this channel.
We also note that the $Z$ boson dipole operator $\Gamma_Z^e$  typically has   the largest cut efficiency. Although the cross section is small compared to other
SMEFT operators, the large cut efficiency implies that the EIC will impose relatively strong constraints on the $Z$ dipoles. 
$C_{GG}$ and $c_{L\varphi}\equiv c^{(1)}_{L\varphi} + c^{(3)}_{L\varphi}$ show a comparable cut efficiency. However, the cross section from the gluonic operators is very small, $\mathcal{O}(10^{-5})~{\rm pb}$, so that we do not expect very strong constraints on these operators. 
Based on Figs. \ref{Fig:kinF}--\ref{Fig:cutL}, we suggest the following kinematic acceptance cuts to suppress the background for the three classes of decay modes:
\begin{itemize}
\item $\tau^-\to e^-\bar\nu_e{\nu}_\tau$: at least one electron, one jet and
\beq
p_{T}^{e}>10~{\rm GeV},\quad p_{T}^{j1}>20~{\rm GeV},\quad \met>15~{\rm GeV},\quad |\eta_e|, |\eta_{j1}|<3.
\label{eq:cute}
\eeq
\item $\tau^-\to \mu^-\bar\nu_\mu{\nu}_\tau$: at least one muon, one jet and 
\beq
p_{T}^{\mu}>10~{\rm GeV},\quad p_{T}^{j1}>20~{\rm GeV},\quad \met>15~{\rm GeV},\quad |\eta_\mu|,|\eta_{j1}|<3.
\eeq
A rejection on electrons is also applied if $p_{T}^{e}>10~{\rm GeV}$.
\item $\tau^-\to {\nu}_\tau+X_h$: no leptons and at least two jets with, 
\beq
p_T^{j1}>20~{\rm GeV},\quad p_T^{j2}>15~{\rm GeV},\quad \met>15~{\rm GeV}, \quad |\eta_{j1}|,|\eta_{j2}|<3.
\label{eq:cuth}
\eeq
Here $\eta_i = -\ln\tan(\theta_i/2)$ is the pseudorapidity of the particle $i$ with respect to the $p$ direction, with $i=e,\mu,j_{1,2}$.
\end{itemize}
In the electronic and hadronic modes, the typical cut efficiency of the SM background after we include the cuts in Eqs.~\eqref{eq:cute} and ~\eqref{eq:cuth} is $\mathcal{O}(10^{-4})$. 
Combining the inclusive production cross section with the background cut efficiency, the background cross section after the cuts is around $\mathcal{O}(1)~{\rm pb}$, which is still much larger than the signals. To get sensitive bounds in these channels, it is therefore necessary to further refine the analysis.
In the hadronic mode, this could be done by including jet-substructure information to single out the jet emerging from $\tau$ decay, which is expected to
be displaced from the primary vertex, have small hadron multiplicity and to be correlated with the missing energy \cite{Zhang}. We will pursue this direction in future work.
Here we will focus on the muonic channel, which is essentially background-free and thus allows for strong constraints on the CLFV coefficients.  The distributions of $p_T^\mu,~p_T^{j1}$ and $\met$ from $ep\to \tau (\rightarrow  \mu \bar\nu_\mu \nu_\tau)+X$ are shown in Fig.~\ref{Fig:event}, with Wilson coefficients set to one. 
Most of our results do not change notably if we extend the rapidity cuts in Eqs.~\eqref{eq:cute}--\eqref{eq:cuth} into the more forward/backward regions $\abs{\eta}<4$ or 5.
Tracking and particle identification performance at EIC, however, will vary over rapidity regions.  We assumed uniform identification parameters for muons and electrons in our rudimentary study. It will be interesting in future studies to study the performance particularly for forward and backward rapidities. As a preliminary example, we compare the muon pseudorapidity $\eta_\mu$ distributions for several possible signals and the DIS background in Fig.~\ref{Fig:rap}. We note that $\eta_\mu$ from most of the signals would favor the central rapidity region, although the background falls a bit faster for forward rapidity than most of the signals. This is especially true of the dipole $\Gamma_\gamma^e$ signal, which peaks significantly in the forward region. The distinct $\eta_\mu$ distributions between signals and background will be interesting in future studies to further optimize EIC sensitivity, although we may need to consider smaller $p_T$ triggers, especially if we want to consider forward jets.

\begin{table}
\begin{center}
\begin{tabular}{||c||c|c|c|c|c|c|c|}
\hline
&  $(C_{LQ,U})_{uu}$  & $(C_{LQ,U})_{cc}$ & $(C_{LQ,D})_{dd}$ & $(C_{LQ,D})_{ss}$ & $(C_{LQ,D})_{bb}$ & $c_{L \varphi}^{(1)} + c_{L \varphi}^{(3)}$\\
\hline
$\epsilon_{\rm cut} (\%)$& $9.9$ & $2.6$ & $5.8$ & $3.2$ & $0.91$ & $4.9$ \\
\hline
\hline
& $(C_{eu})_{uu}$ & $(C_{eu})_{cc}$ & $(C_{ed})_{dd}$ & $(C_{ed})_{ss}$ & $(C_{ed})_{bb}$ & $C_{e\varphi}$\\
\hline
$\epsilon_{\rm cut}(\%)$& $9.6$ & $2.5$ & $5.6$ & $3.1$ & $0.85$ & $3.3$ \\
 \hline
  \hline
&  $(C_{GG})_{\tau e}$ & $(\Gamma_\gamma^e)_{\tau e}$  & $(\Gamma_Z^e)_{\tau e}$  &  $(C_{GG})_{e \tau}$ & $(\Gamma_\gamma^e)_{e\tau }$  & $(\Gamma_Z^e)_{e\tau}$\\
\hline
$\epsilon_{\rm cut}(\%)$& $6.8$  & $0.15$ & $19$ &$6.4$ & $0.15$&$18$\\
\hline
\end{tabular} 
\end{center}
\vspace{-1em}
\caption{Cut efficiency in the muonic channel, in units of $10^{-2}$, for  various SMEFT operators at the EIC with energy $E_e=20~{\rm GeV}$ and $E_p=250~{\rm GeV}$ ($\sqrt{S}=141$ GeV). There is no background after including the kinematic cuts.}
\label{tab:eff}
\end{table}

The cut efficiency $\epsilon_{\rm cut}$ (i.e. percentage of events left intact by the cuts) for different SMEFT operators is shown in  Table~\ref{tab:eff}. 
Notice that, as in Eq. \eqref{poissonmean}, $\epsilon_{\rm cut}$ is defined after factoring out the branching ratio in a specific channel. 
For four-fermion operators,
$\epsilon_{\rm cut}$ is only sensitive to the flavor of the  initial state quark, and does not depend on the Lorentz structure and on the flavor of the quark in the final state. We can therefore use the  $\epsilon_{\rm cut}$ shown in Table~\ref{tab:eff} for $C_{LQ, U}$, $C_{LQ, D}$, $C_{eu}$ and
$C_{ed}$ for the other four-fermion operators
in  our basis. In the muonic channel, after combining all the cuts,  $\epsilon_{\rm cut}^{\rm BG}=0$, that is, we obtain a background-free process.
The typical $\epsilon_{\rm cut}$ for four-fermion operators with valence quarks is around $\sim 6\%$--$10\%$, while it reduces to $\sim 1\%$--$3\%$ for operators with heavy quarks. Notice however that we have not imposed additional selection criteria, e.g. $b$ tagging in the final state, which could further suppress the background with more moderate cuts, thus increasing $\epsilon_{\rm cut}$ for heavy quarks.
$c_{e\varphi}$, $c^{(1)}_{L\varphi} + c^{(3)}_{L\varphi}$ and the gluonic operators also have a sizable $\epsilon_{\rm cut}$, from $3\%$ to $7\%$. 
$\Gamma^e_Z$ has the biggest efficiency, around 20\%. For the photon dipole, on the other hand, $\epsilon_{\rm cut}$ is very small, $\epsilon_{\rm cut} \sim 0.1\%$, as expected from Fig. \ref{Fig:kinL}.  The cut efficiency is not sensitive to the $\tau$ polarization, the difference between 
operators with left-handed and right-handed $\tau$, such as $C_{LQ,U}$ and $C_{eu}$, being about few percent.

\begin{table}
\begin{center}
\begin{tabular}{||c | c || c| c ||c  | c ||c |c||}     
\hline
 $|(C_{L Q,\, U}^{} )_{uu}    |$        & 0.6  &  $|(C_{L u})_{uu}   |$  & 0.8 & $\left|(C^{(1)}_{L e Q u})_{uu}   \right|$ &  1.9 & $\left|(C^{(3)}_{L e Q u})_{uu}   \right|$ &0.2\\
 $|(C_{L Q, U}^{} )_{cu}      |$        & 0.6  &  $|(C_{L u})_{cu}   |$   & 0.9 &  $\left|(C^{(1)}_{L e Q u})_{cu}   \right|$ & 2.0 & $\left|(C^{(3)}_{L e Q u})_{cu}   \right|$ & 0.2\\
 $|(C_{L Q,\, U}^{} )_{uc}    |$        & 3.2  &  $|(C_{L u})_{uc}   |$  & 2.6 & $\left|(C^{(1)}_{L e Q u})_{uc}   \right|$ &8.1 & $\left|(C^{(3)}_{L e Q u})_{uc}   \right|$ &0.8\\
 $|(C_{L Q, U}^{} )_{cc}      |$        & 3.6  &   $|(C_{L u})_{cc}   |$  & 3.6 & $\left|(C^{(1)}_{L e Q u})_{cc}   \right|$ &10.0 & $\left|(C^{(3)}_{L e Q u})_{cc}   \right|$ &1.0\\
 $|(C_{L Q,\, D }^{})_{dd}    |$        & 1.0  &   $|(C_{L d})_{dd}   |$  & 1.3 & $\left|(C_{L e d Q})_{dd}   \right|$&3.2 & $|(C_{Qe})_{dd}   |$ & 0.8\\ 
 $|(C_{L Q,\, D}^{} )_{sd}    |$        & 1.1  &  $|(C_{L d})_{sd}   |$  & 1.7 & $\left|(C_{L e d Q})_{sd}   \right|$ &3.7 & $|(C_{Qe})_{sd}   |$ &1.0\\ 
 $|(C_{L Q, D}^{} )_{bd}      |$        & 1.1  &  $|(C_{L d})_{bd}   |$  & 1.8 & $\left|(C_{L e d Q})_{bd}   \right|$ &3.8 & $|(C_{Qe})_{bd}   |$  &1.7\\ 
 $|(C_{L Q, D}^{} )_{ds}      |$        & 3.2  &  $|(C_{L d})_{ds}   |$  & 2.2 & $\left|(C_{L e d Q})_{ds}   \right|$ &7.3 & $|(C_{Qe})_{ds}   |$ & 1.2\\ 
 $|(C_{L Q, D}^{} )_{ss}      |$        & 4.6  &  $|(C_{L d})_{ss}   |$  & 4.6 & $\left|(C_{L e d Q})_{ss}   \right|$ &12.3 & $|(C_{Qe})_{ss}   |$ & 1.8\\  
 $|(C_{L Q,\, D}^{} )_{bs}    |$        & 5.0  &  $|(C_{L d})_{bs}   |$  & 6.1 & $\left|(C_{L e d Q})_{bs}   \right|$&14.5 & $|(C_{Qe})_{bs}   |$ &5.4\\ 
 $|(C_{L Q,\, D }^{} )_{db}   |$        & 6.7  &  $|(C_{L d})_{db}   |$  & 4.3 & $\left|(C_{L e d Q})_{db}   \right|$ &14.6 &$|(C_{Qe})_{db}   |$ & 4.1\\   
 $|(C_{L Q, D}^{} )_{sb}      |$        & 11.5  &  $|(C_{L d})_{sb}   |$  & 9.5 & $\left|(C_{L e d Q})_{sb}   \right|$ &27.2 & $|(C_{Qe})_{sb}   |$ & 8.6\\   
 $|(C_{L Q,\, D}^{} )_{bb}    |$        & 13.6  &  $|(C_{L d})_{bb}   |$   & 13.6 & $\left|(C_{L e d Q})_{bb}   \right|$  &33.5 & $|(C_{Qe})_{bb}   |$ &13.6\\   
 \hline
 $\left|(c_{L \varphi}^{(1)} + c_{L \varphi}^{(3)} ) \right|$ &1.9 &$|c_{e \varphi} |    $ &2.7 &$\left|(\Gamma^e_\gamma)_{\tau e/e\tau} \right |$   &4.7 &$\left|(\Gamma^e_Z)_{\tau e/e\tau} \right|$ &5.0\\
  $\left|Y^\prime_{\tau e/e\tau} \right|$ &941 & $\left|(C_{GG})_{\tau e/e\tau} \right|$& 1075& $\left|(C_{G\tilde{G}})_{\tau e/e\tau} \right|$& 1075&&\\
 \hline
 \end{tabular}
 \end{center}
 \vspace{-1em}
\caption{EIC sensitivity, in units of $10^{-2}$, to CLFV  operators  at 90\% CL  with $E_e=20~{\rm GeV}$, $E_p=250~{\rm GeV}$ ($\sqrt{S}=141$ GeV) and $\mathcal{L}=100~{\rm fb}^{-1}$. Bounds on the right-handed operators  $C_{eu}$ and $C_{ed}$ are almost the same as $C_{LQ, U}$ and $C_{LQ, D}$.
}\label{EIC_sens}
 \end{table}

For the background-free channels, we can use the Bayesian posterior probability method to determine the upper limits on the CLFV coefficients; see Eq.~\eqref{eq:CL} with $n_b =0$.
The 90\% CL upper limits on the CLFV operators at the EIC, assuming  $E_e=20~{\rm GeV}$, $E_p=250~{\rm GeV}$ and $\mathcal{L}=100~{\rm fb}^{-1}$, are given in Table~\ref{EIC_sens}.  The EIC can put very strong constraints on the light quark components of four-fermion operators, ranging from 0.2\% to few percent
in dependence of the Lorentz  and quark-flavor structures of the operators. 
With our cuts, the small $\epsilon_{\rm cut}$ causes  the heavy quark components to be relatively less well constrained,
at the 10\% level.  The limits on $Z$ boson CLFV couplings and dipole operators are comparable to the four-fermion operators. 
Finally, it will be difficult to give  useful constraints on the Yukawa and gluonic operators, because of the small production cross sections at the EIC.

The polarization of the electron beam will be very useful to single out the chiral structure of SMEFT operators.
Since the cut efficiencies of CLFV operators are not sensitive to the $\tau$ polarization, the limits on CLFV coefficients with  $\lambda_e\neq 0$ can be written as
\beq
|C_i(e_L,\lambda_e)|=\frac{1}{\sqrt{1-\lambda_e}}|C_i(e_L,\lambda_e=0)|,\quad 
|C_i(e_R,\lambda_e)|=\frac{1}{\sqrt{1+\lambda_e}}|C_i(e_R,\lambda_e=0)|. 
\eeq
Here $e_{L,R}$ is the helicity of the incoming electron in $e^-p\to \tau^-X$. It is clear that a negative $\lambda_e$ would improve the limits of the operators with left-handed electron, while it would weaken the results for the right-handed electron operators and vice versa.

\section{Complementary high energy limits on CLFV operators}
\label{sect:LHC}

CLFV interactions have been probed at other high-energy collider experiments.
In particular, LEP and the LHC have searched for CLFV decays of the Higgs boson \cite{Aad:2019ugc}, $Z$ boson \cite{Akers:1995gz,Aad:2020gkd}, and $t$ quark \cite{ATLAS:2018avw,Gottardo:2676841,Ruina:2653340}.
The relevant scales for these processes are the decaying particles' masses, well within the regime of validity of SMEFT.
The ATLAS experiment has also looked for the process $p p \rightarrow \tau e$ \cite{Aaboud:2018jff}. In this case, the invariant mass of the 
$e \tau$ pair can reach values larger than 3 TeV, and the comparison of the LHC and projected EIC limits requires to make sure that one is working in the regime of validity of the EFT.

\subsection{$Z$, Higgs and $t$  decays}
\label{ZHT}

The OPAL collaboration at the LEP experiment constrained the branching ratio of the $Z$ boson into $\tau e$ to be $\textrm{BR}(Z \rightarrow e \tau) < 9.8 \cdot 10^{-6}$  ($95\%$ CL) \cite{Akers:1995gz}. This limit was recently superseded by the ATLAS collaboration \cite{Aad:2020gkd}, which found
\begin{eqnarray}\label{Zobs}
\textrm{BR}(Z \rightarrow e \tau) < 8.1 \cdot 10^{-6} \quad (95\%\; {\rm CL}).
\end{eqnarray}
This branching ratio is mostly sensitive to the operators $c_{e\varphi}$ and $c^{(1,3)}_{L\varphi}$, which induce CLFV $Z$ vertices, and to the dipole operator
$\Gamma^e_Z$. Their contributions to the branching ratio are
\begin{equation}\label{Zpre}
\textrm{BR}(Z\rightarrow e \tau) = \frac{1}{\widehat\Gamma_Z} \left(  \frac{1}{4} \left| \left[c^{(1)}_{L\varphi} + c^{(3)}_{L\varphi}\right]_{\tau e}\right|^2 
+ \frac{1}{4} |\left[c_{e\varphi}\right]_{\tau e}|^2  + \frac{m_Z^2}{2 v^2} \left( |\left[ \Gamma^e_Z\right]_{e\tau} |^2 + \left[ \Gamma^e_Z\right]_{\tau e} |^2\right)
\right),
\end{equation}
where the branching ratio includes both $e^+ \tau^-$ and $e^- \tau^+$ channels, and we used 
\begin{eqnarray}
\Gamma_Z = \frac{G_F m_Z^3}{3 \sqrt{2} \pi}   \widehat{\Gamma}_Z.
\end{eqnarray}
The dimensionless number $\widehat \Gamma_Z$ is, at leading order in QCD and EW corrections,
\begin{equation}
\widehat{\Gamma}_Z  = \sum_f  N^f_{c} (z^2_{f_L} + z^2_{f_R}),
\end{equation}
with $N^f_c = 1$ for leptons and $N^f_c = N_c$ for quarks. In terms of the observed $Z$ width, $\widehat\Gamma_Z = 3.76$.
From Eqs. \eqref{Zobs} and \eqref{Zpre} we get the $90\%$ CL limits
\begin{eqnarray}\label{Zbound}
|c_{e\varphi}| < 1.0 \cdot 10^{-2}, \quad  |c^{(1)}_{L\varphi} + c^{(3)}_{L\varphi}| < 1.0 \cdot 10^{-2}, \quad 
|\left[\Gamma^e_{Z}\right]_{e\tau, \, \tau e}| < 1.9 \cdot 10^{-2}.
\end{eqnarray}

The Higgs decay width into $\tau e$ is given by   \cite{Harnik:2012pb}
\begin{equation}
\Gamma(H_0\rightarrow e^- \tau^+ + \tau^- e^+) = \frac{m_H}{8\pi} \left( \left[Y_e^\prime \right]_{\tau e}^2 + \left[Y_e^\prime \right]_{e \tau}^2  \right).
\end{equation}
Using the bounds on the branching ratio \cite{Aad:2019ugc} 
\begin{equation}
\textrm{BR}(H_0\rightarrow e^- \tau^+ + \tau^- e^+) \equiv \mathcal B_e <  4.7 \cdot 10^{-3}  \quad (95\%  {\rm CL}),
\end{equation}
and the relation:
\begin{equation}
\left( \left[Y_e^\prime \right]_{\tau e}^2 + \left[Y_e^\prime \right]_{e \tau}^2  \right) = \frac{8 \pi}{m_H}  \frac{\mathcal B_e}{1 - \mathcal B_e} \, \Gamma_{\rm SM},
\end{equation}
where the SM Higgs width is $\Gamma_{\rm SM} = 4.07 \cdot 10^{-3}$ GeV,
one gets the strong constraint \cite{Aad:2019ugc}
\begin{eqnarray}
\left[Y_e^\prime \right]_{\tau e, \, e \tau} < 2.0 \cdot 10^{-3}. 
\end{eqnarray}

The ATLAS experiment has put bounds on the top branching ratio ${\rm BR}(t \rightarrow q \ell \ell^\prime) < 1.86 \cdot 10^{-5}$ ($95\%$ CL) \cite{ATLAS:2018avw}. The analysis 
is sensitive to the $e \tau$, $\mu \tau$ and $e\mu$ channels, putting the strongest constraints on the latter.
To obtain a constraint on the $e \tau$ channel, we first of all 
get the yield and shape of the $t \rightarrow q e \tau$ and $t \rightarrow q \mu \tau$  signal distributions by subtracting the signal histograms with and without $\tau$ vetos in Fig. 3 of Ref. \cite{ATLAS:2018avw}. We then estimate the $t \rightarrow q e \tau$ fraction of signal events by accounting for the different electron versus muon acceptance,  obtained from the yields of the two validation regions given in Ref. \cite{Gottardo:2676841}. We then used signal and background events in a likelihood analysis using \texttt{pyhf} \cite{pyhf_joss,pyhf,ATL-PHYS-PUB-2019-029},   
obtaining\footnote{We thank C. A. Gottardo for illustrating the procedure for the extraction of bounds on $t \rightarrow q e \tau$ from Ref. \cite{ATLAS:2018avw}, and for checking the limit in Eq. \eqref{topBR}. }
\begin{equation}\label{topBR}
 \textrm{BR}(t \rightarrow q e \tau) \le 2.2 \cdot 10^{-4}. 
\end{equation}
Dedicated analyses in the $\tau$ channels are in progress, and preliminary results for  $\textrm{BR}(t \rightarrow q \mu \tau)$  show bounds at the $10^{-4}$ level \cite{Ruina:2653340}.
The BR for the decay $t\to q e^+\tau^-$ is \cite{Davidson:2015zza}
\begin{eqnarray}
& & \textrm{BR}(t\to q e^+\tau^-)=\frac{1}{6 \widehat \Gamma_t}\! \left(\frac{m_t}{4 \pi v}\right)^{\!2} \! \! \bigg[4\Bigl(\left| \left[C_{LQ, U}\right]_{\tau e q t}\right|^2 \!+\left| \left[C_{Lu}\right]_{\tau e q t}\right|^2 \!
+\left|\left[C_{Qe}\right]_{\tau e q t}\right|^2  \! +\left|\left[C_{e u}\right]_{\tau e q t}\right|^2 \Bigr) \nn \\
& & \qquad\quad + \left| \left[C_{LeQu}^{(1)}\right]_{\tau e q t} \right|^2
+ \left| \left[C_{LeQu}^{(1)}\right]_{e \tau t q} \right|^2 
+ 48 \left| \left[C_{LeQu}^{(3)}\right]_{\tau e q t} \right|^2 + 48 \left| \left[C_{LeQu}^{(3)}\right]_{e \tau t q} \right|^2
\bigg],
\end{eqnarray}
where we expressed the SM top width as 
\begin{equation}
\Gamma(t \rightarrow W b) = \frac{m_t^3}{16 \pi v^2} \widehat{\Gamma}_t, 
\end{equation}
with $\widehat{\Gamma}_t$ a dimensionless function of $V_{tb}$, $m_t$ and $m_W$. In terms of the measured top width, $\widehat{\Gamma}_t = 1.01^{+0.14}_{-0.11}$ \cite{Zyla:2020zbs}.
The resulting constraints on top CLFV operators are 
\begin{equation}
\left[C_{LQ,U} \right]_{\tau e qt}< 0.35,\qquad
\left[C^{(1)}_{LeQu} \right]_{\tau e qt}< 0.7,\qquad
\left[C^{(3)}_{LeQu} \right]_{\tau e qt}< 0.1,
\end{equation}
where the limit on $C_{Lu},~C_{Qe}$ and $C_{eu}$ is the same as the one on $C_{LQ,U}$.

\subsection{CLFV Drell-Yan}

\begin{figure}
\vspace{-1em}
\center
 \includegraphics[width=12cm]{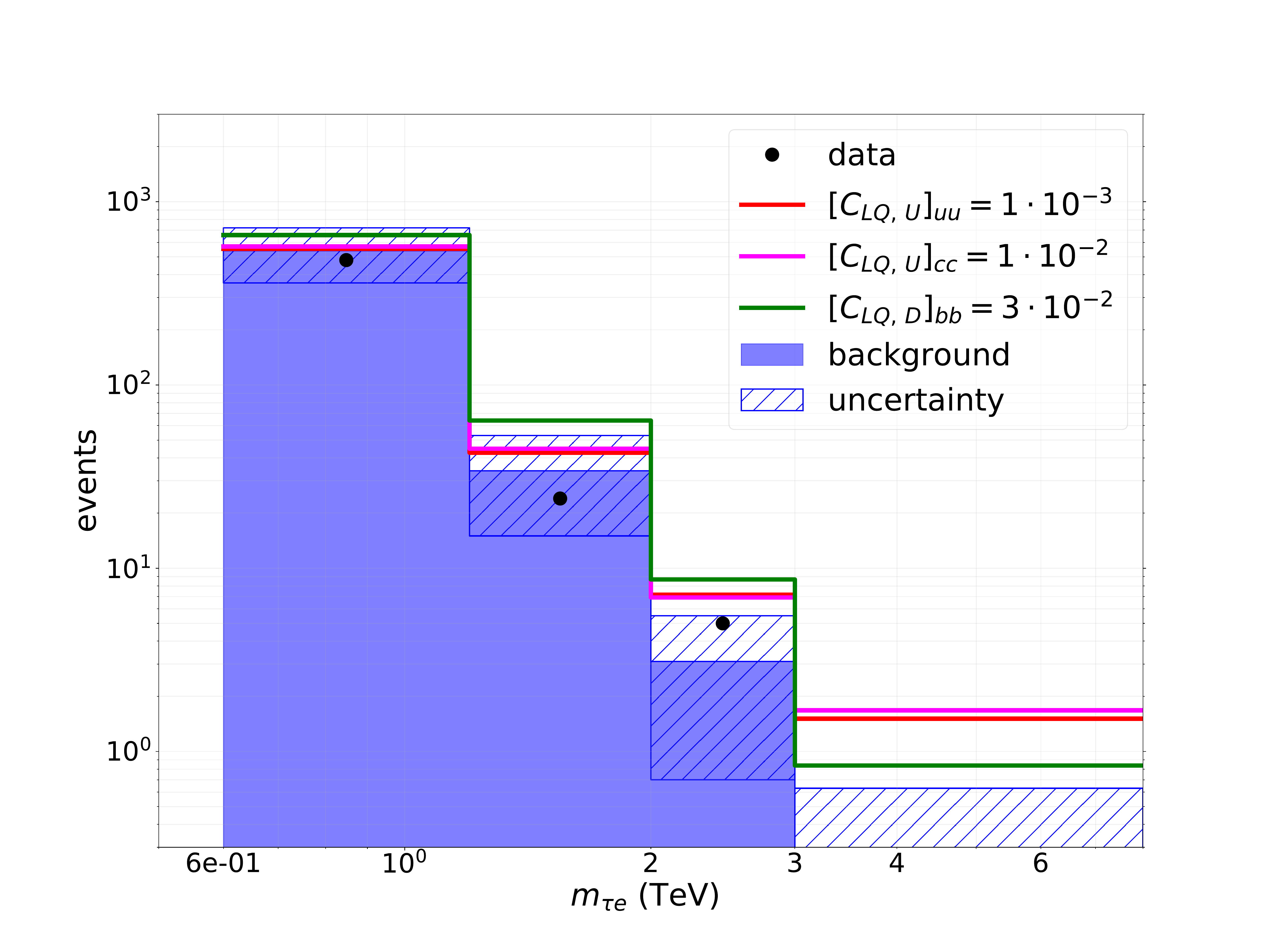}
\vspace{-1em}
\caption{Observed and background events in $p p \rightarrow e \tau$, at $\sqrt{S} = 13$ TeV with luminosity of 36.1 fb$^{-1}$,
as a function of the $\tau$-electron invariant mass $m_{\tau e}$ \cite{Aaboud:2018jff}.
The dashed area denotes the uncertainty on the backgrounds. The red, magenta and green lines denote the sum of the background and signal events induced by SMEFT operators.
}\label{LHCnum}
\end{figure}

The SMEFT operators in Eqs. \eqref{eq:dipole} and \eqref{eq:fourfermion} can also affect the process $p p \rightarrow e \tau$, which has been studied in Refs. \cite{Aaboud:2016hmk,Aaboud:2018jff}. These analyses look for $e\tau$, $e\mu$ and $\mu \tau$ pairs in several invariant mass bins, and they provide the strongest constraints at high invariant mass, where the SM background is highly suppressed. They are thus most sensitive to four-fermion operators \cite{Han:2010sa}. 
In the $e\tau$ channel, Ref. \cite{Aaboud:2018jff} considered 6 invariant mass bins, from $m_{e\tau} < 300$ GeV to $m_{e\tau} > 3$ TeV.
The number of observed and background events in the four invariant mass bins we consider are shown in Fig. \ref{LHCnum}.

\begin{table}
\center
\begin{tabular}{||c || c | c || c |c  || c |c ||}
\hline
                            &              &  90\% CL  &  & 90\% CL &  & 90\% CL  \\ \hline \hline
$C_{L Q,\, U}^{}$, $C_{eu}$ &  $uu$        & $9.8 \cdot 10^{-4}$  &  $uc$  & $1.9 \cdot 10^{-3}$ & $cc$ & $9.6 \cdot 10^{-3}$  \\
$C_{L u}^{}      $          &  $uu$        & $9.9 \cdot 10^{-4}$  &  $uc$  & $2.0 \cdot 10^{-3}$ & $cc$ & $9.7 \cdot 10^{-3}$  \\
\hline
\hline
$C_{L Q,\, D}^{}$, $C_{ed}$ &  $dd$        & $1.4 \cdot 10^{-3}$  &  $ds$  & $2.8 \cdot 10^{-3}$ & $db$ & $4.9 \cdot 10^{-3}$  \\
                            &  $ss$        & $6.3 \cdot 10^{-3}$  &  $sb$  & $1.2 \cdot 10^{-2}$ & $bb$ & $2.9 \cdot 10^{-2}$  \\
$C_{L d}$                   &  $dd$        & $1.4 \cdot 10^{-3}$  &  $ds$  & $2.8 \cdot 10^{-3}$ & $db$ & $5.0 \cdot 10^{-3}$  \\
                            &  $ss$        & $6.3 \cdot 10^{-3}$  &  $sb$  & $1.2 \cdot 10^{-2}$ & $bb$ & $2.9 \cdot 10^{-2}$  \\
$C_{Qe}$                    &  $dd$        & $8.3 \cdot 10^{-4}$  &  $ds$  & $1.5 \cdot 10^{-3}$ & $db$ & $5.0 \cdot 10^{-3}$  \\
                            &  $ss$        & $4.0 \cdot 10^{-3}$  &  $sb$  & $1.2 \cdot 10^{-2}$ & $bb$ & $3.0 \cdot 10^{-2}$  \\                          
\hline \hline
$C_{LedQ}$                  &  $dd$        & $1.2 \cdot 10^{-3}$  &  $ds$  & $3.3 \cdot 10^{-3}$ & $db$ & $5.8 \cdot 10^{-3}$  \\
                            &  $ss$        & $5.2 \cdot 10^{-3}$  &  $sb$  & $1.5\cdot 10^{-2}$ & $bb$ & $2.5 \cdot 10^{-2}$  \\                          
$C_{Le Qu}^{(1)}$           &  $uu$        & $8.1 \cdot 10^{-4}$  &  $uc$  & $2.3 \cdot 10^{-3}$ & $cc$ & $8.0 \cdot 10^{-3}$  \\
                            &  $tt$ & $2.5\cdot 10^{-2}$ & & & &\\
$C_{Le Qu}^{(3)}$           &  $uu$        & $3.5 \cdot 10^{-4}$  &  $uc$ & $9.9 \cdot 10^{-4}$ & $cc$ & $3.4 \cdot 10^{-3}$  \\
 \hline
 \hline 
 $\Gamma_{\gamma}^e$  & & 0.10& & &&\\
 \hline
 \end{tabular}
\caption{$90\%$ CL bounds on four-fermion LFV operators from the LHC.
The coefficients are evaluated at the scale $\mu= 1$ TeV. For quark-flavor-changing operators, the bounds on the $q_j q_i$ and $q_i q_j$ components are identical.}\label{LHC_bounds}
\end{table}

We generate CLFV Drell-Yan events from SMEFT operators with a trivial modification of the \texttt{POWHEG} implementation of Ref. \cite{Alioli:2018ljm}.
We include NLO QCD corrections, which, as shown in Ref. \cite{Alioli:2018ljm}, can give a $\sim 30\%$ correction in the high invariant mass bins, and 
the parton-level events are showered with \texttt{Pythia8}, 
which we also use for the decays of the $\tau$ lepton. We apply the selection cuts described in Ref. \cite{Aaboud:2018jff},
in particular the electron and the jet from hadronic $\tau$ decays are required to have  $p_T > 65$ GeV and $|\eta |< 2.5$.
We simulate the detector and $\tau$ tagging  with \texttt{Delphes}.
The effect of selecting hadronic decays, of the cut on the electron and jet $p_T$, and of the efficiency of $\tau$-tagging 
combine to give a selection efficiency between $\epsilon = 0.24$ and $\epsilon = 0.27$
in the four different invariant mass bins. The efficiencies do not show a strong dependence on the Lorentz or flavor structure of the four-fermion operators.
We also simulated $Z$ vertex corrections and dipole operators, but, for coefficients compatible with the bounds in Eq.  \eqref{Zbound}, they are negligible. 
The top scalar operators $\left[C^{(1)}_{LeQu}\right]_{\tau e t t}$ and $\left[C^{(1)}_{LeQu}\right]_{e \tau t t}$ contribute to 
CLFV Drell-Yan via the gluon fusion process $g g \rightarrow \tau e$ at  the loop level. We parametrize the finite one-loop corrections as form factors that are function of the external momenta. The form factors are implemented as effective new vertices in a dedicated UFO model file, which is then used in \texttt{MadGraph5}.  
We have compared the cross section with the amplitude in Appendix \ref{GGatLHC} to the \texttt{MadGraph5} code and found excellent agreement.  
QCD corrections are taken into account by introducing a constant $\kappa$ factor in our simulation, i.e. $\kappa\simeq3$~\cite{Dittmaier:2011ti}. 
We checked that this is a reasonable assumption by simulating off-shell Higgs production via gluon fusion in the relevant invariant mass bins with  
\texttt{MCFM}~\cite{Boughezal:2016wmq,Campbell:2019dru}. Non-standard Yukawa couplings would contribute via the same mechanism, but the constraints from off-shell Higgs production are much weaker than those shown in Section \ref{ZHT}.

In Table \ref{LHC_bounds} we show the 90\% CL bounds on the coefficients of effective operators, evaluated at the renormalization scale $\mu= 1$ TeV.
To obtain the bounds, we use a generalization of Eq. \eqref{eq:CL} to multiple bins \cite{Favara:1997ui}.
Since the uncertainties on the background are non-negligible, 
we generate a large number of pseudoexperiments, assuming the number of signal and background events in each bin to follow a Poisson distribution.
The mean $\mu_i$ of the distributions of signal events is given by Eq. \eqref{poissonmean}, generalized to several bins. For each value of the operator coefficient $C$,
the mean $\mu_{b_i}$ is picked randomly in the $1\sigma$ intervals shown  in Figure \ref{LHCnum}. 
Each pseudoexperiment is characterized by a number of signal and background events, $n_{s_i}$ and $n_{b_i}$.
We consider only the pseudoexperiments with $n_{b_i} \le n_i$, where $n_i$ is the number of observed events, 
and we construct the confidence level by counting the ratio of pseudoexperiments for which $n_{s_i} + n_{b_i} < n_i$. If this ratio is less than $10\%$, $C$ is excluded.

The bounds in Table \ref{LHC_bounds} are dominated by the last two bins, and our results agree well with  Ref. 
\cite{Angelescu:2020uug}, which also recasts the analysis of Ref. \cite{Aaboud:2018jff} in terms of SMEFT operators.
The LHC puts very strong constraints on operators with two $u$ or two $d$ quarks. The bounds deteriorate to the few percent level in the case of operators with heavy flavors.
Converting into a new physics scale, vector operators with valence quarks give $\Lambda \gtrsim 6.5$ TeV, while operators with one valence and one sea quark give $\Lambda \gtrsim$ 4 TeV.
These scales are larger than the probed $m_{e\tau}$, and the SMEFT analysis is thus justified. For operators with two sea quarks $\Lambda \gtrsim  1.5$--$2$ TeV, and in this case it might be more appropriate to consider explicit BSM degrees of freedom. The bound on top scalar operators are also at the few-percent level, 
of similar size as other heavy flavors. 
The bound on the photon dipole $\Gamma^e_\gamma$ is at the 10\% level, much weaker than from $\tau$ decays. 

We have so far assumed that the SMEFT is valid up to scales of a few TeV.
For  BSM physics contributing at tree level in the $s$-channel, Ref. \cite{Aaboud:2018jff} found comparable limits on the masses 
of new CLFV degrees of freedom, in the range of 4--5 TeV. The limits in Table \ref{LHC_bounds} can be weakened if BSM particles are exchanged in the $t$-channel, as for example in the case of scalar leptoquarks discussed in Section \ref{sect:leptoquarks}.
At LO in QCD, we can study this scenario by replacing the coefficients of SMEFT four-fermion operators with    
\begin{eqnarray}\label{tchan}
C \rightarrow C \frac{M^2}{M^2 -t },
\end{eqnarray}
where $M$ denotes the mass of the exchanged particle.
We find that the bounds on the light-quark components of the four-fermion operators in Table \ref{LHC_bounds} worsen by a factor of 5 (2) for $t$-channel exchange of a particle of mass $M = 1$ TeV ($2$ TeV).

\section{Low-energy observables}
\label{low-energy}

\begin{table}[t]
\centering
\begin{tabular}{c c}
\hline
Decay mode & Upper limit on BR ($90\%$ C.L.) \\ 
\hline\hline
$\tau^- \to e^- \gamma$ & $<3.3\times 10^{-8}$    \\
$\tau^- \to e^- e^+ e^+$ & $<2.7\times 10^{-8}$   \\
$\tau^- \to e^- \mu^+\mu^-$ & $<2.7\times 10^{-8}$   \\
$\tau^- \to e^- \pi^0$ & $<8.0\times 10^{-8}$    \\
$\tau^- \to e^- \eta$ & $<9.2\times 10^{-8}$    \\
$\tau^- \to e^- \eta^{\prime}$ & $<1.6\times 10^{-7}$    \\
$\tau^- \to e^- K_S^0$ & $<2.6\times 10^{-8}$   \\
$\tau^- \to e^- \pi^+\pi^-$ & $<2.3\times 10^{-8}$   \\
$\tau^- \to e^- \pi^+ K^-$ & $<3.7\times 10^{-8}$   \\
$\tau^- \to e^- \pi^- K^+$ & $<3.1\times 10^{-8}$   \\
\hline
$B^0 \to e^{\pm}\tau^{\mp} $ & $<2.8\times 10^{-5}$   \\
$B^+ \to \pi^{+} e^{+} \tau^{-} $ & $<7.4\times 10^{-5}$    \\
$B^+ \to \pi^{+} e^{-} \tau^{+} $ & $<2.0\times 10^{-5}$    \\
$B^+ \to K^{+} e^{+} \tau^{-} $ & $<4.3\times 10^{-5}$    \\
$B^+ \to K^{+} e^{-} \tau^{+} $ & $<1.5\times 10^{-5}$    \\
\hline 
\end{tabular}
\caption{Summary of the low-energy decay modes and current experimental limits on their branching ratios \cite{Zyla:2020zbs}.}
\label{experiment_limit}
\end{table}

We next discuss CLFV low-energy observables. 
The relatively heavy mass of the $\tau$ lepton compared to light hadrons offers a rich array of channels to search for  CLFV $\tau$ decays
including $\tau\to e\gamma$, the purely leptonic channels $\tau \rightarrow 3e$ and $\tau \rightarrow e \mu \mu$,
and semileptonic decays such as $\tau \to e\pi^0, \eta^{(\prime)}$ and $\tau \to e\pi^+\pi^-$. Table \ref{experiment_limit} summarizes the LFV decay modes that we consider and the current experimental upper limits on each branching ratio (BR) at $90\%$ C.L. While most of the $\tau$ decays are associated with the CLFV quark-flavor-conserving interactions, the decay modes $\tau \rightarrow e K^0_S$
and $\tau \rightarrow e K^\pm \pi^{\mp}$ can probe the LFV quark-flavor-violating interactions. In Table \ref{low-energy_operator}, we present a tabulation of which operators contribute to each decay channel. The parentheses indicate decays that are induced only at 1- and/or 2-loop level. For example, the LFV Yukawa interaction originating from $\psi^2\varphi^3$ can induce $\tau \to e\gamma$ through 1- and 2-loop diagrams. The semileptonic four-fermion operators denoted as $\psi^4$  contribute to the leptonic $\tau$ decays via renormalization group running.

\begin{table}[t]
\centering
\begin{tabular}{c c c c c}
\hline
Decay mode  & $\psi^2X\varphi$ & $\psi^2\varphi^2D$  & $\psi^2\varphi^3$ & $\psi^4$  \\ 
\hline \hline
$\tau \to e \gamma$   & $\checkmark$ &   & $(\checkmark)$  & $(\checkmark)$ \\
$\tau \to e e^+ e^-$ & $\checkmark$  & $\checkmark$  & $\checkmark$  & ($\checkmark$)   \\
$\tau \to e \mu^+\mu^-$ & $\checkmark$  & $\checkmark$  & $\checkmark$ & ($\checkmark$)   \\
$\tau \to e \pi^0$        &      & $\checkmark$ & & $\checkmark$    \\
$\tau \to e \eta$          &  & $\checkmark$   & & $\checkmark$   \\
$\tau \to e \eta^{\prime}$ & & $\checkmark$  &  & $\checkmark$    \\
$\tau \to e K_S^0$ & & & & $\checkmark$   \\
$\tau \to e \pi^+\pi^-$ & $\checkmark$ & $\checkmark$ & $\checkmark$ & $\checkmark$   \\
$\tau \to e  K^{\pm} \pi^{\mp} $ & $ $ & $ $ & $ $ & $\checkmark$   \\
\hline
$B^0 \to e^{\pm}\tau^{\mp} $ & &  & & $\checkmark$  \\
$B^+ \to \pi^{+} e^{\pm} \tau^{\mp} $ & & & &  $\checkmark$   \\
$B^+ \to K^{+} e^{\pm} \tau^{\mp} $ &  & & &   $\checkmark$  \\  
\hline 
\end{tabular}
\caption{Illustration of the contributions from six different types of gauge-invariant CLFV operators to low-energy decay modes. The parentheses imply that the operator induces the decay only at 1- or 2-loop level. 
\label{low-energy_operator}
}
\end{table}

Heavy $D$ and $B$ mesons, $J/\psi$ and $\Upsilon$ and other quarkonia can decay into electrons and $\tau$ leptons, offering additional handles on CLFV interactions.
$D$ and $B$ decays probe flavor-changing couplings. At the moment, there are no bounds on $D_0 \rightarrow \tau^{\pm} e^{\mp}$.
This decay would put interesting constraints on the $uc$ and $cu$ components of the flavor matrices introduced in Section \ref{basis}, which, as we will see, are otherwise unconstrained at low energy. $B$ decays put strong constraints on the $bd$, $db$, $bs$ and $sb$ elements. Quarkonium decays constrain the $cc$ and $bb$ components, but the limits are weaker than those from $\tau$ decays.  

We start this section by introducing the low-energy basis in Section \ref{low-energy_basis}. We then discuss quark-flavor-conserving $\tau$
and quarkonium decays in Section \ref{QFCO} and quark-flavor-violating observables in Section \ref{QFVO}. Additional low-energy observables that indirectly probe CLFV interactions are studied in Section \ref{sect:indirect}.

\subsection{The low-energy basis}
\label{low-energy_basis}

In order to study the low-energy observables, we first map the LFV operators listed in Section \ref{basis} onto 
a low-energy $SU(3)_c \times U(1)_{\rm em}$ EFT (LEFT). The matching can be done more immediately in the basis of Ref. \cite{Jenkins:2017jig,Jenkins:2017dyc,Dekens:2019ept}, from which we differ only in the fact that we factorize dimensionful parameters so that the Wilson coefficients of the LEFT operators become dimensionless.

At dimension five, we consider leptonic dipole operators \begin{equation}
\mathcal L_5 = - \frac{e}{2 v}  \bar e_{L}^{p} \, \sigma^{\mu \nu} \left[ \Gamma^{e}_\gamma\right]_{p r} {e_{R}^{r}} F_{\mu \nu}  + {\rm h.c.},
\end{equation}
where $p,r$ are leptonic flavor indices.

At dimension six, there are several semileptonic four-fermion operators.
Those relevant for direct LFV probes have two charged leptons. There are eight vector-type operators
\begin{align}\label{eq:fourfermion2}
\mathcal L_{6} = -\frac{4 G_F}{\sqrt{2}} & \bigg( 
 C^{eu}_{\rm VLL}\, \bar e_L \gamma^\mu e_L \, \bar u_L \gamma_\mu u_L
+ C^{ed}_{\rm VLL}\, \bar e_L \gamma^\mu e_L \, \bar d_L \gamma_\mu d_L 
+ C^{eu}_{\rm VRR}\, \bar e_R \gamma^\mu e_R \, \bar u_R \gamma_\mu u_R \nn \\
&  + C^{ed}_{\rm VRR}\, \bar e_R \gamma^\mu e_R \, \bar d_R \gamma_\mu d_R 
+     C^{ue}_{\rm VLR}\, \bar e_R \gamma^\mu e_R \, \bar u_L \gamma_\mu u_L
+ C^{de}_{\rm VLR}\, \bar e_R \gamma^\mu e_R \, \bar d_L \gamma_\mu d_L \nonumber \\ 
&  + C^{eu}_{\rm VLR}\, \bar e_L \gamma^\mu e_L \, \bar u_R \gamma_\mu u_R 
 + C^{ed}_{\rm VLR}\, \bar e_L \gamma^\mu e_L \, \bar d_R \gamma_\mu d_R 
\bigg), 
\end{align}
and six scalar-tensor type operators
\begin{align}\label{eq:fourfermion3}
\mathcal L_{6} = -\frac{4 G_F}{\sqrt{2}} & \bigg( 
  C^{eu}_{\rm SRR}\, \bar e_L  e_R \, \bar u_L  u_R
+ C^{ed}_{\rm SRR}\, \bar e_L  e_R \, \bar d_L  d_R 
+ C^{eu}_{\rm TRR}\, \bar e_L \sigma^{\mu\nu} e_R \, \bar u_L \sigma_{\mu \nu} u_R \\
& + C^{ed}_{\rm TRR}\, \bar e_L \sigma^{\mu\nu} e_R \, \bar d_L \sigma_{\mu \nu} d_R 
+     C^{eu}_{\rm SRL}\, \bar e_L  e_R \, \bar u_R  u_L
+ C^{ed}_{\rm SRL}\, \bar e_L  e_R \, \bar d_R d_L  
\bigg) + \textrm{h.c.} \nn
\end{align}
There are in addition four purely leptonic operators
\begin{align}
\label{eq:fourlepton}
{\cal L}_{6}=-\frac{4G_F}{\sqrt{2}} & \Bigl[C_{\rm VLL}^{ee}\bar{e}_L\gamma^{\mu}e_L\bar{e}_L\gamma_{\mu}e_L +C_{\rm VRR}^{ee}\bar{e}_R\gamma^{\mu}e_R\bar{e}_R\gamma_{\mu}e_R +C_{\rm VLR}^{ee}\bar{e}_L\gamma^{\mu}e_L\bar{e}_R\gamma_{\mu}e_R  \nonumber\\
&
+\bigl(C_{\rm SRR}^{ee}\bar{e}_Le_R\bar{e}_Le_R +{\rm h.c.} \bigr) \Bigr].
\end{align}

LFV operators can also affect probes with one or two neutrinos, in which the neutrino flavor is not observed.
There are four operators with two neutrinos, which will affect rare meson decays,
\begin{align}\label{eq:fourfermion4}
\mathcal L_{6} = -\frac{4 G_F}{\sqrt{2}} &  \bigg( 
  C^{\nu u}_{\rm VLL}\, \bar \nu_L \gamma^\mu \nu_L \, \bar u_L \gamma_\mu u_L
+ C^{\nu d}_{\rm VLL}\, \bar \nu_L \gamma^\mu \nu_L \, \bar d_L \gamma_\mu d_L 
 + C^{\nu u}_{\rm VLR}\, \bar \nu_L \gamma^\mu \nu_L \, \bar u_R \gamma_\mu u_R \nn \\
&   + C^{\nu d}_{\rm VLR}\, \bar \nu_L \gamma^\mu \nu_L \, \bar d_R \gamma_\mu d_R 
\bigg),
\end{align}
and five charged-current operators 
\begin{align}\label{eq:fourfermion5}
\mathcal L_{6} = -\frac{4 G_F}{\sqrt{2}} &  \bigg(
  C^{\nu e d u}_{\rm VLL}\, \bar \nu_L \gamma^\mu e_L \, \bar d_L \gamma_\mu u_L
+ C^{\nu e d u}_{\rm VLR}\, \bar \nu_L \gamma^\mu e_L \, \bar d_R \gamma_\mu u_R \\ 
& + C^{\nu e d u}_{\rm TRR}\, \bar \nu_L \sigma^{\mu \nu} e_R \, \bar d_L \sigma_{\mu \nu} u_R
 + C^{\nu e d u}_{\rm SRR}\, \bar \nu_L  e_R \, \bar d_L  u_R 
 + C^{\nu e d u}_{\rm SRL}\, \bar \nu_L  e_R \, \bar d_R  u_L
\biggr) + {\rm h. c.} \nn
\end{align}
The coefficients of the operators in Eqs.~\eqref{eq:fourfermion2}, \eqref{eq:fourfermion3}, \eqref{eq:fourfermion4} and \eqref{eq:fourfermion5} are not all independent, if one matches from SMEFT.
For example, the four-fermion contributions to the semi-leptonic vector operators with charged leptons in \eq{fourfermion2} are given by:
\begin{subequations}
\begin{eqnarray}
\Big[ C^{eu}_{\rm VLL} \Big]_{\tau e j i} &=& \Big[ C_{LQ, U}\Big]_{\tau e j i} + \delta_{i j} \Big[c^{(1)}_{L\varphi} + c^{(3)}_{L\varphi}\Big]_{\tau e} z_{u_L}, \\
\Big[ C^{ed}_{\rm VLL} \Big]_{\tau e j i} &=& \Big[ C_{LQ, D}\Big]_{\tau e j i} + \delta_{i j} \Big[c^{(1)}_{L\varphi} + c^{(3)}_{L\varphi}\Big]_{\tau e} z_{d_L}, \\ 
\Big[ C^{eu}_{\rm VRR} \Big]_{\tau e j i} &=& \Big[ C_{eu}\Big]_{\tau e j i}    + \delta_{i j} \Big[c_{e\varphi}\Big]_{\tau e} z_{u_R}, \\  
\Big[ C^{ed}_{\rm VRR} \Big]_{\tau e j i} &=& \Big[ C_{ed}\Big]_{\tau e j i}    + \delta_{i j} \Big[c_{e\varphi}\Big]_{\tau e} z_{d_R}, \\
\Big[ C^{eu}_{\rm VLR} \Big]_{\tau e j i} &=& \Big[ C_{Lu}\Big]_{\tau e j i}    + \delta_{i j} \Big[c^{(1)}_{L\varphi} + c^{(3)}_{L\varphi}\Big]_{\tau e} z_{u_R}, \\
\Big[ C^{ed}_{\rm VLR} \Big]_{\tau e j i} &=& \Big[ C_{Ld}\Big]_{\tau e j i}    + \delta_{i j} \Big[c^{(1)}_{L\varphi} + c^{(3)}_{L\varphi}\Big]_{\tau e} z_{d_R}, \\  
\Big[ C^{ue}_{\rm VLR} \Big]_{\tau e j i} &=& \Big[ V_{\rm CKM} C_{Qe} V^{\dagger}_{\rm CKM}\Big]_{\tau e j i} + \delta_{i j} \Big[ c_{e\varphi}\Big]_{\tau e} z_{u_L}, \\ 
\Big[ C^{de}_{\rm VLR} \Big]_{\tau e j i} &=& \Big[ C_{Qe} \Big]_{\tau e j i}   + \delta_{i j} \Big[ c_{e\varphi}\Big]_{\tau e} z_{d_L}.
\end{eqnarray}
\end{subequations}
The coefficients of the leptonic operators in \eq{fourlepton} are given by:
\begin{subequations}
\begin{eqnarray}
\Big[C_{\rm VLL}^{ee} \Big]_{p r st}&=&   \left[C_{LL}\right]_{p r s t}   + 
\frac{z_{e_L}}{4} \left[ \left(c^{(1)}_{L\varphi}+c^{(3)}_{L\varphi} \right)_{pr}\delta_{st} + \left(c^{(1)}_{L\varphi}+c^{(3)}_{L\varphi} \right)_{p t}\delta_{sr} \right]  \\
& & + \frac{z_{e_L}}{4} \left[ \left(c^{(1)}_{L\varphi}+c^{(3)}_{L\varphi} \right)_{st}\delta_{pr} + \left(c^{(1)}_{L\varphi}+c^{(3)}_{L\varphi} \right)_{s r}\delta_{pt } \right] \,, \nn\\
\Big[C_{\rm VRR}^{ee} \Big]_{p r st}&=& \left[C_{ee}\right]_{p r s t} +   \frac{z_{e_R}}{4} \left[
\left(c_{e\varphi}\right)_{p r}\delta_{st} + \left(c_{e\varphi}\right)_{p t}\delta_{s r}  \right]  \\
& &  +   \frac{z_{e_R}}{4} \left[
\left(c_{e\varphi}\right)_{s t}\delta_{p r} + \left(c_{e\varphi}\right)_{s r}\delta_{p t}  \right]  \nn \\
\Big[C_{\rm VLR}^{ee} \Big]_{prst}&=& \left[C_{Le}\right]_{pr st}
+
z_{e_R}\left(c^{(1)}_{L\varphi}+c^{(3)}_{L\varphi}\right)_{pr}\delta_{st}+z_{e_L}\left(c_{e\varphi}\right)_{st}\delta_{pr}
,\\
\Big[C_{\rm SRR}^{ee} \Big]_{p r st}&=&- \frac{v^2}{2m^2_h}\left( \left(Y_e^{\prime} \right)_{pr }\left(Y_e \right)_{st}\delta_{st}
+  \left(Y_e^{\prime} \right)_{s t}\left(Y_e \right)_{p r}\delta_{pr}
\right).
\end{eqnarray}
\end{subequations}
The coefficients of the vector charged-current operators in \eq{fourfermion5} are given by:
\begin{subequations}
\begin{eqnarray}
\Big[C^{\nu e d u}_{\rm VLL}\Big]_{\nu_\tau e j i}  &=& \Big[ C_{LQ, D} V_{\rm CKM}^{\dagger} - V^{\dagger}_{\rm CKM} C_{LQ, U}\Big]_{\tau e j i} 
- \Big[c^{(3)}_{L\varphi}\Big]_{\nu_\tau e} \Big[ V_{\rm CKM}^{\dagger} \Big]_{j i}, \\
\Big[ C^{\nu e d u}_{\rm VLR} \Big]_{\nu_\tau e j i} &=& 0,
\end{eqnarray}
\end{subequations}
while the neutrino operators in \eq{fourfermion4} are 
\begin{subequations}
\begin{eqnarray}
\Big[C^{\nu u}_{\rm VLL}  \Big]_{\nu_\tau \nu_e j i}  &=&  \Big[V_{\rm CKM } C_{LQ, D}  V^{\dagger}_{\rm CKM}\Big]_{\tau e j i} +  \delta_{i j} \Big[c^{(1)}_{L\varphi} - c^{(3)}_{L\varphi}\Big]_{\tau e} z_{u_L}, \\
\Big[C^{\nu d}_{\rm VLL}  \Big]_{\nu_\tau \nu_e j i}  &=&  \Big[V^{\dagger}_{\rm CKM} C_{LQ, U} V_{\rm CKM}\Big]_{\tau e j i}   +  \delta_{i j} \Big[c^{(1)}_{L\varphi} - c^{(3)}_{L\varphi}\Big]_{\tau e} z_{d_L}, \\
\Big[C^{\nu u}_{\rm VLR}  \Big]_{\nu_\tau \nu_e j i}  &=&  \Big[C_{Lu}\Big]_{\tau e j i}  + \delta_{i j} \Big[c^{(1)}_{L\varphi} - c^{(3)}_{L\varphi}\Big]_{\tau e} z_{u_R} , \\
\Big[ C^{\nu d}_{\rm VLR} \Big]_{\nu_\tau \nu_e j i}  &=&  \Big[C_{Ld}\Big]_{\tau e j i}  + \delta_{i j} \Big[c^{(1)}_{L\varphi} - c^{(3)}_{L\varphi}\Big]_{\tau e} z_{d_R}.  
\end{eqnarray}
\end{subequations}

The scalar and tensor operators, $C^{(1)}_{LeQu}$, $C^{(3)}_{LeQu}$ and $C^{}_{LedQ}$, and the LFV Yukawa $Y^\prime_e$ match onto scalar and tensor operators \eq{fourfermion3} at low energy. In the neutral current sector one finds
\begin{subequations}
\begin{eqnarray}
\Big[ C^{eu}_{\rm SRR} \Big]_{\tau e j i} &=& - \Big[C^{(1)}_{LeQu}\Big]_{\tau e j i} - \delta_{i j} \frac{v^2}{2 m_H^2} \Big[Y^{\prime}_e\Big]_{\tau e} Y_u , \\
\Big[ C^{ed}_{\rm SRR} \Big]_{\tau e j i} &=& - \delta_{i j} \frac{v^2}{2 m_H^2} \Big[Y^{\prime}_e\Big]_{\tau e} Y_d, \\
\Big[ C^{eu}_{\rm SRL} \Big]_{\tau e j i} &=& - \delta_{i j} \frac{v^2}{2 m_H^2} \Big[Y^{\prime}_e\Big]_{\tau e} Y_u \\
\Big[ C^{ed}_{\rm SRL} \Big]_{\tau e j i} &=& + \Big[C_{LedQ}\Big]_{\tau e j i} - \delta_{i j} \frac{v^2}{2 m_H^2} \Big[Y^{\prime}_e\Big]_{\tau e} Y_d \\
\Big[ C^{eu}_{\rm TRR} \Big]_{\tau e j i} &=& - \Big[ C^{(3)}_{LeQu}\Big]_{\tau e j i}, \\
\Big[ C^{ed}_{\rm TRR} \Big]_{\tau e j i} &=& 0,
\end{eqnarray}
\end{subequations}
while the charged-current operators in \eq{fourfermion5} are 
\begin{subequations}
\begin{eqnarray}
\Big[C^{\nu e d u}_{\rm TRR}\Big]_{\nu_\tau e j i}   &=& \Big[V^{\dagger}_{\rm CKM} C^{(3)}_{LeQu}\Big]_{\tau e j i}, \\ 
\Big[C^{\nu e d u}_{\rm SRR}\Big]_{\nu_\tau e j i}   &=& \Big[V^{\dagger}_{\rm CKM}C^{(1)}_{LeQu} \Big]_{\tau e j i}\\ 
\Big[C^{\nu e d u}_{\rm SRL}\Big]_{\nu_\tau e j i}   &=& \Big[C^{}_{LedQ} V^{\dagger}_{\rm CKM} \Big]_{\tau e j i}.
\end{eqnarray}
\end{subequations}
At the $b$ and $c$ thresholds, the scalar operators also induce corrections to the gluonic operators in \eq{Lgluonic}, yielding
\begin{subequations}
\begin{align}
\left[C_{GG} \right]_{\tau e} \!&= \frac{1}{3} \! \sum_{q = b, c} \frac{v}{m_q} \!\left[ C^{eq}_{\rm SRR} + C^{eq}_{\rm SRL} \right]_{\!\tau e qq}, \
\left[C_{GG} \right]_{e \tau}\! =  \frac{1}{3}\! \sum_{q = b, c}   \frac{v}{m_q} \!\left[ C^{eq}_{\rm SRR} + C^{eq}_{\rm SRL} \right]_{\!e \tau qq}\label{matching_GGb}\\
\left[C_{G\widetilde G} \right]_{\!\tau e} \!&= \frac{i}{2} \!\sum_{q = b, c}  \frac{v}{m_q} \!\left[  C^{eq}_{\rm SRR} - C^{eq}_{\rm SRL} \right]_{\!\tau e qq}, \
\left[C_{G\widetilde G} \right]_{e \tau} \!=  \frac{i}{2} \!\sum_{q = b, c}  \frac{v}{m_q}\! \left[  C^{eq}_{\rm SRR} -C^{eq}_{\rm SRL} \right]_{\! e \tau qq}  \label{matching_GGtildeb}
\end{align}
\end{subequations}

The running of the LEFT operators between the electroweak scale and the scales relevant for $\tau$ and $B$ decays was computed in 
Ref.~\cite{Jenkins:2017dyc} and is summarized in Appendix \ref{RGE_low}. The most important effects are the QCD running of the scalar and tensor operators,
and the penguin contributions from operators with $b$ and $c$ quarks onto purely leptonic operators and operators with light quarks. 
The coefficients of LEFT operators, evaluated at the scale $\mu=2$ GeV, as a function of SMEFT operators at the scale $\mu_0 = 1$ TeV are given in Tables \ref{ScalarTensorRGE_low}, \ref{RGE_low_vector} and \ref{RGE_low_lepto}.
In the computation of $\tau$ decay rates we follow very closely Ref.~\cite{Celis:2014asa}, which adopts a different basis for the low-scale operators. 
We provide the appropriate conversion formulae  in Appendix~\ref{app:conversion}.

\subsection{Quark-flavor-conserving decays}\label{QFCO}
We first discuss bounds on $\Gamma^e_{\gamma},~c^{(1,3)}_{L\varphi},~c_{e\varphi}$ and quark-flavor-conserving four-fermion operators from $\tau$ decays. In this section, we give explicitly the full expressions for the decay rates of two decay modes, $\tau \to e \gamma$ and $\tau \to e\pi^+\pi^-$, which lead to many of the strongest limits on these operators. Expressions for all other $\tau$ decay rates we consider, along with relevant input parameters, are collected in Appendix \ref{decay_rate}.
All branching ratios are expressed in terms of LEFT operator coefficients  evaluated at the scale $\mu=2$ GeV. 
These can be expressed in terms of SMEFT coefficients at the high-energy scale $\mu \sim \Lambda$ via the matching formulae given in Section \ref{low-energy_basis} and the RGEs discussed in Sections \ref{RGE_ew} and \ref{RGE_low}.

The branching ratio for $\tau\to e\gamma$ is given by
\begin{align}
{\rm BR}\left(\tau\to e\gamma \right)&=\tau_{\tau}\frac{m^3_{\tau}\alpha_{\rm em}}{4v^2}\Bigl[\bigl\lvert\left( \Gamma^e_{\gamma}\right)_{e\tau} \bigr\rvert^2 +\bigl\lvert\left(\Gamma^e_{\gamma}\right)_{\tau e} \bigr\rvert^2  \Bigr],
\end{align}
where  $\tau_\tau$ is the $\tau$ lifetime, given in Table \ref{input_tau_decay}. Writing
\begin{equation}
\tau_\tau = \Bigl( 5 \frac{G_F^2 m_\tau^5}{192 \pi^3} \widehat{\Gamma}_\tau \Bigr)^{-1},
\end{equation}
with the dimensionless factor $\widehat{\Gamma}_\tau  = 1.12$, we obtain
\begin{equation}
{\rm BR}\left(\tau\to e\gamma \right)= \frac{96 \pi^3 \alpha_{\rm em}}{5 \widehat{\Gamma}_\tau} \frac{v^2}{m_\tau^2}  \Bigl[\bigl\lvert\left(\Gamma^e_{\gamma}\right)_{e\tau} \bigr\rvert^2 +\bigl\lvert\left(\Gamma^e_{\gamma}\right)_{\tau e} \bigr\rvert^2  \Bigr]\simeq7.4\times 10^{4}\Bigl[\bigl\lvert\left(\Gamma^e_{\gamma}\right)_{e\tau} \bigr\rvert^2 +\bigl\lvert\left(\Gamma^e_{\gamma}\right)_{\tau e} \bigr\rvert^2  \Bigr].
\end{equation}
The $\tau\to e\gamma$ branching ratio is thus enhanced with respect to other modes by the two-body phase space, and by the dipole operator appearing at dimension five at low energy.  We notice that $\tau \to e \gamma$ also receives contributions from the tensor operators \cite{Dekens:2018pbu}, which shift the original contribution as
\begin{subequations}
\bea 
(\Gamma^e_\gamma)_{e \tau}  &\to  &(\Gamma^e_\gamma)_{e \tau}   -    \frac{16}{3}   \left(  \frac{i \Pi_{VT} (0)}{v} \right)   \left[C_{\rm TRR}^{eu}\right]_{e \tau uu}  
\\
(\Gamma^e_\gamma)_{ \tau e} ^* &\to  &(\Gamma^e_\gamma)_{ \tau e}^*   - \frac{16}{3}   \left(  \frac{i \Pi_{VT} (0)}{v} \right)   \left[C_{\rm TRR}^{eu}\right]^*_{\tau e uu}   ~, 
\eea
\end{subequations}
with the non-perturbative parameter $(i \Pi_{VT} (0)/v) \approx 1.6 \times 10^{-4}$ at $\mu = 2$~GeV (see Appendix~\ref{sect:tensor2} for details). 
As we will show, this is mostly relevant for global analyses, because in a single operator analysis
$\tau \to e  \pi \pi$  provides a bound  on  the tensor Wilson coefficient that is  four times stronger than the one from $\tau \to e \gamma$.

\begin{figure}
\vspace{-1em}
\centering
\includegraphics[width=.445\textwidth]{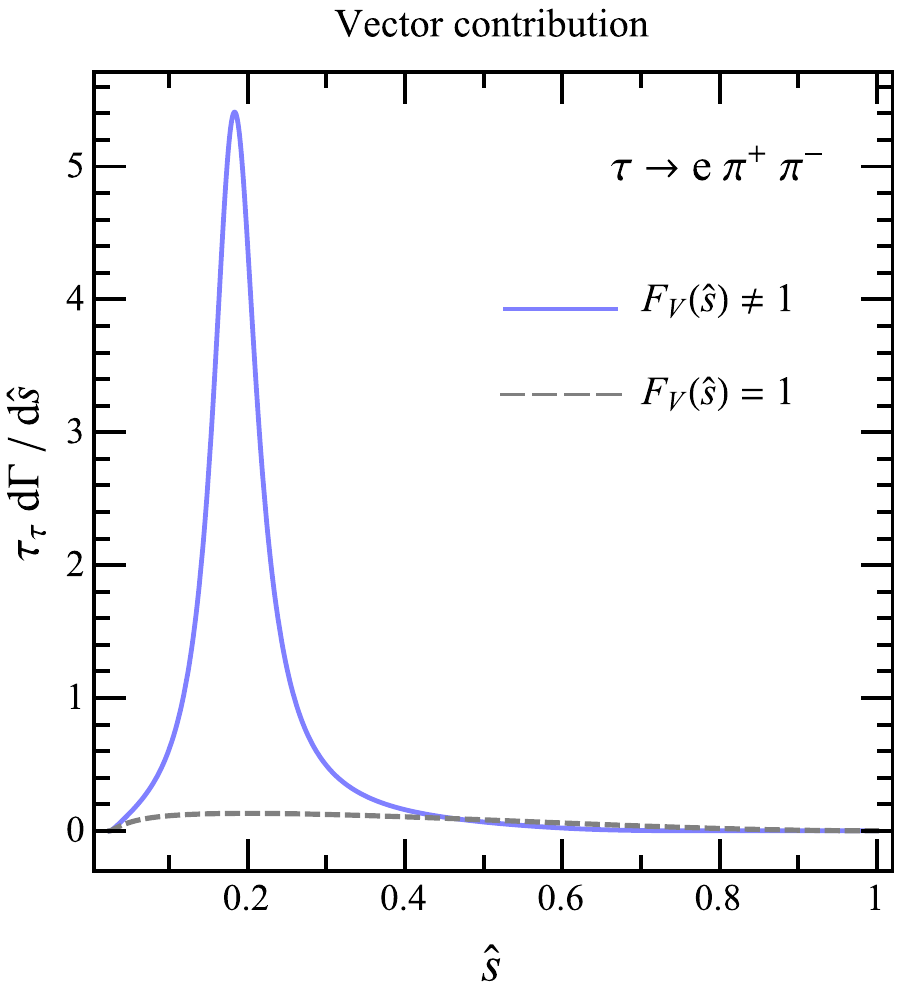}\quad 
\includegraphics[width=0.46\textwidth]{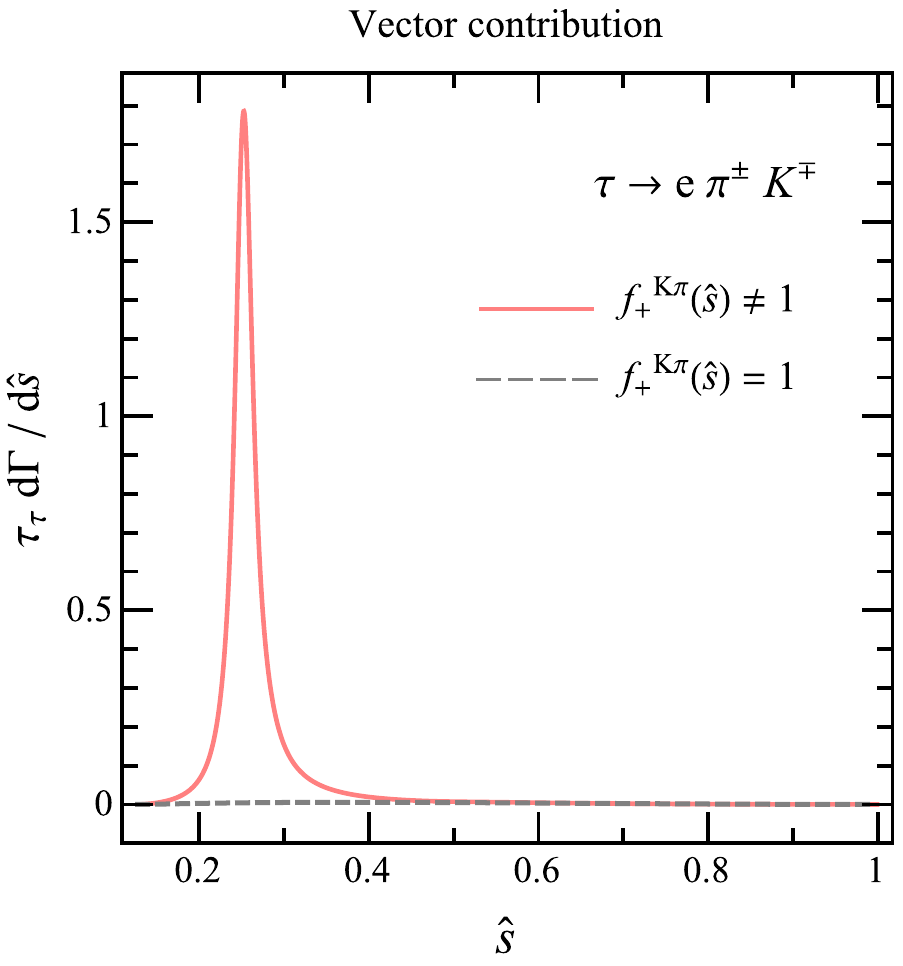}
\vspace{-1em}
\caption{Vector contributions to the differential decay rate of $\tau\to e\pi^+\pi^-$ (left) and $\tau\to e\pi^{\pm}K^{\mp}$ (right). The solid lines include vector form factors as a function of $\hat{s}$, while they are fixed at $1$ in the dashed lines. The Wilson coefficients are set to unity.  }
\label{FormFactors}
\end{figure}

In the case of $\tau \to e \pi^+\pi^-$, the differential decay width is given by
\begin{eqnarray}
\tau_\tau \frac{d\Gamma}{d \hat{s}}&=& \frac{1}{40 \widehat\Gamma_\tau } \left(1-\frac{\rho_\pi}{ \hat{s}} \right)^{1/2} \left(1- \hat{s} \right)^2 
\bigg\{ \frac{6}{m^2_{\tau}} \left(\left|Q_L^{\prime} \right|^2+\left|Q_R^{\prime} \right|^2 \right)  \\ 
& &  + 8\left(1 - \frac{\rho_\pi}{\hat{s}} \right)\left|F_V(s) \right|^2 
\bigg[\frac{2 + \hat{s}}{4 \hat{s}}\left(\left|A_{L} \right|^2+\left|A_{R} \right|^2 \right) +  \left(B_L+B_R\right)  \bigg]\bigg\}, \nn
\end{eqnarray}
where $s$ is the invariant mass of the charged pions, $s=(p_{\pi^+}+p_{\pi^-})^2$ and 
we define the dimensionless quantities  $\hat{s} = s/m_\tau^2$ and $\rho_\pi =4  m_\pi^2/m_\tau^2$.
$\hat{s}$ is kinematically allowed to be in the range $\rho_\pi \leq \hat s \leq 1$.
Here we follow the expression in Ref. \cite{Celis:2014asa}, where the Wilson coefficients are assumed to be real.
$Q^{\prime}_{L,R}$, $A_{L,R}$ and $B_{L,R}$ are combinations of Wilson coefficients and form factors
\begin{subequations}
\begin{align}
Q_L^{\prime}&=\frac{2}{9v}\left(\theta_{\pi}(s)-\Gamma_{\pi}(s)-\Delta_{\pi}(s) \right)\left(C_{GG} \right)_{\tau e}+\frac{\Delta_{\pi}}{m_s}\left(C^{ed}_{\rm SRR}+C^{ed}_{\rm SRL} \right)_{\tau ess} \label{QL1}\\
&\hspace{0.5cm}
+\frac{1}{2}\Gamma_{\pi}(s)\left\{\frac{1}{\hat{m}}\left(C^{eu}_{\rm SRR}+C^{eu}_{\rm SRL} \right)_{\tau e uu}+\frac{1}{\hat{m}}\left(C^{ed}_{\rm SRR}+C^{ed}_{\rm SRL} \right)_{\tau e dd} \right\}, \nn \\
Q_R^{\prime}&=\frac{2}{9v}\left(\theta_{\pi}(s)-\Gamma_{\pi}(s)-\Delta_{\pi}(s) \right)\left(C_{GG} \right)_{e\tau}^*+\frac{\Delta_{\pi}}{m_s}\left(C^{ed}_{\rm SRR}+C^{ed}_{\rm SRL} \right)^*_{e\tau ss}\\
&\hspace{0.5cm}
+\frac{1}{2}\Gamma_{\pi}(s)\left\{\frac{1}{\hat{m}}\left(C^{eu}_{\rm SRR}+C^{eu}_{\rm SRL} \right)^*_{e\tau  uu}+\frac{1}{\hat{m}}\left(C^{ed}_{\rm SRR}+C^{ed}_{\rm SRL} \right)^*_{e\tau dd} \right\}, \nn
\end{align}
\end{subequations}
\begin{subequations}
\begin{align}
A_L&= 4 \pi\alpha_{\rm em} \frac{v}{m_\tau} \left(\Gamma^e_{\gamma}\right)_{\tau e}+\frac{ \hat{s} }{\sqrt{\rho_\pi}}B_T^{\pi,u}(0)\left\{\left(C^{ed}_{\rm TRR} \right)_{\tau e dd}-\left(C^{eu}_{\rm TRR} \right)_{\tau e uu} \right\},\\
A_R&=4\pi \alpha_{\rm em}  \frac{v}{m_{\tau}} \left(\Gamma^e_{\gamma}\right)^*_{e\tau}+\frac{ \hat{s} }{\sqrt{\rho_\pi}}B_T^{\pi,u}(0)\left\{\left(C^{ed*}_{\rm TRR} \right)_{e\tau dd}-\left(C^{eu*}_{\rm TRR} \right)_{e\tau uu} \right\},
\end{align}
\end{subequations}
\begin{subequations}
\begin{align}
B_L=&\left\{\left(C^{ed}_{\rm VLL}+C^{ed}_{\rm VLR} \right)_{\tau e dd}-\left(C^{eu}_{\rm VLL}+C^{eu}_{\rm VLR} \right)_{\tau e uu} \right\} \\
&\times
\bigg[3 A_L +(2 \hat{s} + 1)\left\{\left(C^{ed}_{\rm VLL}+C^{ed}_{\rm VLR} \right)_{\tau e dd}-\left(C^{eu}_{\rm VLL}+C^{eu}_{\rm VLR} \right)_{\tau e uu} \right\}
 \bigg], \nn \\
 B_R=&\left\{\left(C^{ed}_{\rm VRR}\right)_{\tau e dd}+\left(C^{de}_{\rm VLR} \right)_{dd\tau e}-\left(C^{eu}_{\rm VRR}\right)_{\tau e uu}-\left(C^{ue}_{\rm VLR} \right)_{uu\tau e} \right\}\label{BL1}\\
&\times
\bigg[3 A_R +(2 \hat{s}+1)\left\{\left(C^{ed}_{\rm VRR}\right)_{\tau e dd}+\left(C^{de}_{\rm VLR} \right)_{dd\tau e}-\left(C^{eu}_{\rm VRR}\right)_{\tau e uu}-\left(C^{ue}_{\rm VLR} \right)_{uu\tau e} \right\}\bigg],  \nn
 \end{align}
\end{subequations}
and $\hat{m}=(m_u+m_d)/2$.
$Q^{\prime}_{L,R}$ depend on the Wilson coefficients of the scalar and gluonic operators, and on the scalar form factors
$\Gamma_{\pi}(s),~\Delta_{\pi}(s),~\theta_{\pi}(s)$, for which we follow the conventions and determinations in Ref. \cite{Celis:2013xja} (for related work see \cite{Arganda:2008jj,Daub:2012mu,Petrov:2013vka}).
$A_{L,R}$ encode the contributions of dipole and tensor operators, with the value of $B_T^{\pi,u}(0)$  taken from  Ref. \cite{Hoferichter:2018zwu}.
Finally, $B_{L,R}$ encode the contributions of vector operators and their interference with dipole and tensor operators.
For the vector form factor $F_V(s)$ we use the extraction in Ref. \cite{Celis:2013xja}. In the left panel of Fig. \ref{FormFactors}, the solid line depicts the vector contributions to the differential decay rate of $\tau \to e\pi^+\pi^-$. Compared to the dashed line that assumes $F_V(s)=1$, the blue line has a peak around $\hat{s}\sim 0.2$ originating from the $\rho(770)$ resonance. Analogous enhancements are also seen in the scalar contributions as discussed in \cite{Celis:2014asa}. Because of the resonance contribution, the branching ratio in this mode is relatively large.
%
%
Neglecting the interference terms, the following useful  expression for the BR can be obtained
\begin{align}
{\rm BR}\left(\tau \to e \pi^+\pi^- \right)
\simeq&~ 1.9\times 10^{2}\bigl\lvert\Gamma^e_{\gamma} \bigr\rvert^2_{\tau e}+ 1.0\times 10^{-8}\bigl\lvert C_{GG}\bigr\rvert^2+0.13\bigl\lvert C^{ed}_{\rm SRR}+C^{ed}_{\rm SRL} \bigr\rvert^2_{ss}\\
&
+ \Bigl(0.17\Bigl\lvert \big[C^{eq}_{\rm SRR} +C^{eq}_{\rm SRL} \big]_{\tau e (qq)^{(0)}} \Bigr\rvert^2
+ 0.5 \Bigl\lvert \big[C^{eq}_{\rm VLL}+C^{eq}_{\rm VLR} \big]_{\tau e (qq)^{(1)}}     \Bigr\rvert^2\Bigr)
\nonumber\\
&+1.0  \Bigl\lvert \big[ C^{ed}_{\rm TRR}\big]_{\tau e dd} -  \big[C^{eu}_{\rm TRR} \big]_{\tau e uu}  \Bigr\rvert^2 ~,  \nn
\label{tautoepp}
\end{align}
where the notation $(qq)^{(0),(1)}$ indicates that the  isoscalar or isovector ($uu \pm dd$) combination of Wilson coefficients has to be taken. 
A similarly large branching ratio, due to the $\phi(1020)$ resonance, can be seen in the $\tau \rightarrow e K^+ K^-$ mode \cite{Beloborodov:2019uql,Husek:2020fru}, which we discuss in Appendix \ref{taueKK}. 
Because the scalar, tensor and gluonic contributions are affected by larger theoretical uncertainties \cite{Celis:2013xja,Husek:2020fru}, we do not use this channel in our main analysis, and remark on its impact in the multiple operator scenario discussed in Section \ref{global}.

\begin{table}[t]
\centering
\begin{tabular}{||c c c || c c c ||}
\hline
 $\left(\Gamma^e_{\gamma}\right)_{\tau e}$  & $\left(\Gamma^e_Z\right)_{\tau e}$ & $\left(Y_e^{\prime}\right)_{\tau e}$ & $\left(c^{(1)}_{L\varphi} \right)$  & $\left(c_{L\varphi}^{(3)}\right)_{\tau e}$ & $\left(c_{e\varphi}\right)_{\tau e}$\\[4pt]
\hline \hline
 $6.7\times 10^{-7}$  & $3.4\times 10^{-5}$ & $1.1\times 10^{-2}$  &  $4.0\times 10^{-4}$  & $4.0\times 10^{-4}$ & $4.0\times 10^{-4}$ \\
\hline
\end{tabular}
\caption{
90\% C.L. upper
 limits on lepton bilinear operators, assuming a single operator is turned on at the scale $\Lambda =1$ TeV.
 The bounds on the the dipole and Yukawa operators are dominated by $\tau\to e\gamma$, while those
 on the $Z$ couplings $c^{(1,3)}_{L\varphi}$ and $c_{e\varphi}$ by $\tau\to e\pi^+\pi^-$. 
 }
\label{limit_dipole_yukawa}
\end{table}

In order to compare sensitivities across various CLFV  $\tau$ decays, 
we present the numerical results for the remaining decay modes, neglecting interference between operators with different Lorentz structure. 
For leptonic $\tau$ decays we have\footnote{Here, we neglect contributions from scalar four-lepton operators to leptonic decays as they do not give relevant limits on CLFV operators of our interest.}
\begin{subequations}
\begin{align}
{\rm BR}\left(\tau\to e e^+e^- \right)\simeq& ~7.82\times 10^{2} \bigl\lvert\Gamma^e_{\gamma} \bigr\rvert_{\tau e}^2 
+0.36\left(\left|C^{ee}_{\rm VLL} \right|_{\tau eee}^2+0.5\bigl\lvert C^{ee}_{\rm VLR} \bigr\rvert_{\tau eee}^2 \right)\,,\\
{\rm BR}\left(\tau\to e \mu^+\mu^- \right)\simeq&~39\bigl\lvert\Gamma^e_{\gamma} \bigr\rvert_{\tau e}^2+0.17\left(\bigl\lvert C^{ee}_{\rm VLL} \bigr\rvert^2_{\tau e \mu\mu}+\bigl\lvert C^{ee}_{\rm VLR} \bigr\rvert^2_{\tau e \mu\mu} \right) \,, 
\end{align}
\end{subequations}
while for semileptonic $\tau$  decays we have
\begin{subequations}
\begin{align}
{\rm BR}\left(\tau\to e\pi^0 \right)\simeq&~
\Bigl(5.6\times 10^{-2}\Big|   \big[C^{eq}_{\rm VLR}-C^{eq}_{\rm VLL} \big]_{\tau e (qq)^{(1)}} \Big|^2 
+0.14\Big| \big[C^{eq}_{\rm SRR}-C^{eq}_{\rm SRL}\big]_{e \tau (qq)^{(1)}}  \Big|^2\Bigr)
,\\
{\rm BR}\left(\tau\to e\eta \right)\simeq&
 \Bigl([3.3\times 10^{-2}\Big|   \big[C^{eq}_{\rm VLR}-C^{eq}_{\rm VLL} \big]_{\tau e (qq)^{(0)}} \Big|^2 
+1.9 \times 10^{-2} \Big| \big[C^{eq}_{\rm SRR}-C^{eq}_{\rm SRL}\big]_{e \tau (qq)^{(0)}}  \Big|^2\Bigr) \nonumber\\
&+6.6\times 10^{-2}\Big|C^{ed}_{\rm VLR}-C^{ed}_{\rm VLL} \Big|^2_{\tau ess}+0.15\Big|C^{ed}_{\rm SRR}-C^{ed}_{\rm SRL}\Big|_{e\tau ss}^2\nonumber\\
&+1.4\times 10^{-8}\bigl\lvert C_{G\tilde G} \bigr\rvert^2_{\tau e}, \label{tautoeeta}\\
{\rm BR}\left(\tau\to e\eta^\prime \right)\simeq&
 \Bigl([1.3 \! \times\! 10^{-2}\Big|   \big[C^{eq}_{\rm VLR}\!-\! C^{eq}_{\rm VLL} \big]_{\tau e (qq)^{(0)}} \Big|^2 
+1.1 \!\times\! 10^{-2} \Big| \big[C^{eq}_{\rm SRR}\!-\! C^{eq}_{\rm SRL}\big]_{e \tau (qq)^{(0)}}  \Big|^2\Bigr) \nonumber\\
&+6.1\times 10^{-2}\Big|C^{ed}_{\rm VLR}-C^{ed}_{\rm VLL} \Big|^2_{\tau ess}+0.14\Big|C^{ed}_{\rm SRR}-C^{ed}_{\rm SRL} \Big|_{e\tau ss}^2\nonumber\\
&+5.5\times 10^{-8}\bigl\lvert C_{G\tilde G} \bigr\rvert^2_{\tau e}. \label{tautoeetap}
\end{align}
\end{subequations}
Note that  Wilson coefficients corresponding to operators with opposite chiralities ($L \leftrightarrow R$)   contribute to  each decay  mode with the same prefactors.

With the above results at hand, we can get a reasonable picture of the constraints imposed by   $\tau$ decays on various CLFV operators. 
 Table \ref{limit_dipole_yukawa} shows the upper limits on lepton bilinear operators. 
 Starting with  photon-dipole operator, we see that $\tau \to e\gamma$ gives the strongest limit. 
 The bound on $\Gamma^e_Z$ is obtained by considering operator mixing between the $Z$- and $\gamma$-dipole operators. The running effect is given by $\Gamma^e_{\gamma}(m_t)=-2.0\times 10^{-2}~\Gamma^e_{Z}(\Lambda)$ with $\Lambda=1~$TeV, yielding $\left(\Gamma^e_{Z}\right)_{\tau e}<3.4\times 10^{-5}$. Moreover, the $\tau$-$e$ LFV Yukawa coupling induces the photon-dipole operator at 1- and 2-loop level (the expressions are given in Appendix \ref{dipole_yukawa}). The resulting limit is $1.1\times 10^{-2}$, which is consistent with the result in  \cite{Harnik:2012pb}.

\begin{table}[t]
\centering
\begin{tabular}{||c|| c| c|| c | c || c | c |}
\hline
$C_{LQ,U}$ & $uu$ & $2.1\times 10^{-4*}$ & $cc$ & $1.1\times 10^{-2*}$ & $tt$ & $3.4\times 10^{-3*}$  \\
$C_{eu}$ & $uu$ & $2.1\times 10^{-4*}$ & $cc$ & $1.1\times 10^{-2*}$ & $tt$ & $4.1\times 10^{-3*}$  \\
$C_{Lu}$ & $uu$ & $2.1\times 10^{-4*}$ & $cc$ & $1.1\times 10^{-2*}$ & $tt$ & $4.1\times 10^{-3*}$  \\
\hline \hline
$C_{LQ,D}$ & $dd$ &$2.1\times 10^{-4*}$ & $ss$ & $1.2\times 10^{-3\S}$ & $bb$ & $2.5\times 10^{-2*}$  \\
$C_{ed}$ & $dd$ & $2.1\times 10^{-4*}$& $ss$ & $1.2\times 10^{-3\S}$ & $bb$ & $2.5\times 10^{-2*}$ \\
$C_{Ld}$ & $dd$ & $2.1\times 10^{-4*}$ & $ss$ & $1.2\times 10^{-3\S}$ & $bb$ & $2.5\times 10^{-2*}$ \\
$C_{Qe}$ & $dd$ & $8.5\times 10^{-4\S}$ & $ss$ & $1.2\times 10^{-3\S}$ & $bb$ & $3.9\times 10^{-3*}$  \\
\hline\hline
$C_{LedQ}$ & $dd$ & $2.1\times 10^{-4*}$ & $ss$ & $2.3\times 10^{-4*}$ & $bb$ & $3.6\times 10^{-2\sharp}$  \\
$C_{LeQu}^{(1)}$ & $uu$ & $2.0\times 10^{-4*}$ & $cc$ & $9.0\times 10^{-3\sharp}$ & $tt$ & $1.0\times 10^{-3\dagger}$  \\
$C_{LeQu}^{(3)}$ & $uu$ & $1.8\times 10^{-4*}$ & $cc$ & $8.5\times 10^{-5\dagger}$ & $tt$ & $1.5\times 10^{-6\dagger}$  \\
 \hline
\end{tabular}
\caption{90\% C.L. upper
 limits on the quark-flavor-conserving semileptonic operators, assuming a single operator is turned on at the scale $\Lambda =1$ TeV. The superscripts represent that the strongest limit is imposed by decay modes ($^*$) $\tau\to e\pi^+\pi^-$, ($^\dag$) $\tau\to e\gamma$, ($^{\S}$) $\tau\to e\eta$ and ($^\sharp$) $\tau\to e\eta^{\prime}$. For the scalar and tensor operators, the bounds apply to both the $\tau e$ and  $e\tau$ components.}
\label{limit_semileptonic}
\end{table}

A noteworthy feature of CLFV $\tau$ decay phenomenology 
 is a somewhat large contribution of the vector operators to $\tau \to e\pi^+\pi^-$ compared to other $\tau$ decay channels.   This is caused by a resonant effect in the pion vector form factor as seen from the left panel of Fig.~\ref{FormFactors}. The bounds on the $\psi^2\varphi^2D$-type operators, $c_{L\varphi}^{(1,3)}$ and $c_{e\varphi}$,  in Table \ref{limit_dipole_yukawa} are predominantly given by the $\tau \to e \pi^+\pi^-$ channel.  The contributions stem from the tree-level $Z$-exchange process as listed in Section \ref{low-energy_basis}. Similarly, in a single operator analysis, most of the semileptonic vector operators receive the strongest bounds from the $\tau \to e \pi^+\pi^-$ mode. In Table \ref{limit_semileptonic}, we show the upper limits on the four-fermion operators, where the symbol $``*"$ indicates that $\tau \to e \pi^+\pi^-$ provides the most stringent bound. For the vector operators, we consider the RGEs for the heavy quarks $(q=t,b,c)$ from 1~TeV to 2 GeV. For light-quark operators, running effects are negligible. The details of the RGEs are given in Appendix \ref{RGEs}. 
The isoscalar  $\left(C_{Qe} \right)_{\tau e dd}$ 
and the strange components of vector operators are not constrained by  $\tau \rightarrow e \pi\pi$. In this case, among the observables
in Table \ref{experiment_limit}, the strongest bounds arise from $\tau\to e \eta$, marked with $``\S"$ in Table \ref{limit_semileptonic}. As discussed in Appendix \ref{taueKK},  $\tau \rightarrow e K^+ K^-$
imposes stronger constraints on strange operators,  $|\left[C_{LQ,\, D}]_{\tau e s s} \right| < 2.4 \times 10^{-4}$,
and very similar constraints on $ss$ components of $C_{Ld}$, $C_{ed}$ and $C_{Qe}$.

The last three rows in Table \ref{limit_semileptonic} correspond to the limits on the quark-flavor-conserving scalar and tensor operators. Here, we take into account the QCD self-running of these operators. In addition, there are several paths in the RGEs: (1) threshold corrections of the heavy quarks to $C_{GG}$ and $C_{G\widetilde G}$ as in Eqs~(\ref{matching_GG}) and  (\ref{matching_GGtilde}); (2) mixing between $C^{(1)}_{LeQu}$ and $C^{(3)}_{LeQu}$; and (3) mixing from $C^{(3)}_{LeQu}$ to $\Gamma^e_{\gamma}$. These paths enable us to constrain the operators from $\tau \to e\gamma$, yielding the predominant bounds on the top-quark operators. The induced $C_{GG}$ and $C_{G\widetilde G}$ are not large enough to compensate for the suppression factor of roughly ${\cal O}(10^{-8})$ as seen in Eqs.~(\ref{tautoepp}), (\ref{tautoeeta}) and (\ref{tautoeetap}). The mixing to the dipole operator is proportional to the Yukawa coupling, while the threshold corrections are enhanced by the inverse of the coupling in the lighter-quark case. For the charm-quark scalar operator, although each related decay channel gives the comparable limit of ${\cal O}(10^{-2})$, $\tau \to e\eta^{\prime}$ provides the slightly stronger bound. Apart from the heavy up-type quarks, since no mixing is present, it is straightforward to examine $(C_{LedQ})_{bb}$, whose bound results from the contribution of $C_{G\tilde G}$ to $\tau \to e \eta^{\prime}$. The rest of the light-quark operators are primarily constrained by $\tau \to e\pi^+ \pi^-$.

Finally, we comment on LFV quarkonium decays such as $\Upsilon(2S)\to \tau e$ and $J/\psi\to\tau e$.  The current experimental bounds on BRs of these decay modes are $\mathcal O(10^{-6})$. Based on the analysis in \cite{Hazard:2016fnc}, we find that the resulting limit on the four-fermion operators is roughly  $\mathcal O(0.1)$, which is weaker than those from $\tau$ decays. Therefore, we do not include the quarkonium decays in our analysis.

\subsection{Quark-flavor-violating observables}\label{QFVO}

We now turn to the quark-flavor-violating operators that can be constrained by $B$ meson decays as well as $\tau$ decay involving strange mesons. As in the previous section, below we only give a rough sketch of each BR to have an idea of which decay modes are relevant. All the expressions of BRs are listed in Appendix \ref{Bdecays}.

The channels $\tau^- \to e^-K_S$ and $\tau^-\to e^-\pi^{\pm}K^{\mp}$ put bounds on the $sd$ and $ds$ components of the LFV down-type operators:
\begin{subequations}
\begin{align}
{\rm BR}(\tau^-\to e^-K_S)&\simeq ~6.9\times 10^{-2}\left|\left[C^{ed}_{\rm VLR}-C^{ed}_{\rm VLL}\right]_{\tau e ds} - (d\leftrightarrow s) \right|^2
 \\ &\quad
+0.14\left|\left[C^{ed}_{\rm SRR}-C^{ed}_{\rm SRL}\right]_{\tau eds} - (d\leftrightarrow s)  \right|^2, \nn \\
{\rm BR}\left(\tau^-\to e^- \pi^{+}K^-\right) &\simeq~0.17\left|C^{ed}_{\rm VLL}+C^{ed}_{\rm VLR} \right|^2_{\tau e ds}+0.16\left|C^{ed}_{\rm SRR}+C^{ed}_{\rm SRL} \right|_{\tau e ds}^2.
\label{eq:kpi}
\end{align}
\end{subequations}
Wilson coefficients with opposite lepton chirality  contribute to  each decay  mode with the same prefactors.
Compared to  $\tau^-\to e^-K_S$, 
the  $\tau^-\to e^- \pi^{+}K^-$  decay has a stronger sensitivity to Wilson coefficients of the vector semi-leptonic operators. 
This enhancement stems from the $K^*(892)$ resonance, which is seen in the right panel of Fig.~\ref{FormFactors}.\footnote{In this panel, we only plot the vector contribution from $V(s)$ in Eq. (\ref{piK_formfactor}).}  On the other hand, the scalar contribution is comparable between the two decay modes.

The $bd$ and $db$ elements of the LFV down-type operators contribute to $B_d\to \tau e$ and $B^+\to\pi^+ \tau e$ modes:
\begin{subequations}
\begin{align}
{\rm BR}\left(B_d\to\tau^-e^+ \right) &\simeq~6.0\left|\Big[C^{ed}_{\rm VLR}- C^{ed}_{\rm VLL}\Big]_{\tau e bd} \right|^2 +84.8\left|\Big[C^{ed}_{\rm SRR}-\Big[C^{ed}_{\rm SRL}\Big]_{\tau e bd} \right|^2 \\
{\rm BR}\left(B^+\to \pi^+ \tau^-e^+ \right)&\simeq~5.72\left|\Big[C^{ed}_{\rm VLL} +C^{ed}_{\rm VLR} \Big]_{\tau ebd} \right|^2 +8.9\left|\Big[C^{ed}_{\rm SRR} + C^{ed}_{\rm SRL} \Big]_{\tau e bd} \right|^2.
\end{align}
\end{subequations}
Similarly,  $B_d\to \tau^+e^-$ and $B^+\to \pi^+ \tau^+e^-$ are described via the interchange of $b\leftrightarrow d$, and, as usual, we are showing only one lepton chirality. 
For the $bd$ components of the vector operators, although both decay modes give similar bounds, $B_d\to \tau^-e^+$  gives the somewhat stronger  bound due to its slightly stronger experimental limit. The opposite situation can be seen in the $db$ components, which are most strongly restricted by $B^+\to \pi^+\tau^+ e^-$. On the other hand, in the case of the scalar operator, the prefactor in $B_d\to \tau^-e^+$ is enhanced by roughly $(m_{B^0}/m_{\tau})^2$ compared to the vector operator, making this  the most restrictive decay channel.

\begin{table}[t]
\centering
\begin{tabular}{||c|| c| c|| c | c || c | c |}
\hline
$C_{LQ,D}$ & $ds$ & $4.6\times 10^{-4\diamondsuit}$ & $sb$ & $1.2\times 10^{-3\natural}$ & $db$  & $1.9\times 10^{-3\flat}$ \\
                    & $sd$ &$4.2\times 10^{-4\diamondsuit}$ & $bs$ & $2.1\times 10^{-3\natural}$  & $bd$ & $2.2\times10^{-3\star}$ \\
$C_{ed}$ & $ds$ & $4.6\times 10^{-4\diamondsuit}$& $sb$ & $1.2\times 10^{-3\natural}$ & $db$ & $1.9\times 10^{-3\flat}$\\
                & $sd$ & $4.2\times 10^{-4\diamondsuit}$& $bs$ & $2.1\times 10^{-3\natural}$ & $bd$ & $2.2\times10^{-3\star}$  \\
$C_{Ld}$ & $ds$ & $4.6\times 10^{-4\diamondsuit}$ & $sb$ & $1.2\times 10^{-3\natural}$ & $db$ & $1.9\times 10^{-3\flat}$\\
                & $sd$ & $4.2\times 10^{-4\diamondsuit}$ & $bs$ & $2.1\times 10^{-3\natural}$ & $bd$ & $2.2\times10^{-3\star}$  \\
$C_{Qe}$ & $ds$ & $4.6\times 10^{-4\diamondsuit}$ & $sb$ & $1.2\times 10^{-3\natural}$ & $db$ & $1.9\times 10^{-3\flat}$ \\
                  & $sd$ & $4.2\times 10^{-4\diamondsuit}$ & $bs$ & $2.1\times 10^{-3\natural}$ & $bd$ & $2.2\times10^{-3\star}$ \\
\hline\hline
$C_{LedQ}$ & $ds$ & $4.3\times 10^{-4\ddag}$ & $sb$ & $1.1\times 10^{-3\natural}$ & $db$ & $5.7\times 10^{-4\star}$  \\
                    & $sd$ & $4.3\times 10^{-4\ddag}$ & $bs$ & $1.9\times 10^{-3\natural}$  & $bd$ & $5.7\times 10^{-4\star}$ \\
 \hline
\end{tabular}
\caption{90\% C.L. upper limits on the down-type quark-flavor-violating semileptonic operators, assuming a single operator is turned on at the scale $\Lambda =1$ TeV. The superscripts denote that the limits come from $\tau$ decay modes ($^\diamondsuit$)  $\tau\to e\pi K$ or ($^\ddag$) $\tau \to e K_S$, or $B$ meson decay modes ($^\star$) $B_d\to \tau e$, ($^\flat$) $B^+\to\pi^+\tau e$ or ($^\natural$) $B^+\to K^+\tau e$. The limit on the scalar operators is applicable to both the $\tau e$ and  $e\tau$ elements.}
\label{limit_semileptonic_vio}
\end{table}

The last elements are $bs$ and $sb$, which are restricted by $B^+\to K^+e^{\pm}\tau^{\mp}$:
\begin{align}
{\rm BR}\left(B^+ \! \to\! K^+e^+\tau^- \right)&\simeq  9.92\left|\big[C^{ed}_{\rm VLL} + C^{ed}_{\rm VLR} \big]_{\tau ebs} \right|^2 +12.24\left|\big[C^{ed}_{\rm SRR} + C^{ed}_{\rm SRL} \big]_{\tau e bs} \right|^2.
\end{align}

The resulting upper limits on the four-fermion operators are summarized in Table \ref{limit_semileptonic_vio}. Overall, these limits are less than or equal to $O(10^{-3})$. The third column represents those of $ds$ and $sd$ components and their bounds originate from $\tau \to e \pi K$ for the vector operators and $\tau \to eK_S$ for the scalar operators. These decay modes are represented by $``\diamondsuit"$ and $``\ddag"$, respectively. The fifth column corresponds to the bounds on the $bs$ and $sb$ elements from the $B^+\to K^+\tau e$ channel symbolized by $``\natural"$. The constraints on the $bd$ and $db$ elements from $B_d\to \tau e~(\star)$ and $B^+\to \pi^+ \tau e~(\flat)$ are in the rightmost column.

\section{Indirect bounds: charged current and neutrino processes}
\label{sect:indirect}

\begin{table}
\center
\begin{tabular}{||c |c  | c | c||}
\hline
                                   & BR (90\% CL) & & BR (90\% CL)   \\
\hline   
$\pi \rightarrow e \nu$            &  $\left(1.230 \pm 0.004 \right) \times 10^{-4}$ & $K^+ \rightarrow \pi^+ \nu \bar\nu$   &  $ < 1.78  \times 10^{-10}$    \\ 
$K \rightarrow e \nu$ &  $ (1.582 \pm 0.007 )\times 10^{-5}$  &  $K_L \rightarrow \pi^0 \nu \bar \nu$   &  $ < 3.0   \times 10^{-9}$    \\ 
$D \rightarrow e \nu$   &  $ < 8.8 \times 10^{-6}$  & $B^+ \rightarrow \pi^+ \nu \bar \nu$   &  $ < 1.4   \times 10^{-5}$    \\
$D \rightarrow \tau \nu$   &  $  (1.20 \pm 0.27 )\times 10^{-3}$  & $B^+ \rightarrow K^+ \nu \bar \nu$   &  $ < 1.6   \times 10^{-5}$     \\
$D_s \rightarrow e \nu$   &  $ < 8.3 \times 10^{-5}$  &  &   \\
$D_s \rightarrow \tau \nu$   &  $ (5.48 \pm 0.23) \times 10^{-2}$  & &   \\
$B \rightarrow e \nu$   &  $ < 9.8 \times 10^{-7}$  & &   \\
$B \rightarrow \mu \nu$   &  $ \left(6.46 \pm  2.74\right) \times 10^{-7}$  & &   \\
$B \rightarrow \tau \nu$   &  $  (1.09 \pm  0.24) \times 10^{-4}$  &  &    \\
\hline
\hline
\end{tabular}
\caption{Charged current and neutrino processes sensitive to CLFV operators.
All limits are taken from Ref. \cite{Zyla:2020zbs}, with the exception of $K^+\rightarrow \pi^+ \nu \bar\nu$, for which we use the more recent result in Ref.  \cite{CortinaGil:2020vlo}. \label{ChargedAndNeutrinos}
}
\end{table}

Invariance under the $SU(2)_L$ gauge group implies that some of the SMEFT four-fermion operators in \eq{fourfermion}  induce LFV operators with one or two neutrinos rather than charged leptons. These can mediate meson or nuclear $\beta$ decays with a  $e \nu_\tau$ or $\tau \nu_e$ in the final state,
or flavor-changing-neutral-current meson decays such as $K^+ \rightarrow \pi^+ \nu_e \bar \nu_{\tau}$.
These observables probe LFV  indirectly, since the flavor of the neutrino is not identified. However, the 
agreement between experiment and SM predictions for these processes can put severe constraints on the coefficient of LFV four-fermion operators and provide useful information on their flavor structure.
The branching ratios that we use in this section are summarized in Table \ref{ChargedAndNeutrinos}.

Leptonic decays of charged pseudoscalar mesons are particularly sensitive to new scalar interactions.
CLFV interactions can contribute to these processes, since the flavor of the outgoing neutrino is not determined.
The branching ratio is given by
\begin{align}
{\rm BR}(P^+_{u_i d_j} \rightarrow \ell^+ \nu_{\ell^\prime}) &= 
\frac{G_F^2 |V_{i j}|^2}{8 \pi} \tau_P \, f^2_{P} m_P m^2_\ell  \left(1 - \frac{m^2_\ell}{m^2_P}\right)^2 \\
\times &  \left( \delta_{\ell \ell^\prime} + \frac{1}{|V_{i j}|^2} \left|  \left[ C^{\nu e d u}_{\rm VLL}\right]_{\nu_{\ell^\prime} \ell ji} +   \frac{m^2_P}{m_{\ell} ( m_{u_i} + m_{d_j})}
\left[C^{\nu e d u}_{\rm SRR} -  C^{\nu e d u}_{\rm SRL}\right]_{\nu_{\ell^\prime} \ell ji}\right|^2 \right) \nonumber,
\end{align}
where the CLFV operators do not interfere with the SM contribution. Here, the indices $u_i$ and $d_j$ correspond to constituent quarks of pseudoscalar meson $P$.

In the case of light pseudoscalar mesons, the ratios $R_P = \Gamma(P \rightarrow e \nu)/\Gamma(P \rightarrow \mu \nu)$,
with $P = \pi, K$, are very well determined. The general expression in the SMEFT (extended with light sterile neutrinos) is given in 
Ref.~\cite{Cirigliano:2013xha}.
Neglecting flavor-conserving operators, and considering only CLFV in the $\tau$-$e$ sector, the ratios $R_{\pi}$ and $R_K$ are
\begin{eqnarray}
\frac{R_{\pi}}{R^{SM}_\pi} &=& 1 + \frac{1}{|V_{u d}|^2}\left|  \left[ C^{\nu e d u}_{\rm VLL}\right]_{\nu_{\tau} e d u} +   \frac{m^2_\pi}{m_e ( m_{u} + m_{d})}
\left[C^{\nu e d u}_{\rm SRR} -  C^{\nu e d u}_{\rm SRL}\right]_{\nu_{\tau} e d u}\right|^2 , \\
 \frac{R_{K}}{R^{SM}_\pi} &=& 1 + \frac{1}{|V_{u s}|^2} \left|  \left[ C^{\nu e d u}_{\rm VLL}\right]_{\nu_{\tau} e s u} +   \frac{m^2_K}{m_e ( m_{u} + m_{s})}
\left[C^{\nu e d u}_{\rm SRR} -  C^{\nu e d u}_{\rm SRL}\right]_{\nu_{\tau} e s u}\right|^2 .
\end{eqnarray}
Comparing theory and experiment (see  \cite{Cirigliano:2013xha} and reference therein)  one obtains
\begin{eqnarray}
\frac{R_{\pi}}{R^{SM}_\pi} = 0.996 \pm 0.005, \qquad \frac{R_{K}}{R^{SM}_K} = 1.0048 \pm 0.0048,
\end{eqnarray}
which, because of the enhancement of $1/m_e$, leads to strong bounds on the scalar operators. 

For the $D$ mesons, we can look at the ratio between the $\tau$ and $\mu$ or $e$ leptonic decays. Using the input in Table \ref{ChargedAndNeutrinos}, we obtain
\begin{eqnarray}
R^\mu_{D}= \frac{\Gamma(D \rightarrow \mu \nu)}{\Gamma(D \rightarrow \tau \nu)}     &=&   0.312 \pm 0.072, \qquad 
R^e_{D}= \frac{\Gamma(D \rightarrow e \nu)}{\Gamma(D \rightarrow \tau \nu)} <  0.013,  \\
R^{\mu}_{D_s}=\frac{\Gamma(D_s \rightarrow \mu \nu)}{\Gamma(D_s \rightarrow \tau \nu)} &=&  0.100 \pm 0.005, \qquad 
R^{e}_{D_s} =\frac{\Gamma(D_s \rightarrow e \nu)}{\Gamma(D_s \rightarrow \tau \nu)} <   1.5 \times 10^{-3}.  
\end{eqnarray}
Similarly, for $B$ mesons the ratio of $B\rightarrow e \nu$ and $B \rightarrow \tau \nu$ is constrained to be
\begin{eqnarray}
 R^e_{B}= \frac{\Gamma(B \rightarrow e \nu)}{\Gamma(B \rightarrow \tau \nu)} <  0.9 \times 10^{-2}.
\end{eqnarray}
The expressions for these ratios are
\begin{eqnarray}
R^\mu_D &=& \left( \frac{m_D^2 - m^2_\mu}{m^2_D - m^2_\tau} \right)^2  \frac{m_\mu^2}{m^2_\tau}
\frac{1}{
 1 + |V_{c d}|^{-2}  \left|  \left[ C^{\nu e d u}_{\rm VLL}\right]_{\nu_{e} \tau d c} +   \frac{m^2_D}{m_{\tau} ( m_{c} + m_{d})}
\left[C^{\nu e d u}_{\rm SRR} -  C^{\nu e d u}_{\rm SRL}\right]_{\nu_e \tau d c}\right|^2 } \nonumber\\ \\
R^e_D &=& \left( \frac{m_D^2 - m^2_e}{m^2_D - m^2_\tau} \right)^2  \frac{m_e^2}{m^2_\tau}
\frac{1 + |V_{c d}|^{-2} \left|  \left[ C^{\nu e d u}_{\rm VLL}\right]_{\nu_{\tau} e d c} +   \frac{m^2_D}{m_{e} ( m_{c} + m_{d})}
\left[C^{\nu e d u}_{\rm SRR} -  C^{\nu e d u}_{\rm SRL}\right]_{\nu_\tau e d c}\right|^2 }{ 1 + |V_{c d}|^{-2}  \left|  \left[ C^{\nu e d u}_{\rm VLL}\right]_{\nu_{e} \tau d c} +   \frac{m^2_D}{m_{\tau} ( m_{c} + m_{d})}
\left[C^{\nu e d u}_{\rm SRR} -  C^{\nu e d u}_{\rm SRL}\right]_{\nu_e \tau d c}\right|^2 },  \nonumber\\
\end{eqnarray}
with $d \rightarrow s$ for $D_s$ decays, and $m_D\rightarrow m_B$, $d \rightarrow b$, $c \rightarrow u$ for $B$ decays.

\begin{table}[t]
\centering
\begin{tabular}{||c|| c| c|| c | c || c | c || c | c||}
\hline
 $(C^{(1)}_{LeQu})_{\tau e}$ & $uu$ & $1.3\times 10^{-5\, *}$ & $uc$ &  $1.7 \times 10^{-3\, \S }$  & $cu$  &  $5.1\times 10^{-6\, \dagger}$  & $cc$ &   $4.3\times 10^{-3\, \S}$   \\
 & $t u$ &  $8.5 \times 10^{-5\, \sharp}$ & & & & & & \\
 $(C^{(1)}_{LeQu})_{e \tau}$ & $cu$ & $4.2\times 10^{-2\,\sharp}$ &   $t u$ &  $1.7 \times 10^{-3\,\sharp}$  & & & & \\
 \hline
$ (C^{}_{LedQ})_{\tau e}$ & $dd$ & $1.3 \times 10^{-5\, *}$ & $ds$ & $5.7 \times 10^{-5\, *}$ & $db$ & $3.4 \times 10^{-3\, *}$ & $sd$ & $5.1 \times 10^{-6\, \dagger}$ \\
 & $ss$ & $2.3 \times 10^{-5\, \dagger}$ & $sb$ & $1.4 \times 10^{-3\, \dagger}$ & $bd$ & $8.3 \times 10^{-5\, \sharp}$ & $bs$ & $3.6 \times 10^{-4\, \sharp}$ \\
 & $b b $& $2.2 \times 10^{-2\, \sharp}$ & & & & & &  \\
 $ (C^{}_{LedQ})_{e\tau}$ &  $bd$ & $1.8 \times 10^{-3\,\sharp}$ & $bs$ & $7.6 \times 10^{-3\,\sharp}$ & & & &  \\
 \hline
$C_{LQ,U}$ & $uu$ & $3.3 \times 10^{-5\,\flat}$ & $uc$ &  
$1.0 \times 10^{-5\,\flat}$ & $ut$ & $1.4 \times 10^{-3\,\star}$ & $cc$ & $3.2 \times 10^{-5\,\flat}$ \\ &  $ct$ & $1.0 \times 10^{-3\,\star}$ & && &&&\\
$C_{Ld}$ & $ds$ & $1.0 \times 10^{-5\,\flat}$ & $sb$ & $1.0 \times 10^{-3\,\star}$ & $db$ &$1.4 \times 10^{-3\,\star}$ & & \\
\hline
\end{tabular}
\caption{90\% C.L. limits from charged-current leptonic decays and neutrino processes, assuming a single operator is turned on at the scale $\Lambda =1$ TeV.
The superscripts denote limits from decay ratios ($^*$) $R_\pi/R^{SM}_{\pi}$, ($^\dag$) $R_K/R_K^{SM}$, ($^\S$) $R^e_D$ and $R^e_{D_s}$, and  ($^\sharp$) $R^e_B$ and $R^\mu_B$, and from decay modes ($^\flat$) $K \rightarrow \pi \nu \nu$ and ($^\star$) $B\to (K,\pi)\nu\nu$.
Purely leptonic decays of pseudoscalar mesons
constrain the $\tau e$ component of the scalar operators $C^{(1)}_{LeQu}$ and $C^{}_{LedQ}$.
Limits on the $e\tau$ components of scalar operators are much weaker. For example,  $|(C^{(1)}_{LeQu})_{e\tau cc}| \lesssim 0.22$. 
On these components, we only quote bounds that are better than 0.1.
$C_{LQ,\, U}$ and $C_{Ld}$ are constrained by $K \rightarrow \pi \nu \bar\nu$ and $B\to (K,\pi)\nu\nu$.
In this case, the bounds on the $j i$ and $i j$ components are the same, and we only show one flavor combination.  }
\label{limit_cc}
\end{table}

The operators $C_{LQ,\, U}$, $C_{LQ,\, D}$, $C_{Lu}$ and $C_{Ld}$ also induce effective interactions with two neutrinos of different flavor.
In these cases, strong constraints can arise from bounds on $K \rightarrow \pi \nu \bar{\nu}$, 
$B \rightarrow K  \nu \bar \nu$ and $B \rightarrow \pi  \nu \bar \nu$. For kaon decays, the differential decay rate can be expressed as
\cite{Cirigliano:2004pv,Cirigliano:2011ny}
\begin{align}
\frac{d\Gamma(K^+ \rightarrow \pi^+ \nu_i \bar \nu_j)}{d z d y} &= \frac{G_F^2 m_K^5}{128 \pi^3} \left|f_+^{K\pi}(0)\right|^2 \rho(y,z) \left| \left[C^{\nu d}_{\rm VLL}
+ C^{\nu d}_{\rm VLR}\right]_{\nu_i \nu_j s d} \right|^2 \\
\frac{d\Gamma(K_L \rightarrow \pi^0 \nu_i \bar \nu_j)}{d z d y} &= \frac{G_F^2 m_K^5}{128 \pi^3} \left|f_+^{K\pi}(0)\right|^2 \rho(y,z)
\frac{1}{2}
\nonumber \\
& \quad \times  \left| \left[C^{\nu d}_{\rm VLL}
+ C^{\nu d}_{\rm VLR}\right]_{\nu_i \nu_j s d} -
\left[C^{\nu d}_{\rm VLL}
+ C^{\nu d}_{\rm VLR}\right]_{\nu_i \nu_j d s}
\right|^2, 
\end{align}
where $y=2 p_{\nu_i}\cdot p_K/m_K^2$, $z=  p_\pi \cdot p_K/m_K^2$.
The function $\rho$ is given by
\begin{eqnarray}
\rho(y,z) = 4 (z+y-1)(1-y) - 4 r_\pi, 
\end{eqnarray}
with $r_\pi = m_\pi^2/m_K^2$, and the limits of integrations
\begin{eqnarray}
0 < y <  1 - r_\pi, \qquad   1 - y + \frac{r_\pi}{1-y} < z < 1 + r_\pi.
\end{eqnarray}
Integrating over the phase space, we get
\begin{eqnarray}
\textrm{BR}(K^+ \rightarrow \pi^+ \nu_i \bar \nu_j)  &=& 1.76  \left| \left[C^{\nu d}_{\rm VLL}
+ C^{\nu d}_{\rm VLR}\right]_{\nu_i \nu_j s d} \right|^2, \\
\textrm{BR}(K_L \rightarrow \pi^0 \nu_i \bar \nu_j)  &=& 3.63  \left| \left[C^{\nu d}_{\rm VLL}
+ C^{\nu d}_{\rm VLR}\right]_{\nu_i \nu_j s d} -
\left[C^{\nu d}_{\rm VLL}
+ C^{\nu d}_{\rm VLR}\right]_{\nu_i \nu_j d s}
\right|^2.
\end{eqnarray}
From $B \rightarrow K \nu_i \bar \nu_j$ and 
$B \rightarrow \pi \nu_i \bar \nu_j$ we can use the expressions for $B \rightarrow K e \tau$ reported in Appendix \ref{Bdecays}, and take the limit of zero lepton masses.

The limits on the $C_{LQ,U}$ and $C_{Ld}$ operators from processes with two neutrinos are shown in Table \ref{limit_cc}.
Here we neglect the SM contributions and assume  that the bound is saturated by SMEFT operators. 
For $K \rightarrow \pi \nu \bar{\nu}$, this approximation leads to a weaker, and hence more conservative,  bound.

\section{Interim summary---constraints on SMEFT operators}
\label{sect:summary}

In this Section we summarize and display our results so far on constraints from low- and high-energy experiments on coefficients of CLFV SMEFT operators. First in \ssec{singlesummary} we summarize results for turning on one SMEFT operator at a time. In Sec.~\ref{global} we preview what a more global analysis might look like by considering two scenarios of constraints on multiple CLFV operators that contribute simultaneously. A full global analysis is deferred to future work.

\subsection{Single-operator dominance hypothesis}
\label{ssec:singlesummary}

\begin{figure}[t]
\centering
\includegraphics[width=1\textwidth]{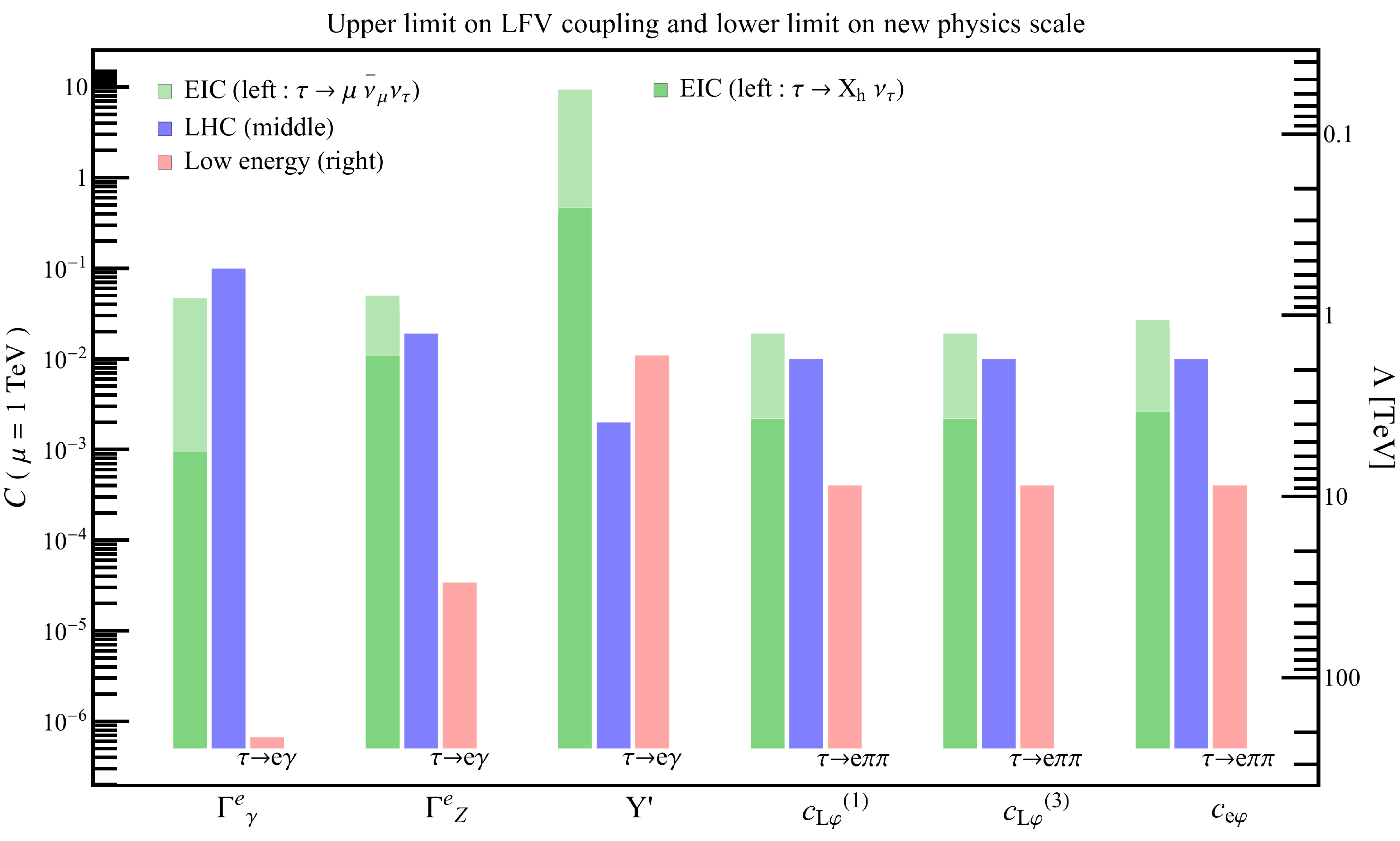}
\caption{Upper limits on $\Gamma^e_{\gamma,Z},~Y_e^{\prime},~c^{(1,3)}_{L\varphi}$ and $c_{e\varphi}$ from the EIC (light green, left), LHC (blue, middle) and low-energy observables (pink, right). The rightmost vertical axis depicts the lower limit on the scale of new physics. The darker green bar overlaid on the light green one is the expected sensitivity in hadronic $\tau$ decays at the EIC assuming the efficiency is $100\%$ with no SM backgrounds.  }
\label{barchart_dipole}
\end{figure}

\begin{figure}[t]
\centering
\includegraphics[width=0.89\textwidth]{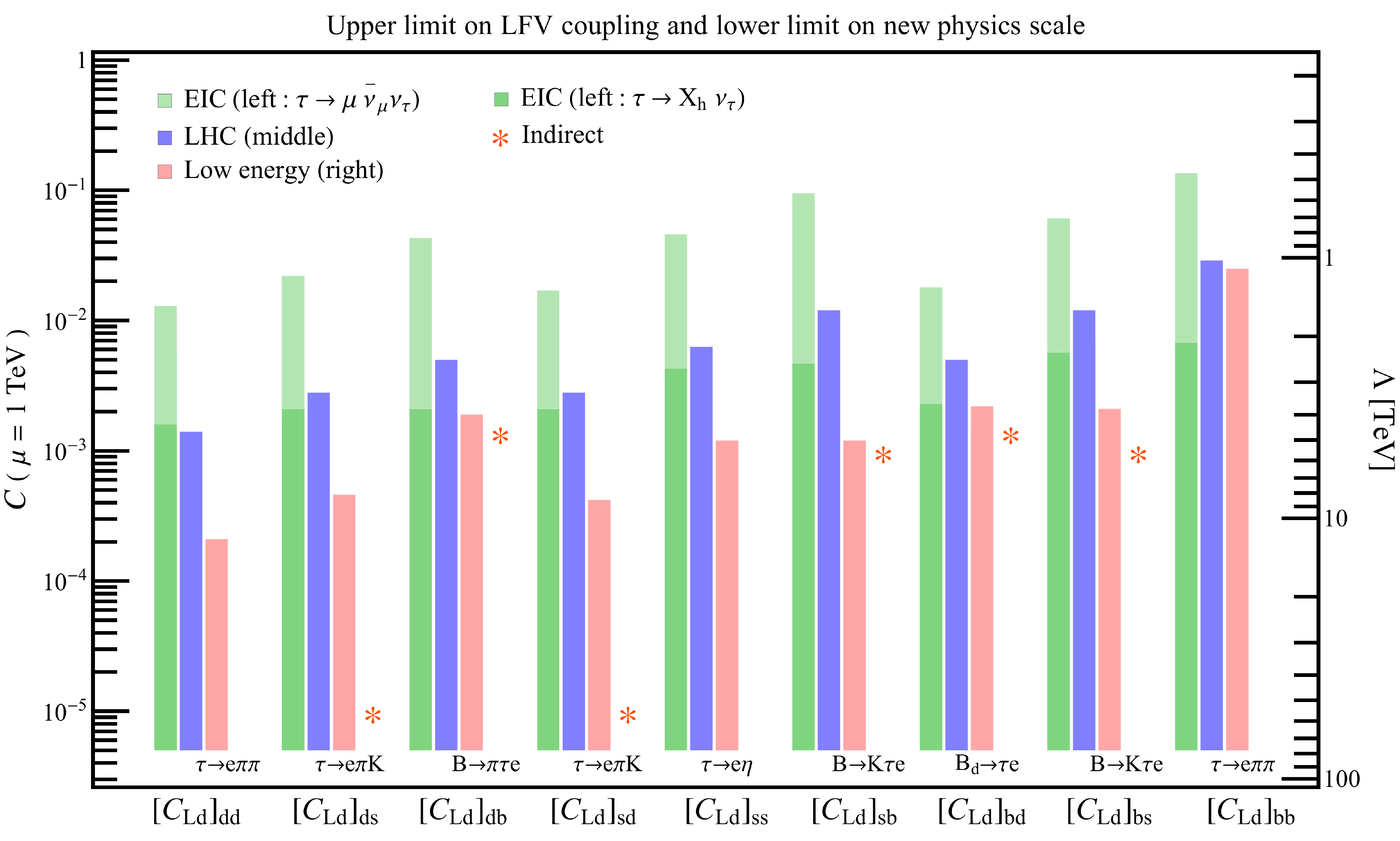}
\caption{Upper limit on $C_{Ld}$ (leftmost axis) and lower limit on new physics scale $\Lambda$ (rightmost axis) from the EIC (left), LHC (middle) and low-energy observables (right). The symbol $``*"$ indicates indirect bounds discussed in Sec. \ref{sect:indirect}. For the EIC expected sensitivity, the light green bar corresponds to the result in Table \ref{EIC_sens}, while the dark green one represents the case in hadronic tau decay mode assuming $\epsilon_{n_b}=1$ with $n_b=0$ in Tables \ref{EIC_sens_LL} -- \ref{EIC_sens_S}.  }
\label{barchart_CLd}
\end{figure}

\begin{figure}[h]
\centering
\includegraphics[width=0.89\textwidth]{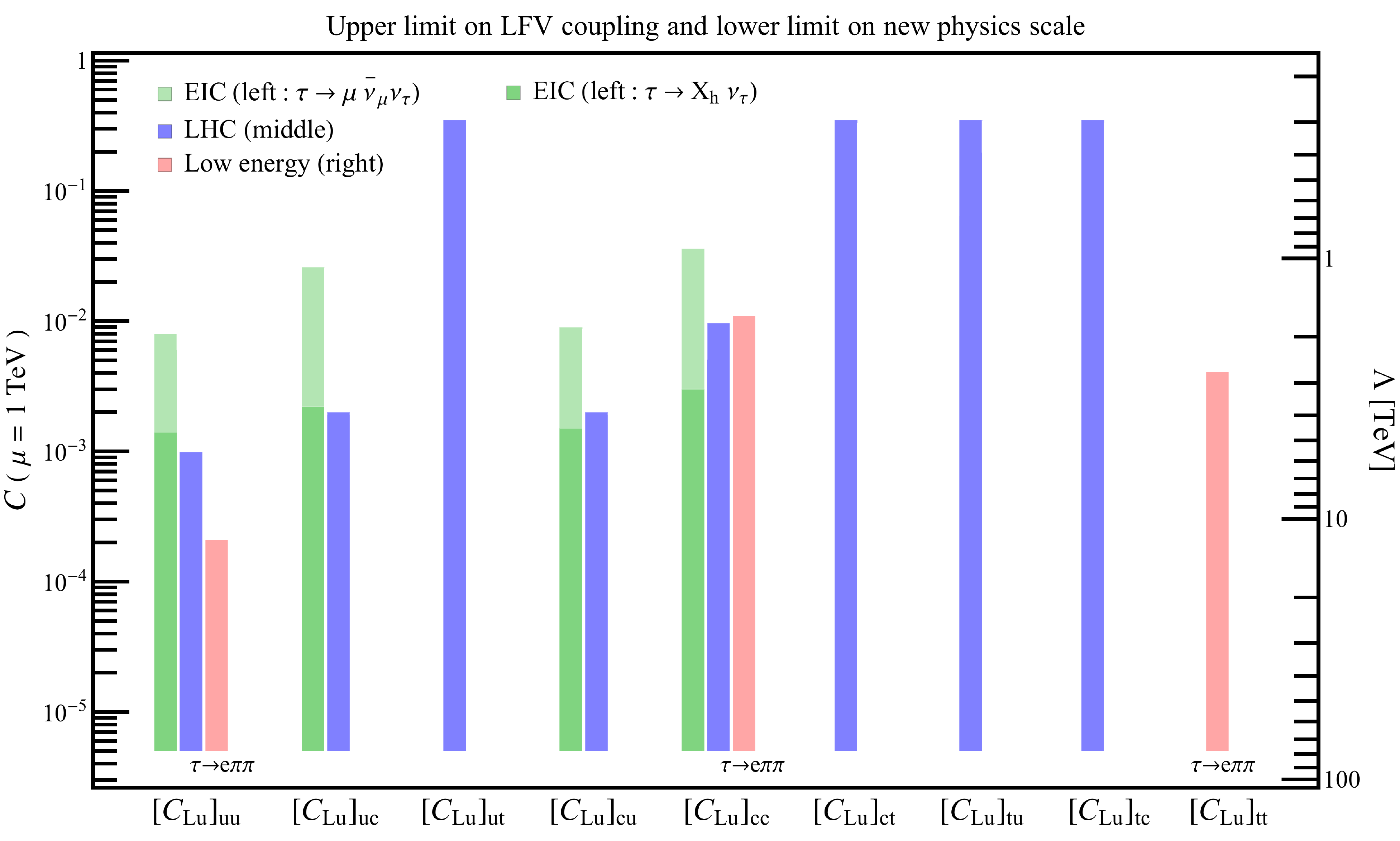}
\caption{Upper limit on $C_{Lu}$ (leftmost axis) and lower limit on new physics scale $\Lambda$ (rightmost axis). For the EIC expected sensitivity, the light green bar corresponds to the result in Table \ref{EIC_sens}, while the dark green one represents the case in hadronic tau decay mode assuming $\epsilon_{n_b}=1$ with $n_b=0$ in Tables. \ref{EIC_sens_LL} -- \ref{EIC_sens_S}.  }
\label{barchart_CLu}
\end{figure}

\begin{figure}[h]
\centering
\includegraphics[width=0.89\textwidth]{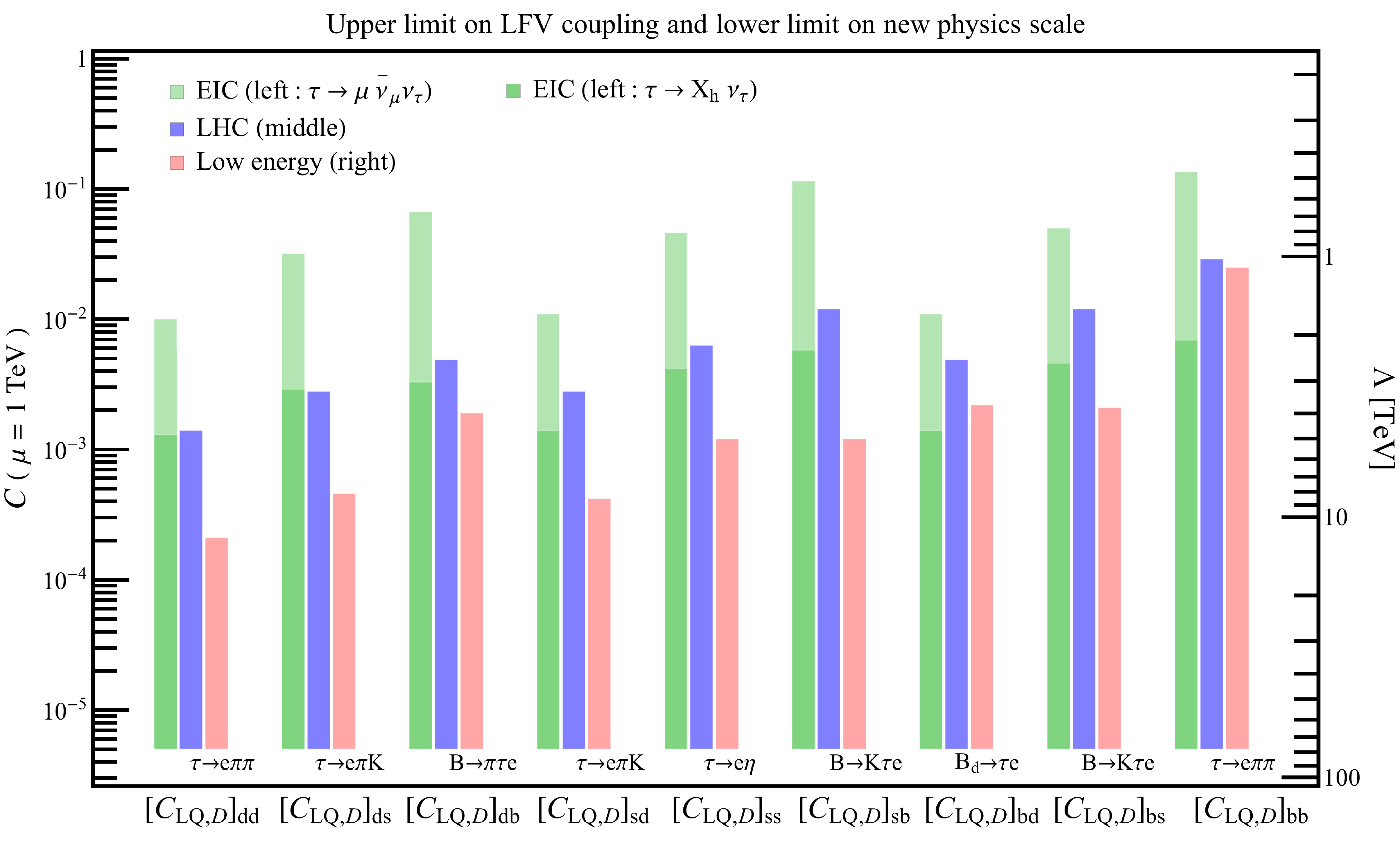}
\caption{Upper limit on $C_{LQ,D}$ (leftmost axis) and lower limit on new physics scale $\Lambda$ (rightmost axis). For the EIC expected sensitivity, the light green bar corresponds to the result in Table \ref{EIC_sens}, while the dark green one represents the case in hadronic tau decay mode assuming $\epsilon_{n_b}=1$ with $n_b=0$ in Tables \ref{EIC_sens_LL} -- \ref{EIC_sens_S}.}
\label{barchart_CLQD}
\end{figure}

\begin{figure}[h]
\centering
\includegraphics[width=0.89\textwidth]{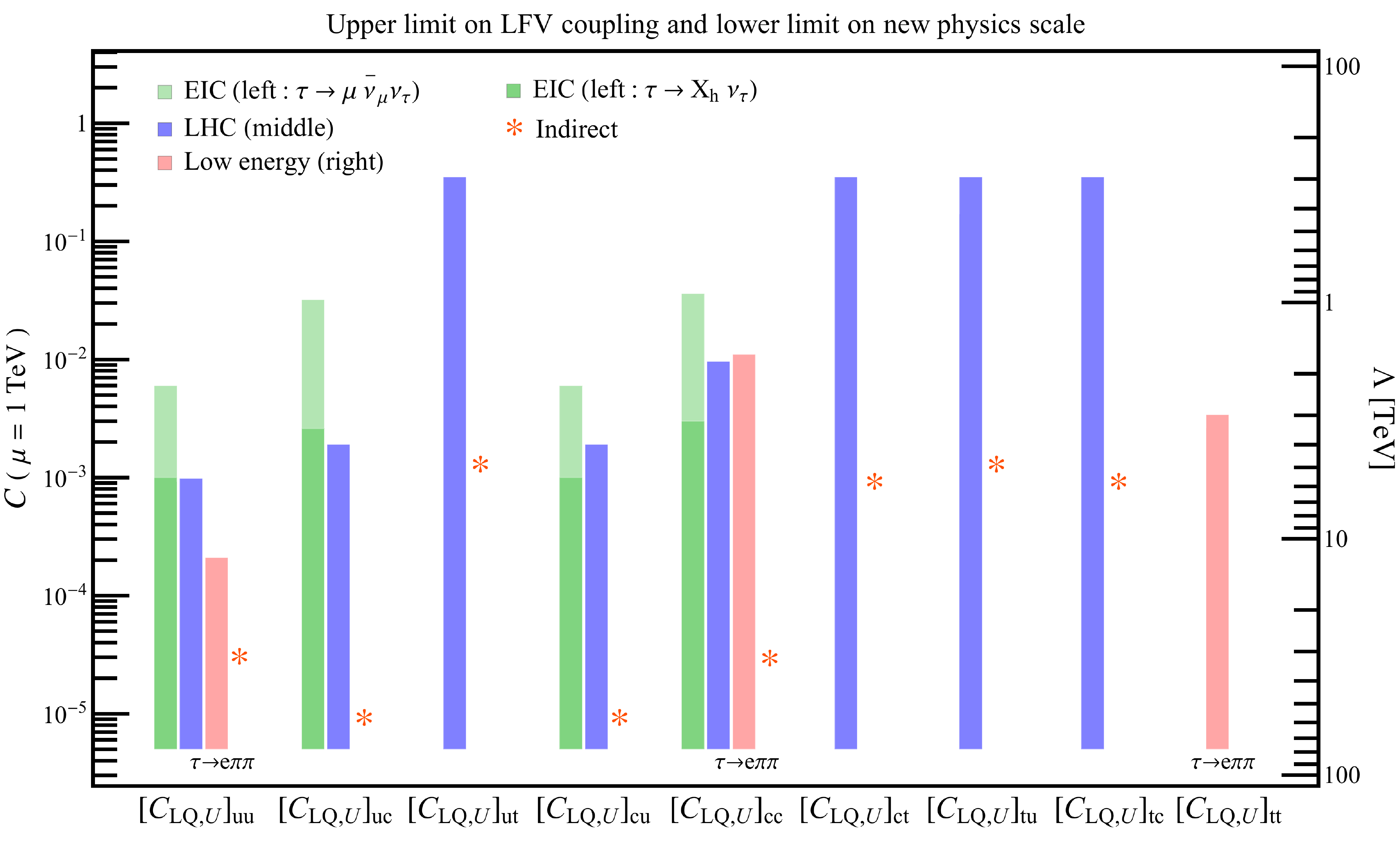}
\caption{Upper limit on $C_{LQ,U}$ (leftmost axis) and lower limit on new physics scale $\Lambda$ (rightmost axis). For the EIC expected sensitivity, the light green bar corresponds to the result in Table \ref{EIC_sens}, while the dark green one represents the case in hadronic tau decay mode assuming $\epsilon_{n_b}=1$ with $n_b=0$ in Tables \ref{EIC_sens_LL} -- \ref{EIC_sens_S}.}
\label{barchart_CLQU}
\end{figure}

\begin{figure}[h]
\centering
\includegraphics[width=0.89\textwidth]{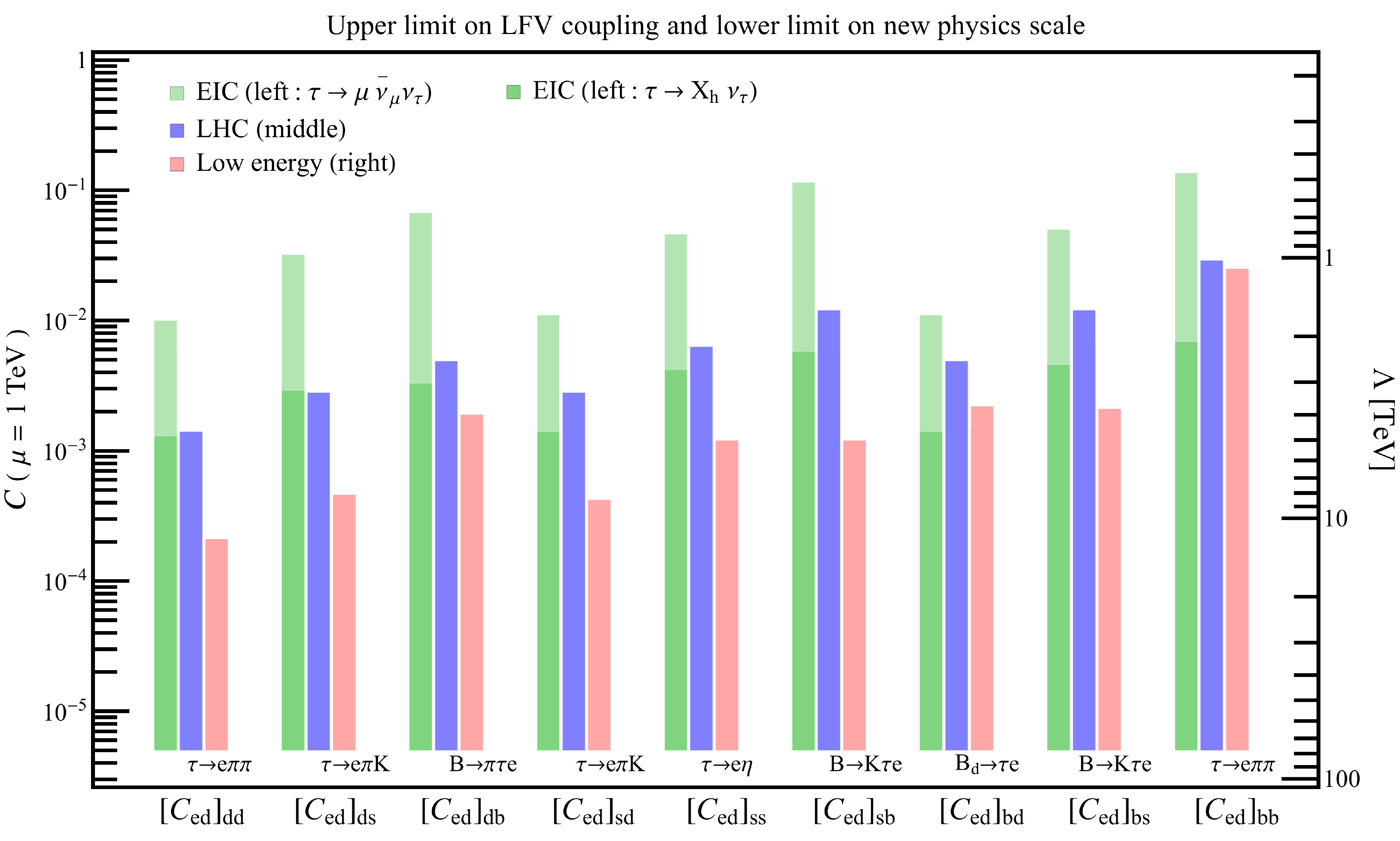}
\caption{Upper limit on $C_{ed}$ (leftmost axis) and lower limit on new physics scale $\Lambda$ (rightmost axis). For the EIC expected sensitivity, the light green bar corresponds to the result in Table \ref{EIC_sens}, while the dark green one represents the case in hadronic tau decay mode assuming $\epsilon_{n_b}=1$ with $n_b=0$ in Tables \ref{EIC_sens_LL} -- \ref{EIC_sens_S}.}
\label{barchart_Ced}
\end{figure}

\begin{figure}[h]
\centering
\includegraphics[width=0.89\textwidth]{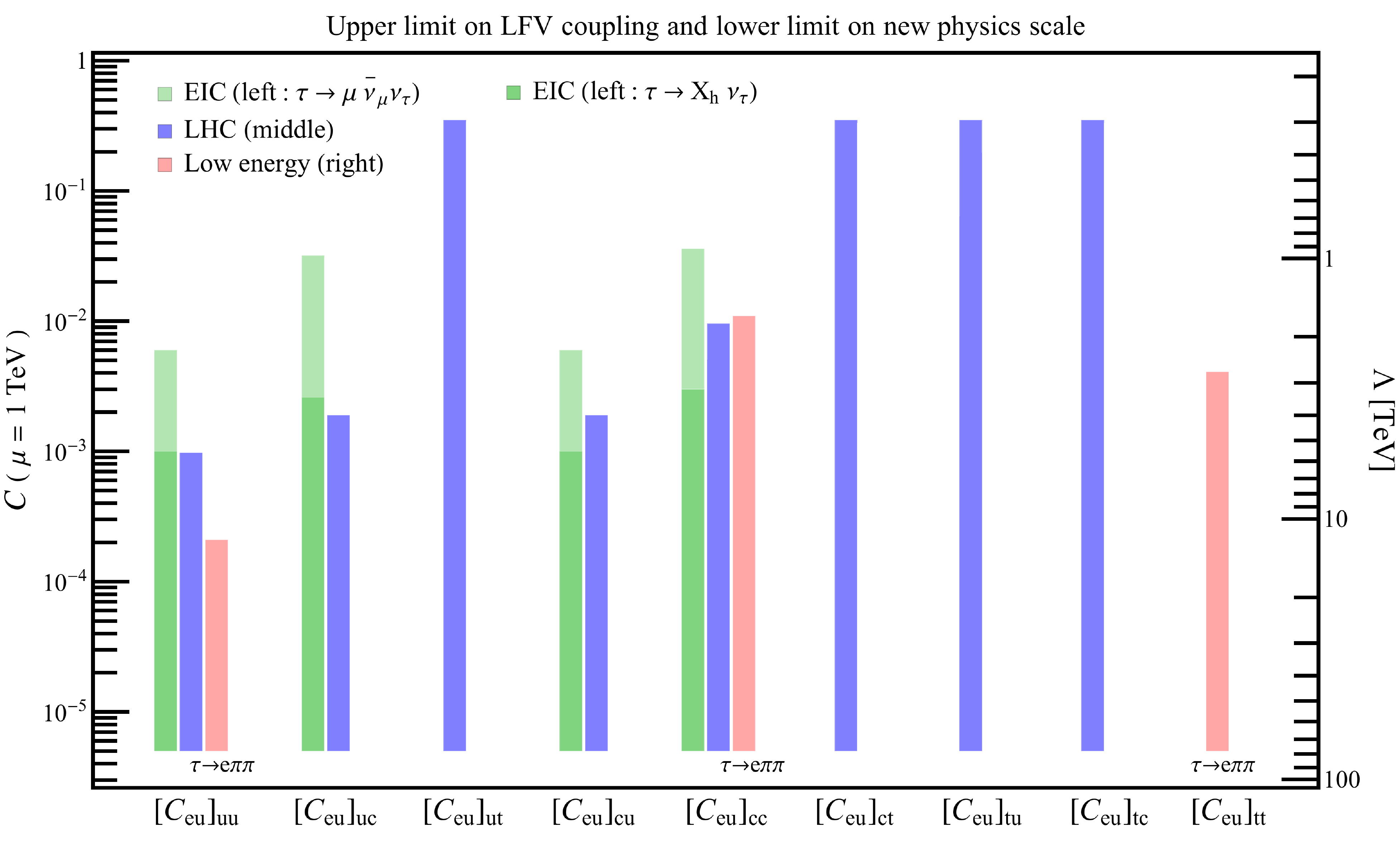}
\caption{Upper limit on $C_{eu}$ (leftmost axis) and lower limit on new physics scale $\Lambda$ (rightmost axis). For the EIC expected sensitivity, the light green bar corresponds to the result in Table \ref{EIC_sens}, while the dark green one represents the case in hadronic tau decay mode assuming $\epsilon_{n_b}=1$ with $n_b=0$ in Tables \ref{EIC_sens_LL} -- \ref{EIC_sens_S}.}
\label{barchart_Ceu}
\end{figure}

\begin{figure}[h]
\centering
\includegraphics[width=0.89\textwidth]{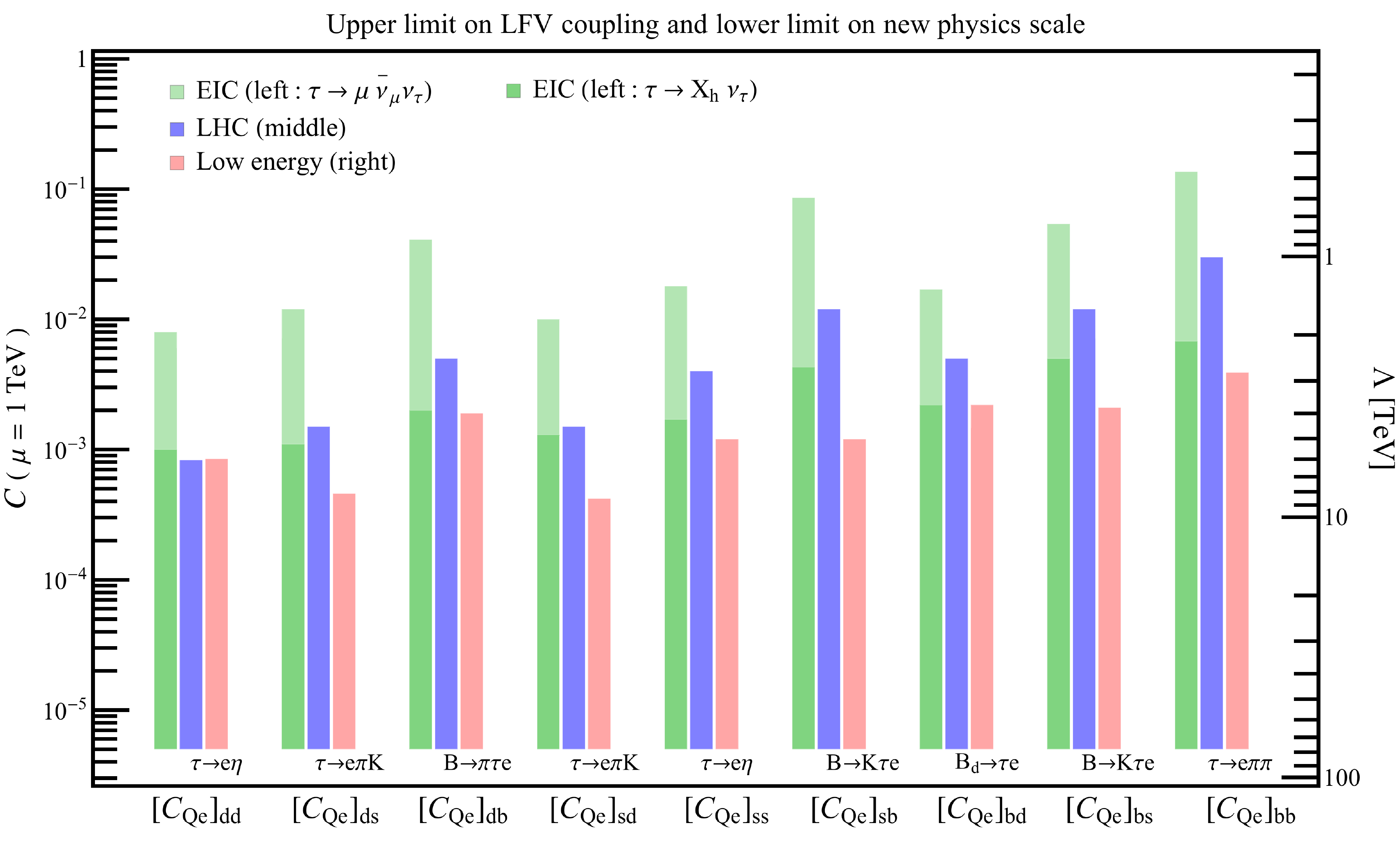}
\caption{Upper limit on $C_{Qe}$ (leftmost axis) and lower limit on new physics scale $\Lambda$ (rightmost axis). For the EIC expected sensitivity, the light green bar corresponds to the result in Table \ref{EIC_sens}, while the dark green one represents the case in hadronic tau decay mode assuming $\epsilon_{n_b}=1$ with $n_b=0$ in Tables \ref{EIC_sens_LL} -- \ref{EIC_sens_S}.}
\label{barchart_CQe}
\end{figure}

\begin{figure}[h]
\centering
\includegraphics[width=0.89\textwidth]{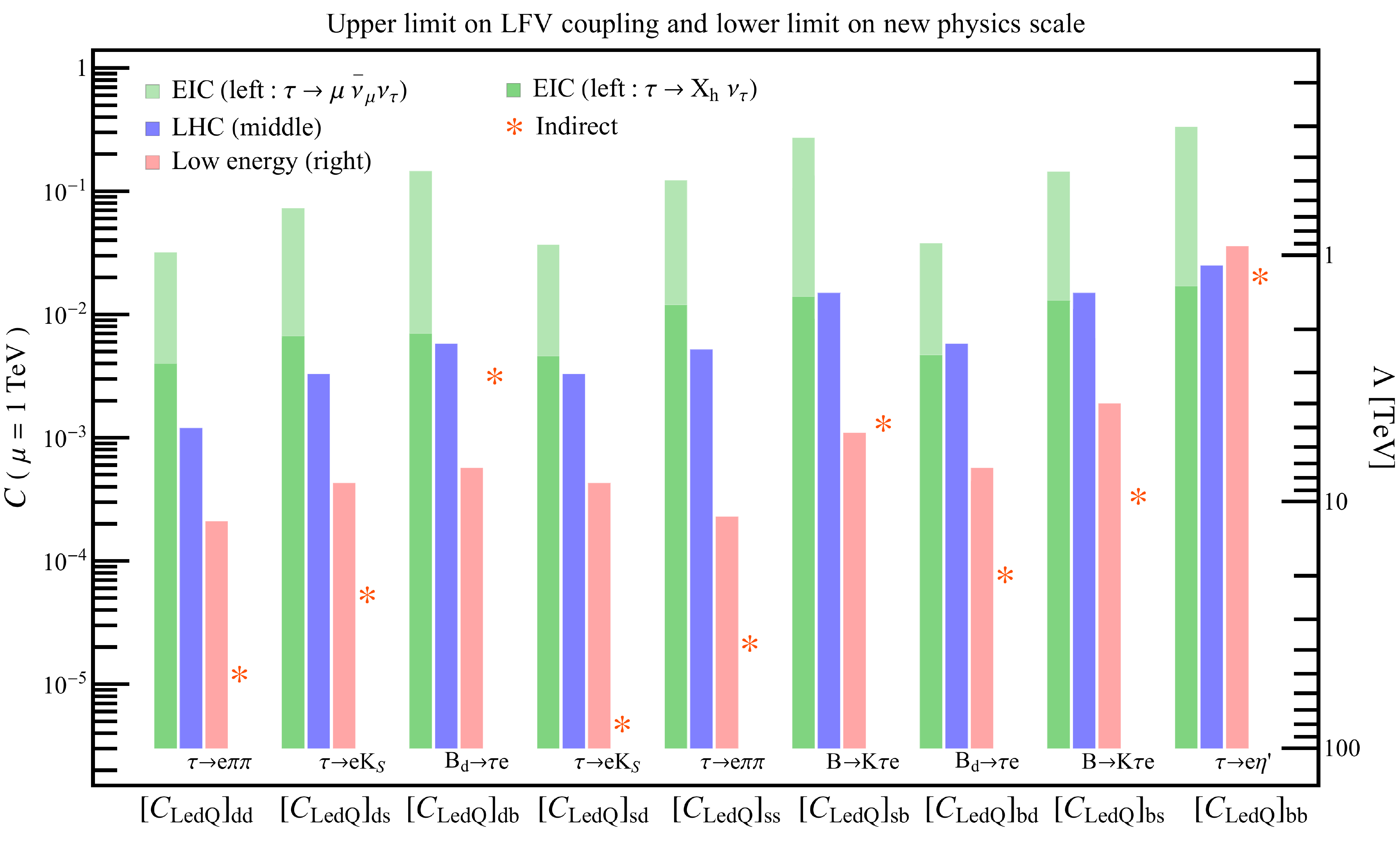}
\caption{Upper limit on $C_{LedQ}$ (leftmost axis) and lower limit on new physics scale $\Lambda$ (rightmost axis). For the EIC expected sensitivity, the light green bar corresponds to the result in Table \ref{EIC_sens}, while the dark green one represents the case in hadronic tau decay mode assuming $\epsilon_{n_b}=1$ with $n_b=0$ in Tables \ref{EIC_sens_LL} -- \ref{EIC_sens_S}.}
\label{barchart_CLedQ}
\end{figure}

\begin{figure}[h]
\centering
\includegraphics[width=0.89\textwidth]{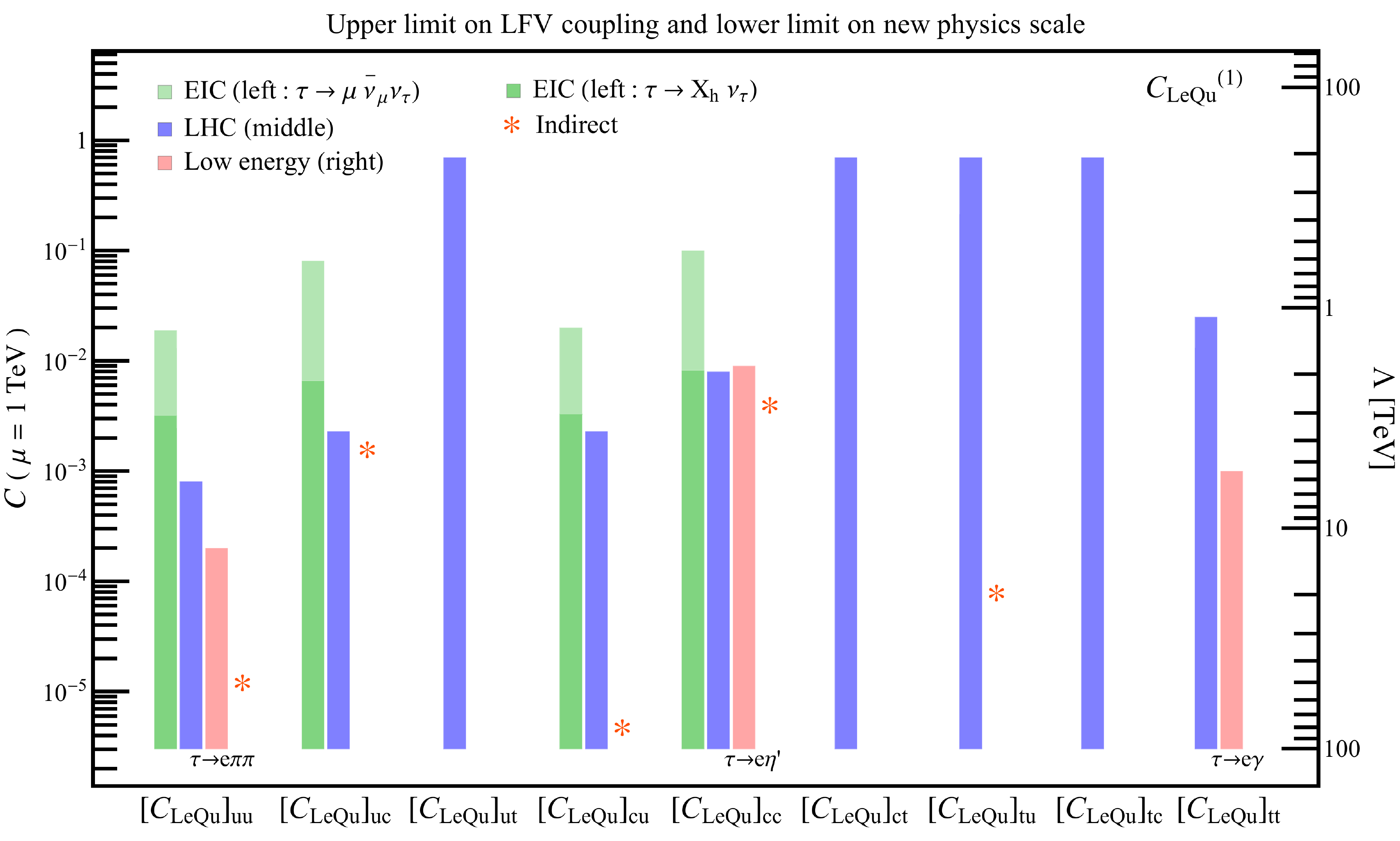}
\caption{Upper limit on $C_{LeQu}^{(1)}$ (leftmost axis) and lower limit on new physics scale $\Lambda$ (rightmost axis). For the EIC expected sensitivity, the light green bar corresponds to the result in Table \ref{EIC_sens}, while the dark green one represents the case in hadronic tau decay mode assuming $\epsilon_{n_b}=1$ with $n_b=0$ in Tables. \ref{EIC_sens_LL} -- \ref{EIC_sens_S}.}
\label{barchart_CLeQu1}
\end{figure}

\begin{figure}[h]
\centering
\includegraphics[width=0.89\textwidth]{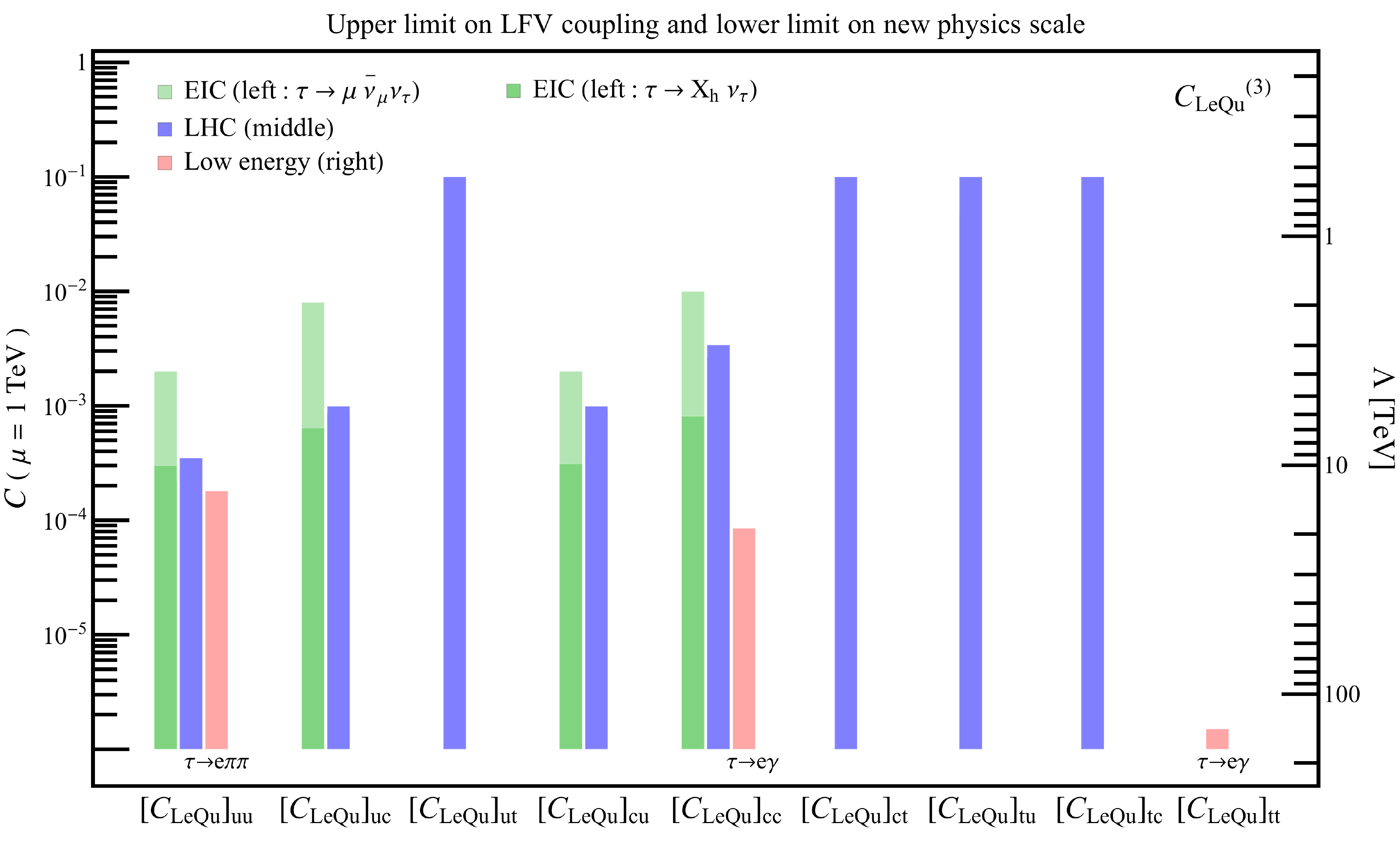}
\caption{Upper limit on $C_{LeQu}^{(3)}$ (leftmost axis) and lower limit on new physics scale $\Lambda$ (rightmost axis). For the EIC expected sensitivity, the light green bar corresponds to the result in Table \ref{EIC_sens}, while the dark green one represents the case in hadronic tau decay mode assuming $\epsilon_{n_b}=1$ with $n_b=0$ in Tables. \ref{EIC_sens_LL} -- \ref{EIC_sens_S}.}
\label{barchart_CLeQu3}
\end{figure}

We summarize here the upper limits on the LFV couplings discussed in Sections \ref{efficiency}, \ref{sect:LHC} and \ref{low-energy}, 
obtained by assuming that a single operator at a time is turned on at the high scale $\Lambda$. 
While not necessarily reflecting the pattern of Wilson coefficients in  concrete extensions of the SM, this analysis nonetheless provides a good 
guidance on the relative sensitivity of various probes of $e$-$\tau$ CLFV.  
Our findings are summarized in Figs. \ref{barchart_dipole}--\ref{barchart_CLeQu3}. 
The leftmost and rightmost vertical axes in Figs. \ref{barchart_dipole}--\ref{barchart_CLeQu3} present the bounds on the dimensionless Wilson coefficient $C(\mu=1~{\rm TeV})$ and the scale $\Lambda$, respectively. The value of $\Lambda$ is obtained by taking $4G_FC/\sqrt{2} \equiv1/\Lambda^2$. The blue and pink bars represent 
existing $90\%$ C.L. limits from the LHC and low-energy observables, respectively. The pink bars are labeled  by the decay mode that gives the strongest limit as in Tables \ref{limit_semileptonic} and \ref{limit_semileptonic_vio}.
The green bars show the EIC sensitivity, assuming $\sqrt{S}=141~$GeV and an integrated luminosity  ${\cal L} =$100 fb$^{-1}$ (the bound on  
 the Wilson coefficient scales as $1/\sqrt{{\cal L}}$). The light green bars are 
 based on the analysis with muonic $\tau$ decay,  
  for which the cuts discussed in Section \ref{efficiency} allow to reduce the SM background to a negligible level. The cut efficiencies are given in Table \ref{tab:eff}, and vary between $10\%$ and $1\%$, depending on whether the SMEFT operators include valence or sea quarks.
The darker green bar overlaid on the lighter one depicts the maximally optimistic scenario utilizing hadronic $\tau$ decay channels,  
and assuming that the SM background can be reduced to  $n_b=0$ with $\epsilon_{n_b}=1$. The indirect bounds discussed in Sec. \ref{sect:indirect} are
 indicated by a mark $``*"$ in orange. 


The bar charts in  Figs.~\ref{barchart_dipole}--\ref{barchart_CLeQu3} contain several interesting messages. 
For the `vertex correction' operators (dipoles, gauge-fermion, Higgs-fermion) the bounds are depicted in  Fig.~\ref{barchart_dipole}. 
The main take-away points are: 

\begin{itemize}
\item  
The photon dipole  $\Gamma^e_{\gamma}$ receives by far the strongest constraint from $\tau \rightarrow e\gamma$, 
corresponding to the effective new physics scale $\Lambda \gtrsim 200~$TeV. This is the highest scale currently probed 
by $e$-$\tau$ LFV transitions. 
High-invariant-mass Drell-Yan is not very sensitive to this operator, leading to weak limits from the LHC. The EIC can in principle provide better constraints,
but, even in the most optimistic scenario, they would be three orders of magnitude weaker than from $\tau \rightarrow e\gamma$.

\item 
Similarly, the $Z$ dipole $\Gamma^e_{Z}$ is most strongly constrained by $\tau \rightarrow e \gamma$, via RGE running. The second best limit is currently from $Z \rightarrow e \tau$ at the LHC. To be competitive with $\tau \rightarrow e \gamma$, however, the branching ratio ${\rm BR}(Z \rightarrow e \tau)$ 
needs to reach the prohibitive level of $2 \cdot 10^{-11}$.  

\item The most severe limit on non-standard Yukawa couplings  $[Y_e^{\prime}]_{\tau e}$  originates  from the ATLAS search for $h \to \tau  e$~\cite{Aad:2019ugc}.   The strongest low-energy limit on $[Y_e^{\prime}]_{\tau e}$ comes from $\tau \rightarrow e \gamma$, which is roughly a factor of five weaker than the LHC. The EIC  can at best probe Yukawa couplings of order one.

\item The constraints on the $Z$ couplings $c^{(1)}_{L\varphi} + c^{(3)}_{L\varphi}$ and $c_{e\varphi}$ are dominated by $\tau \rightarrow e \pi^+ \pi^-$,
which limits these couplings to be less than $4 \cdot 10^{-4}$, corresponding to a new physics scale of 10 TeV. 
High-invariant mass Drell-Yan is not sensitive to these couplings, since the cross section shows the same dependence on $\sqrt{S}$ as the SM. The best LHC limit therefore comes from $Z\rightarrow e \tau$. A measurement of the $Z\rightarrow e \tau$ branching ratio at the $10^{-8}$ level will be competitive with low-energy constraints. At the EIC, these couplings can be probed at the few permill level.

\end{itemize}

Bounds on the  four-fermion operators with vector/axial and scalar/tensor  Lorentz structure are reported in 
Figs.~\ref{barchart_CLd}--\ref{barchart_CQe}
and \ref{barchart_CLedQ}--\ref{barchart_CLeQu3},  respectively.  We note that: 

\begin{itemize} 

\item The $uu$ component of the vector-like operators $C_{LQ, U}$, $C_{eu}$ and $C_{Lu}$,
and the $dd$ component of the $C_{LQ, D}$, $C_{ed}$, $C_{Ld}$ are very well constrained by hadronic $\tau$ decays,
in particular $\tau \rightarrow e \pi^+ \pi^-$. The LHC limits are currently weaker by a factor of five, and the EIC, especially with improvements in the hadronic channel, can reach levels comparable to the LHC. 

\item The $dd$ component of the isoscalar operator $C_{Qe}$ is constrained at low-energy by $\tau \rightarrow e \eta$. Current LHC limits are already comparable with low-energy. The dominant constraint on the $ss$ components of vector-like operators is also from  $\tau \rightarrow e \eta$.

\item The sensitivities to the $cc$ and $bb$ elements of $C_{Ld/Lu},~C_{LQ,U/D}$ and $C_{ed/eu}$ are comparable among the EIC, LHC and low-energy observables.
At low energy, these operators are constrained via mixing with leptonic operators and semileptonic operators with light quarks,
with the weak loop causing a $\sim 10^{-3}$ suppression in the amplitude. High energy processes are relatively less suppressed.    

\item The top component of the vector operators has a large mixing with $c^{(1)}_{L\varphi} + c^{(3)}_{L\varphi}$ and $c_{e\varphi}$. As a consequence,
these operators are constrained by $\tau \rightarrow e \pi^+ \pi^-$ at the few permill level.

\item The $uu$, $dd$ and $ss$ components of the scalar operators $C^{(1)}_{LeQu}$ and $C_{LedQ}$ receive their dominant direct constraints from $\tau \rightarrow e \pi^+ \pi^-$.
LHC limits are a factor of five to ten weaker. The $cc$ and $bb$ components are equally well constrained by high- and low-energy experiments,
while the top component runs at two loop onto $\Gamma^{e}_\gamma$, which dominates the bound. 
Low energy and collider constraints on the $uu$ component of the tensor operator $C^{(3)}_{LeQu}$ are similar, while low-energy dominates on the $cc$
and $tt$ components, due to the mixing of the tensor operator onto the dipole.

\item Concerning quark-flavor-changing decays, $\tau \rightarrow K_S e$, $\tau \rightarrow e K \pi$, $B \rightarrow \tau e$, $B \rightarrow \pi \tau e$
and $B \rightarrow K \tau e$ allow to constrain the off-diagonal components of $d$-type vector and scalar operators. 
A bound on $B_s \rightarrow e \tau$ at the same level as the recent LHCb limit on  $B_s \rightarrow \mu \tau$ \cite{Aaij:2019okb,LHCb-PUB-2018-009} would provide complementary information.

\end{itemize}

Finally, we note that the results from the indirect observables  (orange $``*"$ in the plots), when available, provide 
limits that are comparable to or stronger than those from the direct observables.

We conclude this survey with some  considerations on the current and future impact of  LHC and  EIC searches for 
$e$-$\tau$ CLFV:  

\begin{itemize}

\item Collider searches play a crucial role in bounding off-diagonal elements of up-type four-fermion operators, while the low-energy observables are insensitive to them. For the $t$-$q$ components, weak running onto flavor-diagonal operators is very suppressed by small Yukawa and CKM elements, so that top decays provide the only sensitive probe.  The $uc$ and $cu$ components could be constrained by $D \rightarrow e \tau$, which will be investigated at LHCb 
\cite{Gisbert:2020vjx} \footnote{We thank M. Fontana, D. Mitzel and M. Williams for communications on this point.}. For both pseudoscalar and axial operators, however, the Drell-Yan limits imply the prohibitive BR$(D\rightarrow e \tau)\sim 10^{-7}-10^{-8}$. 

\item Inclusion of hadronic $\tau$ decays in the EIC analysis provides a great opportunity to improve the sensitivity by a factor of 10 depending on LFV operators.

\end{itemize}

\subsection{Towards a global analysis}\label{global}

\begin{table}[t]
\small
\centering
\begin{tabular}{c c c c c c }
\hline
Decay mode  & $C^{eq}_{\rm VLL + VLR}$ & $C^{eq}_{\rm VLL - VLR}$  & $C^{eq *}_{\rm SRR + SRL}$ & $C^{eq *}_{\rm SRR - SRL}$ & $C^{eq *}_{\rm TRR}$ \\ 
\hline \hline
$\tau \to e \gamma$   &  &   &   & & $(uu)$, $(cc)$ \\
$\tau \to e \ell^+ \ell^-$ & $(cc)$, $(bb)$  &  &   &    \\
$\tau \to e \pi^0$        &      & $uu-dd$ & & $uu-dd$ &    \\
$\tau \to e \eta^{(\prime)}$          &  & $uu+dd$, $ss$   & & $uu+dd$, $ss$   \\
                                      &  &                 & & $(cc)$, $(bb)$    \\
$\tau \to e \pi^+\pi^-$ & $uu-dd$,  &  & $uu+dd$, $ss$ & & $uu$   \\
&$(cc)$, $(bb)$ &  & $(cc)$, $(bb)$\\ 
$\tau \to e K^+ K^-$ & $uu+dd$, $uu-dd$  &  & $uu+dd$, $uu-dd$ & &  $uu$   \\
&$ss$, $(cc)$, $(bb)$ &  & $ss$, $(cc)$, $(bb)$ \\
    \hline
$\tau \to e K_S^0$ & & $sd -ds$ & &  $sd-ds$    \\
$\tau^- \to e^- K^+ \pi^+ $ & $sd$ & & $ds$ &    \\
$\tau^- \to e^- K^- \pi^+ $ & $ds$ & & $sd$ &    \\
$B^0 \to e^{\pm}\tau^{\mp} $ & & $db$, $bd$ &  & $bd$, $db$  &   \\
$B^+ \to \pi^{+} e^{-} \tau^{+} $ & $db$ & & $bd$  &     \\
$B^+ \to \pi^{+} e^{+} \tau^{-} $ & $bd$ & & $db$ &     \\
$B^+ \to K^{+} e^{-} \tau^{+} $   & $sb$ & & $bs$ &     \\  
$B^+ \to K^{+} e^{+} \tau^{-} $   & $bs$ & & $sb$ &     \\  
\hline 
\end{tabular}
\caption{Dependence of low-energy decay channels on the coefficients of semileptonic LEFT operators at the matching scale $\mu \sim v$.
We focus here on operators with left-handed electrons, an analogous table can be made for operators with right-handed electrons. 
The parentheses imply that the operator induces the decay mode at the loop level, either in perturbation theory, e.g. via the RGE running of $\left(C^{ed}_{\rm VLL + VLR}\right)_{\tau e b b}$ onto four-lepton operators or the matching of $\left(C^{ed *}_{\rm SRR - SRL}\right)_{e\tau bb}$ onto $C_{G \widetilde G}$ at the $m_b$ threshold, or via hadronic loops, e.g. the contribution of $\left(C^{eu *}_{\rm TRR}\right)_{e \tau u u}$ to $\tau \rightarrow e \gamma$. 
We ignore $d$-type tensor operators, which are not induced by matching onto SMEFT.
}
\label{low-energy_operator_2}
\end{table}

The discussion has so far focused on a single coupling analysis. 
In most extensions of the SM this is not a realistic scenario, as  several operators are generated at the matching scale $\Lambda$. 
`Switching on' more than one coupling at the high scale could in principle result in cancellations 
that weaken the bounds reported in previous sections. 
We next discuss  the extent to which this is possible,  showing that 
complementary information from colliders in general and EIC in particular becomes very relevant. 
Our discussion below is exploratory  and we  refrain  from a global analysis that is beyond the scope of this work. 

To facilitate the identification of directions in parameter space that are unconstrained by low-energy probes, in Table \ref{low-energy_operator_2}
we summarize the dependence of the $\tau$ and $B$ branching ratios used in our analysis on LEFT semileptonic operators, defined at the matching scale between the SMEFT and the LEFT. For exclusive channels, the contributions are more easily organized by constructing combinations in which the quark bilinears
have well defined parity transformations \cite{Celis:2014asa,Celis:2013xja}. Since the interference between operators with left- and right-handed electrons is suppressed by the electron mass and always negligible, in Table \ref{low-energy_operator_2} we only show operators with left-handed electrons, similar conclusions can be drawn for operators with right-handed electrons.

For down-type operators,
assuming the presence of a single operator structure with all the flavor entries 
simultaneously turned on does not  entail a significant weakening of the constraints. Consider for example
the operator $C_{Ld}$. From the summary in Fig. \ref{barchart_CLd} we can see that only two components receive the strongest constraint from the same process, namely $dd$ and $bb$, which are both limited by $\tau \rightarrow e \pi^+ \pi^-$ (the $sd$ and $ds$ components contribute to $\tau^- \rightarrow e^- K^- \pi^+$ and $\tau^- \rightarrow e^- K^+ \pi^-$, respectively, for which there are two independent constraints; similar considerations apply to the $sb$ and $bs$ components). When we simultaneously turn on $\left[C_{Ld}\right]_{dd}$ and $\left[C_{Ld}\right]_{bb}$, $\tau \rightarrow e \pi$ and $\tau \rightarrow e \ell \ell$ become relevant and the limits only slightly deteriorate, $\left| \left[C_{Ld}\right]_{dd} \right| < 5.3 \cdot 10^{-4}$
and $\left| \left[C_{Ld}\right]_{bb} \right| < 4.8 \cdot 10^{-2}$, at the 90\% CL.

The situation changes  if we simultaneously turn on two or more operators at the same time.
As an example, we consider two scenarios, in which we turn on: A) the left-handed $Z$-coupling operator $c^{(1)}_{L\varphi} + c^{(3)}_{L\varphi}$  and the  light-quark components of all operators with two left-handed leptons and B) $c^{(1)}_{L\varphi} + c^{(3)}_{L\varphi}$ and 
all flavor-diagonal components of the two down-type operators with two left-handed leptons, $\left[C_{Ld}\right]$ and $\left[C_{LQ,\, D}\right]$. The nonzero coefficients in the two scenarios are summarized in Table \ref{multioperator}. Including operators with two right-handed leptons would not further weaken the limits, since, as we already noted, the interference of vector operators with leptons with different chirality is suppressed by the electron mass.

\begin{table}
\centering
\begin{tabular}{c|c}
\hline
Scenario & Operators \\
\hline\hline
A & $c^{(1,3)}_{L\varphi},~[C_{LQ,U}]_{uu},~[C_{LQ,D}]_{dd,ss},~[C_{Lu}]_{uu},~[C_{Ld}]_{dd,ss}$ \\
B& $c^{(1,3)}_{L\varphi},~[C_{LQ,D}]_{dd,ss,bb},~[C_{Ld}]_{dd,ss,bb}$ \\
\hline
\end{tabular}
\caption{Multi-operator scenarios A and B.}
\label{multioperator}
\end{table}

\begin{figure}
\center
\includegraphics[width=0.48\textwidth]{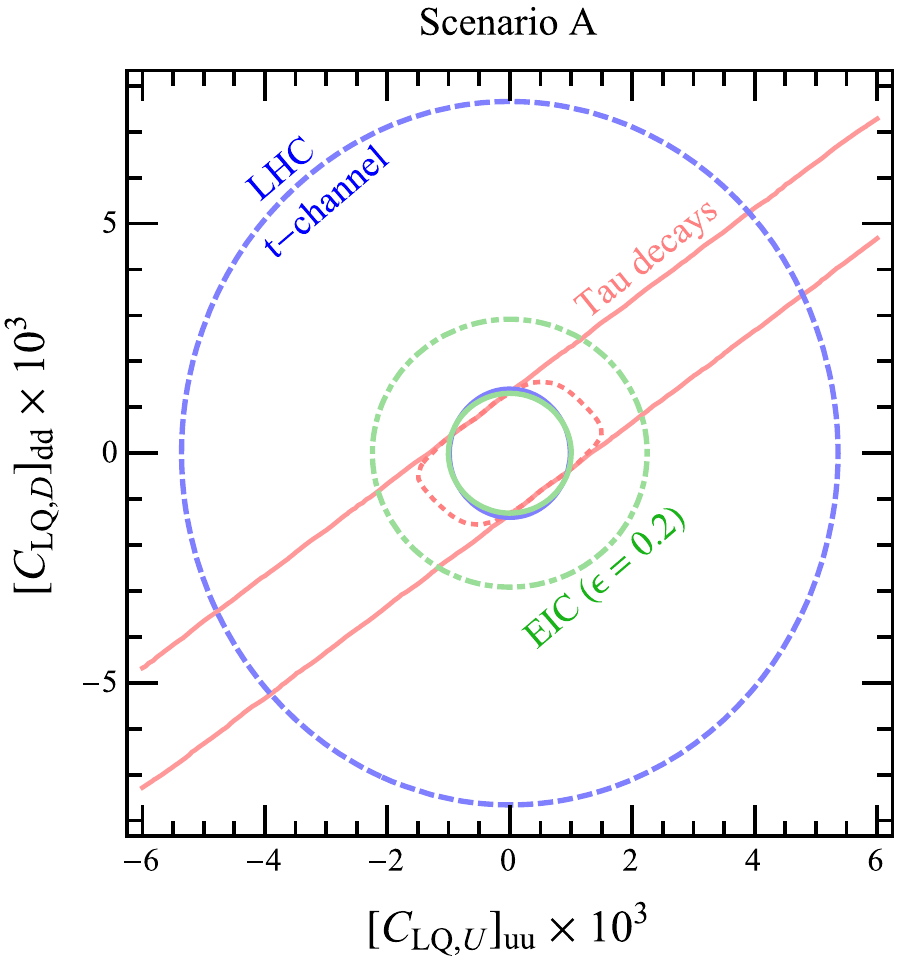}\quad 
\includegraphics[width=0.48\textwidth]{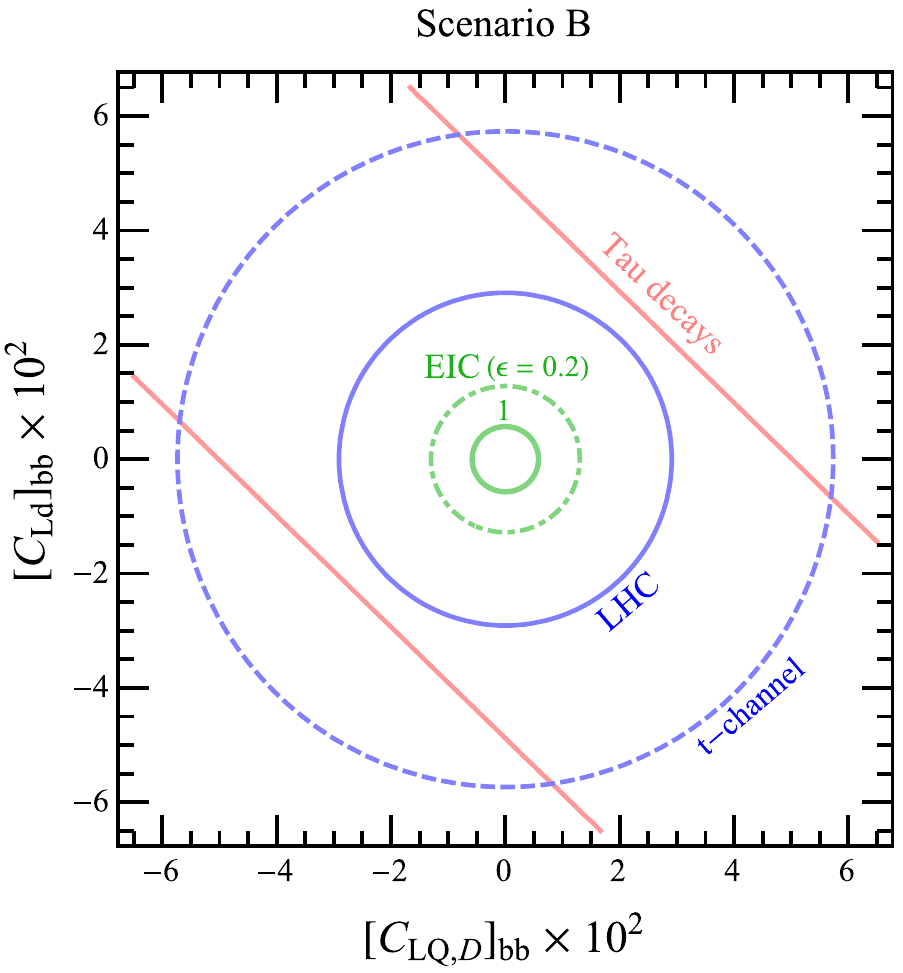}
\vspace{-1ex}
\caption{The $90\%$ C.L. limits in $[C_{LQ,U}]_{uu}-[C_{LQ,D}]_{dd}$ (scenario A, left) and $[C_{LQ,D}]_{bb}-[C_{Ld}]_{bb}$  planes (scenario B, right). The pink lines are limits from $\tau$ decays, while the pink dotted contour presents the case incorporating $\tau \to e K^+K^-$ channel.
The blue and green solid lines are bounds from the LHC and EIC ($\epsilon_{n_b}=1$ with $n_b=0$), respectively. The blue dashed line assumes a $t$-channel exchange of a particle with $M=1$ TeV at the LHC, and the green dash-dotted line assumes $\epsilon_{n_b}=0.2$ in the EIC sensitivity.  }
\label{fig:marginalized}
\end{figure}

In the left panel of Fig. \ref{fig:marginalized}, we show the 90\% C.L. limits on $\left[C_{LQ, U}\right]_{\tau e uu}$ and $\left[C_{LQ, U}\right]_{\tau e dd}$ in the scenario A, marginalized over the six remaining couplings.  The region between the two pink lines is allowed by low-energy experiments. The blue solid and dashed lines correspond to the limits from the LHC. The solid line corresponds to the EFT analysis of Section \ref{sect:LHC}, while the dashed line is obtained by assuming that the effective operators are induced by the $t$-channel exchange of a new particle with mass $M=1$ TeV, see Eq. \eqref{tchan}. The green solid (dash-dotted) line represents the projected EIC sensitivity in hadronic $\tau$ decay mode assuming the efficiency is 1 (0.2) with zero SM background. We see that now there are enough couplings to engineer cancellations in the leading hadronic channel, $\tau \rightarrow e \pi^+ \pi^-$. In the axial direction there are still enough constraints from $\tau \rightarrow e \pi$,  $\tau \rightarrow e \eta$ and $\tau \rightarrow e \eta^\prime$. The isoscalar combination of vector couplings,
$\left[C_{LQ, U} + C_{Lu}\right]_{\tau e uu} + \left[C_{LQ, D} + C_{Ld}\right]_{\tau e dd}$,
is however unconstrained by the observables we consider in Sec.~\ref{low-energy}, leading to the appearance of a free direction. Including the $\tau \to e K^+K^-$ channel closes this free direction, since the process receives contributions from 
both isovector and isoscalar operators. The fit including the $\tau \to e K^+K^-$ mode is presented by the pink dotted contour.
Even with the inclusion of this mode, colliders are very competitive with low-energy.

The right panel of Fig. \ref{fig:marginalized} presents the bounds on $[C_{LQ,D}]_{\tau e bb}$ and $[C_{Ld}]_{\tau e bb}$ in the scenario B, where the rest of the operators are marginalized in the same way as in scenario A. 
As can be seen from Table \ref{RGE_low_vector}, modulo a small component induced by the $b$ Yukawa, the purely leptonic and the semileptonic operators with light quarks receive a contribution that is proportional to the vector combination 
$\left[C_{LQ,\, D} +  C_{Ld}\right]_{\tau e bb}$, leaving the axial combination $\left[C_{LQ,\, D} -  C_{Ld}\right]_{\tau e bb}$ unconstrained by low-energy processes.
The free direction can be closed using LHC data, which currently impose percent level constraints. In this case, assuming that the effective operators are induced by the $t$-channel exchange of a mediator with $M=1$ TeV (dashed blue line) only weakens the bound by a factor of two. 
The EIC can potentially do much better and improve the bounds by a factor of five. In addition, 
while high-invariant-mass Drell-Yan is sensitive to the sum of all quark flavors, the EIC could clearly identify the CLFV mechanism
by tagging the $b$ quark in the final state.  
Similar considerations hold for off-diagonal couplings. 
While $B \rightarrow \tau^+ e^- + \tau^- e^+$ and $B\rightarrow \pi \tau e$ are sufficient to constrain both the vector and axial 
combinations $\left[C_{LQ,\, D} \pm  C_{Ld}\right]_{\tau e bd,\, db}$, $B\rightarrow K \tau e$ 
and $\tau \rightarrow e K \pi$ only constrain the vector combinations $\left[C_{LQ,\, D} +  C_{Ld}\right]_{\tau e bs,\, sb}$
and $\left[C_{LQ,\, D} +  C_{Ld}\right]_{\tau e sd, ds}$, and $\tau \rightarrow e K_S$ the linear combination  $\left[C_{LQ,\, D} -  C_{Ld}\right]_{\tau e sd} - \left[C_{LQ,\, D} -  C_{Ld}\right]_{\tau e ds}$.  Collider information is thus always necessary to complement the strong constraints from low-energy.

For scalar and pseudoscalar operators, Table \ref{low-energy_operator_2} suggests that the isovector $uu-dd$ component
of scalar operators and the $bs$ and $sb$ components of pseudoscalar operators are unconstrained at low energy. In this case, however, 
the $SU(2)_L \times U(1)_Y$ invariance of the SMEFT implies that the scalar and pseudoscalar linear combinations are not independent, 
and the observables included in our analysis are sufficient to fully constrain all the diagonal components of $C^{(1)}_{LeQu}$ 
and all the components of $C^{}_{LedQ}$.

\section{Leptoquark models}
\label{sect:leptoquarks}

To illustrate the EFT framework we consider three simplified models involving scalar leptoquarks (LQ).
In the notation of Ref.~\cite{Davidson:1993qk}, we consider the leptoquarks 
$S_{1/2}$, and $\widetilde{S}_{1/2}$, which are color (anti)triplets and  weak isospin doublets, 
with weak hypercharge  $-7/3$ and $-1/3$, respectively~\footnote{These fields correspond via charge conjugation to $R_2$ and $\tilde R_2$ in the notation of Ref.~\cite{Buchmuller:1986zs}.}.
We further restrict the interactions  of $S_{1/2}$ by requiring that it 
couples only to L-handed  leptons ($S_{1/2}^L$) or R-handed leptons ($S_{1/2}^R$).
Apart from the LQ  gauge-kinetic term and mass terms, 
the SM Lagrangian density is extended by:
\bea
{\cal L}_{S_{1/2}^L} &=& \lambda_L^{\alpha a}  \ \bar u_R^\alpha \, \ell_L^a \, S_{1/2}^{L \dagger} + {\rm h.c.}~,
\\
{\cal L}_{S_{1/2}^R} &=&  \lambda_R^{\alpha a}  \ \bar q_R^\alpha \, i \tau_2 e_R^a \, S_{1/2}^{R \dagger} + {\rm h.c.}~, 
\\
{\cal L}_{\widetilde S_{1/2}} &=&  \tilde{\lambda}^{\alpha a}  \ \bar d_R^\alpha \, \ell_L^a \, \widetilde S_{1/2}^{\dagger} + {\rm h.c.}~.
\eea
In the above equations we have denoted by $\alpha$  and $a$   the quark and lepton generation indices, respectively. 
In what follows we will continue to use greek letters for quark generation and latin letters for lepton generation indices.  

Assuming the LQ masses to be considerably above the electroweak scale (consistently with LHC phenomenology \cite{Khachatryan:2015qda,Khachatryan:2016jqo,Sirunyan:2018btu,Aad:2020iuy,Aad:2021rrh}), 
we integrate out the LQ and match onto the SMEFT effective Lagrangian. 
Each of the above models matches at tree level onto a single  four-fermion operator in SMEFT at dimension six. 
At loop level one can generate more operators. However, 
for the purposes of studying lepton flavor violation the most  relevant one is the photon dipole operator 
mediating $\tau \to e \gamma$.  For the one-loop  matching coefficient we will use the results of Refs.~\cite{Davidson:1993qk} and 
\cite{Gonderinger:2010yn}.
For the three models we find:

\begin{itemize}
\item Integrating out $S_{1/2}^L$ generates $O_{Lu}$ and the dipole, with coefficients 
\begin{subequations}
\label{mLQ1}
\begin{align}
\left[C_{Lu}\right]_{ab \alpha \beta} 
&=  \frac{v^2}{4 M_{LQ}^2}   \, (\lambda_L^\dagger)^{a \beta} (\lambda_L)^{\alpha b}, 
\\
\left[ \Gamma_\gamma^e \right]_{e\tau} 
 &= - \frac{3}{64 \pi^2}  \frac{vm_{\tau}}{M_{LQ}^2} \, \sum_\alpha  (\lambda_L^\dagger)^{e \alpha} 
(\lambda_L)^{\alpha \tau} = -\frac{3}{16\pi^2}\frac{m_{\tau}}{v}\sum_{\alpha} \Big[C_{Lu}\Big]_{\tau e\alpha\alpha} ,
\\
\left[ \Gamma_\gamma^e \right]^{*}_{\tau e} 
 &\propto   \, Y_e \approx 0~, 
\end{align}
\end{subequations}
where $Y_\ell = m_\ell/v$ is the charged  lepton Yukawa coupling and we have set the electron mass to zero in the last equation.
Hermiticity implies 
\beq
\Big[C_{Lu}\Big]_{ab \alpha \beta}  = \Big[ C_{Lu} \Big]^*_{ba  \beta \alpha}~.
\eeq
LFV $\tau$ decays probe 
$\left[  \Gamma^e_\gamma \right]_{e \tau}$ and $[C_{Lu}]_{e \tau \alpha \beta}$, 
while EIC processes probe the complex conjugate of these coefficients.

\item Integrating out $S_{1/2}^R$ generates $O_{Qe}$ and the dipole, with coefficients
\begin{subequations}
\begin{align}
\Big[C_{Qe}\Big]_{\alpha \beta ab} &=  \frac{v^2}{4 M_{LQ}^2}   \, (\lambda_R^\dagger)^{a \beta} (\lambda_R)^{\alpha b} ,
\\
\left[ \Gamma_\gamma^e \right]_{\tau e}^{*} 
 &= - \frac{3}{64 \pi^2}  \frac{vm_{\tau}}{M_{LQ}^2} \,   \sum_\alpha  (\lambda_R^\dagger)^{e \alpha} 
(\lambda_R)^{\alpha \tau} = -\frac{3}{16\pi^2}\frac{m_{\tau}}{v}\sum_{\alpha} \Big[C_{Qe}\Big]_{\alpha\alpha\tau e} \,,
\\
\left[ \Gamma_\gamma^e \right]_{e \tau} 
 &\propto   \, Y_e \approx 0~.
\end{align}
\end{subequations}

\item Integrating out $\widetilde S_{1/2}$ generates $O_{Ld}$  and no dipole operator 
due to a cancellation between the photon emission from internal quark and LQ lines~\cite{Davidson:1993qk,Gonderinger:2010yn}. 
For the four-fermion operator we find 
\bea
\Big[C_{Ld}\Big]_{\alpha \beta ab} &=&  \frac{v^2}{4 M_{LQ}^2}   \, (\tilde \lambda^\dagger)^{a \beta} (\tilde \lambda)^{\alpha b} ~.
\eea

\end{itemize}

Introducing two vectors that express the LQ couplings, e.g. $(v_\tau)^\alpha \equiv  \lambda^{\alpha \tau}$ and $(v_e)^\alpha \equiv \lambda^{\alpha e}$ (the index $\alpha$ runs over the three quark generations), we can express the induced LFV couplings as an outer product of the two vectors,  $[C_M]_{\tau e\alpha \beta}=v_{\tau}^{\alpha}v_e^{\beta}$,
where $M$ labels four-fermion operator. Our analysis assumes that each Wilson coefficient is real, which enables all the coefficients to be expressed by only five independent parameters. 
For example, if $[C_M]_{\tau e11,21,31}$ are chosen as  three independent parameters, the rest of the components are described by a product of one of the three elements and a ratio, $r_2=v_e^2/v_e^1$ or $r_3=v_e^3/v_e^1$. 

\begin{figure}[t]
\vspace{-1em}
\centering
\includegraphics[width=0.45\textwidth]{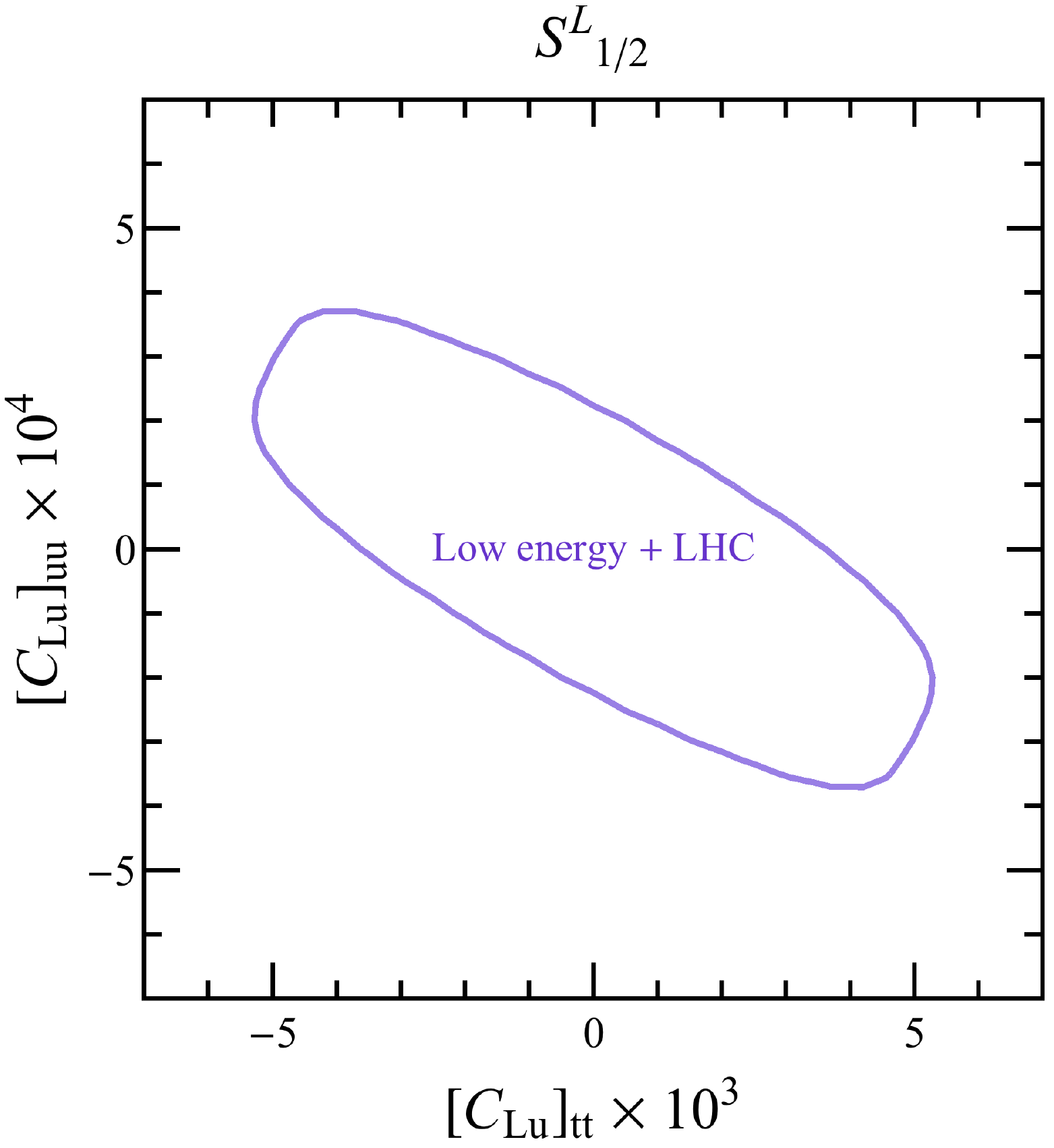}\quad
\includegraphics[width=0.455\textwidth]{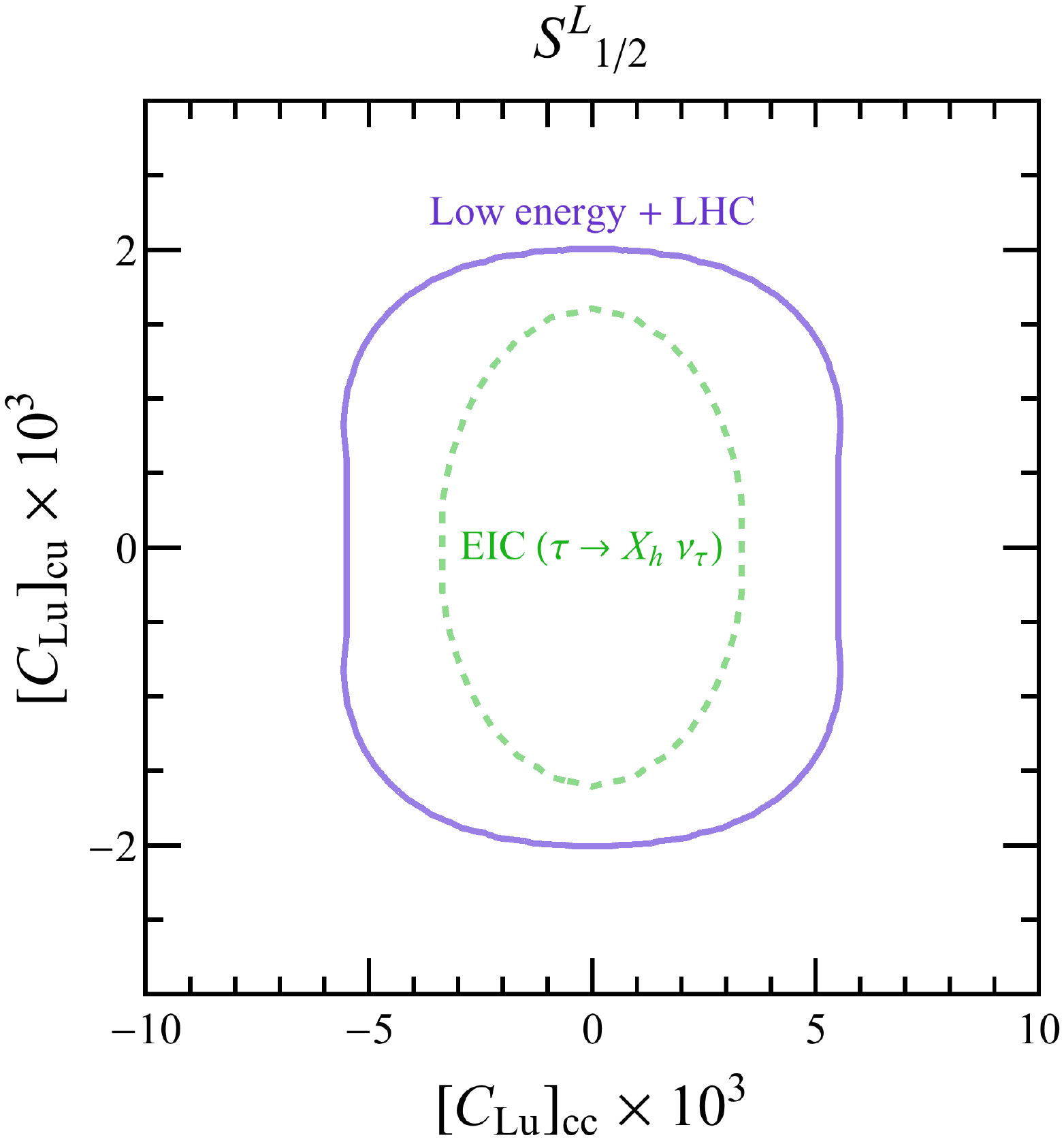}
\vspace{-1em}
\caption{The  region of $\Delta\chi^2<2.71$ in $[C_{Lu}]_{tt}-[C_{Lu}]_{uu}$~(left) and $[C_{Lu}]_{cc}-[C_{Lu}]_{cu}$~(right) planes. While the purple contour represents existing limits from low-energy experiments and the LHC, the dashed green line corresponds to the EIC expected sensitivity in hadronic $\tau$ decays under the assumption of $\epsilon_{n_b}=1$ with $n_b=0$.  The projection of these regions onto each axis corresponds to the  $90\%$ C.L. allowed region for that coupling.}
\label{LQ_SL}
\end{figure}

In what follows,  we determine the allowed regions in parameter space by minimizing a $\chi^2$ function  which includes LFV $\tau$ decays, $B$ meson decays and LHC searches. 
We present our results in terms of two-dimensional plots marginalizing over the remaining three free parameters in each model. 
The regions we obtain correspond to $\Delta\chi^2<2.71$, which gives a  $90\%$ C.L.  limit on single operator couplings  when we project the obtained confidence regions onto one dimension.\footnote{The resulting contour in two dimensions corresponds to the allowed region at $74.2\%$ C.L. } Below, we present our fitting results in several scenarios.
\begin{figure}[t]
\vspace{-1em}
\centering
\includegraphics[width=0.48\textwidth]{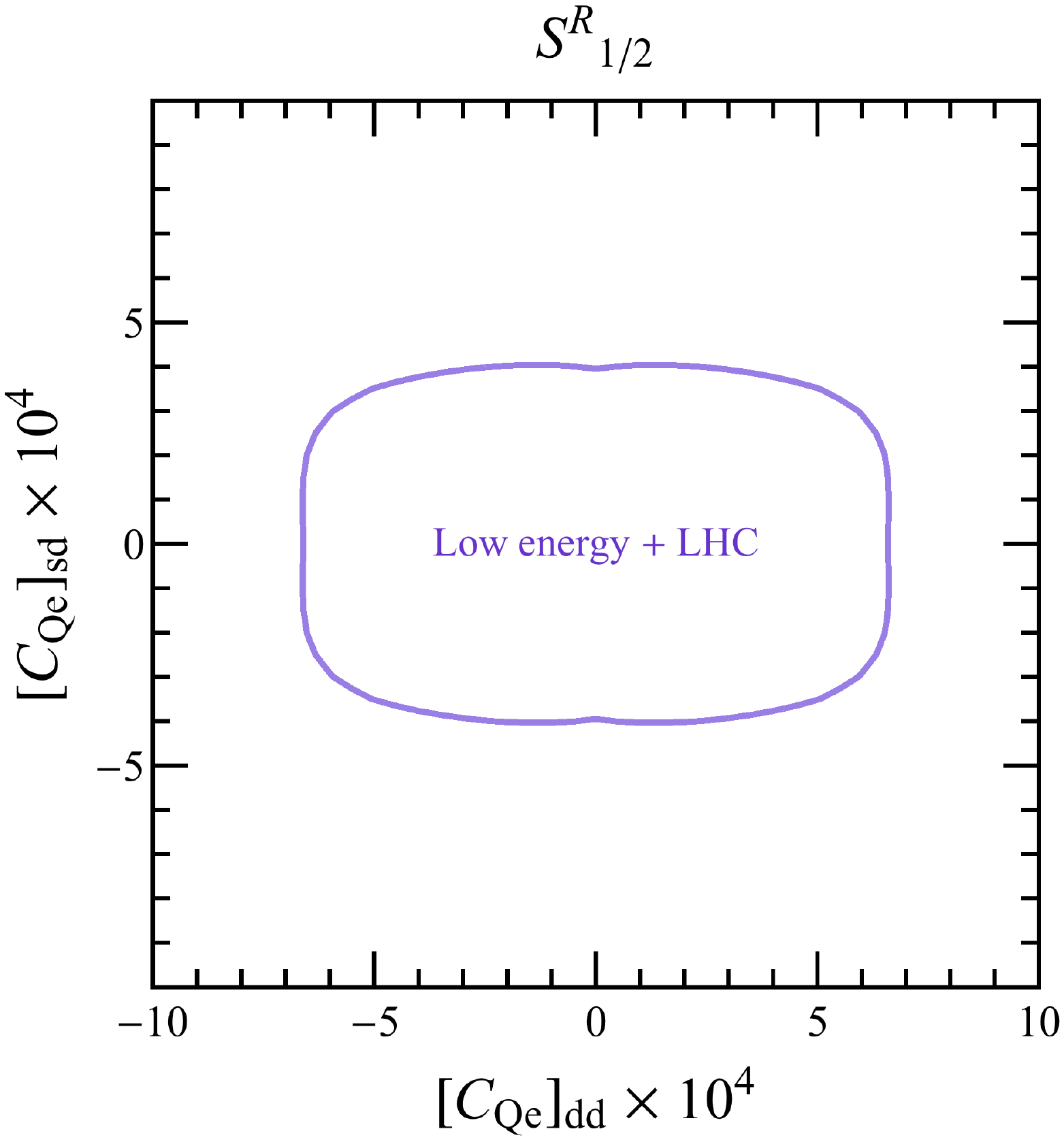}\quad
\includegraphics[width=0.48\textwidth]{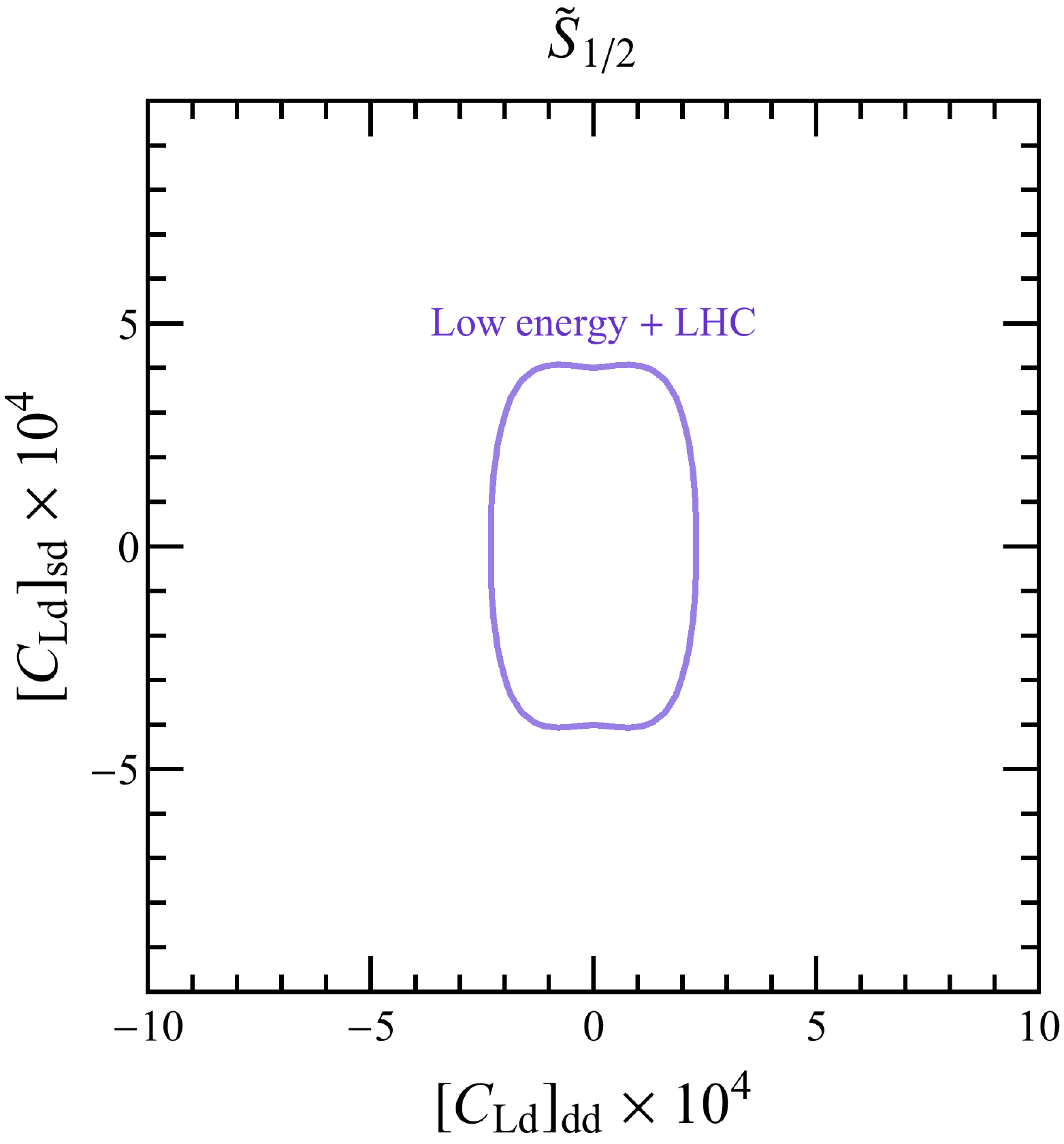}
\includegraphics[width=0.48\textwidth]{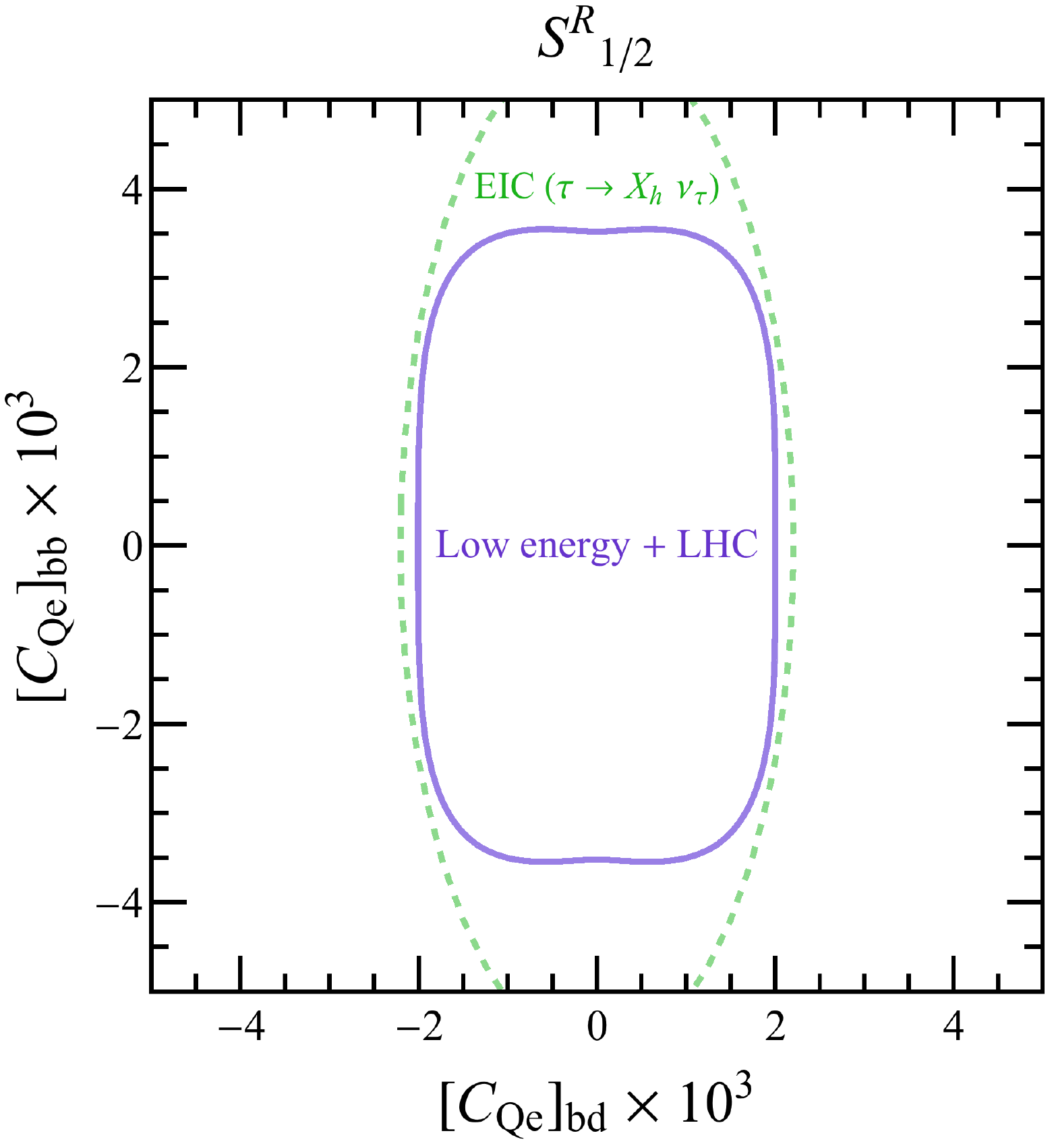}\quad
\includegraphics[width=0.48\textwidth]{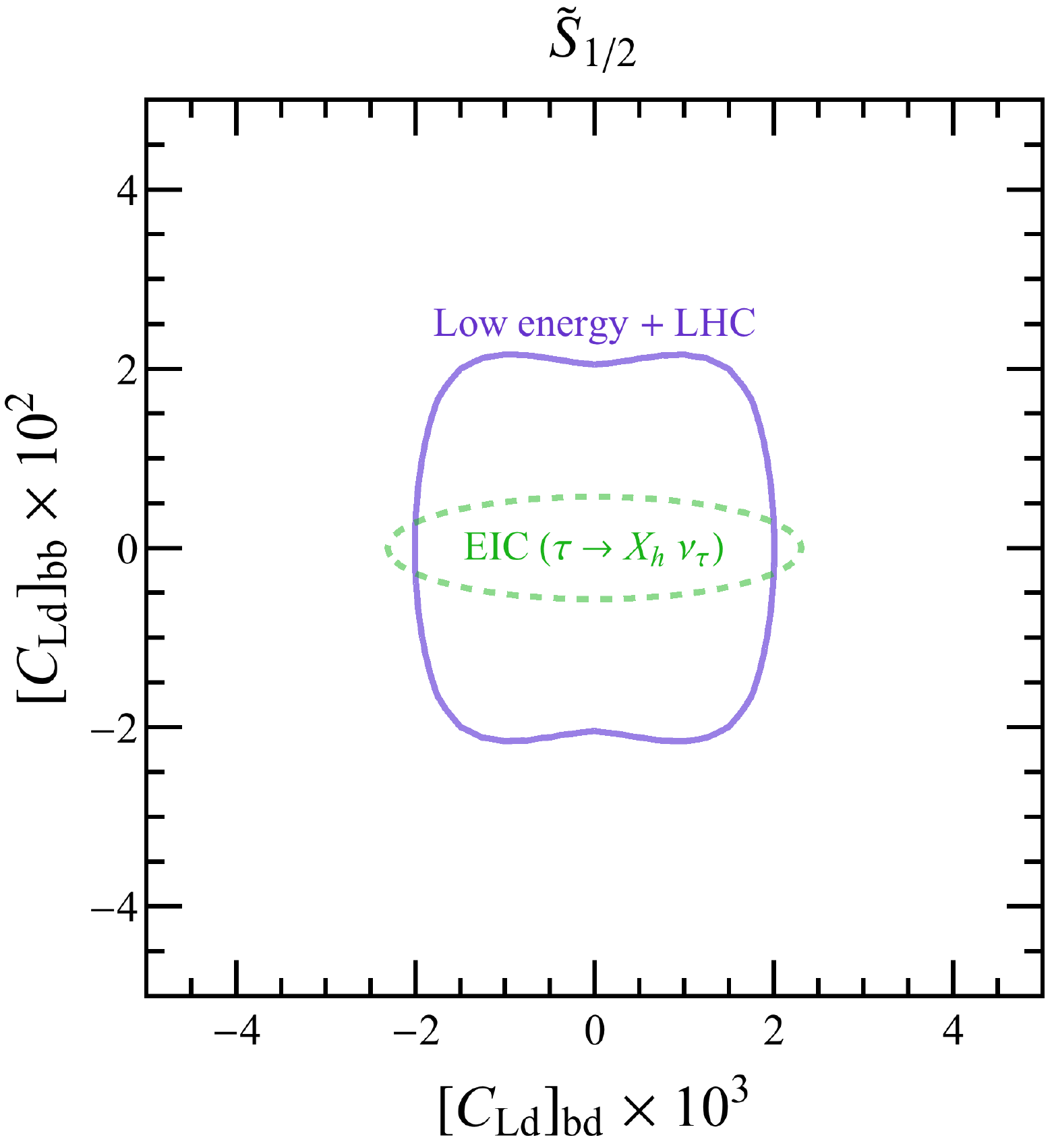}
\vspace{-1em}
\caption{[Upper] the contour that satisfies $\Delta\chi^2<2.71$ in $[C_{Qe}]_{dd}-[C_{Qe}]_{sd}$~(left) and $[C_{Ld}]_{dd}-[C_{Ld}]_{sd}$~(right) planes. [Lower] the same contour as the top two panels but in $[C_{Qe}]_{bd}-[C_{Qe}]_{bb}$~(left) and $[C_{Ld}]_{bd}-[C_{Ld}]_{bb}$~(right) planes. Current limits from low-energy experiments and the LHC are depicted by the purple contour. The EIC expected sensitivity in hadronic $\tau$ decays is described by the dashed green line under the assumption of $n_b=0$ and $\epsilon_{n_b}=1$. }
\label{LQ_SRStilde}
\end{figure}

In the case of the $S^L_{1/2}$ LQ model, since the induced operators are those of up-type quarks, the LFV $\tau$ decays can only restrict quark-flavor-conserving elements, namely, $[C_{Lu}]_{uu,cc,tt}$. On the other hand, the LFV  searches at the LHC 
play a significant role in bounding off-diagonal components $[C_{Lu}]_{uc/cu}$ and $[C_{Lu}]_{tu/ut}$ as well as the first- and second-diagonal elements. Figure \ref{LQ_SL}  shows the results of $\chi^2$ fitting in $[C_{Lu}]_{tt}-[C_{Lu}]_{uu}$ and $[C_{Lu}]_{cc}-[C_{Lu}]_{cu}$ planes. In the left panel, the bound on $[C_{Lu}]_{tt}$ is determined by the low-energy observables via the RGEs as the collider searches cannot constrain the flavor-diagonal  top-quark operator. Conversely, in the right panel, the width of the contour along the vertical direction is controlled by the LHC limit. While the single-operator analyses presented in Tables  \ref{EIC_sens}, \ref{LHC_bounds} and \ref{limit_semileptonic} show the constraint on $[C_{Lu}]_{cc}$ is $O(10^{-2})$, the contour indicates the relatively strong limit $\sim 6\times 10^{-3}$, which originates from $\tau \to e\gamma$ contribution.  
This happens because in this particular model there exists a correlation between four-quark and dipole operators, 
as shown by the matching conditions in Eqs.~\eqref{mLQ1}.

Unlike the $S^L_{1/2}$ LQ case, in the $S^R_{1/2}$ and $\widetilde{S}_{1/2}$ models, all the elements of the induced operators can be constrained by
both    low-energy observables and   LHC searches. The allowed regions in these models are depicted in Fig. \ref{LQ_SRStilde}. In the upper two panels, the contours in the vertical direction are controlled by $\tau \to e \pi^{\pm}K^{\mp}$ and $\tau \to e K_S$. On the other hand, the LHC search contributes to the bound on $[C_{Qe}]_{dd}$ due to the comparable limit to that from $\tau$ decays  as seen from the single-operator analyses in Table \ref{LHC_bounds} and \ref{limit_semileptonic}. 
In the lower two panels, an order of magnitude difference between the $S^R_{1/2}$ and $\widetilde{S}_{1/2}$ models is found in the width of the contours in the $[C_{M}]_{bb}$ direction. This is due to the fact that the RGEs of $[C_{Qe}]_{bb}$ involve top-quark Yukawa coupling, resulting in relatively large corrections to LEFT operators as seen from Table \ref{RGE_low_vector}.

In Figs.~\ref{LQ_SL} and \ref{LQ_SRStilde}, we also report  the prospective reach of the EIC with  ${\cal L} =$100 fb$^{-1}$, 
in  the ideal scenario  in which the  $\tau$ is reconstructed through the hadronic decay channels,  
and the SM background can be reduced to  $n_b=0$ with $\epsilon_{n_b}=1$  
(this corresponds to the dark green bands in Figs.~\ref{barchart_dipole}--\ref{barchart_CLeQu3}).
While for the couplings involving light quarks and top quark  the EIC is not competitive (almost the entire plotted region is allowed), 
the EIC can be quite competitive for couplings involving the charm and beauty quarks, both flavor diagonal and off-diagonal. 
These simple models illustrate a general lesson emerging from our study:  the  discovery window for CLFV at the EIC comes 
mostly from semileptonic  interactions that involve one or two heavy flavors. 

\begin{table}[t]
\centering
\begin{tabular}{||c|| c| c|| c | c || c | c |}
\hline
              & $[C_{Lu}]_{uu}$ & $[-0.37,~0.37]$ & $[C_{Lu}]_{uc}$ & $[-2.0,~2.0]$ & $[C_{Lu}]_{ut}$ & $[-348,~348]$  \\
$S^L_{1/2}$ & $[C_{Lu}]_{cu}$ & $[-2.0,~2.0]$ & $[C_{Lu}]_{cc}$ & $[-5.6,~5.6]$ & $[C_{Lu}]_{ct}$ & $[-348,~348]$  \\
                 & $[C_{Lu}]_{tu}$ & $[-348,~348]$ & $[C_{Lu}]_{tc}$ & $[-348,~348]$ & $[C_{Lu}]_{tt}$ & $[-5.3,~5.3]$  \\
\hline \hline
	 & $[C_{Qe}]_{dd}$ & $[-0.66,~0.66]$ & $[C_{Qe}]_{ds}$ & $[-0.44,~0.44]$ & $[C_{Qe}]_{db}$ & $[-1.6,~1.6]$ \\
$S^R_{1/2}$ & $[C_{Qe}]_{sd}$ & $[-0.40,~0.40]$ & $[C_{Qe}]_{ss}$ & $[-1.1,~1.1]$ & $[C_{Qe}]_{sb}$ & $[-1.2,~1.2]$ \\
	 & $[C_{Qe}]_{bd}$ & $[-2.0,~2.0]$ & $[C_{Qe}]_{bs}$ & $[-2.1,~2.1]$ & $[C_{Qe}]_{bb}$ & $[-3.7,~3.7]$ \\
\hline\hline
	 & $[C_{Ld}]_{dd}$ & $[-0.23,~0.23]$ & $[C_{Ld}]_{ds}$ & $[-0.44,~0.44]$ & $[C_{Ld}]_{db}$ & $[-1.6,~1.6]$  \\
$\tilde{S}_{1/2}$ & $[C_{Ld}]_{sd}$ & $[-0.41,~0.41]$ & $[C_{Ld}]_{ss}$ & $[-1.1,~1.1]$ & $[C_{Ld}]_{sb}$ & $[-1.2,~1.2]$  \\
		 & $[C_{Ld}]_{bd}$ & $[-2.0,~2.0]$ & $[C_{Ld}]_{bs}$ & $[-2.1,~2.1]$ & $[C_{Ld}]_{bb}$ & $[-21.6,~21.6]$  \\
 \hline
\end{tabular}
\caption{$90\%$ C.L.  ranges for the Wilson coefficients (in units of $10^{-3}$), in the three leptoquark models considered here. 
The ranges are obtained after marginalizing over all  other couplings. }
\label{LQ_limits}
\end{table}

The currently allowed  $90\%$ C.L.  ranges for each coupling are summarized in Table \ref{LQ_limits}. 
When comparing our leptoquark analysis to previous studies  in Refs.~\cite{Gonderinger:2010yn,Chekanov:2005au,Aktas:2007ji}, 
 several remarks are in order: 
\begin{itemize}
\item 
We improve  the bounds on the first-generation quark-flavor diagonal couplings by including $\tau \to e\pi^+\pi^-$ and the LHC searches. 
This leads to constraints that are an order of magnitude stronger  than the expected sensitivity at the EIC.
\item As discussed in \cite{Gonderinger:2010yn}, $\tau \to e \gamma$ constrains the quark-flavor diagonal components of the four-fermion operators 
 in the $S^L_{1/2}$ and $S^R_{1/2}$ models, yielding  a somewhat stronger limit   on $[C_{Lu}]_{cc}$ than those from the  LHC and other low-energy decay channels. For this coupling, prospective EIC limits are quite competitive. 
On the other hand, $\tau \to e\eta, \tau \to e\pi^+\pi^-$ and the LHC searches, which are newly incorporated into our analyses, give the most stringent bounds on the rest of the second- and third-generation diagonal elements.
\item Concerning the quark-flavor changing couplings,  the LHC searches 
currently provide the strongest bounds on the $uc$ and $cu$ elements , but the EIC can be quite competitive in the future. 
 In addition, the recent ATLAS search for LFV top-quark decays enables us to put bounds of $O(0.1)$ on the flavor-violating operators involving top quark.
\item For the strangeness-changing couplings, the inclusion of $\tau \to e \pi^{\pm} K^{\mp}$ improves  the limits on $sd$ and $ds$ elements by a factor 10 compared to previous analyses.\footnote{For $[C_{Ld}]_{sd/ds}$, if we include indirect bounds from kaon decays, they are superior to other low-energy limits as also discussed in \cite{Chekanov:2005au,Aktas:2007ji}.} The bounds on the $sb$ and $bs$ components are improved by incorporating the latest experimental results of $B^{\pm}\to K^{\pm} \tau e$, yielding stronger limits than the LHC and future EIC  sensitivities.

\item As illustrated by Figs.~\ref{LQ_SL} and \ref{LQ_SRStilde}, 
we find that after inclusion of low-energy constraints the CLFV discovery potential    at the EIC 
arises mostly  for  LQ couplings  involving the charm and beauty quarks, both flavor diagonal and off-diagonal.

\end{itemize}

Although our analysis focuses on LFV $\tau-e$ couplings, with emphasis on the EIC discovery potential,  the above LQ models have several intriguing connections to other interesting 
phenomenology,  such as neutrino mass \cite{FileviezPerez:2008dw, Dekens:2020ttz} and $B$ physics \cite{Tanaka:2012nw, Sakaki:2013bfa, Dorsner:2013tla,Crivellin:2020mjs}. 
This would open a number of additional observables to probe LQ couplings. We defer the analysis to future work.

\section{Conclusions}
\label{sect:concl}

It has been long recognized that searches  for CLFV processes are a very promising tool to probe new physics beyond the SM. 
In this paper  we have performed,  in the framework of the SMEFT,  a first comprehensive analysis of the CLFV sensitivity at the EIC 
in the channel $ep\to \tau X$. 
The SMEFT is particularly appealing  because it  captures a large class of new physics models originating at energies above the electroweak scale 
and allows for a systematic comparison of  all  probes of CLFV in the $\tau$-$e$ sector. 
We considered all the dimension-six CLFV operators in the SMEFT, including CLFV $Z$ and Higgs couplings, photon and $Z$ dipole interactions,
and ten semileptonic four-fermion operators, with different Lorentz and completely general quark-flavor structures.

For  the DIS cross section $ep\to \tau X$ we 
found that, for all operators except Yukawa and electron-gluon operators, 
the unpolarized cross sections at $\sqrt{S}=141~$GeV are in the 1--10 pb range 
for SMEFT coefficients of order one\footnote{Recall we have written the dimensionful couplings for the dimension-six  operators  as $\sim C/v^2$, 
where $v=246$~GeV is the electroweak scale and $C$ are dimensionless SMEFT coefficients. } 
(see Tables~\ref{CrossSection:AV}--\ref{CrossSection:ST}). 
Operators with sea quarks in the initial state give rise to somewhat smaller cross sections, as expected from the 
suppression of the  corresponding PDFs. 
In order to account for the decay of the $\tau$ lepton, and to realistically assess the sensitivity of the EIC, 
we simulated SMEFT events in \texttt{Pythia8}, using the \texttt{Delphes} package to simulate the detector smearing effects 
(see Figs.~\ref{Fig:kinF}--\ref{Fig:event}).
We found the muonic  reconstruction  channel $\tau \to \mu \bar{\nu}_{\mu} \nu_{\tau}$ to be very promising, since moderate cuts on the muon $p_T$ and on 
the missing energy allow one  to eliminate all SM background  without excessively suppressing the signal. The signal efficiency depends 
strongly on the flavor of the SMEFT operators, since operators with heavy quarks in the initial state give rise to distributions peaked at smaller $p_T$, which are more affected by the cuts to suppress the SM background. The efficiency is on the other hand rather insensitive to the Lorentz  structure of the SMEFT operators. In the electron channel $\tau \to e \bar{\nu}_e \nu_{\tau}$, the background from neutral and charged-current DIS is always very large. 
In the hadronic channels $\tau \to X_h \nu_{\tau}$, the naive cuts we imposed in Section \ref{efficiency}, where we vetoed leptons with $p_T > 10$ GeV and
asked for two jets with $p_T$ larger than 15 and 20 GeV, are not sufficient to fully suppress the SM background. 
We however did not use additional information on the jet that emerges from $\tau$ decay, such as the presence of a secondary vertex, the hadron multiplicity 
or the correlation with $\met$, which could provide more efficient ways to tag hadronic $\tau$ events \cite{Zhang}. 
At $\sqrt{S}= 141$ GeV and ${\cal L}=100~$fb$^{-1}$, the EIC expected sensitivity 
for the dimensionless SMEFT coefficients 
reaches $C \sim O(10^{-(3-2)})$  for light-quark four-fermion, dipole and $Z$-coupling operators.
Bounds on heavy-quark operators result in 
$C \sim O(10^{-(2-1)})$, 
while it is more challenging to constrain LFV Yukawa and electron-gluon operators as their cross sections are strongly suppressed.
\looseness-1

To assess the discovery potential of the EIC in  $\tau$-$e$ transitions, we have 
compared its sensitivity to other probes of the same interactions, across a broad range of energy scales, 
ranging from other collider processes to decays of  $\tau$ lepton and $B$ meson. 
In Sec.~\ref{sect:sensitivities} we have provided simple order of magnitude estimates, 
substantiated by a detailed analysis in  Secs.~\ref{sect:LHC} , \ref{low-energy} , \ref{sect:indirect}, and ~\ref{sect:summary}.
We summarize our main findings below,  starting with the LHC and going down in energy.

The LHC can probe LFV by studying the decays of the $Z$ and Higgs bosons
and of the top quark. In addition, if the scale of new physics is larger than a few TeV, the same semileptonic four-fermion operators that induce CLFV DIS can be studied in high-invariant mass Drell-Yan $pp \to e \tau X$. The bounds we obtain are discussed in Sec.~\ref{ZHT} and Table \ref{LHC_bounds}.
While the LHC has a clear edge in measuring Higgs and top quark-flavor-changing couplings, we found that the EIC could competitively probe $Z$ couplings and four-fermion interactions with light quarks, especially if the efficiency in the hadronic channel can be improved with respect to our simple analysis.
Four-fermion operators with two heavy quarks are somewhat more suppressed in Drell-Yan compared to DIS, because of the presence of two heavy quark PDFs. Here  the EIC could have a larger impact, provided analysis strategies  are devised in order to improve the signal efficiency.  
When comparing the EIC with the LHC, it is worth keeping in mind  that the formalism of the SMEFT might not be applicable at LHC energies. The two colliders could thus be probing complementary regions in parameter space, and are both necessary to fully constrain CLFV. 
\looseness-1

We then carried out a comprehensive comparison of the EIC and LHC sensitivity with current bounds from $\tau$ and $B$ decays,
including the radiative decay $\tau \to e \gamma$, purely leptonic channels, $\tau \rightarrow e \ell^+ \ell^-$, and semileptonic channels. 
The limits obtained under the hypothesis that a single SMEFT operator is present at the new physics scale are shown in Figs.~\ref{barchart_dipole}--\ref{barchart_CLeQu3}.
$\tau \to e \gamma$  gives the most severe limits,  at least a factor of 100 stronger than those expected from the EIC, on dipole and $Z$-coupling operators. 
In the single coupling analysis, quark-flavor-diagonal four-fermion operators with light quarks are very well constrained by $\tau\to e \pi^+\pi^-$,
$\tau\to e \eta$, $\tau\to e \eta^\prime$ and $\tau\to e \pi$. In particular, the constraints from  $\tau\to e \pi^+\pi^-$ on operators with valence quarks
are currently a factor of five better than high-invariant mass Drell-Yan  and a factor of fifty better than the EIC in the muonic channel. 
In the case of heavy quarks, however, contributions to $\tau$ decay only arise at the loop level, and 
the EIC  sensitivity on these operators is very competitive with  LHC and low-energy observables. 
While the muonic  reconstruction  channel is rather clean, the full potential of the EIC is better represented by the hadronic channel, whose BR is a factor of 4 larger than the muonic channel. Assuming, optimistically, that all the SM background can be suppressed without losing any signal
event, we find that the EIC sensitivity to four-fermion operators can exceed that  of the current LHC and low-energy experiments.
It will therefore be very important to more thoroughly explore the hadronic reconstruction channels and devise analysis strategies to maximize the signal/background ratio.

Due to the prominence of $\tau \rightarrow e \pi \pi$ in the single coupling analysis, it is rather easy to weaken the low-energy bounds by turning on several operators 
at the matching scale. In this paper we refrained from a global fit to CLFV observables, but explored the impact of multiple operators in two scenarios,
turning on two down-type vector operators with generic quark flavors, left-handed leptons and left- or right-handed quarks, and turning on 
five operators with left-handed leptons and couplings only to the light $u$, $d$ and $s$ quarks. In both cases, low-energy observables are not sufficient to constrain all operator coefficients and free directions appear.  As illustrated in Fig.~\ref{fig:marginalized},  collider experiments are crucial to close these free directions 
and discovery windows arise for the EIC in these more general scenarios. 
\looseness-1

Our analysis  applies to any new physics originating at  energies higher than the electroweak scale. 
In Sec.~\ref{sect:leptoquarks} we have applied our framework to study three different leptoquark scenarios,  which 
yield more than one LFV operator at the matching scale. 
Leptoquarks provide   interesting extensions of the Standard Model  motivated  both by model building and by several phenomenological puzzles. 
We  improve upon the current literature  by including state-of-the art analyses of $\tau$ decays, in particular semi-leptonic modes. 
As expected from the single-operator analysis, most of the LFV leptoquark couplings are constrained quite  
severely by the LHC,  $\tau$ and $B$ decays. 
We find that after inclusion of low-energy constraints the CLFV discovery potential   at the EIC 
arises mostly  for  LQ couplings  involving the charm and beauty quarks (see Figs.~\ref{LQ_SL} and \ref{LQ_SRStilde}).
We leave to future work  a more comprehensive analysis of leptoquark models and their implications for lepton flavor violation and  beyond. 
 \looseness-1

To fully explore the EIC potential to probe CLFV physics, our analysis needs to be extended in several directions:  
\begin{itemize}
\item The LO DIS cross sections for processes initiated by a strange or heavy quark are affected by significant theoretical errors, as shown in Tables \ref{CrossSection:AV} and  \ref{CrossSection:ST}. 
To reduce the error, and to have a more robust assessment of the theory uncertainties, it is necessary to consider NLO QCD corrections. These corrections for the SM DIS inclusive cross section range from $\mathcal{O}(10\%)$ for light flavor contributions to $\mathcal{O}(20-50\%)$ for heavy flavors, and we may expect similar corrections to SMEFT contributions.

\item As shown in Figs.~\ref{barchart_dipole} to \ref{barchart_CLeQu3}, improving the analysis in the  hadronic channels $e p \rightarrow \tau X \rightarrow \nu_\tau X_h X$, which has the largest branching ratio, could highly impact the EIC reach. 
It will be important to take advantage of the distinctive features of the jets emerging from the hadronic $\tau$ decay 
in order to devise a robust and efficient $\tau$ tagger, to suppress the SM background without losing signal events.

\item One of the most promising directions for the EIC is to probe CLFV operators with heavy $c$ and $b$ quarks, whose effects are suppressed 
by one electroweak loop in $\tau$ decays and by two heavy quark PDFs at the LHC. 
However, the missing energy, lepton and jet $p_T$ distributions induced by heavy-flavor operators are peaked at small $p_T$, and thus severely suppressed by the cuts imposed in Section \ref{efficiency}.
It will be important to explore whether tagging  $b$ and $c$ jets allows to achieve the same background suppression with looser $p_T$ cuts, thus boosting the efficiency for heavy flavor operators.  Higher-order perturbative QCD corrections and resummation may be particularly important to predict accurately the dependence of cross sections with jet $p_T$ cuts.

\item New data on $\tau$, $B$ and $D$ meson decays at Belle II, LHCb and  BESIII, combined with increased luminosity at the LHC and, in the future, with data from the EIC, will help paint a complete picture of CLFV in the  $\tau$ sector. 
To fully exploit this wealth of data, a global analysis (beyond  single-coupling)  is highly desirable.  
In this context, the inclusion of more observables will help eliminate flat directions that emerged already in our discussion. 
For light quarks, additional constraints can be obtained by including  
the decays $\tau \rightarrow e K^+ K^-$,
$\tau \rightarrow e K_S K_S$ and $\tau \rightarrow e \pi \pi \pi$, whose branching ratios are bounded at the   few $\!\times 10^{-8}$
level. 
For flavor changing interactions,   $D \rightarrow e \tau$ 
could be measured at BESIII and LHCb, while $B_s \rightarrow e \tau$ will be in reach of LHCb and Belle II.
At colliders, heavy flavor tagging at the EIC (e.g. \cite{Li:2020sru}) could provide unambiguous probes of the operator flavor structure, while angular distributions 
in high-invariant mass Drell-Yan and helicity fractions in top decays can pinpoint the Lorentz structure of the contributing operators. 
Because of the similar sensitivity, we expect that, in case of observation, by correlating observables at low and high energy it will be possible to remove degeneracies and clearly identify the dominant CLFV mechanism.
\looseness-1

\end{itemize}

\acknowledgments
We thank Carlo Alberto Gottardo for illustrating the procedure for the extraction of bounds on $t \rightarrow q e \tau$ from Ref. \cite{ATLAS:2018avw}, and for checking the limit on the branching ratio we obtained. We acknowledge interesting discussion with Marianna Fontana, Dominik Mitzel, and Mark Williams on the LHCb sensitivity to $\tau$ CLFV in $D$ and $B_s$ decays. We thank Miguel Arratia for discussion on the EIC Delphes card used in the analysis of Section \ref{efficiency}. 
We are indebted to Emilie Passemar for sharing numerical data files on the $K\pi$ form factors, and to Konstantin Beloborodov for providing the $K^+ K^-$ form factors. We thank Jure Zupan for early discussions on a SMEFT analysis of CLFV at the EIC. 
We also acknowledge stimulating discussions with Sacha Davidson and Krishna Kumar. 
This work is supported  by the US Department of Energy through  
the Office of Nuclear Physics,   an Early Career Research Award,  and  the  
LDRD program at Los Alamos National Laboratory. Los Alamos National Laboratory is operated by Triad National Security, LLC, for the National Nuclear Security Administration of U.S.\ Department of Energy (Contract No. 89233218CNA000001).

\appendix

\section{Renormalization group equations and their solutions}\label{RGEs}

In this section we consider the renormalization group evolution of SMEFT operators between $\Lambda$ and the electroweak scale,
and then between the EW scale and the low-energy scale $\mu \sim 2$ GeV.
We focus in particular on the mixing of operators that cannot be probed at tree level (at the EIC or low-energy) onto operator that can. 
Our  analysis is mostly limited to leading logarithmic accuracy, but it also includes important threshold corrections, such as the 
contribution of CLFV Yukawa couplings to dipole operators.
In Section \ref{RGE_ew} we discuss the evolution from $\Lambda$ to the electroweak scale in the SMEFT, 
adding threshold corrections in Section \ref{dipole_yukawa}, and, finally, in Section \ref{RGE_low} we consider the evolution in the LEFT.

\subsection{Running between $\Lambda$ and the electroweak scale}\label{RGE_ew}

\begin{table}
\small
\centering
\begin{tabular}{||c | c || c | c || c | c||}
\hline
$\alpha_s(m_Z) $      &  0.118                     &  $g_2$ & 0.65                                      & $g_1$ &  0.36 \\          
$m_u( 2\, {\rm GeV})$ & $2.16^{+0.46}_{-0.26}$ MeV & $m_d( 2\, {\rm GeV})$ & $4.67^{+0.48}_{-0.17}$ MeV & $m_s( 2\, {\rm GeV})$ & $93^{+11}_{-5}$ MeV \\
$m_c( m_c)$           & $1.27 \pm 0.02$ GeV & $m_b( m_b)$ & $4.18^{+0.03}_{-0.02}$ GeV & $m_t( m_t)$ & $162.5^{+2.1}_{-1.5}$  GeV\\
\hline
\end{tabular}
\caption{Standard Model parameters used in the solution of the RGEs}\label{Tab:SM}
\end{table}

The RGEs in the SMEFT can be found in Refs. \cite{Jenkins:2013zja,Jenkins:2013wua,Alonso:2013hga}. We report them here for completeness, in a slightly different choice of basis.
We work in a basis in which both the $u$ and $d$ quark mass matrices are diagonal, and define the SM Yukawa coupling as
\begin{align}
Y_f=\frac{m_f}{v}. 
\end{align}
The values of the masses, in the $\overline{\textrm{MS}}$ scheme, are given in Table \ref{Tab:SM}.
The hypercharge assignments are
\begin{align}\label{hypc}
{\rm y_q}=\frac{1}{6}, \hspace{0.5cm}{\rm y_u}=\frac{2}{3}, \hspace{0.5cm}{\rm y_d}=-\frac{1}{3}, \hspace{0.5cm}{\rm y_l}=-\frac{1}{2},\hspace{0.5cm}{\rm y_e}=-1.
\end{align}
The running of the SM couplings is given by
\begin{align}
\mu\frac{d}{d\mu}g_s(\mu)&=-  \left( \frac{11}{3} N_c - \frac{4}{3} T_f n_f \right)  \frac{\left(g_s(\mu)\right)^3}{(4\pi)^2}, \label{SM_rge1}\\
\mu\frac{d}{d\mu}g_2(\mu)&=-  \left( \frac{43}{6}     - \frac{4}{3} n_G \right) \frac{\left(g_2(\mu)\right)^3}{(4\pi)^2} , \label{SM_rge2}\\
\mu\frac{d}{d\mu}g_1(\mu)&= \frac{5}{3} \left( \frac{1}{10} + \frac{4}{3} n_G   \right) \frac{\left(g_1(\mu)\right)^3}{(4\pi)^2} , \label{SM_rge3}\\
\mu\frac{d}{d\mu}Y_t(\mu)&=\frac{1}{(4\pi)^2}Y_t(\mu)\left(9Y_t(\mu)^2-8g^2_s(\mu)-\frac{9}{4}g^2_2(\mu)-\frac{17}{12}g_1^2(\mu) \right), \label{SM_rge4}
\end{align}
where $N_c =3$, $T_f = 1/2$, $n_f$ is the number of active quarks, $n_f = 6$ above the EW scale, and $n_G$ the number of fermion generations.
The RGEs in \eqref{SM_rge1}-\eqref{SM_rge4} are not affected by the dimension-six CLFV SMEFT operators we are considering.

\begin{table}
\center
\begin{tabular}{|| c || c c c c c  || }
\hline 
                                & $\left({C}_{LeQu}^{(1)}\right)_{\tau ett}$ & $\left({C}_{LeQu}^{(3)}\right)_{\tau ett}$  & $\left({C}_{LeQu}^{(1)}\right)_{\tau e cc}$ & $\left({C}_{LeQu}^{(3)}\right)_{\tau ecc}$
                                    & $\left({C}_{LedQ}^{}\right)_{\tau e bb}$ 
                                    \\
\hline
$\Gamma^e_\gamma$                     &                  $6.5\cdot  10^{-4}$   & $-0.46$   &$ 2.9 \cdot 10^{-6}$ & $-2.0 \cdot 10^{-3}$         &      --                          \\
$\Gamma^e_Z$                          &                  $9.3\cdot  10^{-5}$   & $-0.07$   & $4.2 \cdot 10^{-7}$ & $-2.9 \cdot 10^{-4}$         &    --                           \\
$\left({C}_{LeQu}^{(1)}\right)_{\tau e cc}$  &   --                                   &  --       &  $1.12$              & $-0.13$  &  -- \\
$\left({C}_{LeQu}^{(3)}\right)_{\tau ecc}$   &   --                                   &  --       & $-2.8 \cdot 10^{-3}$ & $0.96$   & --\\
$\left({C}_{LedQ}^{}\right)_{\tau e bb}$  &   --                                   &  --       &   --             & -- &  $1.12$     \\
\hline\hline
$\left( C_{GG} \right)_{\tau e}$      &   $-0.57$                               & $0.07$   &      $-127$ & $29$                                 &     $32$     \\
$\left( C_{G\tilde G} \right)_{\tau e}$    &   $-0.85$                               & $0.1$   &      $-190$ & $44$                                  &     $-47$     \\
\hline
\end{tabular}
\caption{The Wilson coefficients of dipole, scalar and tensor operators at $\mu=m_t$ induced by a nonzero top-, bottom- and charm-quark scalar and tensor 
operators through operator mixing. The starting point of the running is taken at $\Lambda = 1$~TeV. 
The threshold corrections to $C_{GG}$ and $C_{G\tilde G}$ are given in Eqs. \eqref{matching_GG}, 
\eqref{matching_GGtilde}, \eqref{matching_GGb}, 
\eqref{matching_GGtildeb}. We evaluate them at $\mu=m_t~ (m_b)$ for the top (bottom) quark while $\mu=2~$GeV for the charm quark.}
\label{ScalarTensorRGE}
\end{table}

We consider the QCD running of scalar and tensor operators. In addition, because of the very strong low-energy limits on the CLFV dipole operator $\Gamma^{e}_{\gamma}$, we take into
account the mixing of the $Z$ dipole and of the tensor operators onto $\Gamma^{e}_{\gamma}$. For the tensor operator, the mixing is proportional to the SM Yukawa coupling and thus is most relevant
in the case of operators involving the top quark. The bounds on $\Gamma^{e}_{\gamma}$ are so stringent that they also constrain the top component 
of $C^{(1)}_{LeQu}$, which mixes onto the tensor operator $C^{(3)}_{LeQu}$ at one loop.  We thus solve the RGEs
\begin{eqnarray}
\frac{d}{d\log \mu} \left[\Gamma^{e}_\gamma \right]_{\tau e} &=& \frac{6 g_2^2}{(4\pi)^2} \left(1 - \tan^2\theta_W \right)\, \left[ \Gamma^e_Z \right]_{\tau e}
+ \frac{32}{(4\pi)^2} N_c Q_t \left[ Y_u\right]_{i i } \left[ C^{(3)}_{Le Qu} \right]_{\tau e i i },  \label{rge_st0} \\
\frac{d}{d\log \mu} \left[\Gamma^{e}_Z \right]_{\tau e} &=&  \frac{16}{(4\pi)^2} N_C (z_{u_L} + z_{u_R}) \left[ Y_u\right]_{i i } \left[ C^{(3)}_{Le Qu} \right]_{\tau e i i }, \label{rge_st1}\\
\frac{d}{d\log \mu}C^{(1)}_{LeQu}&=&\frac{\alpha_s}{4\pi}\gamma_S^{(0)}C_{LeQu}^{(1)} -\frac{1}{(4\pi)^2}\bigg[24\left({\rm y}_q+{\rm y}_u \right)\left(2{\rm y}_e-{\rm y}_q+{\rm y}_u \right)g^2_1-18g^2_2 \bigg]C^{(3)}_{LeQu},\nonumber\\
\label{rge_st2}
\\
\frac{d}{d\log \mu}C^{(3)}_{LeQu} &=& \frac{\alpha_s}{4\pi}\gamma_T^{(0)}C_{LeQu}^{(3)} + \frac{1}{8(4\pi)^2}\bigg[-4\left({\rm y}_q+{\rm y}_u \right)\left(2{\rm y}_e-{\rm y}_q+{\rm y}_u \right)g^2_1+3g^2_2 \bigg]C^{(1)}_{LeQu} \nonumber,\\
\label{rge_st3} \\
\frac{d}{d\log\mu}C_{LedQ}&=&\frac{\alpha_s}{4\pi}\gamma_S^{(0)}C_{LedQ},\label{rge_st4}
\end{eqnarray}
where
\begin{align}
\gamma_S^{(0)}&=-6C_F,\hspace{1cm}
\gamma_T^{(0)}=2C_F,
\end{align}
with $C_F=4/3$.  
The solutions of these RGEs take the form,
\begin{equation}
\label{eq:RGEmatrix}
\mathbf{C}_\text{low}^i(\mu) = \mathbb{A}_{ij} \mathbf{C}_\text{hi}^j(\Lambda)\,,
\end{equation}
where $\mathbf{C}_\text{low}$ is a vector of the coefficients on the LHS of Eqs.~\eqref{rge_st0}--\eqref{rge_st4}, evaluated at $\mu= m_t$, and $\mathbf{C}_\text{hi}$ is the vector of coefficients at the scale $\Lambda = 1$ TeV on the RHS.
The coefficients $\mathbb{A}_{ij}$ that solve these equations are given in Table \ref{ScalarTensorRGE}.

Considering now vector-like operators, four-quark operators involving heavy quarks run into the $Z$ couplings $c^{(1,3)}_{L\varphi}$ and $c_{e\varphi}$
via the first diagram in Fig. \ref{Penguin1}.  The RGE has a piece proportional to the quark Yukawas, and a gauge component
\begin{eqnarray}
\frac{d}{d\log \mu}  \left( c^{(1)}_{L\varphi} + c^{(3)}_{L\varphi} \right) &=& -\frac{16 N_c}{(4\pi)^2} \Bigg\{ \frac{1}{2}\left[Y_u\right]^2_{i i}   \left( - C_{LQ,U} + C_{Lu} \right)_{ii}
- \frac{1}{2}\left[Y_d\right]^2_{j j}   \left( - C_{LQ,D} + C_{Ld} \right)_{jj} \nn   \\
& & 
+ \frac{1}{3}\left(\frac{g_2}{2 c_w}\right)^2 \left(  \left( z_{u_L} C_{LQ, U} + z_{u_R} C_{L u}\right)_{ii} +  \left( z_{d_L} C_{LQ, D} + z_{d_R} C_{L d}\right)_{jj} \right) 
\Bigg\}, \nonumber\\ \label{cLphi} \\
\frac{d}{d\log \mu}  c_{e\varphi} &=& -\frac{16 N_c}{(4\pi)^2} \Bigg\{ \frac{1}{2} \left[Y_u\right]^2_{i i} \left( C_{eu} \right)_{i i}
- \frac{1}{2} \left[Y_d\right]^2_{j j} \left( C_{ed} \right)_{j j } \nn \\ & & + \frac{1}{2} \left( \left[Y_d\right]^2_{jj} \delta_{j k}  -  V^*_{i k} \left[Y_u\right]^2_{ii} V_{i j}  \right) (C_{Qe})_{j k} 
 \nn  \\
& & 
+ \frac{1}{3}\left(\frac{g_2}{2 c_w}\right)^2 \left(  z_{u_R} \left (C_{e u} \right)_{ii}
+ z_{d_R} \left (C_{e d} \right)_{jj} - 2 {\rm y_q} s_w^2 \left(C_{Qe}\right)_{jj} 
\right)
\Bigg\},\label{cephi}
\end{eqnarray}
where $V_{i j}$ denotes elements of the CKM matrix. We use $i$ to denote $u$-type indices, $i\in \{u,c,t\}$
and $j,k$ to denote $d$-type indices, $j, k \in \{d,s,b\}$, and a sum over repeated indices is understood in Eqs. \eqref{cLphi} and \eqref{cephi}.

The penguin contributions to the semileptonic four-fermion operators are 
\begin{eqnarray}
\frac{d}{d\log \mu} (C_{LQ, U})_{ll} &=&  \frac{4}{3} N_c \frac{g_1^2}{(4\pi)^2} \Bigg\{ 
  {\rm y^2_q}  (C_{LQ, U})_{ii}  + {\rm y^2_q}(C_{LQ, D})_{jj}  +{\rm y_q} {\rm y_u}  (C_{Lu})_{ii}  +{\rm y_q} {\rm y_d}  (C_{Ld})_{jj} \nn \\
 & &  + \frac{g_2^2}{4 g_1^2} \left(  (C_{LQ, U})_{ii}    -   (C_{LQ, D})_{jj}\right)     \Bigg\}  \label{peng1}\\
\frac{d}{d\log \mu} (C_{LQ, D})_{kk} &=&  \frac{4}{3} N_c \frac{g_1^2}{(4\pi)^2} \Bigg\{  {\rm y^2_q} (C_{LQ, U})_{ii} +  {\rm y^2_q} (C_{LQ, D})_{jj}
+  {\rm y_q} {\rm y_u} (C_{Lu})_{ii}  +   {\rm y_q} {\rm y_d} (C_{Ld})_{jj} \nn \\
& & -         \frac{g_2^2}{4 g_1^2} \left((C_{LQ, U})_{ii} - (C_{LQ, D})_{jj}\right)  
 \Bigg\} \label{peng2} \\
\frac{d}{d\log \mu} (C_{Lu})_{ll} &=&  \frac{4}{3} N_c \frac{g_1^2}{(4\pi)^2} \Bigg\{   {\rm y_u} {\rm y_q} \left(  (C_{LQ, U})_{ii} + (C_{LQ, D})_{jj}   \right)
+  {\rm y^2_u} (C_{Lu})_{ii} + {\rm y_u} {\rm y_d} (C_{Ld})_{jj} \Bigg\} \nn \\ \label{peng3} \\
\frac{d}{d\log \mu} (C_{Ld})_{kk} &=&  \frac{4}{3}N_c \frac{g_1^2}{(4\pi)^2} \Bigg\{   {\rm y_d}{\rm y_q}  \left( (C_{LQ, U})_{ii} + (C_{LQ, D})_{jj} \right)   
+   {\rm y_d} {\rm y_u} (C_{Lu})_{ii} + {\rm y^2_d} \left(C_{Ld} \right)_{jj} \Bigg\} \nn \\ \label{peng4} \\
\frac{d}{d\log \mu} (C_{eu})_{ll} &=&  \frac{4}{3}N_c \frac{g_1^2}{(4\pi)^2} \Bigg\{  {\rm y^2_u}  (C_{eu})_{ii}  + {\rm y_u}{\rm y_d}  (C_{ed})_{jj}  
+  2  {\rm y_u} {\rm y_q} (C_{Qe})_{jj} \Bigg\} \label{peng5} \\
\frac{d}{d\log \mu} (C_{ed})_{kk} &=&  \frac{4}{3}N_c \frac{g_1^2}{(4\pi)^2} \Bigg\{ {\rm y_d} {\rm y_u}  (C_{eu})_{ii} + {\rm y^2_d}  (C_{ed})_{jj}  
+  2 {\rm y_d} {\rm y_q} (C_{Qe})_{jj} \Bigg\} \label{peng6} \\
\frac{d}{d\log \mu} (C_{Qe})_{kk} &=&  \frac{4}{3}N_c \frac{g_1^2}{(4\pi)^2} \Bigg\{  {\rm y_q} {\rm y_u} \left(C_{eu} \right)_{ii} + {\rm y_q} {\rm y_d} \left(C_{ed} \right)_{jj}+ 2 {\rm y^2_q} \left(C_{Qe}\right)_{jj}  \Bigg\}, \label{peng7} 
\end{eqnarray}
where, as before, summation over $u$ and $d$-type flavor indices $i$ and $j$ on the r.h.s of Eqs. \eqref{peng1}--\eqref{peng7}  is understood. 

In addition to the penguin diagrams, there are also current-current contributions shown in the last three diagrams in Fig.~\ref{Penguin1}. Neglecting again the operator self-renormalization, we have:
\cite{Jenkins:2013wua}
\begin{eqnarray}
\frac{d}{d\log \mu} C_{LQ,\, U} &=&   - \frac{2}{(4\pi)^2} V_{\rm CKM}  Y_d \, C_{L d} \, Y_d V^{\dagger}_{\rm CKM}, \label{fc1} \\
\frac{d}{d\log \mu} C_{LQ,\, D} &=&   - \frac{2}{(4\pi)^2} V^{\dagger}_{\rm CKM}  Y_u \, C_{L u} \, Y_u V^{}_{\rm CKM},  \\
\frac{d}{d\log \mu} C_{Lu} &=&   - \frac{2}{(4\pi)^2} Y_u   V_{\rm CKM}   \, C_{L Q,\, D} \,  V^{\dagger}_{\rm CKM} Y_u,  \\
\frac{d}{d\log \mu} C_{Ld} &=&   - \frac{2}{(4\pi)^2}  Y_d  V^{\dagger}_{\rm CKM}  \, C_{L Q,\, U}  V^{}_{\rm CKM} Y_d,   \\
\frac{d}{d\log \mu} C_{eu} &=&   - \frac{4}{(4\pi)^2}  Y_u  V^{}_{\rm CKM}  \, C_{Q e}  V^{\dagger}_{\rm CKM} Y_u,   \\
\frac{d}{d\log \mu} C_{Qe} &=&   - \frac{2}{(4\pi)^2}    V^{\dagger}_{\rm CKM}  Y_u \, C_{e u} \, Y_u V^{}_{\rm CKM}. \label{fc6}
\end{eqnarray}
Because of the CKM and Yukawa factors, these RGE always induce negligible effects.

\begin{table}[t]
\small
\centering
\begin{tabular}{||c|| c c c || c c c c ||}
\hline
                        & $({C}_{LQ,U})_{tt}$ & $({C}_{Lu})_{tt}$ &  $({C}_{eu})_{tt}$  & $({C}_{LQ,D})_{bb}$ & $({C}_{Ld})_{bb}$ &  $({C}_{ed})_{bb}$  & $(C_{Qe})_{bb}$ \\
\hline \hline
$c^{(1)}_{L\varphi}+c^{(3)}_{L\varphi}$ & $-102$        &  $106$        & --            & $-10$               & $1.9$ & $-$     &  --\\
$c_{e\varphi}$                          & --            &   --          & $106$         & --                  & $-$  & $1.9$    &  $-112$ \\
\hline \hline 
                        & $({C}_{LQ,U})_{cc}$ & $({C}_{Lu})_{cc}$ &  $({C}_{eu})_{cc}$  & &  &   & \\
\hline \hline 
$c^{(1)}_{L\varphi}+c^{(3)}_{L\varphi}$ & $8.5$        &  $-3.9$        & --            & &  &    & \\
$c_{e\varphi}$                          & --            &   --          & $-3.9$         & &  &    &  \\
\hline \hline 
                        & $({C}_{LQ,U})_{ll}$ & $({C}_{Lu})_{ll}$ &  $({C}_{eu})_{ll}$  & $({C}_{LQ,D})_{kk}$ & $({C}_{Ld})_{kk}$ &  $({C}_{ed})_{kk}$  & $(C_{Qe})_{kk}$ \\
\hline
$\left({ C}_{LQ,U}\right)_{ii}$     & $-4.9$        &  $-0.65$      &   --          &  $4.5$              & $0.34$  & $-$   &  -- \\
$\left({ C}_{LQ,D}\right)_{jj}$     & $ 4.5$        &  $-0.69$      &   --          & $-4.9$              & $0.34$  & $-$   &  -- \\
$\left({ C}_{Lu}\right)_{ii}$       & $-0.67$       &  $-2.7 $      &   --          & $-0.67$             & $1.3$  & $-$    &  -- \\
$\left({ C}_{Ld}\right)_{jj}$       & $ 0.34$       &  $1.3$        &   --          & $0.34$              & $-0.67$  & $-$  &  --   \\
$\left({ C}_{eu}\right)_{ii}$       &   --          &   --          &  $-2.7$       & $-$                 & $-$  & $1.3$    & $-1.3$ \\
$\left({ C}_{ed}\right)_{jj}$       &   --          &   --          &  $1.3$        & $-$                 & $-$  & $-0.67$  & $0.67$\\
$\left({ C}_{Qe}\right)_{jj}$       &   --          &   --          &  $-0.67$      & $-$                  & $-$  & $0.34$  & $-0.34$  \\
\hline 
$\left(C_{LL}\right)_{\tau e \mu\mu}$     & $-2.1$  & $1.0$ & --       & $ 2.6$ & $-0.5$  & $-$     & $-$ \\
$\left(C_{LL}\right)_{\tau \mu \mu e }$   & $4.7 $  & --     & --       & $-4.7$ & --       & $-$     & $-$  \\
$\left(C_{LL}\right)_{\tau e ee}$         & $2.6 $  & $1.0$ & --       & $-2.1$ & $-0.5$  & $-$     & $-$  \\
$\left(C_{ee}\right)_{\tau e \ell \ell}$  & --      & --     & $2.0$    & --     & $-$      & $-1.0$ & $1.0$ \\
$\left(C_{Le}\right)_{\tau e \ell \ell}$  & $1.0$  & $4.0$  & --       & $1.0$ & $-2.0$   & $-$     & --      \\
$\left(C_{Le}\right)_{\ell \ell \tau e}$  &  --     &  --    & $2.0$    & $-$    & $-$      & $-1.0$ & $1.0$ \\
\hline
\end{tabular}
\caption{The Wilson coefficients at $\mu=m_t$ induced by nonzero top-, bottom- and charm-quark operators at the scale $\Lambda =1$ TeV, in units of $10^{-3}$. The indices  $i$, $l$ and $j$, $k$ denote, 
respectively, $u$- and $d$-type flavor indices, and we consider mixing onto operators of different flavor, $i \neq l$, $j \neq k$.}
\label{topRGE}
\end{table}

The penguin diagrams also induce the following leptonic operators that contribute to $\tau \to e \mu \mu$ and $\tau \to 3 e$ \cite{Alonso:2013hga,Jenkins:2013wua}.
The RGEs for the left-handed operator $C_{LL}$ are
\begin{eqnarray}   \label{peng8} 
\frac{d}{d \log\mu} \left( C_{LL} \right)_{\tau e \mu \mu} &=&    - \frac{1}{6} N_c \frac{g_2^2}{(4\pi)^2}  \left(  - (C_{LQ, U})_{\tau e ii}   + (C_{LQ, D})_{\tau e jj} \right) + \frac{2}{3} N_c \frac{g_1^2}{(4\pi)^2}  \\ & & \times\bigg\{ 
  {\rm y_l} {\rm y_q}  \left(  (C_{LQ, U})_{\tau e ii}   + (C_{LQ, D})_{\tau e jj} \right)   
  + {\rm y_l} {\rm y_u} ( C_{L u}) _{\tau e ii}  + {\rm y_l} {\rm y_d} ( C_{L d}) _{\tau e jj}     \bigg\} 
\nn \\
\frac{d}{d \log\mu} \left( C_{LL} \right)_{\tau \mu \mu e} &=&   \frac{1}{3} N_c \frac{g_2^2}{(4\pi)^2}  \left(  - (C_{LQ, U})_{\tau e ii}   + (C_{LQ, D})_{\tau e jj} \right)
   \label{peng9}  \\ 
    \label{peng10} 
   \frac{d}{d \log\mu} \left( C_{LL} \right)_{\tau e e e} &=&    + \frac{1}{6} N_c \frac{g_2^2}{(4\pi)^2}  \left(  - (C_{LQ, U})_{\tau e ii}   + (C_{LQ, D})_{\tau e jj} \right)+ \frac{2}{3} N_c \frac{g_1^2}{(4\pi)^2}\\ & & \times \bigg\{ 
  {\rm y_l} {\rm y_q}  \left(  (C_{LQ, U})_{\tau e ii}   + (C_{LQ, D})_{\tau e jj} \right)   
  + {\rm y_l} {\rm y_u} ( C_{L u}) _{\tau e ii}   + {\rm y_l} {\rm y_d} ( C_{L d}) _{\tau e jj}     \bigg\} \nn 
\end{eqnarray}
The RGEs for the $\ell \ell \tau e$ and $\ell e \tau \ell$ components are the same as Eqs. \eqref{peng8} -- \eqref{peng10}.
\begin{eqnarray}
\frac{d}{d \log\mu} \! \left( C_{ee} \right)_{\tau e \ell \ell} &=&     \frac{2}{3} N_c \frac{g_1^2}{(4\pi)^2} \Big\{ 
  2 {\rm y_e} {\rm y_q}  (C_{Q e})_{\tau e jj}    
  + {\rm y_e} {\rm y_u} ( C_{e u}) _{\tau e ii}   + {\rm y_e} {\rm y_d} ( C_{e d}) _{\tau e jj}     \Big\}\,,\quad
   \label{peng11} 
\end{eqnarray}   
and again the $\ell \ell \tau e$ component has the same RGE.
Finally, the LR operator 
\begin{eqnarray}
\frac{d}{d \log\mu} \left( C_{Le} \right)_{\tau e \ell \ell} &=&     \frac{4}{3} N_c \frac{g_1^2}{(4\pi)^2} \Bigg\{ 
    {\rm y_e} {\rm y_q}  \left(  (C_{LQ, U})_{\tau e ii} + (C_{LQ, D})_{\tau e jj} \right)     \nn \\&&
  + {\rm y_e} {\rm y_u} ( C_{L u}) _{\tau e ii}   + {\rm y_e} {\rm y_d} ( C_{L d}) _{\tau e jj}     \Bigg\}, \label{peng12}\\
\frac{d}{d \log\mu} \left( C_{Le} \right)_{\ell \ell \tau e } &=&     \frac{4}{3} N_c \frac{g_1^2}{(4\pi)^2} \Bigg\{ 
    2 {\rm y_l} {\rm y_q}    (C_{Qe})_{\tau e jj}   
  + {\rm y_l} {\rm y_u} ( C_{e u}) _{\tau e ii}   + {\rm y_l} {\rm y_d} ( C_{e d}) _{\tau e jj}     \Bigg\}. \quad \label{peng13}
\end{eqnarray}   
The solutions of the RGEs in Eqs. \eqref{cLphi}--\eqref{peng13} take the form \eq{RGEmatrix}, and the solution coefficients are given in Table \ref{topRGE}.

\subsection{Dipole contributions induced  by the LFV Yukawa interaction}\label{dipole_yukawa}  
 The $\tau -e$ LFV Yukawa coupling contributes to $\tau\to e\gamma$ through one- and two-loop diagrams. For the one-loop diagram, the expression is given by \cite{Harnik:2012pb}
\begin{align}\label{Yone}
\left[\Gamma^e_{\gamma}\right]_{e\tau}&=-\frac{vm_{\tau}}{16\pi^2 m^2_h}\left(Y_e' \right)_{e\tau}\left(Y_e \right)_{\tau\tau}\left(\log\frac{m^2_h}{m^2_{\tau}}-\frac{4}{3} \right),\\
\left[\Gamma^{e}_{\gamma}\right]^*_{\tau e}&=-\frac{vm_{\tau}}{16\pi^2 m^2_h}\left(Y_e^{'} \right)^*_{\tau e}\left(Y_e^* \right)_{\tau\tau}\left(\log\frac{m^2_h}{m^2_{\tau}}-\frac{4}{3} \right),\label{Ytwo}
\end{align}
with  the Higgs mass $m_h=125~$GeV.

The two-loop contribution consists of several diagrams, known as Barr-Zee diagrams \cite{Barr:1990vd, Abe:2013qla}, where not only top quark but also $W$ boson runs in the loop. They are given by
\begin{align}
&\left[\Gamma^e_{\gamma}\right]_{e\tau}=\frac{N_cQ_t\alpha_{\rm em}}{8\pi^3 }\frac{v}{m_t}\left(Y'_e \right)_{e\tau}\bigg[Q_t\left\{{\rm Re}\left(Y_u\right)_{tt}f(x_{th})+i~{\rm Im}\left(Y_u \right)_{tt}g(x_{th}) \right\}\nonumber\\
&-\frac{1}{4s^2_w c^2_w}\left(z_{\tau_L}+z_{\tau_R} \right)\left(z_{t_L}+z_{t_R} \right)\left\{{\rm Re}\left(Y_u\right)_{tt}\tilde{F}_H(x_{th},x_{tZ})+i~{\rm Im}\left(Y_u \right)_{tt}\tilde{F}_A(x_{th},x_{tZ}) \right\}
 \bigg]\nonumber\\
& -\frac{\alpha_{\rm em}}{32\pi^3}\left(Y'_e \right)_{e\tau}\bigg[{\cal J}^{\gamma}_W(m_h)-\frac{1}{2s^2_w}\left(z_{\tau_L}+z_{\tau_R} \right){\cal J}^Z_W(m_h) \bigg],\label{BZ1}\\
& \left[\Gamma^e_{\gamma}\right]_{\tau e}^*=\frac{N_cQ_t\alpha_{\rm em}}{8\pi^3 }\frac{v}{m_t}\left(Y'_e \right)_{\tau e}^*\bigg[Q_t\left\{{\rm Re}\left(Y_u\right)_{tt}f(x_{th})-i~{\rm Im}\left(Y_u \right)_{tt}g(x_{th}) \right\}\nonumber\\
&-\frac{1}{4s^2_w c^2_w}\left(z_{\tau_L}+z_{\tau_R} \right)\left(z_{t_L}+z_{t_R} \right)\left\{{\rm Re}\left(Y_u\right)_{tt}\tilde{F}_H(x_{th},x_{tZ})-i~{\rm Im}\left(Y_u \right)_{tt}\tilde{F}_A(x_{th},x_{tZ}) \right\}
 \bigg]\nonumber\\
& -\frac{\alpha_{\rm em}}{32\pi^3}\left(Y'_e \right)_{\tau e}^*\bigg[{\cal J}^{\gamma}_W(m_h)-\frac{1}{2s^2_w}\left(z_{\tau_L}+z_{\tau_R} \right){\cal J}^Z_W(m_h) \bigg], \label{BZ2}
\end{align}
where $x_{ij}=m^2_i/m^2_j$, and the $Z$ couplings $z_{f_L}$ and $z_{f_R}$ are given in Eq. \eqref{zcouplings}. The loop functions are given by
\begin{align}
f(x)&=\frac{x}{2}\int^1_0dy~\frac{1-2y(1-y)}{y(1-y)-x}\ln\left(\frac{y(1-y)}{x} \right),\\
g(x)&=\frac{x}{2}\int^1_0dy~\frac{1}{y(1-y)-x}\ln\left(\frac{y(1-y)}{x} \right),\\
\tilde{F}_A(a,b)&=\frac{1}{a-b}\left[ag(b)-bg(a) \right],\\
\tilde{F}_H(a,b)&=\frac{1}{a-b}\left[af(b)-bf(a) \right],
\end{align}
\begin{align}
{\cal J}^V_W(m_h)&=\frac{2m^2_W}{m^2_h-m^2_V}\bigg[-\frac{1}{4}\left\{\left(6-\frac{m^2_V}{m^2_W}\right)+\left(1-\frac{m^2_V}{2m^2_W}\right)\frac{m^2_h}{m^2_W} \right\}\nonumber\\
&\hspace{4cm}\times
\left(I_1(m_W,m_h)-I_1(m_W,m_V) \right)\nonumber\\
&+\left\{\left(-4+\frac{m^2_V}{m^2_W}\right)+\frac{1}{4}\left(6-\frac{m^2_V}{2m^2_W}\right)+\frac{1}{4}\left(1-\frac{m^2_V}{2m^2_W}\right)\frac{m^2_h}{m^2_W} \right\}\nonumber\\
&\hspace{1cm}\times
\left(I_2(m_W,m_h)-I_2(m_W,m_V) \right)\bigg],\\
I_1(m_1,m_2)&=-2\frac{m^2_2}{m^2_1}f\left(\frac{m^2_1}{m^2_2} \right),\\
I_2(m_1,m_2)&=-2\frac{m^2_2}{m^2_1}g\left(\frac{m^2_1}{m^2_2} \right).
\end{align}
\begin{table}
\center
\begin{tabular}{|| c || c c c  || }
\hline 
                                         & $\Gamma^{e}_Z$ & $\left( Y^{\prime}_e\right)_{\tau e}$  &     \\
\hline                                    
$\Gamma^e_\gamma$                                                           &   $-2.0 \cdot 10^{-2}$   & $2.7 \cdot 10^{-7}$                      &                                 \\
$\left( C_{GG} \right)_{\tau e}$           &   --                               &  $-2.59$                &                                                 \\
$\left( C_{G\tilde G} \right)_{\tau e}$     &   --                               & $0$              &                                               \\                                    
$C^{eq}_{\rm SRR}$, $C^{eq}_{\rm SRL}$     &   --                               &  $-1.94$ $Y_q$                       &                                            \\                                    
$C^{eq}_{\rm TRR}$                       &   --                               &  $7.0 \cdot 10^{-3}$ $Q_qY_q$        &                                           \\                                   
\hline
\end{tabular}
\caption{The Wilson coefficients of dipole, gluonic, scalar, and tensor operators at $\mu=2$ GeV induced by a $Z$ dipole or a CLFV Yukawa through operator mixing. The Yukawa contribution to $\Gamma^e_\gamma$ does not include
the numerically larger effects from top  quarks and weak bosons, given in Eq. \eqref{BZnum}.
$q$ denotes a light quark, $q  \in \{ u,\, d,\, s\}$, and the quark Yukawa couplings $Y_q$ is evaluated at the scale $\mu$.}
\label{ScalarTensorRGE_low}
\end{table}
In Eqs. \eqref{BZ1} and \eqref{BZ2} we consider only the top quark. At leading log, the contributions of $b$ and $c$ quarks are accounted for 
by first matching onto LEFT scalar operators, $C^{eq}_{\rm SRR}$ and $C^{eq}_{\rm SLR}$, which then run into tensor and dipole operators, as discussed in the next section. These contributions are shown in Table \ref{ScalarTensorRGE_low} and are negligible. 
Using the couplings and masses of SM particles in Eqs. \eqref{Yone}--\eqref{BZ2}, we obtain
\begin{align}
\left[\Gamma^e_{\gamma}\right]_{e\tau/\tau e}^{\rm 1-loop}&=-9.2\times 10^{-6}\left(Y^{\prime}_e\right)_{e\tau/\tau e}, \label{BZnum}\\
\left[\Gamma^e_{\gamma}\right]_{e\tau/\tau e}^{\rm 2-loop}&=-5.1\times 10^{-5}\left(Y^{\prime}_e\right)_{e\tau/\tau e}.
\end{align}
One- and two-loop contributions are thus of similar size.

\subsection{Running below the electroweak scale}\label{RGE_low}
The RGEs below the electroweak scale are listed in Ref. \cite{Jenkins:2017dyc}. For the QCD and QED couplings, the one-loop running is given by
\begin{align}
\mu\frac{d}{d\mu}{g}_s(\mu)&=-\frac{1}{(4\pi)^2}\left[\frac{11}{3}N_c-\frac{2}{3}\left(n_u+n_d\right)\right] \left(g_s(\mu)\right)^3,\\
\mu\frac{d}{d\mu}{e}(\mu)&=\frac{4}{3}\frac{1}{(4\pi)^2}\left(n_eQ_e^2+n_dN_cQ_d^2+n_uN_cQ_u^2 \right)\left(e(\mu)\right)^3,
\end{align}
where $n_u, n_d$ and $n_e$ are the number of active up-, down-type quarks and charged leptons. For example, up to $m_b$ scale, $n_u=2, n_d=3$ and $n_e=3$.

The anomalous dimension of the dipole, scalar and tensor operators are 
\begin{eqnarray}
\mu\frac{d}{d\mu}\left({\Gamma}^e_{\gamma} \right)_{\tau e}&=& +\frac{1}{(4\pi)^2}\bigg[ 10Q_e^2 
e^2 \, \left(\Gamma^e_{\gamma} \right)_{\tau e}   -32N_c  \sum_q  Q_q\left(Y_q\right)_{ww}\left(C^{eq}_{\rm TRR}\right)_{\tau e ww}\bigg], \\
\mu\frac{d}{d\mu} C^{eq}_{\rm SRR} &=& - \frac{1}{(4\pi)^2}\bigg[\left(6e^2\left(Q_e^2+Q_q^2 \right)+6g^2_s C_F\right) C^{eq}_{\rm SRR}
+96e^2 Q_e Q_q\, C^{eq}_{\rm TRR} \bigg],\\
\mu\frac{d}{d\mu} C^{eq}_{\rm TRR} &=& + \frac{1}{(4\pi)^2}\bigg[
\left(2e^2\left(Q_e^2+Q_q^2 \right)+2g^2_sC_F \right) C^{eq}_{\rm TRR} 
-2e^2Q_e Q_q  C^{eq}_{\rm SRR}  \bigg],\\
\mu\frac{d}{d\mu} C^{eq}_{\rm SRL} &=&-\frac{6}{(4\pi)^2}\bigg[e^2\left(Q_e^2+Q_q^2\right)+g_s^2C_F \bigg]  C^{eq}_{\rm SRL},
\end{eqnarray}
where, in the last three lines, we have omitted the quark and lepton flavor indices $\tau e s t$ or $e  \tau s t$. $q$ here denotes both $u$ and $d$-type quark, $q \in \{ u, d\}$,
while $w \in \{u,c\}$ for $u$-type operators and $w \in \{ d, s, b \}$ for $d$-type operators. A summation over repeated flavor indices is understood. 
We integrate out the bottom quark at the scale $\mu = m_b$, while the charm quark at the scale $\mu = 2$ GeV. The solutions of the RGEs for $\Gamma^e_Z$ and $(Y^{\prime}_e)_{\tau e}$ are given in Table  \ref{ScalarTensorRGE_low}.

Purely leptonic operators at low-energy are induced by photon penguin diagrams.
The RGEs for the left-handed leptonic operators are 
\begin{eqnarray}
\mu\frac{d}{d\mu}\left({C}_{\rm VLL}^{ee}\right)_{\tau e \mu\mu}&=&\frac{e^2}{3(4\pi)^2}Q_e\bigg[N_c
\sum_{q} Q_q\left(C_{\rm VLL}^{eq}+C_{\rm VLR}^{eq}\right) +Q_e\left(4C_{\rm VLL}^{ee}+C_{\rm VLR}^{ee} \right)\bigg]_{\tau e ww} \nn \label{lowpenguin1}\\
& & 
+\frac{12}{(4\pi)^2}e^2Q_e^2\left(C_{\rm VLL}^{ee}\right)_{\tau e \mu\mu},\\
\mu\frac{d}{d\mu}\left({C}_{\rm VLL}^{ee}\right)_{\tau \mu \mu e}&=&\frac{e^2}{3(4\pi)^2}Q_e\bigg[N_c \sum_q Q_q\left(C_{\rm VLL}^{eq}+C_{\rm VLR}^{eq}\right)
+Q_e\left(4C_{\rm VLL}^{ee}+C_{\rm VLR}^{ee} \right)\bigg]_{\tau e ww} \nn \\
& & +\frac{12}{(4\pi)^2}e^2Q_e^2\left(C_{\rm VLL}^{ee}\right)_{\tau \mu \mu e}, \\ 
\mu\frac{d}{d\mu}\left({C}_{\rm VLL}^{ee}\right)_{\tau e ee}&=&\frac{2}{3(4\pi)^2}e^2Q_e\bigg[N_c \sum_q Q_q\left(C_{\rm VLL}^{eq}+C_{\rm VLR}^{eq}\right)
+Q_e\left(4C_{\rm VLL}^{ee}+C_{\rm VLR}^{ee} \right)\bigg]_{\tau e ww} \nn \\ & &  +\frac{12}{(4\pi)^2}e^2Q_e^2\left(C_{\rm VLL}^{ee}\right)_{\tau e ee},
\end{eqnarray}
while those for right-handed operators are
\begin{eqnarray}
\mu\frac{d}{d\mu}\left({C}_{\rm VRR}^{ee}\right)_{\tau e \mu\mu}&=& \frac{12}{(4\pi)^2}e^2Q_e^2\left(C_{\rm VRR}^{ee}\right)_{\tau e \mu\mu} + \frac{e^2Q_e}{3(4\pi)^2} \left\{  
   \left[ N_c \sum_q Q_q C_{\rm VLR}^{qe} + Q_e C_{\rm VLR}^{ee} \right]_{ww\tau e} \right.  \nonumber \\ & & \left.
+  \left[ N_c \sum_q Q_q C_{\rm VRR}^{eq} + 4 Q_e C_{\rm VRR}^{ee} \right]_{\tau e w w} \right\}\\
\mu\frac{d}{d\mu}\left({C}_{\rm VRR}^{ee}\right)_{\tau \mu \mu e}&=& \frac{12}{(4\pi)^2}e^2Q_e^2\left(C_{\rm VRR}^{ee}\right)_{\tau \mu \mu e} + \frac{e^2Q_e}{3(4\pi)^2} \left\{  
   \left[ N_c \sum_q Q_q C_{\rm VLR}^{qe} + Q_e C_{\rm VLR}^{ee} \right]_{ww\tau e} \right.  \nonumber \\ & & \left.
+  \left[ N_c \sum_q Q_q C_{\rm VRR}^{eq} + 4 Q_e C_{\rm VRR}^{ee} \right]_{\tau e w w} \right\}\\
\mu\frac{d}{d\mu}\left({C}_{\rm VRR}^{ee}\right)_{\tau e e e}&=& \frac{12}{(4\pi)^2}e^2Q_e^2\left(C_{\rm VRR}^{ee}\right)_{\tau \mu \mu e} + \frac{2 e^2Q_e}{3(4\pi)^2} \left\{  
   \left[ N_c \sum_q Q_q C_{\rm VLR}^{qe} + Q_e C_{\rm VLR}^{ee} \right]_{ww\tau e} \right.  \nonumber \\ & & \left.
+  \left[ N_c \sum_q Q_q C_{\rm VRR}^{eq} + 4 Q_e C_{\rm VRR}^{ee} \right]_{\tau e w w} \right\}.
\end{eqnarray}
The RGEs for $LR$ operators are given by
\begin{eqnarray}
\mu\frac{d}{d\mu}\left({C}_{\rm VLR}^{ee}\right)_{\tau e \ell \ell}&=&\frac{4}{3(4\pi)^2}e^2Q_e\bigg[N_c \sum_q Q_q\left(C_{\rm VLL}^{eq}+C_{\rm VLR}^{eq}\right)
+Q_e\left(4C_{\rm VLL}^{ee}+C_{\rm VLR}^{ee} \right)\bigg]_{\tau e ww} \nn \\
& & 
-\frac{12}{(4\pi)^2}e^2Q_e^2\left(C_{\rm VLR}^{ee}\right)_{\tau e ll},\\
\mu\frac{d}{d\mu}\left({C}_{\rm VLR}^{ee}\right)_{\ell \ell \tau e }&=& -\frac{12}{(4\pi)^2}e^2Q_e^2\left(C_{\rm VLR}^{ee}\right)_{ll\tau e} + \frac{4 e^2 Q_e }{3(4\pi)^2} \left\{
  \left[ N_c \sum_q Q_q  C_{\rm VLR}^{qe} + C_{\rm VLR}^{ee}  \right]_{ww\tau e}    \right.  \nn \\ & & \left.
+ \left[ N_c \sum_q Q_q  C_{\rm VRR}^{eq} +4 C_{\rm VRR}^{ee} \right]_{\tau e w w}   \right\}
\end{eqnarray}

\begin{table}[t]
\centering
\begin{tabular}{||c| c c c c || c | c c c ||}
\hline
                        & \! $({C}_{LQ,U})_{tt}$\! &\! $({C}_{Lu})_{tt}$\! & \!$({C}_{LQ,D})_{bb}$\! & \!$({C}_{Ld})_{bb}$\! &  &\!$(C_{eu})_{tt}$\! &  \!$({C}_{ed})_{bb}$\!  & \!$(C_{Qe})_{bb}$\!   \\
\hline \hline
$C^{eu}_{\rm VLL}$      &  $-39 $            &  $ 35$             &   $ 3.0$         & $3.0$             &  $C^{eu}_{\rm VRR}$ & $-19$   & $3.0$                 & $18$\\
$C^{ed}_{\rm VLL}$      &  $ 46 $            &  $-44$             &   $-1.5$         & $-1.5$            &  $C^{ed}_{\rm VRR}$ & $9.5$   & $-1.5$                & $-9$\\ 
$C^{eu}_{\rm VLR}$      &  $15 $             &  $-19$             &    $3.0$         & $3.0$              &  $C^{ue}_{\rm VLR}$ & $ 35$   & $3.0$                 & $-35$\\
$C^{ed}_{\rm VLR}$      &  $-7.5$            &  $ 9.5$           &   $-1.5$         & $-1.5$              &  $C^{de}_{\rm VLR}$ & $-44$   & $-1.5$                & $44$ \\
\hline \hline 
                        & \!$({C}_{LQ,U})_{cc}$\! &\! $({C}_{Lu})_{cc}$\! &   \!$c^{(1)}_{L \varphi} + c^{(3)}_{L \varphi}$\! &  & &\! $({C}_{eu})_{cc}$\!    & \!$c_{e\varphi}$ \! &   \\
                        \hline \hline
$C^{eu}_{\rm VLL}$      &  $-6.8$             & $-6.8$            & $342$       &   & $C^{eu}_{\rm VRR}$  & $-6.8$  &  $-158$ &  \\
$C^{ed}_{\rm VLL}$      &  $3.4$              & $3.4$             & $-421$      &   & $C^{ed}_{\rm VRR}$  &$3.4$    &  $79$    &\\
$C^{eu}_{\rm VLR}$      &  $-6.8$             & $-6.8$            & $-158$      &  & $C^{ue}_{\rm VLR}$   &$-6.8$   &  $342$ & \\
$C^{ed}_{\rm VLR}$      &  $3.4$              & $3.4$             & $79$        &   & $C^{de}_{\rm VLR}$  & $3.4$   & $-421$ &\\
\hline \hline
\end{tabular}
\caption{The Wilson coefficients (in units of $10^{-3}$) at $\mu=2$ GeV induced by nonzero top-, bottom- and charm-quark vector-like operators at the scale $\Lambda = 1$ TeV,
and by $Z$ CLFV vector and axial couplings. The $u$-type operators have flavor indices $uu$, while the $d$-type $dd$ or $ss$. }
\label{RGE_low_vector}
\end{table}

\begin{table}[t]
\centering
\small
\begin{tabular}{||c| c c c c || c| c c c ||}
\hline
                                                   & $({C}_{LQ,U})_{tt}$  \!\!\!\!  & \!\!\!\! $({C}_{Lu})_{tt}$  \!\!\!\! &  \!\!\!\!  $({C}_{LQ,D})_{bb}$  \!\!\!\! & \!\!\!\! $({C}_{Ld})_{bb}$  \!\!\!\! &  & \!\!\!\! $({C}_{eu})_{tt}$   \!\!\!\! & \!\!\!\!  $({C}_{ed})_{bb}$   \!\!\!\! & \!\!\!\!  $(C_{Qe})_{bb}$  \!\!\!\!  \\
\hline \hline
$\left[C^{ee}_{\rm VLL}\right]_{\tau e\, e e}$      &  $ {15} $           &  $ {-13}$           &    ${-2.1}$        &   ${-2.1}$      &  $\left[C^{ee}_{\rm VRR}\right]_{\tau e\, e e}$      & ${14}$   & ${-2.1}$                    & ${-13}$\\
$\left[C^{ee}_{\rm VLL}\right]_{\tau e\, \mu\mu}$   &  ${4} $           &  ${-6}$           &    ${2.5} $        &   ${-1.3}$      &  $\left[C^{ee}_{\rm VRR}\right]_{\tau e\, \mu \mu}$  & ${8}$   & ${-1.5}$                  & ${-6}$\\ 
$\left[C^{ee}_{\rm VLL}\right]_{\tau \mu \,\mu e}$  &  ${11} $           &  ${-7}$           &    ${-4.6}$        &   ${-0.8}$      &  $\left[C^{ee}_{\rm VRR}\right]_{\tau \mu\, \mu e}$  & ${6}$   & ${-0.6}$ &
${-7}$\\ 
$\left[C^{ee}_{\rm VLR}\right]_{\tau e\, \ell \ell}$      &  $-22.5 $         &   $28$            &     $-4.5$       &   $-4.5$      & $\left[C^{ee}_{\rm VLR}\right]_{\ell \ell \, \tau e }$    & $-25$  & $-4.5$ & $27$ \\                    
\hline \hline 
                        & $({C}_{LQ,U})_{cc}$  \!\!\!\! & \!\!\!\!  $({C}_{Lu})_{cc}$  \!\!\!\! & \!\!\!\!   $c^{(1)}_{L \varphi} + c^{(3)}_{L \varphi}$   \!\!\!\! & &  &\!\!\!\!  $({C}_{eu})_{cc}$   \!\!\!\! & \!\!\!\! $c_{e\varphi}$  \!\!\!\! &    \\
                        \hline \hline
$\left[C^{ee}_{\rm VLL}\right]_{\tau e\, e e}$      &  ${4.8}$             &  ${4.8}$             &     ${-132}$       &             &      $\left[C^{ee}_{\rm VRR}\right]_{\tau e\, e e}$         &  $4.6$  & ${118}$ &\\
$\left[C^{ee}_{\rm VLL}\right]_{\tau e\, \mu\mu}$   &  ${-0.9}$            &  ${2.8}$             &     ${-66}$       &             &      $\left[C^{ee}_{\rm VRR}\right]_{\tau e\, \mu \mu}$     &  $3.3$  & ${59}$ & \\ 
$\left[C^{ee}_{\rm VLL}\right]_{\tau \mu \,\mu e}$  &  ${5.7}$             &  ${1.9}$             &     ${-66}$       &             &      $\left[C^{ee}_{\rm VRR}\right]_{\tau \mu\, \mu e}$     & $1.3$   & ${59}$ & \\ 
$\left[C^{ee}_{\rm VLR}\right]_{\tau e\, \ell \ell}$ &  $10$              &   $10$             &     $237$        &             &      $\left[C^{ee}_{\rm VLR}\right]_{\ell \ell \, \tau e }$ &  $10$   & $-263$ &\\                    
\hline \hline
\end{tabular}
\caption{The Wilson coefficients of purely leptonic operators  (in units of $10^{-3}$) at $\mu=2$ GeV induced by nonzero top-, bottom- and charm-quark vector-like operators at the scale $\Lambda = 1$ TeV,
and by $Z$ CLFV vector and axial couplings. Note that $[C_{\rm VLL}^{ee}]_{ee\tau e}$,$[C_{\rm VLL}^{ee}]_{\mu\mu\tau e}$ and $[C_{\rm VLL}^{ee}]_{\mu e\tau \mu}$ are also generated with the same contributions as the operators listed above.} 
\label{RGE_low_lepto}
\end{table}

Finally, the anomalous dimensions of the semileptonic operators are
\begin{eqnarray}
\mu\frac{d}{d\mu}\left({C}_{\rm VLL}^{eq}\right)_{\tau e st}&=&\frac{4}{3(4\pi)^2}e^2Q_{q}\delta_{st}\bigg[N_c \sum_{q^\prime }Q_{q^\prime}\left(C_{\rm VLL}^{eq^\prime}+C_{\rm VLR}^{eq^\prime}\right)  +Q_e\left(4C_{\rm VLL}^{ee}+C_{\rm VLR}^{ee} \right) \bigg]_{\tau e ww} \nonumber\\
&&
+\frac{12}{(4\pi)^2}e^2Q_eQ_q\left(C_{\rm VLL}^{eq}\right)_{\tau e st}, \\
\mu\frac{d}{d\mu}\left({C}_{\rm VRR}^{e q}\right)_{\tau est}&=&\frac{4}{3(4\pi)^2}e^2Q_q\delta_{st}\, \left( \bigg[N_c
\sum_{q^\prime} Q_{q^\prime }C_{\rm VLR}^{q^\prime e}  +   Q_e  C_{\rm VLR}^{ee} \bigg]_{ww \tau e}  \right. \\ && \left. 
+ \bigg[N_c \sum_{q^\prime} Q_{q^\prime }C_{\rm VRR}^{e q^\prime}  +   4 Q_e  C_{\rm VRR}^{ee} \bigg]_{\tau e w w} \right)  +\frac{12}{(4\pi)^2}e^2Q_eQ_u\left(C_{\rm VRR}^{eu}\right)_{\tau e st},\nonumber\\
\mu\frac{d}{d\mu}\left({C}_{\rm VLR}^{eq}\right)_{\tau est}&=&\frac{4}{3(4\pi)^2}e^2Q_q\delta_{st}\bigg[N_c \sum_{q^\prime }Q_{q^\prime}\left(C_{\rm VLL}^{e q^\prime}+C_{\rm VLR}^{e q^\prime}\right) 
+Q_e\left(4C_{\rm VLL}^{ee}+C_{\rm VLR}^{ee} \right)\bigg]_{\tau eww}\nonumber\\
& &-\frac{12}{(4\pi)^2}e^2Q_eQ_q\left(C_{\rm VLR}^{eq}\right)_{\tau est},\\
\mu\frac{d}{d\mu}\left({C}_{\rm VLR}^{qe}\right)_{st \tau e}&=&\frac{4}{3(4\pi)^2}e^2Q_q\delta_{st} \left( \bigg[
N_c \sum_{q^\prime} Q_{q^\prime}  C_{\rm VLR}^{ q^\prime e}  + Q_e C_{\rm VLR}^{ee} \bigg]_{w w \tau e} \right. \\  & & \left. + 
\bigg[N_c \sum_{q^\prime} Q_{q^\prime}  C_{\rm VRR}^{e q^\prime}  + 4 Q_e C_{\rm VRR}^{ee} \bigg]_{\tau e w w } \right) -\frac{12}{(4\pi)^2}e^2Q_e Q_q\left(C_{\rm VLR}^{ue}\right)_{st\tau e },\nonumber
\label{lowpenguin2}
\end{eqnarray}
We give the solution of the RGEs in Eqs. \eqref{lowpenguin1}-\eqref{lowpenguin2} in Tables \ref{RGE_low_vector} and \ref{RGE_low_lepto}.

\section{Partonic cross sections for CLFV processes}\label{partonic}

\subsection{DIS}

Here we complete the collection of  expressions for the partonic cross sections, as defined in Eq. \eqref{eq:sigma}, induced by the CLFV SMEFT operators, some of which we gave in \ssec{CLFV_partonic}.
The prefactors $F_Z$, $F_{\rm dip}$, $F_S$ and $F_G$ are defined in Eqs. \eqref{FZ}, \eqref{Fdip}, \eqref{FS} and \eqref{FG}.

\paragraph{Vertex corrections and vector-axial four-fermion operators}

In \eq{up1} we gave the $u$-type quark contributions to the partonic cross sections for the $Z$ coupling and vector-axial four-fermion operators.
Here we give corresponding results for $\bar u,d,\bar d$-type quark and antiquark contributions.

For $\bar u$ antiquarks: 
\begin{eqnarray}
\hat\sigma^{\bar u_i}_{LR} &=& F_Z (1-y)^2   \left\{  \left|\left[c^{(1)}_{L\varphi}  +  c^{(3)}_{L\varphi}\right]_{\tau e} z_{u_L} +  \hat{\rho}_Z 
\left[C^{}_{LQ,\, U}\right]_{\tau e u_i u_i}   \right|^2  +  \sum_{j \neq i }\left|   \hat{\rho}_Z 
\left[C^{}_{LQ,\, U}\right]_{\tau e u_i u_j}    \right|^2 \right\}  \nn \\
\hat\sigma^{\bar u_i}_{RL} &=& F_Z (1-y)^2  \left\{  \left|  \left[c_{e\varphi}\right]_{\tau e} z_{u_R} +  \hat{\rho}_Z \left[C_{e u} \right]_{\tau e u_i u_i} \right|^2  + \sum_{j \neq i} \left|   \hat{\rho}_Z 
 \left[C_{e u} \right]_{\tau e u_i u_j}   \right|^2 \right\} \nn \\
\hat\sigma^{\bar u_i}_{LL} &=& F_Z    \left\{  \left| \left[ c^{(1)}_{L\varphi}  +  c^{(3)}_{L\varphi}\right]_{\tau e} z_{u_R} + \hat{\rho}_Z \left[C_{Lu}\right]_{\tau e u_i u_i}   \right|^2  +  \sum_{j \neq i }\left|   \hat{\rho}_Z 
\left[C^{}_{Lu}\right]_{\tau e u_i u_j}    \right|^2 \right\}  \nn \\
\hat\sigma^{\bar u_i}_{RR} &=& F_Z    \left\{  \left|  \left[ c_{e \varphi}\right]_{\tau e} z_{u_L}  + \hat{\rho}_Z \left[C_{Q e}\right]_{\tau e u_i u_i}  \right|^2  + \sum_{j \neq i} \left|   \hat{\rho}_Z 
 \left[C_{Q e}\right]_{\tau e u_i u_j}   \right|^2 \right\}, 
 \end{eqnarray}
where $\hat{\rho}_Z =  (m_Z^2 + Q^2)/m_Z^2$,  $u_i = \{u,\, c\}$,  and $\left[C_{Q e}\right]_{u_j u_i}$ includes factors of the CKM matrix as in Eq. \eqref{CQe}.

For $d$ type quarks, the partonic cross sections are 
\begin{eqnarray}
\hat\sigma^{d_i}_{LL} &=& F_Z   \left\{  \left|\left[c^{(1)}_{L\varphi}  +  c^{(3)}_{L\varphi}\right]_{\tau e} z_{d_L} +  \hat{\rho}_Z 
\left[C^{}_{LQ,\, D}\right]_{\tau e d_i d_i}   \right|^2  +  \sum_{j \neq i }\left|   \hat{\rho}_Z 
\left[C^{}_{LQ,\, D}\right]_{\tau e d_j d_i}    \right|^2 \right\}  \nn \\
\hat\sigma^{d_i}_{RR} &=& F_Z   \left\{  \left|  \left[c_{e\varphi}\right]_{\tau e} z_{d_R} +  \hat{\rho}_Z \left[C_{e d} \right]_{\tau e d_i d_i} \right|^2  + \sum_{j \neq i} \left|   \hat{\rho}_Z 
 \left[C_{e d} \right]_{\tau e d_j d_i}   \right|^2 \right\} \nn \\
\hat\sigma^{d_i}_{LR} &=& F_Z  (1 -y)^2  \left\{  \left| \left[ c^{(1)}_{L\varphi}  +  c^{(3)}_{L\varphi}\right]_{\tau e} z_{d_R} + \hat{\rho}_Z \left[C_{Ld}\right]_{\tau e d_i d_i}   \right|^2  +  \sum_{j \neq i }\left|   \hat{\rho}_Z
\left[C^{}_{Ld}\right]_{\tau e d_j d_i}    \right|^2 \right\}  \nn \\
\hat\sigma^{d_i}_{RL} &=& F_Z  (1-y)^2  \left\{  \left|  \left[ c_{e \varphi}\right]_{\tau e} z_{d_L}  + \hat{\rho}_Z \left[C_{Q e}\right]_{\tau e d_i d_i}  \right|^2  + \sum_{j \neq i} \left|  \hat{\rho}_Z 
 \left[C_{Q e}\right]_{\tau e d_j d_i}   \right|^2 \right\},
\end{eqnarray} 
while for $d$ antiquarks  they are:
\begin{eqnarray}
\hat\sigma^{\bar d_i}_{LR} &=& F_Z (1-y)^2   \left\{  \left|\left[c^{(1)}_{L\varphi}  +  c^{(3)}_{L\varphi}\right]_{\tau e} z_{d_L} +  \hat{\rho}_Z 
\left[C^{}_{LQ,\, D}\right]_{\tau e d_i d_i}   \right|^2  +  \sum_{j \neq i }\left|   \hat{\rho}_Z 
\left[C^{}_{LQ,\, D}\right]_{\tau e d_i d_j}    \right|^2 \right\}  \nn \\
\hat\sigma^{\bar d_i}_{RL} &=& F_Z (1-y)^2  \left\{  \left|  \left[c_{e\varphi}\right]_{\tau e} z_{d_R} +  \hat{\rho}_Z \left[C_{e d} \right]_{\tau e d_i d_i} \right|^2  
+ \sum_{j \neq i} \left|   \hat{\rho}_Z 
 \left[C_{e d} \right]_{\tau e d_i d_j}   \right|^2 \right\} \nn \\
\hat\sigma^{\bar d_i}_{LL} &=& F_Z    \left\{  \left| \left[ c^{(1)}_{L\varphi}  +  c^{(3)}_{L\varphi}\right]_{\tau e} z_{d_R} + \hat{\rho}_Z \left[C_{Ld}\right]_{\tau e d_i d_i}   \right|^2  
+  \sum_{j \neq i }\left|  \hat{\rho}_Z
\left[C^{}_{Ld}\right]_{\tau e d_i d_j}    \right|^2 \right\}  \nn \\
\hat\sigma^{\bar d_i}_{RR} &=& F_Z    \left\{  \left|  \left[ c_{e \varphi}\right]_{\tau e} z_{d_L}  + \hat{\rho}_Z \left[C_{Q e}\right]_{\tau e d_i d_i}  \right|^2  
+ \sum_{j \neq i} \left|   \hat{\rho}_Z 
 \left[C_{Q e}\right]_{\tau e d_i d_j}   \right|^2 \right\}, 
 \end{eqnarray}
where $d_i = \{d,\, s, \, b\}$.

\paragraph{Dipole operators}

For dipole operators, the $u$-type quark contributions were given in \eq{udipole}.
The $\bar{u}$ quark contribution  
is obtained by the replacement,
\beq
\hsigma_{RR}^{\bar u}\leftrightarrow\hsigma_{RL}^u,
\eeq
while the down-type contribution 
is obtained by \eq{dipoleud}.
The $e \tau$ component, meanwhile, where the electron is left-handed, is given by:
\begin{eqnarray}
\hat \sigma_{LL}^{u}&=&F_{\rm dip} (1-y) \left| \left[\Gamma_\gamma^e\right]_{e \tau} Q_u  + \frac{z_{u_L}}{c_w^2 s_w^2} \frac{Q^2}{(Q^2 + m_Z^2)} \left[ \Gamma_Z^e \right]_{e \tau}  \right|^2, \nn \\
\hat \sigma_{LR}^{u}&=& F_{\rm dip} (1-y) \left| \left[\Gamma_\gamma^e\right]_{e \tau} Q_u  + \frac{z_{u_R}}{c_w^2 s_w^2} \frac{Q^2}{(Q^2 + m_Z^2)} \left[ \Gamma_Z^e \right]_{e \tau}  \right|^2, \nn\\
\hat \sigma_{RR}^{u}&=& \hat \sigma_{RL}^{u}= 0. 
\end{eqnarray}
The antiquark and $d$ type components are obtained as before.

\paragraph{Higgs, scalar and tensor four-fermion operators}

The $u$-type quark partonic cross sections induced by the  $\tau e$ component of Yukawa, scalar and tensor operators were given in \eq{HiggsST}, while the $\bar u$-type antiquark contributions are given by:
\begin{eqnarray}
\hsigma_{LL}^{\bar u}&=& \hsigma_{LR}^{\bar u} =0, \nn \\
\hsigma_{RL}^{\bar u_i}&=& F_S y^2 \Bigg\{ \left|\left[C^{(1)}_{LeQu}\right]_{\tau e u_i u_i} - 4 \left( 1- \frac{2}{y}  \right) \left[C^{(3)}_{LeQu}\right]_{\tau e  u_i u_i} 
+  \frac{Y_{u_i}}{2} \left[  Y_e^\prime\right]_{\tau e} \frac{v^2}{m_H^2 + Q^2} \right|^2 \nn \\
& &+  \sum_{j \neq i}  \left|\left[C^{(1)}_{LeQu}\right]_{\tau e u_i u_j} + 4 \left( 1- \frac{2}{y}  \right) \left[C^{(3)}_{LeQu}\right]_{\tau e  u_i u_j} \right|^2 \Bigg\}, \nn \\
\hsigma_{RR}^{\bar u_i}&=& F_S y^2\left| \frac{Y_{u_i}}{2} \left[  Y_e^\prime\right]_{\tau e} \frac{v^2}{m_H^2 + Q^2} \right|^2.
\end{eqnarray}
For down-type quarks, there is no tensor operator and the incoming $d$ is left-handed 
\begin{eqnarray}
\hsigma_{LL}^{d}&=& \hsigma_{LR}^{d} = \hsigma_{LL}^{\bar d}= \hsigma_{LR}^{\bar d} =0 , \nn \\
\hsigma_{RL}^{d_i}&=& F_S y^2 \Bigg\{ \left|\left[C^{}_{LedQ}\right]_{\tau e d_i d_i} -  \frac{Y_{d_i}}{2} \left[  Y_e^\prime\right]_{\tau e} \frac{v^2}{m_H^2 + Q^2} \right|^2 +  \sum_{j \neq i}  \left|\left[C^{}_{LedQ}\right]_{\tau e d_j d_i}  \right|^2 \Bigg\} \nn \\
\hsigma_{RR}^{d_i}&=& F_S y^2\left| \frac{Y_{d_i}}{2} \left[  Y_e^\prime\right]_{\tau e} \frac{v^2}{m_H^2 + Q^2} \right|^2, \nn \\
, \nn \\
\hsigma_{RR}^{\bar d_i}&=& F_S y^2 \Bigg\{ \left|\left[C^{}_{LedQ}\right]_{\tau e d_i d_i} -  \frac{Y_{d_i}}{2} \left[  Y_e^\prime\right]_{\tau e} \frac{v^2}{m_H^2 + Q^2} \right|^2 +  \sum_{j \neq i}  \left|\left[C^{}_{LedQ}\right]_{\tau e d_i d_j}  \right|^2 \Bigg\} \nn \\
\hsigma_{RL}^{\bar d_i}&=& F_S y^2\left| \frac{Y_{d_i}}{2} \left[  Y_e^\prime\right]_{\tau e} \frac{v^2}{m_H^2 + Q^2} \right|^2.
\end{eqnarray}
For $e\tau$ operators, the results are the same, but the electron is left-handed.

\subsection{The squared amplitude of $gg\to e^\pm\tau^\mp$ at the LHC}\label{GGatLHC}

The top component of scalar operator $\left[C_{LeQu}^{(1)}\right]_{e\tau tt}$ contributes to the $p p \rightarrow e \tau$ cross section at one loop,
via the partonic process $gg\to e^\pm\tau^\mp$.
Since in the analysis of Ref. \cite{Aaboud:2018jff}  $m_{e\tau}$ ranges from about $m_Z$ to  $m_{e\tau}\gg 2m_t$, it is here necessary to use the full one-loop results rather than the heavy top quark mass expansion in Eqs.~\ref{matching_GG} and \ref{matching_GGtilde}.  
With a slight abuse of notation, we denote the squared amplitude of $gg\to e^\pm\tau^\mp$,  averaged over gluon polarizations and colors, by
\beq
\overline{|{\cal M}|^2}=\frac{\alpha_s^2s^3}{64\pi^2v^6}\left(\left|C_{GG}\right|_{\tau e}^2+\left|C_{GG}\right|_{e\tau}^2+\left|C_{G\widetilde G}\right|_{\tau e}^2+\left|C_{G\widetilde G}\right|_{e\tau}^2\right),
\eeq
where $s=m_{e\tau}^2$ and the functions $[C_{GG}]_{\tau e/e\tau}$ and $[C_{G\widetilde G}]_{\tau e/e\tau}$ are defined, for $s > 4 m_t^2$, as
\begin{align}
[C_{GG}]_{\tau e/e\tau}  &= -\frac{m_t v}{2s^2} \left[(4m_t^2-s)\ln^2\frac{\sqrt{s(s-4m_t^2)}+2m_t^2-s}{2m_t^2}+4s\right] \left[ C^{(1)}_{L e Q u} \right]_{\tau ett/e\tau tt}, \\
[C_{G\widetilde G}]_{\tau e/e\tau} &= \frac{im_tv}{2s}\ln^2\frac{\sqrt{s(s-4m_t^2)}+2m_t^2-s}{2m_t^2} \left[ C^{(1)}_{L e Q u} \right]_{\tau ett/e\tau tt}.
\end{align}
The dependence of $C_{GG}$ and $C_{G\widetilde G}$ on $s/(2 m_t)^2$ is the same as for gluon fusion into a scalar or pseudoscalar Higgs, see for example Ref.  
\cite{Spira:1997dg}.

\section{Conversion to a non-chiral basis of low-energy operators}\label{app:conversion}

Here we make contact with the basis used in Ref.~\cite{Celis:2014asa}, 
which employs non-chiral quark billinears with good parity quantum number, more convenient for the 
analysis of hadronic $\tau$ decays.
For dipole operators one has
\begin{eqnarray}
C_{\rm DR}  = \frac{\Lambda^2 }{2 v m_\tau} e \left[\Gamma^{e}\right]^*_{\tau e}, \qquad 
C_{\rm DL}  = \frac{\Lambda^2 }{2 v m_\tau} e \left[\Gamma^{e}\right]_{e \tau}.
\end{eqnarray}
The vector/axial couplings to the $u$ quark are given by
\begin{eqnarray}
C^{u}_{\rm VL} &=& \frac{\Lambda^2}{v^2} \left( C^{eu}_{\rm VLR} + C^{eu}_{\rm VLL} \right), \qquad 
C^{u}_{\rm AL} = \frac{\Lambda^2}{v^2} \left( C^{eu}_{\rm VLR} - C^{eu}_{\rm VLL} \right), \\
C^{u}_{\rm VR} &=& \frac{\Lambda^2}{v^2} \left( C^{eu}_{\rm VRR} + C^{ue}_{\rm VLR} \right), \qquad
C^{u}_{\rm AR} = \frac{\Lambda^2}{v^2} \left( C^{eu}_{\rm VRR} - C^{ue}_{\rm VLR} \right). 
\end{eqnarray}
The matching for down-type operators is simply obtained by replacing $u \rightarrow d$.

For scalar and tensor operators, the conversion is
\begin{eqnarray}
C^{u}_{\rm SR} &=&  \frac{\sqrt{2} \Lambda^2}{m_\tau m_u} \left[ C^{eu}_{\rm SRL} + C^{eu}_{\rm SRR} \right]_{\tau e}^*, \qquad
C^{u}_{\rm PR} =  \frac{\sqrt{2} \Lambda^2}{m_\tau m_u} \left[ C^{eu}_{\rm SRL} - C^{eu}_{\rm SRR} \right]_{\tau e}^*, \\
C^{u}_{\rm SL} &=&  \frac{\sqrt{2} \Lambda^2}{m_\tau m_u} \left[ C^{eu}_{\rm SRR} + C^{eu}_{\rm SRL} \right]_{e\tau}, \qquad
C^{u}_{\rm PL} =  \frac{\sqrt{2} \Lambda^2}{m_\tau m_u} \left[ C^{eu}_{\rm SRR} - C^{eu}_{\rm SRL} \right]_{e\tau},
\end{eqnarray}
and $u \rightarrow d$ yields the results for the $d$ quark.
At tree-level, the tensor operator is 
\begin{eqnarray}
C^{u}_{\rm TL} =  \frac{2 \sqrt{2} \Lambda^2}{m_u m_\tau} \left( C^{eu}_{\rm TRR}  \right)_{e\tau},\qquad 
C^{u}_{\rm TR} =  \frac{2 \sqrt{2} \Lambda^2}{m_u m_\tau} \left( C^{eu}_{\rm TRR} \right)^*_{\tau e}.
\end{eqnarray}

\section{Compendium of Decay rates}
\label{decay_rate}

\subsection{$\tau$ decay rates}
Below we report the expressions for LFV  $\tau$-decay rates. 
Most of these results are taken from the existing literature. 
We devote separate subsections to original results on   $\tau \to e K \pi$, the tensor operator contribution to $\tau \to e \gamma$, 
and $\tau \rightarrow e K^+ K^-$. 
\begin{itemize}
\item {$\tau\to e\gamma$  \cite{Celis:2014asa}}
\begin{align}
\Gamma\left(\tau\to e\gamma \right)&=
\frac{m^3_{\tau}\alpha_{\rm em}}{4v^2}\bigg[   \left|   \left(  \Gamma^e_{\gamma}\right)_{e\tau} \right|^2 +\left|\left( \Gamma^e_{\gamma}\right)_{\tau e} \right|^2  \bigg].
\end{align}
Besides, contributions from nonzero tensor semileptonic operators are given by
\bea
( \Gamma^e_\gamma)_{e \tau}  &\to  &(\Gamma^e_\gamma)_{e \tau}   - 4   \left( c_3 + \frac{c_8}{\sqrt{3}} \right)   \frac{i \Pi_{VT} (0)}{v} 
\\
( \Gamma^e_\gamma)_{ \tau e} ^* &\to  &(\Gamma^e_\gamma)_{ \tau e}^*   - 4   \left( \tilde c_3 + \frac{\tilde c_8}{\sqrt{3}} \right)   \frac{i \Pi_{VT} (0)}{v} ~.
\eea
The expressions of $c_{3,8}$ and $\tilde c_{3,8}$ in terms of the tensor semileptonic couplings and 
the non-perturbative parameter $\Pi_{VT} (0)$ are given in Section~\ref{sect:tensor2}. 

\begin{table}[t]
\centering
\begin{tabular}{||c| c || c|  c ||}
\hline
$\tau_{\tau}$ & $290.3\times 10^{-15}~$s \cite{Zyla:2020zbs}& $\alpha_{\rm em}$ & $1/137.036$ \cite{Zyla:2020zbs}\\
$G_F$ & $1.166\times 10^{-5}~$GeV$^{-2}$ \cite{Zyla:2020zbs}& $m_{\tau}$ & $1.78~$GeV \cite{Zyla:2020zbs} \\
$m_{\pi^0}$ & $134.98~$MeV \cite{Zyla:2020zbs}& $m_{\pi^{\pm}}$ & $139.57~$MeV \cite{Zyla:2020zbs}\\
$m_{\eta}$ & $547.862~$MeV \cite{Zyla:2020zbs}& $m_{\eta^{\prime}}$ & $957.78~$MeV \cite{Zyla:2020zbs}\\
$f_{\pi}$ & $130.2~$MeV \cite{Aoki:2019cca} & $B^{\pi,u}_T(0)$  & $0.195$ \cite{Hoferichter:2018zwu} \\
$f_{\eta}^q$ & $0.11~$GeV  \cite{Celis:2014asa} & $f^s_{\eta}$ & $-0.11~$GeV  \cite{Celis:2014asa} \\
$h_{\eta}^q$ & $0.001~$GeV$^3$  \cite{Celis:2014asa}& $h^s_{\eta}$ & $-0.055~$GeV$^3$  \cite{Celis:2014asa}\\
$f_{\eta^{\prime}}^q$ & $0.087~$GeV  \cite{Celis:2014asa}& $f^s_{\eta^{\prime}}$ & $0.135~$GeV  \cite{Celis:2014asa} \\
$h_{\eta^{\prime}}^q$ & $0.001~$GeV$^3$  \cite{Celis:2014asa} & $h^s_{\eta^{\prime}}$ & $0.068~$GeV$^3$  \cite{Celis:2014asa}\\
 $a_{\eta}$ & $0.022~$GeV$^3$  \cite{Celis:2014asa} &  $a_{\eta^{\prime}}$ & $0.056~$GeV$^3$  \cite{Celis:2014asa}\\
 $f_K$ & $155.7~$MeV \cite{Aoki:2019cca}  & $m_{K^0}$ & $497.611~$MeV \cite{Zyla:2020zbs} \\
\hline
\end{tabular}
\caption{Input parameters for $\tau$ decays.}
\label{input_tau_decay}
\end{table}

\item {$\tau\to 3e$  \cite{Crivellin:2017rmk}}
\begin{align}
\Gamma\left(\tau \to 3e\right)=&\frac{\alpha^2_{\rm em}m^3_{\tau}}{48\pi v^2}
\bigg[\left|\left(\Gamma^2_{\gamma} \right)_{e\tau} \right|^2+\left|\left(\Gamma^2_{\gamma} \right)_{\tau e} \right|^2 \bigg]\left\{8\log\left(\frac{m_{\tau}}{m_e} \right)-11 \right\}+X_{\gamma}\nonumber\\
&+\frac{m^5_{\tau}G_F^2}{1536\pi^3}\bigg[\left|\left(C^{ee}_{\rm SRR} \right)_{e\tau ee} \right|^2+16\left|\left(C^{ee}_{\rm VLL} \right)_{\tau e ee} \right|^2+8\left|\left(C^{ee}_{\rm VLR} \right)_{\tau e ee} \right|^2\nonumber\\
&\hspace{2.5cm}
+\left|\left(C^{ee}_{\rm SRR} \right)_{\tau e ee} \right|^2+16\left|\left(C^{ee}_{\rm VRR} \right)_{\tau e ee} \right|^2+8\left|\left(C^{ee}_{\rm VLR} \right)_{ee\tau e }\right|^2 \bigg]\label{taueee}
\end{align}
where $X_{\gamma}$ is the interference term with the dipole operator 
\begin{align}
X_{\gamma}&=-\frac{\sqrt{2}\alpha_{\rm em}}{3(4\pi)^2}\frac{m^5_{\tau}G_F}{vm_{\tau}}{\rm Re}\bigg[\left(\Gamma^e_{\gamma} \right)_{e\tau}^*\left\{\left(C^{ee}_{\rm VLR} \right)^*_{ee\tau e}+2\left(C_{\rm VRR}^{ee} \right)^*_{\tau eee} \right\}\nonumber\\
&\hspace{5cm}
+\left(\Gamma^e_{\gamma} \right)_{\tau e}\left\{\left(C^{ee}_{\rm VLR} \right)^*_{\tau eee}+2\left(C_{\rm VLL}^{ee} \right)^*_{\tau eee} \right\}.
\bigg].
\end{align}
Notice that in Eq. \eqref{taueee} we use a single symbol to denote the contributions of both the $\tau e e e$ and $e e \tau e$ components of LEFT operators, for example
\begin{equation*}
\left( C_{\rm VLL}^{ee} \right)_{\tau eee} \equiv \left( C_{\rm VLL}^{ee} \right)_{\tau eee} + \left( C_{\rm VLL}^{ee} \right)_{ee \tau e}.
\end{equation*}

\item {$\tau\to e\mu^+\mu^-$  \cite{Dassinger:2007ru}}
\begin{align}
&\Gamma\left(\tau\to e\mu^+\mu^- \right)=\int^{X_{\rm max}}_{X_{\rm min}}dX\int^{Y_{\rm max}}_{Y_{\rm min}}dY\nonumber\\
&\times\bigg[\frac{G_F^2}{64\pi^3m^3_{\tau}}\left\{\left| \left(C^{ee}_{\rm VRR}\right)_{\tau e\mu\mu} \right|^2+\left| \left(C^{ee}_{\rm VLL}\right)_{\tau e\mu\mu} \right|^2 \right\}\left\{m^4_{\tau}-(2X-m^2_{\tau}-2m^2_{\mu})^2 \right\}\nonumber\\
&\hspace{0.5cm}
+\frac{G_F^2}{64\pi^3m^3_{\tau}}\left\{\left| \left(C^{ee}_{\rm VLR}\right)_{\tau e\mu\mu} \right|^2+\left| \left(C^{ee}_{\rm VLR}\right)_{\mu\mu\tau e} \right|^2 \right\}\left\{m^4_{\tau}-(2Z-m^2_{\tau}-2m^2_{\mu})^2 \right\}\nonumber\\
&\hspace{0.5cm}
+\frac{\alpha^2_{\rm em}}{16\pi m_{\tau}^3v^2}\left\{\left|\left(\Gamma^e_{\gamma}\right)_{\tau e} \right|^2+\left|\left(\Gamma^e_{\gamma}\right)_{e\tau} \right|^2\right\}\nonumber\\
&\hspace{1.5cm}\times
\bigg\{\frac{m^2_{\mu}}{Y^2}\left(Y-m^2_{\tau} \right)^2+\frac{1}{2Y}\left(X^2+m^4_{13}-2m^4_{\mu} \right)+\frac{1}{2}\left(m^2_{\tau}-Y \right)\bigg\}+X_{\gamma}^{\mu\mu} \bigg],\label{tauemumu}
\end{align}
where the interference term is expressed by
\begin{align}
X^{\mu\mu}_{\gamma}=&-\frac{\alpha_{\rm em}}{16\pi^2v^3}\bigg[{\rm Re}\left\{\left(C^{ee}_{\rm VLL}\right)_{\tau e\mu\mu}\left(\Gamma^e_{\gamma} \right)^*_{\tau e} +\left(C^{ee}_{\rm VRR}\right)_{\tau e\mu\mu}\left(\Gamma^e_{\gamma} \right)_{e\tau} \right\}\left(\frac{1}{m^2_{\tau}}\left(X-2m^2_{\mu} \right)+\frac{m^2_{\mu}}{Y} \right) \nonumber\\
&\hspace{0.5cm}
+{\rm Re}\left\{\left(C^{ee}_{\rm VLR}\right)_{\tau e\mu\mu}\left(\Gamma^e_{\gamma} \right)_{\tau e}^* +\left(C^{ee}_{\rm VLR}\right)_{\mu\mu\tau e}\left(\Gamma^e_{\gamma} \right)_{e\tau} \right\}\left(\frac{1}{m^2_{\tau}}\left(Z-2m^2_{\mu} \right)+\frac{m^2_{\mu}}{Y} \right)\bigg].
\end{align}
As for $\tau \rightarrow e e e$, in Eq. \eqref{tauemumu} we use a single symbol to denote the sum of equivalent contributions, for example
\begin{equation*}
\left( C_{\rm VLL}^{ee} \right)_{\tau e\mu\mu} \equiv \left( C_{\rm VLL}^{ee} \right)_{\tau e\mu\mu} + \left( C_{\rm VLL}^{ee} \right)_{\mu\mu \tau e}
+\left( C_{\rm VLL}^{ee} \right)_{\tau \mu\mu e} + \left( C_{\rm VLL}^{ee} \right)_{\mu e\tau \mu},
\end{equation*}
with the coefficients on the right hand side given in Table \ref{RGE_low_lepto}.
The parameters, $X,Y$ and $Z$, denote invariant masses $m_{ij}^2$ as
\begin{align}
X=m^2_{12}&=\left(p_{e}+p_{\mu^-} \right)^2,\\
Y=m^2_{23}&=\left(p_{\mu^-}+p_{\mu^+} \right)^2,\\
Z=m^2_{13}&=m_{\tau}^2+2m_{\mu}^2-X-Y,
\end{align}
which are kinematically limited by
\begin{align}
\left(m_{e}+m_{\mu} \right)^2\leq X\leq \left(m_{\tau}-m_{\mu} \right)^2,
\end{align}
\begin{align}
Y_{\rm min,max}&=\left(E_{\mu^-}+E_{\mu^+} \right)^2-\bigg[\left(E^2_{\mu^-}-m^2_{\mu}\right)^{\frac{1}{2}}\pm \left(E^2_{\mu^+}-m^2_{\mu} \right)^{\frac{1}{2}} \bigg]^2,
\end{align}
with
\begin{align}
E_{\mu^-}&=\frac{X-m^2_e+m^2_{\mu}}{2m_{12}},\hspace{1cm}
E_{\mu^+}=\frac{m^2_{\tau}-m^2_{\mu}-X}{2m_{12}}.
\end{align}
\end{itemize}

\begin{itemize}
\item{$\tau^+ \to e^+ M,~(M=\pi^0,K_S^0)$}
\begin{align}
\Gamma \left(\tau^+\to e^+ M \right)&=\frac{m_{\tau}^3}{32\pi}\left(1-\frac{m^2_M}{m^2_{\tau}} \right)^2G_F^2f^2_{M}
\bigg[\left|A_L^M \right|^2 + \left|A_R^M \right|^2  \bigg],
\end{align}
where $f_{M}$ is the decay constant. $A_{L,R}^M$ is expressed by
\begin{align}
A_L^{\pi}=&\Big(C^{eu}_{\rm VLR}-C^{eu}_{\rm VLL}\Big)_{\tau e uu} -\Big(C^{ed}_{\rm VLR}- C^{ed}_{\rm VLL} \Big)_{\tau edd}\nonumber\\
&+\frac{m^2_{\pi}}{m_{\tau}\left(m_u+m_d \right)}\left[\Big(C^{eu*}_{\rm SRR}-C^{eu*}_{\rm SRL} \Big)_{e\tau uu} - \Big(C^{ed*}_{\rm SRR}-C^{ed*}_{\rm SRL} \Big)_{e\tau dd}   \right],\\
A_R^{\pi}=&\Big(C^{ue}_{\rm VLR}\Big)_{uu\tau e}-\Big(C^{eu}_{\rm VRR}\Big)_{\tau e uu} -\left[\Big(C^{de}_{\rm VLR}\Big)_{dd\tau e}-\Big(C^{ed}_{\rm VRR} \Big)_{\tau edd}\right]\nonumber\\
&+\frac{m^2_{\pi}}{m_{\tau}\left(m_u+m_d \right)}\left[\Big(C^{eu}_{\rm SRR}-C^{eu}_{\rm SRL} \Big)_{\tau e uu} - \Big(C^{ed}_{\rm SRR}-C^{ed}_{\rm SRL} \Big)_{\tau edd}   \right],
\end{align}
for $\tau^+ \to e^+ \pi^0$,
\begin{align}
A_L^{K}=&\Big(C^{ed}_{\rm VLR}-C^{ed}_{\rm VLL}\Big)_{\tau e ds} +\frac{m^2_{K}}{m_{\tau}\left(m_d+m_s \right)}\Big(C^{ed*}_{\rm SRR}-C^{ed*}_{\rm SRL} \Big)_{e\tau sd}-\left(d \leftrightarrow s\right),\\
A_R^{L}=&\Big(C^{de}_{\rm VLR}\Big)_{ds\tau e}-\Big(C^{ed}_{\rm VRR}\Big)_{\tau e ds}
+\frac{m^2_{K}}{m_{\tau}\left(m_d+m_s \right)}\Big(C^{ed}_{\rm SRR}-C^{ed}_{\rm SRL} \Big)_{\tau  eds} -\left(d \leftrightarrow s\right),
\end{align}
for $\tau^+ \to e^+ K^0_S$.

\item{$\tau \to e \eta^{(\prime)}$ \cite{Celis:2014asa}}
\begin{align}
&\Gamma\left(\tau \to e\eta \right)\nonumber\\
=&\frac{m_{\tau}^3}{32\pi }\left(1-\frac{m^2_{\eta}}{m^2_{\tau}} \right)^2G_F^2\bigg[\left(\frac{\sqrt{2}a_{\eta}}{m_{\tau}v} \right)^2\Big(\left|C_{G\tilde G}\right|_{\tau e}^2 +\left|C_{G\tilde G}\right|_{e\tau }^2  \Big)+\left|A^{\eta}_L\right|^2+\left|A^{\eta}_R\right|^2  \bigg],
\end{align}
with
\begin{align}
A_L^{\eta}=&\sum_{q=u,d}\Big[ f_{\eta}^q\Big(C^{eq}_{\rm VLR}-C^{eq}_{\rm VLL}\Big)_{\tau e qq} +\frac{h_{\eta}^q}{m_{\tau}\left(m_u+m_d \right)}\Big(C^{eq*}_{\rm SRR}-C^{eq*}_{\rm SRL}\Big)_{e\tau qq}\Big]\nonumber\\
&+\sqrt{2}\Big[f_{\eta}^s\Big(C^{ed}_{\rm VLR}-C^{ed}_{\rm VLL}\Big)_{\tau e ss}+\frac{h_{\eta}^s}{2m_{\tau}m_s}\Big(C^{ed*}_{\rm SRR}-C^{ed*}_{\rm SRL}\Big)_{e\tau ss} \Big],\\
A_R^{\eta}=&\sum_{q=u,d}\Big[ f_{\eta}^q\Big\{\Big(C^{qe}_{\rm VLR}\Big)_{qq\tau e}-\Big(C^{eq}_{\rm VRR}\Big)_{\tau e qq}\Big\} +\frac{h_{\eta}^q}{m_{\tau}\left(m_u+m_d \right)}\Big(C^{eq}_{\rm SRR}-C^{eq}_{\rm SRL}\Big)_{\tau e qq}\Big]\nonumber\\
&+\sqrt{2}\Big[f_{\eta}^s\Big\{\Big(C^{de}_{\rm VLR}\Big)_{ss\tau e}-\Big(C^{ed}_{\rm VRR}\Big)_{\tau e ss}\Big\}+\frac{h_{\eta}^s}{2m_{\tau}m_s}\Big(C^{ed}_{\rm SRR}-C^{ed}_{\rm SRL}\Big)_{\tau e ss} \Big].
\end{align}
Here, $f_{\eta}^{q,s},~h^{q,s}_{\eta}$ and $a_{\eta}$ denote decay constants. The BR for $\tau\to e\eta^{\prime}$ can be expressed by the replacement of $\eta\to \eta^{\prime}$.

\item{$\tau \to e \pi^+\pi^-$ \cite{Celis:2014asa}}
\begin{align}
\frac{d\Gamma}{ds}&=\frac{\left(s-4m^2_{\pi} \right)^{1/2} \left(m^2_{\tau}-s \right)^2 }{1536\pi^3 m_{\tau}s^{5/2}}\nonumber\\
&\hspace{0.5cm}\times
\bigg[\frac{6}{m^2_{\tau}}s^2G_F^2\left(\left|Q_L^{\prime} \right|^2+\left|Q_R^{\prime} \right|^2 \right)-4\left(4m^2_{\pi}-s \right)\left|F_V(s) \right|^2 \nonumber\\
&\hspace{1cm}\times
\bigg\{\frac{2m^2_{\tau}+s}{2m^2_{\tau}}\left(\left|A_{L} \right|^2+\left|A_{R} \right|^2 \right) +\sqrt{2}sG_F\left(B_L+B_R\right)  \bigg\}\bigg].
\end{align}
$Q^{\prime}_{L,R}$, $A_{L,R}$ and $B_{L,R}$ are combinations of Wilson coefficients and form factors, and are given in Eqs. \eqref{QL1}--\eqref{BL1}.
\end{itemize}
All the related input parameters are listed in Table \ref{input_tau_decay}.

\tocless\subsubsection{$\tau \to e  \pi K$ modes}
We provide below a detailed expression for the decay rate in  the channel  
$\tau^- \to e^-  \pi^+ K^-$, mediated by operators with structure 
$\bar e \Gamma \tau \ \bar{s} \Gamma d$.  
The decay $\tau^- \to e^-  \pi^- K^+$ has a completely analogous expression, 
in terms of the Wilson Coefficients of the operators  $\bar e \Gamma \tau \ \bar{d} \Gamma s$. 
Similar considerations apply to the decay of $\tau^+$. 
Finally, note that  while  the PDG does not provide  a bound on the mode $\tau^- \to e^- \pi^0 \bar K^0$, 
its theoretical prediction  is   related  $\tau \to e \pi^+ K^-$ by  isospin symmetry, 
\begin{equation}
\Gamma (\tau \to e \pi^0 \bar{K}^0) = \frac{1}{2} \Gamma (\tau \to e \pi^+ K^-)~. 
\end{equation}

To obtain an expression for $\Gamma (\tau \to e \pi^+ K^-)$ 
we note that  isospin symmetry gives  
\begin{equation}
\langle \pi^- \bar {K}^0 | \bar s \Gamma u | 0 \rangle = 
\langle \pi^+  {K}^- | \bar s \Gamma d | 0 \rangle ~,
\end{equation} 
which in turn implies that for our LFV decay we can use the form factors 
$f_{+,0}^{K \pi} (s)$ and $B_T^{K\pi} (s) $  ($s = (p_K + p_\pi)^2$)
appearing in the decay 
$\tau^- \to \nu_\tau \pi^- \bar{K}^0$: 
\begin{align}
 \langle\bar K^0(p_K)\pi^-(p_\pi)|\bar s\gamma^\mu u|0\rangle&=(p_K- p_\pi)^\mu f^{K\pi}_+(s) +(p_K+ p_\pi)^\mu f^{K\pi}_-(s),\notag\\
 \langle\bar K^0(p_K)\pi^-(p_\pi)|\bar s u|0\rangle&=\frac{M_K^2-M_\pi^2}{m_s-m_u}f^{K\pi}_0(s),\notag\\
 \langle\bar K^0(p_K)\pi^-(p_\pi)|\bar s\sigma^{\mu\nu} u|0\rangle&=i\frac{p_K^\mu p_\pi^\nu-p_K^\nu p_\pi^\mu}{M_K}B^{K\pi}_T(s),
\end{align}
where
\beq
f^{K\pi}_-(s)=\frac{M_K^2-M_\pi^2}{s}\big(f^{K\pi}_0(s)-f^{K\pi}_+(s)\big).
\eeq
Moreover, in the limit $m_e \to 0$ the LFV decay rate  $\Gamma (\tau \to e \pi^+ K^-)$  
can be read off the expressions for  $\Gamma (\tau \to \nu_\tau  \pi^-  \bar{K}^0)$  given  in Ref.~\cite{Cirigliano:2017tqn}.
In terms of the effective couplings 
\begin{eqnarray}
c_V &=&  \Big[ C_{\rm VLL}^{ed}\Big]_{e \tau s d} + \Big[C_{\rm VLR}^{ed} \Big]_{e \tau s d} + \Big[C_{\rm VLR}^{de}\Big]_{sd\tau e} + \Big[C_{\rm VRR}^{ed} \Big]_{e \tau s d} ,
\\
{c_A}&=&-\Big[C^{ed}_{\rm VLL} \Big]_{e\tau sd}+\Big[C^{de}_{\rm VLR} \Big]_{sd\tau e}+\Big[C^{ed}_{\rm VRR} \Big]_{e\tau sd}-\Big[C^{ed}_{\rm VLR} \Big]_{e\tau sd}, \\ 
c_S &=& \Big[ C_{\rm SRR}^{ed}\Big]_{e\tau sd} + \Big[C_{\rm SRL}^{ed} \Big]_{e \tau s d} +\Big[C^{ed*}_{\rm SRR}\Big]_{\tau e ds}+\Big[C^{ed*}_{\rm SRL}\Big]_{\tau e ds},\\
c_P &=&-i\left\{\Big[C^{ed}_{\rm SRR}\Big]_{e\tau sd}+\Big[C^{ed}_{\rm SRL}\Big]_{e\tau sd} -\Big[C^{ed*}_{\rm SRR} \Big]_{\tau e ds}-\Big[C^{ed*}_{\rm SRL} \Big]_{\tau e ds} \right\},\\
c_{TR} &=& 2\Big[ C_{\rm TRR}^{ed} \Big]_{e \tau s d} ~,\\
c_{TL} &=& 2\Big[ C_{\rm TRR}^{ed*} \Big]_{\tau e ds},
\end{eqnarray}
we find 
\begin{align}
\label{decay_width}
\frac{d \Gamma}{d s} &(\tau \to e \pi^+ K^-)
=G_F^2    \frac{\lambda^{1/2}_{\pi K}(s)(m_\tau^2-s)^2(M_K^2-M_\pi^2)^2}{512\pi^3 m_\tau s^3}\notag\\
&\times\bigg[\xi(s)\bigg\{|V(s)|^2+|A(s)|^2+\frac{2(m_\tau^2-s)^2}{9s m_\tau^2}\Big(|T_+(s)|^2+|T_-(s)|^2\Big)\bigg\}\
+|S(s)|^2+|P(s)|^2\bigg],
\end{align}
where $\lambda_{\pi K}(s)=\lambda(s,M_\pi^2,M_K^2)$, $\lambda(a,b,c)=a^2+b^2+c^2-2(ab+ac+bc)$,
\beq
\xi(s)=\frac{(m_\tau^2+2s)\lambda_{\pi K}(s)}{3 m_\tau^2(M_K^2-M_\pi^2)^2},
\eeq
and
\begin{align}
 V(s)&=f^{K\pi}_+(s) c_V-T_+(s),\qquad A(s)=f^{K\pi}_+(s) c_A+T_-(s),\notag\\
 S(s)&=f^{K\pi}_0(s)\bigg(c_V+\frac{s}{m_\tau (m_s-m_u)}c_S\bigg),\notag\\
 P(s)&=f^{K\pi}_0(s)\bigg(c_A-  \frac{s}{m_\tau(m_s-m_u)}ic_{P}\bigg),\notag\\
 T_{\pm}(s)&=\frac{3s}{m_\tau^2+2s}\frac{m_\tau}{m_K}\left(c_{TR}\pm c_{TL}\right) B^{K\pi}_T(s). \label{piK_formfactor}
\end{align}

Finally, for the vector and scalar form  factors  $f_{+,0}^{K \pi} (s)/f^{K\pi}_{+,0}(0)$ we  use the numerical results from Ref.~\cite{Antonelli:2013usa} 
and for the normalization we use the lattice QCD input 
$f^{K\pi}_+ (0)=f^{K\pi}_0 (0) = 0.970(3)$~\cite{Aoki:2019cca}.
For the tensor form factor $B_T^{K\pi} (s) $ we use the elastic unitarity relation (accurate in the dominant region of phase space) 
\cite{Cirigliano:2017tqn,Hoferichter:2018zwu}
\begin{equation}
B_T^{K\pi} (s)  = \frac{ B_T^{K\pi} (0)}{f_+^{K \pi} (0)}  \times f_+^{K\pi} (s)
\end{equation}
with the lattice QCD input  $B_T^{K\pi} (0)  = 0.686(25)$~\cite{Baum:2011rm}.

\leavevmode

\tocless\subsubsection{Tensor operator contribution to $\tau \to e \gamma$}
\label{sect:tensor2}

In order to derive the tensor operator contribution to $\tau \to e \gamma$, 
we write the relevant part of  the low-scale effective Lagrangian  as follows 
\bea
{\cal L}   & \supset  &  - \frac{4 G_F}{\sqrt{2}} \, \bar{e}_L \sigma^{\mu \nu} \tau_R \ \bar{q} \sigma_{\mu \nu} \left[ c_0 \, I +  c_3 \, T^3 + c_8 \, T^8 \right] q 
 \ + \  \{ L \leftrightarrow R ,  c_{0,3,8} \to \tilde{c}_{0,3,8} \}
\nonumber \\
&+& e A^\mu J_\mu^{EM}  ~.
\label{eq:LTEM}
\eea
Here $q = (u,d,s)^T$, 
the electromagnetic current is given by 
\beq
J_\mu^{EM} = \bar q \ \gamma_\mu \left[\frac{1}{\sqrt{3}} T^8 + T^3 \right] q~, 
\eeq
and   the matrices $T^a$ are SU(3) flavor generators.  
The tensor couplings are given by 
\bea
c_0 &=& \frac{1}{3}  \left(
\Big[ C_{\rm TRR}^{eu} \Big]_{e \tau uu} + \Big[ C_{\rm TRR}^{ed} \Big]_{e \tau dd} +  \Big[ C_{\rm TRR}^{ed} \Big]_{e \tau ss}
\right)
\\
c_3 &=&  \Big[ C_{\rm TRR}^{eu} \Big]_{e \tau uu} -  \Big[ C_{\rm TRR}^{ed} \Big]_{e \tau dd} 
\\
c_8 &=& \frac{1}{\sqrt{3}}
\left(
\Big[ C_{\rm TRR}^{eu} \Big]_{e \tau uu} + \Big[ C_{\rm TRR}^{ed} \Big]_{e \tau dd} -2  \Big[ C_{\rm TRR}^{ed} \Big]_{e \tau ss}
\right)
\eea
and 
\bea
\tilde c_0 &=& \frac{1}{3}  \left(
 \Big[ C_{\rm TRR}^{eu} \Big]_{ \tau e uu }^*   +  \Big[ C_{\rm TRR}^{ed} \Big]_{ \tau e dd }^* + \Big[ C_{\rm TRR}^{ed} \Big]_{ \tau e ss}^ *
\right)
\\
\tilde c_3 &=&   \Big[ C_{\rm TRR}^{eu} \Big]_{ \tau e uu }^*   -   \Big[ C_{\rm TRR}^{ed} \Big]_{ \tau e dd }^* 
\\
\tilde c_8 &=& \frac{1}{\sqrt{3}}
\left(
 \Big[ C_{\rm TRR}^{eu} \Big]_{ \tau e uu }^*   +  \Big[ C_{\rm TRR}^{ed} \Big]_{ \tau e dd }^* -2 \Big[ C_{\rm TRR}^{ed} \Big]_{ \tau e ss}^ *
\right)~.
\eea

The S-matrix element for the process $\tau (p_\tau) \to e (p_e) \gamma (q)$ is  obtained in second-order  perturbation theory,  by   simultaneously 
inserting  the tensor and electromagnetic interaction from \eqref{eq:LTEM}
\bea
S &=& - i^2 \frac{4 e G_F}{ \sqrt{2}}  \ \bar u_L (p_e) \sigma^{\mu \nu} u_R (p_\tau) 
\ \int d^4y  \, e^{i y \cdot (q + p_e - p_\tau)} 
\nonumber \\
&\times & \int d^4x  \, e^{i q \cdot x}  \ \langle T \Big(  J_\sigma^{EM} (x) \   \bar{q} (0)\sigma_{\mu \nu} \left[ c_0 \, I +  c_3 \, T^3 + c_8 \, T^8 \right] q (0)    \Big) \rangle ~.
\eea
The non-perturbative hadronic contribution to the amplitude is contained in the  correlation function of the 
vector and tensor densities
\beq
V^a_\mu (x) = \bar q (x)  \gamma_\mu  T^a q (x)  \qquad \qquad 
T^a_{\mu \nu} (x) =  \bar q (x)  \sigma_{\mu \nu}  T^a q (x)  ~. 
\eeq
Using the decomposition
\beq
\int d^4x \, e^{i q \cdot x} \, \langle T \left(V_\sigma^a (x) \ T_{\mu \nu}^b (0) \right) \rangle = - i \delta^{ab} \ \Pi_{VT} (q^2) \, \left( g_{\nu \sigma} q_\mu 
- g_{\mu \sigma} q _\nu \right), 
\eeq
and the definition of the amplitude $S \equiv i (2\pi)^4 \delta^4 (p_\tau - p_e - q) \,  {\cal A}$,  one arrives at~\footnote{In the case of matching to SMEFT 
only  $\left[C_{\rm TRR}^{eu}\right]_{e \tau uu} \neq 0$  and one has $c_3 + c_8/\sqrt{3}  = (4/3) \left[C_{\rm TRR}^{eu}\right]_{e \tau uu}$.}
\bea
{\cal A} &=&   \Big[  i e  \bar u_L (p_e) \sigma^{\mu \nu} u_R (p_\tau)  \, (q_\mu \epsilon_\nu^*(q) - q_\nu \epsilon_\mu^* (q)) \Big] \times 
2 \sqrt{2} G_F  \left( c_3 + \frac{c_8}{\sqrt{3}} \right)  
\, i \Pi_{VT} (0)~. 
\eea
The term in the square brackets coincides with the matrix element of the dipole operator, namely  
$ \langle  e \gamma | \bar e_L \sigma^{\mu \nu}  \tau_R \, (e F_{\mu \nu} ) | \tau \rangle$. 
Following Refs.~\cite{Knecht:2001xc,Mateu:2007tr,Cata:2008zc}   we estimate the non-perturbative parameter $\Pi_{VT}(0)$ by using a
large-$N_C$ inspired lowest resonance saturation for the $VT$ correlation function, 
\beq
\Pi_{VT} (q^2 ) = i \frac{\langle \bar q q \rangle}{M_V^2 - q^2} = - i \frac{ B_0 \, F_\pi^2}{M_V^2- q^2}~, 
\eeq
which is also consistent with the high-$q^2$ behavior dictated by the OPE. 
In the $\overline{\rm MS}$ scheme at $\mu = 2$~GeV one has $\langle \bar q q \rangle = - (286(23) {\rm MeV})^3$ or equivalently $B_0 \simeq 2.7$~GeV~\footnote{These numbers are from the FLAG 2019 review~\cite{Aoki:2019cca}, using 2+1+1 dynamical quarks.}.  The pion decay constant $F_{\pi}$ is $92.2~$MeV and we use $\rho$ meson mass $M_V=770~$MeV.  

Based on the above results, the formulae for the $\tau \to e \gamma$ decay rate are modified by the substitutions:
\bea
(\Gamma^e_\gamma)_{e \tau}  &\to  &(\Gamma^e_\gamma)_{e \tau}   - 4   \left( c_3 + \frac{c_8}{\sqrt{3}} \right)   \frac{i \Pi_{VT} (0)}{v} 
\\
(\Gamma^e_\gamma)_{ \tau e} ^* &\to  &(\Gamma^e_\gamma)_{ \tau e}^*   - 4   \left( \tilde c_3 + \frac{\tilde c_8}{\sqrt{3}} \right)   \frac{i \Pi_{VT} (0)}{v} ~.
\eea
The interference between dipole and tensor couplings is controlled by the non-perturbative parameter
\beq
\frac{i \Pi_{VT} (0)}{v} = \frac{B_0}{v} \frac{F_\pi^2}{M_V^2} 
\eeq
which takes the numerical value  $\approx 1.6 \times 10^{-4}$  at $\mu = 2$~GeV.
Since the above estimate is based on large-$N_C$ considerations and a truncation of the spectrum to the lowest lying resonance, 
we assign to it a 50\% uncertainty. Lattice QCD calculations of  $\Pi_{VT}(q^2)$ can reduce the uncertainty in the future. 
Finally, we note that our result is consistent with a similar analysis of the tensor operator to $\mu \to e \gamma$~\cite{Dekens:2018pbu}.

\leavevmode

\tocless\subsubsection{$\tau \rightarrow e K^+ K^-$}\label{taueKK}

We discuss here the contribution of vector operators to $\tau \rightarrow e K^+ K^-$.
Since the scalar and gluonic contributions are affected by large theoretical errors, we do not use this process in the  analysis 
of Section \ref{low-energy}. As discussed in Section \ref{global}, $\tau \rightarrow e K^+ K^-$ can play an important role in global analyses, since it receives contributions from isoscalar combinations of vector couplings, which are otherwise unconstrained at low energy.
In the case of $\tau \to e K^+ K^-$,  the differential decay width induced by vector operators is
\begin{eqnarray}
\tau_\tau \frac{d\Gamma}{d \hat{s}}&=& \frac{1}{5 \bar\Gamma_\tau } \left(1-\frac{\rho_K}{ \hat{s}} \right)^{3/2} \left(1- \hat{s} \right)^2 
(2 \hat{s} + 1) \left( \left|B_L\right|^2+\left|B_R\right|^2\right),
\end{eqnarray}
where $s$ is the invariant mass of the charged kaons, 
and  we define the dimensionless quantities  $\hat{s} = s/m_\tau^2$ and $\rho_K =4  m_K^2/m_\tau^2$.
The kinematically allowed region is $\rho_K \leq \hat{s} \leq 1$. 
$B_{L,R}$ are combinations of Wilson coefficients and form factors
\begin{eqnarray}
B_L&=&\left\{\left(C^{eu}_{\rm VLL}+C^{eu}_{\rm VLR} \right)_{\tau e uu}  \left(F_V^{(8)}(s) + F_V^{(3)}(s) + F_V^{(0)}(s)\right) \nonumber \right. \\
& & \left.
          + \left(C^{ed}_{\rm VLL}+C^{ed}_{\rm VLR} \right)_{\tau e dd}  \left(F_V^{(8)}(s) - F_V^{(3)}(s) + F_V^{(0)}(s)\right) \nonumber 
          \right. \\
          && \left.
          + \left(C^{ed}_{\rm VLL}+C^{ed}_{\rm VLR} \right)_{\tau e ss}  \left(- 2 F_V^{(8)}(s) + F_V^{(0)}(s)\right)
\right\},\\
B_R&=&\left\{\left(C^{eu}_{\rm VRR}+C^{ue}_{\rm VLR} \right)_{\tau e uu}  \left(F_V^{(8)}(s) + F_V^{(3)}(s) + F_V^{(0)}(s)\right) \nonumber \right. \\
& & \left.
          + \left(C^{ed}_{\rm VRR}+C^{de}_{\rm VRL} \right)_{\tau e dd}  \left(F_V^{(8)}(s) - F_V^{(3)}(s) + F_V^{(0)}(s)\right) \nonumber 
          \right. \\
          && \left.
          + \left(C^{ed}_{\rm VRR}+C^{de}_{\rm VRL} \right)_{\tau e ss}  \left(- 2 F_V^{(8)}(s) + F_V^{(0)}(s)\right)
\right\},
\end{eqnarray}
with the form factors defined as
\begin{eqnarray}
& &\frac{1}{2} \langle K^+(p_1) K^-(p_2) |  \left(\bar u \gamma^\mu u - \bar d \gamma^\mu d \right) | 0 \rangle = (p_1 - p_2)_\mu F_V^{(3)}(s) \\
& & \frac{1}{6} \langle K^+(p_1) K^-(p_2) | \left( \bar u \gamma^\mu u + \bar d \gamma^\mu d - 2 \bar s \gamma^\mu s \right) | 0 \rangle = (p_1 - p_2)_\mu F_V^{(8)}(s) \\
& & \frac{1}{3}\langle K^+(p_1) K^-(p_2) |  \left( \bar u \gamma^\mu u + \bar d \gamma^\mu d +  \bar s \gamma^\mu s \right) | 0 \rangle = (p_1 - p_2)_\mu F_V^{(0)}(s) 
\end{eqnarray}
The isoscalar and isovector form factors $F_V^{(8)}$ and $F^{(3)}_{V}$ have been extracted in Ref. \cite{Beloborodov:2019uql}
from data on $e^+ e^-  \rightarrow K^+ K^-$, $e^+ e^-  \rightarrow K_L K_S$ and $\tau  \rightarrow K^+ K^0 \nu_\tau$.
Ref. \cite{Beloborodov:2019uql} used a parametrization in terms of resonances, with the $\rho$ resonance and its excitations contributing to $F_V^{(3)}$
and the $\omega$ and $\phi$ resonances to the $F_V^{(8)}$,
\begin{eqnarray}
F_V^{(3)}(s) &=& \frac{1}{2}\sum_{V=\rho, \rho^\prime, \ldots} c_V \, BW_V(s) \label{FV3}\\
F_V^{(8)}(s) &=& \frac{1}{6}\sum_{V=\omega, \omega^\prime, \ldots} c_V \, BW_V(s)  + \frac{1}{3} \sum_{V=\phi, \phi^\prime, \ldots} c_V \, BW_V(s),\label{FV8}
\end{eqnarray}
with 
\begin{equation}
BW_V(s) = \frac{M_V^2}{M_V^2 -s - i M_V \Gamma_V(s) }.
\end{equation}
We thank K. Beloborodov for providing the energy-dependent widths $\Gamma_V(s)$.
The fit coefficients $c_V$ are given in Ref. \cite{Beloborodov:2019uql}, in two scenarios, Model I and II, with the latter achieving a better description of the data.   
The case $c_{\omega} = c_{\phi} = 1$, with coefficients of the $\omega$ and $\phi$ excitations set to zero, corresponds to the case of single-resonance dominance and ``ideal mixing'', with $\phi$ coupling only to $\bar s \gamma^\mu s$ and $\omega$ to $\bar u \gamma^\mu u + \bar d \gamma^\mu d $.
The actual fits coefficients $c_{\omega} = 1.28 \pm 0.14 $ and $c_\phi =1.038 \pm 0.001$ do not deviate from this expectation very significantly.

The isosinglet component  $F_V^{(0)}$ cannot be directly extracted from data. We will here assume that the tower of $\phi$ resonances couple only 
to the $s$ quarks, while the $\omega$, $\omega^\prime$, $\ldots$ to light $u$ and $d$. This assumption is well justified for the $\omega(782)$
and $\phi(1020)$, which are very close to ideal mixing, and lattice QCD calculations of the meson spectrum find very small mixing between the $\bar s s$ and $\bar u u + \bar d d$ components also for other vector isoscalar excitations \cite{Dudek:2013yja,Johnson:2020ilc}. We thus write 
\begin{eqnarray}\label{model0}
F_V^{(0)}(s) &=& \frac{1}{3} \left(\sum_{V=\omega, \omega^\prime, \ldots} c_V \, BW_V(s)  - \sum_{V=\phi, \phi^\prime, \ldots} c_V \, BW_V(s) \right).
\end{eqnarray}
The coefficients from Ref. \cite{Beloborodov:2019uql} are compatible with $F^{(0)}_V(0) =0$, as expected at NLO in $\chi$PT.  
An alternative model for $F^{(0)}_V$  is provided in Ref. \cite{Husek:2020fru}, and corresponds to considering only the contribution of the lowest resonances. 

Using Eqs. \eqref{FV3}, \eqref{FV3} and \eqref{model0}, the fit coefficients from Model II in Ref. \cite{Beloborodov:2019uql}, and assuming all coefficients to be real, we obtain
\begin{eqnarray}\label{BRKK}
\textrm{BR}(\tau \rightarrow e K^+ K^-) &=& 0.59  \Big|\Big(C^{ed}_{\rm VLL}+C^{ed}_{\rm VLR} \Big)_{\tau e ss}\Big|^2 
+ (1.0 \pm 0.1) \cdot 10^{-2} \Big|\Big(C^{eu}_{\rm VLL}+C^{eu}_{\rm VLR} \Big)_{\tau e uu}\Big|^2  \nn \\
& &+ (0.6 \pm 0.2) \cdot 10^{-3} \Big|\Big(C^{ed}_{\rm VLL}+C^{ed}_{\rm VLR} \Big)_{\tau e dd}\Big|^2 
\nn \\ & &
-(4.6 \pm 0.2) \cdot 10^{-2}\Big(C^{ed}_{\rm VLL}+C^{ed}_{\rm VLR} \Big)_{\tau e ss} \Big(C^{eu}_{\rm VLL}+C^{eu}_{\rm VLR} \Big)_{\tau e uu}
\nn \\ 
& &-(4.3 \pm 1.5) \cdot 10^{-3}\Big(C^{ed}_{\rm VLL}+C^{ed}_{\rm VLR} \Big)_{\tau e ss} \Big(C^{ed}_{\rm VLL}+C^{ed}_{\rm VLR} \Big)_{\tau e dd}
\nn \\ 
& &+ (3.5 \pm 0.8)\cdot 10^{-3} \Big(C^{ed}_{\rm VLL}+C^{ed}_{\rm VLR} \Big)_{\tau e dd} \Big(C^{eu}_{\rm VLL}+C^{eu}_{\rm VLR} \Big)_{\tau e uu}, 
\end{eqnarray}
where the error is obtained by propagating the errors in the fit parameters in Ref. \cite{Beloborodov:2019uql}. 
We can assess at least part of the theoretical error by using the extraction of $c_V$ with Model I in Ref.\cite{Beloborodov:2019uql},
and the one-resonance model for $F_V^{(0)}$ discussed in Ref. \cite{Husek:2020fru}.
While the prefactor of the  $\Big|\Big(C^{ed}_{\rm VLL}+C^{ed}_{\rm VLR} \Big)_{\tau e ss}\Big|^2$ Wilson coefficient in Eq. \eqref{BRKK} barely changes,
the prefactors of $\Big|\Big(C^{eu}_{\rm VLL}+C^{eu}_{\rm VLR} \Big)_{\tau e u u}\Big|^2$ 
and $\Big|\Big(C^{eu}_{\rm VLL}+C^{eu}_{\rm VLR} \Big)_{\tau e dd}\Big|^2$ show a $\sim 40\%$ and $> 100\%$ variation, respectively.
Eq. \eqref{BRKK} shows that the contribution of the $ss$ component of the vector current to the branching ratio is enhanced,
resulting in very strong single-coupling limits on $\left|\left[C_{LQ, D},\, C_{Ld},\,C_{ed},\, C_{Qe}\right]_{\tau e s s}\right| < 2.4 \cdot 10^{-4}$.
The limits on the $uu$ component are weaker by approximately a factor of ten, and affected by larger theoretical uncertainties. 
\\

\subsection{$B$ decays}
\label{Bdecays}
The input parameters relevant for $B$ decays are listed in Table \ref{input_Bmeson}. 
\begin{table}[t]
\centering
\begin{tabular}{||c| c || c|  c || c | c||}
\hline
$\tau_{B^0}$ & $1.519\times 10^{-12}~$s \cite{Zyla:2020zbs}& $f_B$ & $192.0~$MeV  \cite{Aoki:2019cca} & $m_{B^0}$ & $5279.65~$MeV \cite{Zyla:2020zbs} \\

$\tau_{B^{\pm}}$ & $1.638\times 10^{-12}~$s \cite{Zyla:2020zbs}& $m_{B^{\pm}}$ & $5279.34~$MeV  \cite{Zyla:2020zbs} & $m_{K^{\pm}}$ & $493.677~$MeV \cite{Zyla:2020zbs} \\
\hline
\end{tabular}
\caption{Input parameters for $B$ meson decays.}
\label{input_Bmeson}
\end{table}

\tocless\subsubsection{$B_d\to\tau^-e^+$}
The BR of $B_d\to e\tau$ is expressed by
\begin{align}
{\rm BR}\left(B_d\to\tau^-e^+ \right)
&=\tau_{B^0}\frac{G_F^2}{16\pi }\frac{f_{B}^2}{m_{B^0}}\lambda^{\frac{1}{2}}\left(1,\frac{m^2_{\tau}}{m^2_{B^0}},\frac{m^2_{e}}{m^2_{B^0}} \right)\nonumber\\
&\hspace{0.5cm}\times\bigg[\left(m^2_{B^0}-(m_{\tau}+m_e)^2 \right)
\left|\left(m_{\tau}-m_e \right)A+\frac{m^2_{B^0}}{m_b+m_d}C \right|^2\nonumber\\
&\hspace{1.5cm}+\left(m^2_{B^0}-(m_{\tau}-m_e)^2 \right)
\left|\left(m_{\tau}+m_e \right)B+\frac{m^2_{B^0}}{m_b+m_d}D \right|^2\bigg],
\end{align}
with
\begin{align}
A&=\left(C^{ed}_{\rm VLR} \right)_{\tau e bd}-\left(C^{ed}_{\rm VLL} \right)_{\tau e bd}+\left(C^{ed}_{\rm VRR} \right)_{\tau e bd}-\left(C^{de}_{\rm VLR} \right)_{bd\tau e},\\
B&=-\left(C^{ed}_{\rm VLR} \right)_{\tau e bd}+\left(C^{ed}_{\rm VLL} \right)_{\tau e bd}+\left(C^{ed}_{\rm VRR} \right)_{\tau e bd}-\left(C^{de}_{\rm VLR} \right)_{bd\tau e},\\
C&=\left(C^{ed}_{\rm SRR} \right)_{\tau e bd}-\left(C^{ed}_{\rm SRL} \right)_{\tau e bd}+\left(C^{ed*}_{\rm SRL} \right)_{e\tau db}-\left(C^{ed*}_{\rm SRR} \right)_{e\tau db},\\
D&=\left(C^{ed}_{\rm SRR} \right)_{\tau e bd}-\left(C^{ed}_{\rm SRL} \right)_{\tau e bd}-\left(C^{ed*}_{\rm SRL} \right)_{e\tau db}-\left(C^{ed*}_{\rm SRR} \right)_{e\tau db}.
\end{align}

\tocless\subsubsection{$B^+\to K^+ (\pi^{+}) \tau^{\pm} e^{\mp}$}

For the estimation of $B^+\to K^+ \tau^{\pm} e^{\mp}$, we follow the discussion in \cite{Duraisamy:2016gsd} where the effective Lagrangian for $b\to sl^-_il^+_j$ is defined as
\begin{align}
{\cal L}_{\rm eff}=&N_F\bigg[C_V\bar{s}\gamma^{\mu}P_Lb~\bar{l}_i\gamma_{\mu}l_j+C_A\bar{s}\gamma^{\mu}P_Lb~\bar{l}_i\gamma_{\mu}\gamma_5l_j \nonumber\\
&\hspace{1cm}
+C_V^{\prime}\bar{s}\gamma^{\mu}P_Rb~\bar{l}_i\gamma_{\mu}l_j+C_A^{\prime}\bar{s}\gamma^{\mu}P_Rb~\bar{l}_i\gamma_{\mu}\gamma_5l_j\nonumber\\
&\hspace{1cm}
+C_S\bar{s}P_Rb~\bar{l}_il_j+C_P\bar{s}P_Rb~\bar{l}_i\gamma_5l_j+C'_S\bar{s}P_Lb~\bar{l}_il_j+C'_P\bar{s}P_Lb~\bar{l}_i\gamma_5l_j\bigg],
\end{align}
with $N_F=G_F\alpha_{\rm em}V_{tb}V^*_{ts}/(\sqrt{2}\pi)$. The Wilson coefficients are converted into those in our basis
\begin{align}
N_FC_V&=-\sqrt{2}G_F\bigg[\left(C^{ed}_{\rm VLL} \right)_{\tau e sb}+\left(C^{de}_{\rm VLR} \right)_{sb\tau e}  \bigg],\\
N_FC_V^{\prime}&=-\sqrt{2}G_F\bigg[\left(C^{ed}_{\rm VRR} \right)_{\tau e sb}+\left(C^{ed}_{\rm VLR} \right)_{\tau e sb}  \bigg],\\
N_FC_A&=-\sqrt{2}G_F\bigg[-\left(C^{ed}_{\rm VLL}\right)_{\tau esb}+\left(C^{de}_{\rm VLR}\right)_{sb\tau e} \bigg],\\
N_FC_A^{\prime}&=-\sqrt{2}G_F\bigg[\left(C^{ed}_{\rm VRR}\right)_{\tau esb}-\left(C^{ed}_{\rm VLR}\right)_{\tau esb} \bigg],\\
N_FC_S&=-\sqrt{2}G_F\bigg[\left(C_{\rm SRR}^{ed} \right)_{\tau e sb}+\left(C^{ed*}_{\rm SRL} \right)_{e\tau bs} \bigg],\\
N_FC^{\prime}_S&=-\sqrt{2}G_F\bigg[\left(C_{\rm SRR}^{ed*} \right)_{e\tau bs}+\left(C^{ed}_{\rm SRL} \right)_{\tau esb} \bigg],\\
N_FC_P&=-\sqrt{2}G_F\bigg[\left(C_{\rm SRR}^{ed} \right)_{\tau e sb}-\left(C^{ed*}_{\rm SRL} \right)_{e\tau bs} \bigg],\\
N_FC^{\prime}_P&=-\sqrt{2}G_F\bigg[-\left(C_{\rm SRR}^{ed*} \right)_{e\tau bs}+\left(C^{ed}_{\rm SRL} \right)_{\tau esb} \bigg].
\end{align}

\begin{table}[t]
\centering
\begin{tabular}{ |c || c  c | }
\hline
 & $B^+\to K^+$ &  $B^+\to \pi^+$\\
\hline \hline
$a_0^+$     & $0.4696$ &  $0.404$ \\
$a_1^+$     & $-0.73$ & $-0.68$  \\
$a_2^+$     & $0.39$ & $-0.86$ \\
$a_0^0$        & $0.3004$  & $0.490$\\
$a_1^0$        & $0.42$ & $-1.61$ \\
$a_2^0$        & $0.621$  & $1.269$ \\
$a_0^T$     & $0.454$ & $0.393$ \\
$a_1^T$     & $-1.00$ & $-0.65$ \\
$a_2^T$     & $-0.89$ & $-0.6$\\
$a_3^T$     & $-$ &  $0.1$\\
\hline 
\end{tabular}
\caption{Fitting parameters of the form factors, $f_+,f_0$ and $f_T$, for the $B^+\to K^+$ and $B^+\to\pi^+$ modes \cite{Aoki:2019cca}.}
\label{formfactor_BK}
\end{table}

The related form factors are analyzed in \cite{Aoki:2019cca}, in which the following $N=3$ BCL parametrization is used:
\begin{align}
f_{+,T}(q^2)&= 
\frac{1}{P_{+,T}(q^2)}\sum^{N-1}_{n=0}a_n^{+,T}\left[z^n-(-1)^{n-N}\frac{n}{N}z^N \right],\\
f_0(q^2)&=
\frac{1}{P_0(q^2)}\sum^{N-1}_{n=0}a_n^0z^n,
\end{align}
where
\begin{align}
P_{+,0,T}(q^2)&=1-\frac{q^2}{M^2_{+,0,T}},\\
z(q^2,t_0)&=\frac{\sqrt{t_+-q^2}-\sqrt{t_+-t_0}}{\sqrt{t_+-q^2}+\sqrt{t_+-t_0}},\\
t_+&=\left(m_{B^{\pm}}+m_{P} \right)^2,\\
t_0&=\left(m_{B^{\pm}}+m_{P} \right)\left(\sqrt{m_{B^{\pm}}}-\sqrt{m_P} \right)^2,
\end{align}
where $m_P=m_{K^{\pm}}$ or $m_{\pi^{\pm}}$. The pole mass is $M_{+,T}=5.4154~$GeV and $M_0=5.711~$GeV for $B\to K$, $M_{+,T}=5.3252~$GeV and $M_0=\infty$ for $B\to \pi$.
The numerical values of $a_n^{+,0,T}$ are summarized in Table \ref{formfactor_BK}. It should be noted that the parameter $a^0_2$ is obtained by imposing $f_+(q^2=0)=f_0(q^2=0)$.

\bibliographystyle{JHEP} 
\bibliography{bibliography}

\end{document}